\documentclass[a4paper,11pt]{article}
\pdfoutput=1 
\usepackage{jcappub}

\usepackage{amsmath}
\usepackage{bm}
\usepackage{graphicx}
\usepackage{enumitem}
\usepackage{geometry}
\usepackage{xspace}
\usepackage{pdflscape}
\usepackage{placeins}
\usepackage{epstopdf}
\usepackage{multicol}
\usepackage{comment}
\usepackage{caption}
\usepackage{subcaption}

\usepackage{multirow}
\usepackage{array}
\usepackage{longtable}

\makeatletter
\gdef\@fpheader{}
\g@addto@macro\bfseries{\boldmath}
\makeatother

\let\oldsqrt\sqrt
\def\sqrt{\mathpalette\DHLhksqrt}
\def\DHLhksqrt#1#2{%
\setbox0=\hbox{$#1\oldsqrt{#2\,}$}\dimen0=\ht0
\advance\dimen0-0.2\ht0
\setbox2=\hbox{\vrule height\ht0 depth -\dimen0}%
{\box0\lower0.4pt\box2}}

\newcommand{\dd}{\mathrm{d}}

\newcommand{\sss}[1]{{\scriptscriptstyle{#1}}}

\newcommand{\uPl}{\mathrm{Pl}}

\newcommand{\uend}{\mathrm{end}}

\newcommand{\uini}{\mathrm{ini}}

\newcommand{\uT}{\mathrm{T}}

\newcommand{\usssT}{\sss{\uT}}

\newcommand{\usssPl}{\sss{\uPl}}

\newcommand{\nT}{n_\usssT}

\newcommand{\calH}{\mathcal{H}}

\newcommand{\Rea}{\Re \mathrm{e}\,}

\newcommand{\cs}{c_{_\mathrm{S}}}

\newcommand{\Mp}{M_\usssPl}

\newcommand{\beq}{\begin{eqnarray}}
\newcommand{\eeq}{\end{eqnarray}}
\newcommand{\bea}{\begin{equation}\begin{aligned}}
\newcommand{\eea}{\end{aligned}\end{equation}}

\newlength{\wsingfig}
\newlength{\wdblefig}
\newlength{\wquadfig}
\newlength{\wtriplefig}
\newcommand{\Refs}[1]{Refs.~{\cite{#1}}}
\newcommand{\Sec}[1]{Sec.~\ref{#1}}

\newcommand{\JM}[1]{{\color{blue}\ JM: #1} }

\subheader{}

\title{Opening the reheating box in multifield inflation}

\author[a]{J\'er\^ ome Martin,}
\author[a]{Lucas Pinol,}

\affiliation[a]{Institut d'Astrophysique de Paris, UMR 7095-CNRS,
  Sorbonne Universit\'e, 98bis boulevard Arago, 75014
  Paris, France}

\emailAdd{jmartin@iap.fr}
\emailAdd{pinol@iap.fr}

\date{today}

\begin{document}

\sloppy

\abstract{The robustness of multi-field inflation to the physics of reheating is investigated. In order to carry out this study, reheating is described in detail by means of a formalism which tracks the evolution of scalar fields and perfect fluids in interaction (the inflatons and their decay products). This framework is then used to establish the general equations of motion of the background and perturbative quantities controlling the evolution of the system during reheating. Next, these equations are solved exactly by means of a new numerical code.
Moreover, new analytical techniques, allowing us to interpret and approximate these solutions, are developed. As an illustration of a physical prediction that could be affected by the micro-physics of reheating, the amplitude of non-adiabatic perturbations in double inflation is considered. It is found that ignoring the fine-structure of reheating, as usually done in the standard approach, can lead to  differences as big as $\sim 50\%$, while our semi-analytic estimates can reduce this error to $\sim 10\%$. We conclude that, in multi-field inflation, tracking the perturbations through the details of the reheating process is important and, to achieve good precision, requires the use of numerical calculations.} 

\keywords{physics of the early universe, inflation}


\maketitle

\section{Introduction}
\label{sec:intro}

One of the most important result recently obtained in the field of Cosmology is that the physical conditions that prevailed in the very early Universe can be convincingly described by the theory of cosmic inflation~\cite{Starobinsky:1980te, Guth:1980zm, Linde:1981mu,Albrecht:1982wi, Linde:1983gd,Starobinsky:1979ty,Mukhanov:1981xt,Hawking:1982cz,Guth:1982ec,Bardeen:1983qw}. Moreover, and this is even more impressive, detailed pieces of information about the mechanism responsible for this inflationary phase have started to be gathered. So far, all astrophysical data are compatible with single-field models~\cite{Martin:2015dha} and the corresponding potential is known to be of the plateau shape (or close to it)~\cite{Martin:2013tda, Martin:2013nzq, Chowdhury:2019otk}, the prototypical scenario satisfying these physical requirements being the Starobinsky model~\cite{Starobinsky:1980te}.

Does that mean that the final word has been said about the mechanism that drives inflation? There are reasons to believe this is not the case. In particular, since inflation proceeds at very high energies, much higher energies than the ones probed in accelerators, its description must be based on extensions of the standard model of particle physics. Generically, these frameworks predict the presence of several (scalar) fields. So a natural arena for designing well-motivated models seems to be multi-field inflation (see, \textit{e.g.}, Refs.~\cite{Polarski:1992dq,Peter:1994dx,Langlois:1999dw,Gordon:2000hv,GrootNibbelink:2001qt,Tsujikawa:2002qx,Wands:2002bn,Wands:2007bd,Choi:2008et,Achucarro:2010jv,Kaiser:2015usz,Schimmrigk:2017jwa,Braglia:2020fms} for works that are particularly relevant for the present article).
In this sense, a fundamental and so far open problem consists in explaining how, from this multi-field description, an effective single-field scenario emerges, as revealed by the data.

It is also interesting to notice that the above described situation could, maybe, change in the near future. Indeed, the next generation of experiments will probe even further the micro-physics of inflation, for instance by measuring in refined details the properties of the large scale structures and/or of the Cosmic Microwave Background (CMB) anisotropies~\cite{Amendola:2016saw,Dore:2014cca,Matsumura:2013aja}. These experiments will soon come on line. They could reveal that the physical nature of inflation, so far compatible (and, to some extent, hidden) in an effective single-field framework, is in fact multi-field at a deeper level. Given the present efficiency of the single-field description, however, the corresponding observational signatures, if ever detected, are likely to manifest themselves as small deviations. The previous considerations indicate that it is important to know what physical predictions are expected from multi-field inflation and to calculate those predictions with enough accuracy. 

The typical signatures of multi-field inflation, namely  those that would allow us to distinguish multi-field inflation from single-field inflation, have already been studied in great details in the literature. Among the emblematic predictions of multi-field inflation that are not compatible with a single-field description are a violation of consistency relation $r=-8\nT$ ($r$ is the tensor to scalar ratio and $\nT$ is the tensor spectral index)~\cite{Wands:2002bn}, the presence, at a level to be determined, of Non-Gaussianity (NG)~\cite{Allen:2005ye} and of non-adiabatic perturbations~\cite{Huston:2013kgl,Byrnes:2006fr,Schutz:2013fua}. In fact, to be more precise, NG is also compatible with single-field models provided the fields have non-canonical terms but non-canonical and multi-field NG may be of different type and therefore can be, at least in principle, distinguished (see, \textit{e.g.}, Refs.~\cite{Chen:2009we,Chen:2009zp,Baumann:2011nk,Assassi:2012zq,Noumi:2012vr,Gong:2013sma,Arkani-Hamed:2015bza,Lee:2016vti,Garcia-Saenz:2018ifx,Garcia-Saenz:2018vqf,Fumagalli:2019noh,Bjorkmo-Ferreira-Marsh_2019,Wang:2019gok,Garcia-Saenz:2019njm,Ferreira:2020qkf,Pinol:2020kvw} for specific signatures of multi-field inflation in the NG signal). On the other hand, the presence of non-adiabatic modes in the data would establish, beyond any doubt, that the dynamics during inflation involves several degrees of freedom.

At this stage, it ought to be mentioned that there is a fundamental difference between single-field and multi-field inflation: in multi-field inflation, the details of what happens during the reheating epoch~\cite{Albrecht:1982mp,Turner:1983he,Shtanov:1994ce,Bassett:2005xm,Allahverdi:2010xz,Amin:2014eta,Martin:2010kz,Martin:2014nya,Martin:2016oyk} affect the behavior of the perturbations even on super-Hubble scales~\cite{Bassett:1998wg,Bassett:1999mt,Bassett:1999ta,Tsujikawa:2000ab,Tsujikawa:2002nf,Lyth:2003im} (note, however, that even in single-field inflation, the small scales can be affected by the details of metric preheating~\cite{Finelli:1998bu,Jedamzik:2010dq,Martin:2019nuw,Martin:2020fgl}; for interesting references on preheating in multi-field inflation, see Refs.~\cite{Finelli:2000ya,DeCross:2015uza,DeCross:2016fdz,DeCross:2016cbs,Jiang:2018uce,Nguyen:2019kbm,vandeVis:2020qcp}). Technically, this is because, in presence of non-adiabatic perturbations, curvature perturbations can evolve, even on large scales. This implies that calculating the behavior of the system 
during inflation is \textit{a priori} not sufficient to establish the predictions of a multi-field scenario: we need to model the reheating and to follow the perturbations during this phase. What we have just described has important implications for the above discussion. Since the reheating phase is \textit{a priori} complicated, so is an accurate calculation of the corresponding properties of multi-field scenarios. However, as we have already mentioned, we need such accurate estimates in order to optimize the use of cosmological data and to constrain inflationary scenarios beyond the single-field framework.

In the literature, considerable efforts have been made to understanding what happens during the (multi-field) inflationary phase but much less efforts have been devoted to investigating the influence of reheating. This will be the main focus of the present paper. Very often, the dynamics of reheating is simply ignored and the predictions follow from simple assumptions about the continuity of the relevant quantities used to describe  the inflationary scenario under scrutiny. In this article, on the contrary, we will model reheating in detail by carefully following the decay of the inflaton fields during this epoch. This will require the use of a formalism which explicitly includes interactions between scalar fields and perfect fluids (since perfects fluids will be assumed to be a good description of the inflaton fields decay products) both at the background and perturbative levels~\cite{Malik:2002jb,Lyth:2003im,Malik:2004tf}. 
Then, we will follow the fate of the perturbations through this detailed description of reheating and study how it depends on the parameters of the models, for instance the decay rates. This will also allow us to compare approaches where the details of  reheating are ignored to the framework of this article where the dynamics of reheating is taken into account. As a consequence, we will be able to assess how much precision on the final predictions of a model is lost by using a simplified treatment of reheating. As already mentioned, since the curvature perturbation is not necessarily conserved during reheating, the corresponding effect is expected to be relevant.

As noticed above, multi-field inflationary scenarios can be quite complicated and lead to several predictions that differ from those of single-field models. As a simple illustration of how the predictions in multi-field inflation depend on reheating, we will therefore not try to be exhaustive nor to consider only fully realistic scenarios (that is to say, necessarily fully compatible with the most recent astrophysical data) but, rather, we will concentrate on one observable, namely the amplitude of non-adiabatic perturbations. Moreover, when it comes to concrete numbers, we will focus on a specific scenario, namely double inflation~\cite{Polarski:1992dq,Peter:1994dx,Langlois:1999dw,Tsujikawa:2002qx}.
We will calculate exactly the evolution of the system by mean of two numerical codes (in order to check our conclusions). Then, we will compare these exact results with several analytical approaches, some of them being already present in the literature~\cite{Polarski:1992dq,Langlois:1999dw} and some others being introduced here. The previous program will allow us to address the main question of the paper, namely assessing the robustness of the predictions in multi-field inflation to changes in the physics of reheating.

This article is organized as follows. After this foreword, Sec.~\ref{sec:intro}, we discuss how exchanges between fluids can be introduced and modeled, at the background level in Sec.~\ref{sec:background}, and at the perturbed level in Sec.~\ref{sec:pert}. Then, in Sec.~\ref{sec:applic2field}, we consider the previous formalism in the general case of reheating after a model of two-field inflation and, in Sec.~\ref{sec:doubleinf}, we apply this approach to a concrete model, namely two-field inflation. In Sec.~\ref{sec:comparison}, we compare our results to the ones already existing in the literature. Finally, we present our conclusions in Sec.~\ref{sec:conclusions}. We end the paper with an appendix~\ref{app:decay} where we develop new analytical techniques to describe the behavior of the background at the time of the heavy field decay.

\section{Scalar fields and fluids in presence of
energy-momentum exchanges}
\label{sec:background}

\subsection{General equations}
\label{subsec:general}

In this paper, we consider a situation where there are several fluids living in a
space-time described by the four-dimensional metric tensor $g_{\mu
  \nu}(x^\kappa)$; each fluid is characterized by
its own energy-momentum tensor, $T_{\mu \nu}^{(\alpha)}$. The index
``$(\alpha)$'' is written between parenthesis to indicate that it is not a
space-time index but a label identifying the fluid. However, in the
following, when no confusion is possible, we will not write the
parenthesis in order to avoid cluttered notations. Also, the fluid label will indifferently appear
either up or down. Concretely, the different fluids will be
either scalar fields or perfect fluids with constant equations of state
(although one could easily accommodate time-dependent equations of
state). A crucial aspect of the physical situation considered in the present article is
that energy-momentum transfers between the fluids will be possible. This
will be described by the following conservation equations, based on a
detailed balance analysis~\cite{Malik:2002jb,Lyth:2003im,Malik:2004tf}
\begin{align}
  \label{eq:conserv}
\nabla _{\nu}T^{\mu \nu}_{(\alpha)}=\sum_{(\beta)}\left[Q^{\mu}_{(\alpha)\rightarrow (\beta)}
-Q^{\mu}_{(\beta)\rightarrow (\alpha)}\right],
\end{align}
where $\nabla_\nu$ denotes the covariant derivative associated with
the metric $g_{\mu \nu}$. The vector $Q^{\mu}_{(\alpha)\rightarrow
  (\beta)}$ describes the transfer from the fluid ``$(\alpha)$'' to
the fluid ``$(\beta)$'' while, evidently, the vector
$Q^{\mu}_{(\beta)\rightarrow (\alpha)}$ describes the transfer from
the fluid ``$(\beta)$'' to the fluid ``$(\alpha)$''. The vector
$Q^{\mu}_{(\alpha)\rightarrow (\beta)}$ can always be written in a covariant, non-perturbative way, as~\cite{Malik:2002jb,Lyth:2003im,Malik:2004tf}
\begin{align}
\label{eq:defQ}
Q^{\mu}_{(\alpha)\rightarrow (\beta)}=Q_{(\alpha)\rightarrow (\beta)}u^{\mu}
+f^{\mu}_{(\alpha)\rightarrow (\beta)},
\end{align}
with $f^{\mu}_{(\alpha)\rightarrow (\beta)}u_{\mu}=0$ where $u^{\mu}$ is
the total velocity of matter. We see that
$Q^{\mu}_{(\alpha)\rightarrow (\beta)}$ has been decomposed in terms of a
scalar, $Q_{(\alpha)\rightarrow (\beta)}$, and a vector,
$f^{\mu}_{(\alpha)\rightarrow (\beta)}$.

We have just mentioned that individual energy momentum tensors are not conserved
due to the presence of possible transfers between the fluids. But,
clearly, the total energy momentum tensor must be conserved because it
sources Einstein equations 
\begin{align}
G_{\mu \nu}=\frac{1}{\Mp^2}\sum_{(\alpha)} T_{\mu \nu}^{(\alpha)},
\end{align}
where, of course, $G_{\mu \nu}\equiv R_{\mu \nu}-Rg_{\mu \nu}/2$ is the Einstein
tensor ($R_{\mu \nu}$ is the Ricci tensor and $R$ the scalar
curvature) while $\Mp$ is the reduced Planck mass. In order to satisfy
the Bianchi identities, one must have $\nabla_\nu
[\sum_{(\alpha)}T_{(\alpha)}^{\mu \nu}]=0$. Physically, as already
mentioned, this expresses the fact that, if energy momentum transfers
can occur among fluids, the total energy momentum must be conserved and the net sum of all transfers must equal zero.

\subsection{Scalar fields}
\label{subsec:scalar}

In this section, we consider the case where the fluids under consideration are all scalar fields. Therefore, we assume that
we have $N_\mathrm{field}$ scalars with canonical kinetic terms, that we denote $\phi_a$ with $a=1, \cdots, N_\mathrm{field}$ (as it was the case in the last subsection for indices between parenthesis, $a$ is not a space-time index and, in the following, for
notational convenience, will be displayed either up or down). As is well-known, the corresponding stress-energy tensor can be written as
\begin{align}
  T_{\mu \nu}=\sum_{a=1}^{N_\mathrm{field}} \left(\partial_\mu \phi_a\partial_\nu \phi_a
  -\frac12 g_{\mu \nu}g^{\alpha
    \beta}\partial_\alpha \phi_a\partial_\beta\phi_a\right)
  -g_{\mu \nu}V(\phi_1, \cdots, \phi_{N_\mathrm{field}}),
\end{align}
where $V(\phi_a)$ is the potential function that we do not need to
specify at this stage.
A crucial remark is that, because the potential
term $V(\phi_a)$ is \textit{a priori} non-separable, the above stress-energy
tensor cannot be written as the sum of individual stress-energy
tensors, namely cannot be written as $\sum_{a=1}^{N_\mathrm{field}}T_{\mu \nu}^a$ where $T_{\mu \nu}^a$ would be the stress-energy tensor associated with the field $\phi_a$.
However, a collection of $N_\mathrm{field}$ scalar fields can also be viewed as a collection of $N_\mathrm{field}+1$ perfect fluids with constant equations of state~\cite{Malik:2002jb,Lyth:2003im,Malik:2004tf}. In this approach, we have $N_\mathrm{field}$ ``kinetic fluids'' with stress-energy tensor\footnote{In accordance with the remark made in Sec.~\ref{subsec:general}, we have written the stress-energy tensors without a parenthesis around the labels ``$K_a$" and ``$V$" since, in this case, there is no possible confusion with a space-time index.}
\begin{align}
  \label{eq:stressK}
  T_{\mu \nu}^{K_a}=\partial_\mu \phi_a \partial_\nu \phi_a -\frac12 g_{\mu \nu}
  g^{\alpha \beta}\partial _\alpha \phi_a\partial _\beta \phi_a,
  \end{align}
and one ``potential fluid'' with stress-energy tensor
\begin{align}
  \label{eq:stressV}
  T_{\mu \nu}^{V}=-g_{\mu \nu}V(\phi_1, \cdots, \phi_{N_\mathrm{field}}),
  \end{align}
and the total stress-energy tensor can be expressed as the sum of the
stress-energy tensors of the kinetic and potential fluids, $T_{\mu
  \nu}=\sum_{a=1}^{N_\mathrm{field}}T_{\mu \nu}^{K_a}+T_{\mu \nu}^{V}$. More
precisely, the kinetic fluids have energy density and pressure
\begin{align}
  \rho_{K_a}=p_{K_a}
  =-\frac12 g^{\alpha \beta}\partial_\alpha \phi_a\partial_\beta \phi_a,
\end{align}
which shows that each of them has a constant, ``stiff'', equation of
state, $w_{K_a}\equiv p_{K_a}/\rho_{K_a}=1$. The kinetic fluids have
velocity $u_\mu^{K_a}$ defined by
$u_\mu^{K_a}u_\nu^{K_a}=-\partial_\mu\phi_a\partial_\nu
\phi_a/(g^{\alpha \beta}\partial_\alpha \phi_a\partial_\beta \phi_a)$
and, clearly, one verifies that $u_\mu^{K_a}u^{\mu}_{K_a}=-1$ as
expected. On the other hand, the energy density and pressure of the
potential fluid are given by
\begin{align}
\label{eq:rhopV}
  \rho_V=-p_V=V(\phi_1,\cdots , \phi_{N_\mathrm{field}}),
  \end{align}
which means that this fluid has a vacuum equation of state, namely
$w_V=-1$. The velocity of the potential fluid is not defined, which is
not problematic since it will be shown that, in fact, this quantity never appears in the equations of motion and is, therefore, not relevant.

For the previous description to hold, there is however a price to pay:
we must assume that the kinetic and potential fluids interact. This is
indeed necessary in order to recover the correct equations of motion
(namely, the Klein-Gordon equation) of the scalar fields. This can be shown as
follows. Using Eqs.~(\ref{eq:conserv}) and~(\ref{eq:stressV}), the
conservation equation for the potential fluid can be expressed as
\begin{align}
  \nabla_\nu T_V^{\mu \nu}=-g^{\mu\nu}\partial_\nu V
  =-\sum_{a=1}^{N_\mathrm{field}}\partial^\mu \phi_a \frac{\partial V}{\partial \phi_a}=
  \sum_{a=1}^{N_\mathrm{field}}\left(Q^{\mu}_{V\rightarrow K_a}
  -Q^\mu_{K_a\rightarrow V}\right),
  \end{align}
which is satisfied by
\begin{align}
  \label{eq:generalQ}
  Q^\mu_{V\rightarrow K_a}=0, \quad
  Q^{\mu}_{K_a\rightarrow V}=\partial^\mu \phi_a\frac{\partial V}{\partial \phi_a} \, ,
\end{align}
where no sum is meant despite the repeated index ``$a$" (recall that ``$a$" is not a space-time index). Then, if one writes the conservation equation for the kinetic fluids,
using Eq.~(\ref{eq:stressK}), one arrives at
\begin{align}
  \nabla_\nu T_{K_a}^{\mu \nu}
=\partial^\mu \phi_a \nabla_\nu \nabla^\nu\phi_a
=\sum_{b\neq a}\left(Q^\mu_{K_a\rightarrow K_b}-Q^\mu_{K_b\rightarrow K_a}\right)
+Q^\mu_{K_a\rightarrow V}-Q^\mu_{V\rightarrow K_a},
\end{align}
and, recalling Eqs.~(\ref{eq:generalQ}), this reduces to the known
Klein-Gordon equation only if $Q^\mu_{K_a\rightarrow K_b}=0$ for any index $a$ and $b$.

We conclude that a situation with $N_\mathrm{field}$ scalar fields is in fact equivalent to a situation with $N_\mathrm{field}+1$ perfect fluids (with constant equations of state) in interaction~\cite{Malik:2002jb,Lyth:2003im,Malik:2004tf}. This effective interaction is such that the interaction among kinetic fluids vanishes and the energy-momentum transfer only proceeds from the kinetic fluids to the potential one (and not the opposite).

\subsection{Scalar fields and perfect fluids in interaction}
\label{subsec:inter}

In this article, since our goal is to study the reheating in multi-field inflationary models, where, at the end of inflation, the inflaton fields decay in various channels the physical properties of which can be described by means of hydro-dynamical considerations, we are interested in a situation where there are
scalar fields and perfect fluids in the Universe. We have just seen
that scalar fields can in fact be viewed as a collection of perfect
fluids, provided those fluids interact in a specific way. Therefore,
a situation with scalar fields and perfect fluids is in fact equivalent to a situation where there are only perfect fluids, some of them
being ``fictitious'' and some others being ``real''. In the previous
section~\ref{subsec:scalar}, we have written the interactions between the ``fictitious"
kinetic and potential fluids obtained from the requirement that the usual equations of motion
describing the behavior of scalar fields are recovered. In this section, we consider
the ``real'' interaction between a scalar field and a ``real'' fluid.

Firstly, we notice that, on general grounds, the interaction between
two fluids can be characterized in a covariant, non-perturbative way, by the following exchange vector~\cite{Malik:2002jb,Lyth:2003im,Malik:2004tf}
\begin{align}
  \label{eq:Qfluid}
Q^{\mu}_{(\alpha)\rightarrow (\beta)}=\Gamma_{(\alpha)\rightarrow (\beta)}
T^{\mu \nu}_{(\alpha)}u_{\nu}^{(\alpha)},
\end{align}
where $\Gamma_{(\alpha)\rightarrow (\beta)}>0$ is a coefficient that
controls the strength of the interaction and which can also be
interpreted as a decay rate. If the fluid has a perfect fluid form (in this article, we do not consider fluids which have anisotropic stress),
namely $T_{\mu
  \nu}^{(\alpha)}=\left[\rho_{(\alpha)}+p_{(\alpha)}\right]
u_\mu^{(\alpha)}u_{\nu}^{(\alpha)}+p_{(\alpha)}g_{\mu \nu}$ then, it
is easy to show that
\begin{align}
\label{eq:defQperfectfluid}
Q^{\mu}_{(\alpha)\rightarrow (\beta)}=-\Gamma_{(\alpha)\rightarrow (\beta)}
\rho_{(\alpha)}u^{\mu}_{(\alpha)}\,.
\end{align}

Secondly, we can apply the previous considerations to the questions studied in this article. The crucial ingredient is to realize that the interaction
between a scalar field and a ``real'' fluid can in fact be viewed
as an interaction between the corresponding ``fictitious'' kinetic
fluid and the ``real'' fluid. In other words, this interaction will be
characterized by $Q^\mu_{K_a\rightarrow (\beta)}$ given by
Eq.~(\ref{eq:Qfluid}) and $Q^\mu_{(\beta)\rightarrow K_a}=0$. The
``fictitious'' potential fluid will remain decoupled from all the ``real''
fluids present in space-time, namely $Q^\mu_{V\rightarrow (\beta)}=Q^{\mu}_{(\beta)\rightarrow V}=0$.

\subsection{The Homogeneous and Isotropic FLRW Universe}
\label{subsec:fluidFLRW}

The results discussed in the previous subsections are valid for any
metric tensor, namely for any space-time. We now assume that the
Universe is homogeneous and isotropic on large scales and, therefore, is described by
a Friedmann-Lemaitre-Robertson-Walker (FLRW) metric ${\rm d}s^2=-{\rm
  d}t^2+a^2(t)\delta_{ij}{\rm d}x^i{\rm d}x^j$, where $t$ is the
cosmic time and $a(t)$ is the scale factor. The metric can also be
written ${\rm d}s^2=a^2(\eta)(-{\rm d}\eta^2+\delta_{ij}{\rm d}x^i{\rm
  d}x^j)$, where $\eta$ is the conformal time related to cosmic time
$t$ by ${\rm d}t=a(\eta){\rm d}\eta$. For a FLRW Universe (using
conformal time), the total velocity of matter is given by
$u^{\mu}=(1/a,{\bm 0})$, $u_\mu=(-a,{\bm 0})$. On very general
grounds, for any type of fluid, the relation
$f^{\mu}_{(\alpha)\rightarrow (\beta)}u_{\mu}=0$ implies that, in
Eq.~(\ref{eq:defQ}), one has $f^0_{(\alpha)\rightarrow (\beta)}=0$
and, since $f^i_{(\alpha)\rightarrow (\beta)}$ must vanish in an
homogeneous and isotropic background\footnote{Indeed, this quantity can be decomposed into a scalar and a transverse 3-vector, $f^i_{(\alpha)\rightarrow (\beta)}=\partial_i f_{(\alpha)\rightarrow (\beta)} + \tilde{f}^i_{(\alpha)\rightarrow (\beta)}$ with $\partial_i  \tilde{f}^i_{(\alpha)\rightarrow (\beta)}=0$. At the background level, the scalar part must vanish because the universe is homogeneous and the transverse 3-vector must also vanish because the universe is isotropic.}, we reach the conclusion that
$f^{\mu}_{(\alpha)\rightarrow (\beta)}=0$.
Therefore, this quantity simply does
not appear at the background level. As a consequence, for any type of
fluid living in a FLRW Universe, one has (using conformal time)
$Q^0_{(\alpha)\rightarrow (\beta)}=Q_{(\alpha)\rightarrow (\beta)}/a$
and $Q^i_{(\alpha)\rightarrow (\beta)}=0$, with, using Eq.~(\ref{eq:defQ}),
\begin{align}
\label{eq:Qbackground}
Q_{(\alpha)\rightarrow (\beta)}=-\Gamma_{(\alpha)\rightarrow (\beta)}
\rho_{(\alpha)},
\end{align}
since at the background level $u^{\mu}_{(\alpha)}=u^{\mu}$ for any
$(\alpha)$ and, as we have just seen, $f^{\mu}_{(\alpha)\rightarrow (\beta)}=0$.

\subsubsection{Scalar fields in the FLRW Universe} 
\label{subsubsec:sf}

Now, let us see how the formalism described in Secs.~\ref{subsec:general}, \ref{subsec:scalar} and~\ref{subsec:inter} can be applied in an
homogeneous and isotropic Universe filled with $N_\mathrm{field}$ scalar fields. The total energy density and pressure
can be written as
\begin{align}
\rho=\sum _{a=1}^{N_\mathrm{field}}\frac{\phi_a'{}^2}{2a^2}+V(\phi_1,\cdots, \phi_{N_\mathrm{field}}), \quad 
p=\sum_{a=1}^{N_\mathrm{field}}\frac{\phi_a'{}^2}{2a^2}-V(\phi_1,\cdots ,\phi_{N_\mathrm{field}}),
\end{align}
where a prime denotes a derivative with respect to conformal time. As
already discussed, it is not possible to decompose $\rho$ and $p$ as a sum over $N_\mathrm{field}$ individual energy densities and pressures $\rho_a$ and $p_a$, one for each scalar field,
unless the potential is separable, namely unless we deal with the particular case where
$V(\phi_1,\cdots,\phi_{N_\mathrm{field}})=\sum_{a=1}^{N_\mathrm{field}} V_a(\phi_a)$. In general, the
conservation equation leads to the equations of motion for the fields
$\phi_a$, namely
\begin{align}
  \label{eq:sumKG}
\sum_{a=1}^{N_\mathrm{field}}\frac{\phi_a'}{a}\left(\phi_a''+2{\cal H}\phi_a'
+a^2V_{\phi_a}\right)=0,
\end{align}
where the quantity ${\cal H}=a'/a$ is the conformal Hubble parameter,
related to the Hubble parameter $H=\dot{a}/a$ (a dot standing for a
derivative with respect to cosmic time) by ${\cal H}=aH$ and where
$V_{\phi_a}\equiv \partial V/\partial \phi_a$. Eq.~(\ref{eq:sumKG}) is
satisfied if, for each field $\phi_a$, one has
\begin{align}
\label{eq:KG}
\phi_a''+2{\cal H}\phi_a'+a^2V_{\phi_a}=0,
\end{align}
which is the standard form of the Klein-Gordon equations in a FLRW
Universe.

As discussed in the previous sections~\ref{subsec:scalar} and~\ref{subsec:inter}, another way to proceed is to
introduce $N_\mathrm{field}$ kinetic fluids and one potential fluid, which are
perfect fluids with energy density and pressure given by
\begin{align}
\rho_{K_a}=\frac{\phi_a'{}^2}{2a^2}=p_{K_a}, \quad 
\rho_V=V=-p_V.
\end{align}
Then, one can write the total energy density and pressure as
$\rho=\sum _{a=1}^{N_\mathrm{field}}\rho_{K_a}+\rho_V$ and
$p=\sum_{a=1}^{N_\mathrm{field}}p_{K_a}+p_V$. In order to recover the Klein-Gordon
equations, we must introduce interactions between these $N_\mathrm{field}+1$ fluids
which, in an homogeneous and isotropic Universe, take the following form [we recall that all $Q^i_{(\alpha)\rightarrow (\beta)}=0$ at the background level]
\begin{align}
\label{interactions-background}
aQ_{K_a\rightarrow V}&=-\phi_a'V_{\phi_a}, \quad aQ_{V\rightarrow
  K_a}=0, \quad aQ_{K_a\rightarrow K_b}=0.
\end{align}
Indeed, with the above transfer vectors, the most general conservation equation for the
kinetic fluid, namely 
\begin{align}
  \label{eq:KGhomofluid}
\rho_{K_a}'+3{\cal H}\left(\rho_{K_a}+p_{K_a}\right)
&=a\left(Q_{K_a\rightarrow V}-Q_{V\rightarrow K_a}\right)
+a\sum_{b\neq a}\left(Q_{K_a\rightarrow K_b}-Q_{K_b\rightarrow K_a}\right),
\end{align}
simply reproduces the Klein-Gordon equation~(\ref{eq:KG}) for $\phi_a$. On the other hand, the conservation equation for the potential fluid
\begin{align}
\rho_V'=a\sum_{a=1}^{N_\mathrm{field}}\left(Q_{V\rightarrow K_a}-Q_{K_a\rightarrow V}\right)
=\sum_{a=1}^{N_\mathrm{field}}\phi_a'V_{\phi_a},
\end{align}
is identically satisfied.

\subsubsection{Scalar fields and perfect fluids in interactions in the FLRW Universe} 
\label{subsubsec:sfpfinter}

We now take into account the interactions between $N_\mathrm{field}$ scalar fields and $N_\mathrm{fluid}$ ``real'' fluids.
Since we have these ``fictitious" and ``real"  fluids in the Universe, the total energy density and pressure can now be expressed as
\begin{align}
    \rho=\sum _{a=1}^{N_\mathrm{field}}\rho_{K_a}+\rho_V + \sum _{(\alpha)=1}^{N_\mathrm{fluid}} \rho_{(\alpha)}, \quad p=\sum_{a=1}^{N_\mathrm{field}}p_{K_a}+p_V + \sum _{(\alpha)=1}^{N_\mathrm{fluid}} P_{(\alpha)} \,.
\end{align}
Then, one must rewrite the conservation
equation~(\ref{eq:KGhomofluid}) for the kinetic fluid and add the terms describing the exchanges between fields (or ``fictitious" fluids) and
``real'' fluids. This leads to
\begin{align}
\rho_{K_a}'+3{\cal H}\left(\rho_{K_a}+p_{K_a}\right)
&=a\left(Q_{K_a\rightarrow V}-Q_{V\rightarrow K_a}\right)
+a\sum_{b\neq a}\left(Q_{K_a\rightarrow K_b}-Q_{K_b\rightarrow K_a}\right)
\nonumber \\ &
+a\sum_{(\alpha)=1}^{N_\mathrm{fluid}}\left[Q_{K_a\rightarrow (\alpha)}-Q_{(\alpha)\rightarrow K_a}\right],
\end{align}
where $Q_{K_a\rightarrow (\alpha)}$ is given by
Eq.~(\ref{eq:Qbackground}), $Q_{(\alpha)\rightarrow K_a}=0$ and other interactions verify Eq.~\eqref{interactions-background}. As a consequence, one obtains the following modified Klein-Gordon equation
\begin{align}
\label{eq:KGmodified}
\phi_a''+\left[2{\cal H}+\frac{a}{2}\sum_{(\alpha)=1}^{N_\mathrm{fluid}}
  \Gamma_{K_a\rightarrow (\alpha)}\right]\phi_a'+a^2V_{\phi_a}=0 \,,
\end{align}
where it is clear that these new interactions result in extra friction terms for the scalar fields, parameterized by the decay rates $\Gamma_{K_a\rightarrow (\alpha)}$.
On the other hand, the conservation equation for $\rho_V$ is not
modified by any new terms since the potential fluid does not interact
with the ``real'' fluids. As a consequence, it is still identically
satisfied.

If we now consider the ``real'' fluids, the conservation
equation~(\ref{eq:conserv}) leads to only one non-trivial equation,
its time-component. In a homogeneous and isotropic Universe, it reads
\begin{align}
\rho_{(\alpha)}'+3{\cal H}\left[\rho_{(\alpha)}+p_{(\alpha)}\right]
=a\sum_{a=1}^{N_\mathrm{field}}\left[Q_{(\alpha)\rightarrow K_a}-Q_{K_a\rightarrow (\alpha)}
\right]+a\sum_{(\beta)=1}^{N_\mathrm{fluid}}\left[Q_{(\alpha)\rightarrow (\beta)}
  -Q_{(\beta)\rightarrow (\alpha)}\right],
\end{align}
where we recall that $\rho_{(\alpha)}$ and $p_{(\alpha)}$ are the
energy density and pressure of the ``real" fluid $(\alpha)$. Using again
Eq.~(\ref{eq:Qbackground}), one obtains
\begin{align}
\rho_{(\alpha)}'+3{\cal H}\left[\rho_{(\alpha)}+p_{(\alpha)}\right]
-\sum_{a=1}^{N_\mathrm{field}}\frac{\Gamma_{K_a\rightarrow (\alpha)}}{2a}\phi_a'^2
+a\sum_{(\beta)=1}^{N_\mathrm{fluid}}\left[\Gamma_{(\alpha)\rightarrow (\beta)}\rho_{(\alpha)}
  -\Gamma_{(\beta)\rightarrow (\alpha)}\rho_{(\beta)}\right]=0\,,
\end{align}
where the decays of the scalar fields result in an enhancement of the energy densities of the ``real" fluids, as should be the case during reheating, while the self-interactions of the ``real" fluids can add extra complexity in the system with positive and negative contributions.
As already mentioned, the space component of the conservation equation is identically satisfied.

The formalism described above is particularly well suited to describe the transfers of energy that occur, at the background level, from scalar fields to cosmological fluids at the end of inflation, namely during the reheating. This formalism will therefore be very useful to study this epoch of the inflationary scenario which, we recall, is the main target of this paper.
Notice also that the interactions introduced before need not be turned on by hand at the end of inflation: they are negligible but present during inflation and dynamically become relevant only when the decay rates become of the order of the Hubble parameter. This will be exemplified in the following when we study the case of double inflation. Finally, as an additional remark, let us stress that
this formalism can also be used for the warm inflation scenario~\cite{Berera:1995ie,Yokoyama:1998ju}. 

We now turn to the description of linear fluctuations around a FLRW background in the presence of interactions between scalar fields and ``real" fluids. This is indeed crucial in order to establish reliable cosmological predictions in multi-field models since, in that case, and contrary to single-field scenarios, the curvature perturbation can evolve during reheating, even on super-Hubble scales. It is therefore important to track the behavior of the perturbations when the interactions between the inflaton fields and their decay products play an important role in the dynamics of the Universe. This is the goal of the next section.

\section{Theory of cosmological perturbations in presence of
  energy-momentum exchanges}
\label{sec:pert}

\subsection{General equations}
\label{subsec:genepert}

In this section, we consider a Universe which is no longer homogeneous and
isotropic and which is filled with various fluids that can interact
with each others. We assume that the deviations from homogeneity and
isotropy are small and, therefore, can be treated perturbatively~\cite{Mukhanov:1990me}. This
leads to the theory of cosmological perturbations in presence of
energy-momentum exchanges~\cite{Malik:2002jb,Lyth:2003im,Malik:2004tf}. As expected, this only differs from the
standard approach to cosmological perturbations in the fact that the
perturbed conservation equations of the various fluids (``fictitious" ones describing scalar fields,
``real'' fluids, etc.) acquire new terms to describe these
exchanges, in very much the same way as described before for the
background. As a consequence, the corresponding perturbed line element
is written in the standard way, namely ${\rm
  d}s^2=a^2(\eta)\{-(1-2\phi){\rm d}\eta^2 +2(\partial_iB){\rm
  d}x^i{\rm d}\eta+[(1-2\psi)\delta_{ij} +2\partial_i\partial_jE]{\rm
  d}x^i{\rm d}x^j\}$ and is characterized by four functions, $\phi$,
$B$, $\psi$ and $E$ that are time and space dependent. Here, we have
taken into account only scalar perturbations and, in principle, the
perturbed line element should also contain a vector and a tensor
parts. In this article, we focus on the scalar sector only, that is decoupled from vectors and tensors at this order of perturbation theory. In order
to be consistent, the matter sector is also perturbed and we introduce
perturbed scalar fields $\delta\phi_a(\eta,{\bm x})$ and/or perturbed
energy density $\delta \rho_{(\alpha)}(\eta ,{\bm x})$ and pressure,
$\delta p_{(\alpha)}(\eta,{\bm x})$, for the fluids. For the velocity,
one has $\delta u^{\mu}_{(\alpha)}=(-\phi/a,v^i_{(\alpha)}/a)$ [and
  $\delta u_\mu^{(\alpha)}=(-a\phi,av_i^{(\alpha)}+a\partial_i B)$],
where the index of $v_i$ is raised by $\delta_{ij}$ at this order of perturbation theory. One can check that
it preserves the normalization of the four-velocity at the
perturbative level. As usual, not all perturbed quantities are
physically meaningful because of the gauge problem. In the following,
we will deal with this issue by making use of the so-called
gauge-invariant formalism for cosmological perturbations ~\cite{Bardeen:1980kt,Mukhanov:1990me}. The gravity
sector will be described by the Bardeen potentials,
$\Phi=\phi+[a(B-E')]'/a$ and $\Psi=\psi-{\calH}(B-E')$. In the matter
sector, as is well-known~\cite{Bardeen:1980kt}, there are different ways to define gauge-invariant perturbed energy densities. Here, we work in terms of $\delta
\rho_{(\alpha)}^{\mathrm{(gi)}}=\delta
\rho_{(\alpha)}+\rho_{(\alpha)}'(B-E')$. For the pressure, we have
$\delta p_{(\alpha)}^{\mathrm{(gi)}}=\delta
p_{(\alpha)}+p'_{(\alpha)}(B-E')$ and for the velocity,
$v^{(\mathrm{gi})}_{(\alpha)}=v_{(\alpha)}+E'$ with
$v_i^{(\alpha)}=\partial_i v^{(\alpha)}$ (and a similar definition for
the gauge-invariant part) since we consider scalar perturbations
only. Finally, if matter is described by scalar fields, the
inhomogeneous field fluctuations can be described in terms of the quantity $\delta
\phi^{(\mathrm{gi})}_a=\delta \phi_a+\phi_a'(B-E')$.
It can be checked that all these quantities, $\Phi,\Psi$ and all the perturbations with a ``$(\mathrm{gi})$" symbol are gauge-invariant at linear order in cosmological perturbation, and coincide with the corresponding quantities in the longitudinal gauge~\cite{Bardeen:1980kt}.

Let us now examine the equations of motion controlling the behavior
of the perturbations. We obviously have the perturbed Einstein
equations
\begin{align}
\delta G_{\mu \nu}=\frac{1}{\Mp^2}\sum_{(\alpha)}\delta T_{\mu \nu}^{(\alpha)},
\end{align}
and, perturbing the conservation equation~(\ref{eq:conserv}), we also
obtain another set of equations, namely
\begin{align}
  \label{eq:perturbconserv}
\delta\left[\nabla _{\nu}T^{\mu \nu}_{(\alpha)}\right]=\sum_{(\beta)}
\left[\delta Q^{\mu}_{(\alpha)\rightarrow (\beta)}
-\delta Q^{\mu}_{(\beta)\rightarrow (\alpha)}\right].
\end{align}
In this formula, the perturbed stress-energy tensor can be calculated
in the standard way and expressed in terms of the gauge-invariant
quantities introduced before. For the exchange vector, using the covariant, non-perturbative
Eq.~(\ref{eq:defQ}), one has
\begin{align}
\delta Q^{\mu}_{(\alpha)\rightarrow (\beta)}=\delta Q_{(\alpha)\rightarrow (\beta)}u^{\mu}
+Q_{(\alpha)\rightarrow (\beta)}\delta u^{\mu}
+\delta f^{\mu}_{(\alpha)\rightarrow (\beta)}.
\end{align}
As it was the case for the background, we must satisfy the constraint
that the vector $f^\mu_{(\alpha)\rightarrow (\beta)}$ is perpendicular
to the total velocity of matter $u^\mu$. At the perturbative level, this means
that $\delta \left[f^{\mu}_{(\alpha)\rightarrow (\beta)}
  u_{\mu}\right]=0$. This implies that
\begin{align}
\delta f^{\mu}_{(\alpha)\rightarrow (\beta)}u_{\mu}
+f^{\mu}_{(\alpha)\rightarrow (\beta)}\delta u_{\mu}=0.
\end{align}
The second term is vanishing because we have seen that, at the
background level, $f^{\mu}_{(\alpha)\rightarrow (\beta)}=0$. Moreover, $u_0\neq 0$,
$u_i=0$ and, therefore, one has
$\delta f^{0}_{(\alpha)\rightarrow (\beta)}=0$. From the above
considerations, we deduce that the time and space components of
$\delta Q^{\mu}_{(\alpha)\rightarrow (\beta)}$ are given by
\begin{align}
\label{eq:pertQzero}
\delta Q^{0}_{(\alpha)\rightarrow (\beta)}&=\frac{1}{a}
\delta Q_{(\alpha)\rightarrow (\beta)}
-\frac{\phi}{a}Q_{(\alpha)\rightarrow (\beta)}, 
\\
\label{eq:pertQi}
\delta Q^i_{(\alpha)\rightarrow (\beta)}
&=Q_{(\alpha)\rightarrow (\beta)}\delta u^i
+\delta f^i_{(\alpha)\rightarrow (\beta)}.
\end{align}
In the following, since we consider scalar perturbations, we will work
in terms of $v$ and $\delta f_{(\alpha)\rightarrow (\beta)}$ defined
by $v_i=\partial_i v$ where $\delta u^i=v^i/a$ and $\delta f_i^{(\alpha)\rightarrow
  (\beta)}=\partial_i \delta f_{(\alpha )\rightarrow (\beta)}$.
Let us
emphasize again that, in the above expressions, $\delta u^i$ is the
space component of the total velocity and not the space component of
the velocity of some individual fluid. In order to clarify further this
point, let us explain how these quantities are related. The time-space
component of the perturbed Einstein equations can be written as
\begin{align}
\Psi'+{\cal H}\Phi=-\frac{a^2}{2\Mp^2}\sum _{(\alpha)}
\left[\rho_{(\alpha)}+p_{(\alpha)}\right]v_{(\alpha)}^{\rm (gi)}=-\frac{a^2}{2\Mp^2}
\left(\rho+p\right)v^{\rm (gi)},
\end{align}
where $\rho=\sum_{(\alpha)}\rho_{(\alpha)}$ and
$p=\sum_{(\alpha)}p_{(\alpha)}$ are the total energy density and
pressure, respectively. This immediately implies that
\begin{align}
  \label{eq:totalvelocity}
v^{\rm (gi)}=\frac{1}{\rho+p}\sum _{(\alpha)}
\left[\rho_{(\alpha)}+p_{(\alpha)}\right]v_{(\alpha)}^{\rm (gi)},
\end{align}
can be understood as describing the total velocity.

Finally, we must define gauge-invariant exchange vectors or, rather,
define them in terms of gauge-invariant quantities. For the scalar
$\delta Q_{(\alpha)\rightarrow (\beta)}$, we introduce the
gauge-invariant expressions $\delta Q_{(\alpha)\rightarrow
  (\beta)}^{(\mathrm{gi})}=\delta Q_{(\alpha)\rightarrow
  (\beta)}+Q'_{(\alpha)\rightarrow (\beta)}(B-E')$. This 
expression is similar to the definition of the
gauge-invariant scalar field fluctuations, perturbed energy densities, etc. and this is of course due to
the fact that, in all these cases, we deal with scalar quantities. Regarding
$\delta f_{(\alpha)\rightarrow (\beta)}$, the situation is even
simpler: this quantity is already gauge-invariant since it has no background counterpart. 

\subsection{Perturbed conservation equations for fluids}
\label{subsec:pertconserv}

We now consider Eq.~(\ref{eq:perturbconserv}) and write it explicitly
for a perfect
fluid. Let us first start with the time component. It
can be expressed as
\begin{align}
\label{eq:perturbrho}
  \delta \rho_{(\alpha)}^{\rm (gi)}{}'
&+3{\cal H}\left[\delta \rho_{(\alpha)}^{\rm (gi)}
+\delta p_{(\alpha)}^{\rm (gi)}\right]
-3\left[\rho_{(\alpha)}+p_{(\alpha)}\right]\Psi' \nonumber
+\left[\rho_{(\alpha)}+p_{(\alpha)}\right]\nabla ^2\left[v_{(\alpha)}^{\rm (gi)}
\right] \nonumber
\\ &
=\sum_{(\beta)}\biggl[a\delta Q_{(\alpha)\rightarrow (\beta)}^{\rm (gi)}
+aQ_{(\alpha)\rightarrow (\beta)}\Phi
-a\delta Q_{(\beta)\rightarrow (\alpha)}^{\rm (gi)}
-aQ_{(\beta)\rightarrow (\alpha)}\Phi\biggr].
\end{align}
It is interesting to notice that, even if the velocity appears in the left hand side of this
equation, the scalar $\delta f_{(\alpha)\rightarrow (\beta)}$
is absent. Contrary to the background case, the space component of
the conservation equation leads, at the perturbative level, to an
interesting equation which controls the evolution of the perturbed
velocity. It reads
\begin{align}
  \label{eq:perturbv}
\left[\rho_{(\alpha)}+p_{(\alpha)}\right]v_{(\alpha)}^{\rm (gi)}{}'
&+\left[\rho_{(\alpha)}'+p_{(\alpha)}'\right]v_{(\alpha)}^{\rm (gi)}
+4{\cal H}\left[\rho_{(\alpha)}
+p_{(\alpha)}\right]v_{(\alpha)}^{\rm (gi)}
+\left[\rho_{(\alpha)}+p_{(\alpha)}\right]\Phi
+\delta p_{(\alpha)}^{\rm (gi)}
\nonumber \\ &
=
\sum _{(\beta)}\biggl[aQ_{(\alpha)\rightarrow (\beta)}v^{\rm (gi)}
+\delta f_{(\alpha)\rightarrow (\beta)}
-aQ_{(\beta)\rightarrow (\alpha)}v^{\rm (gi)}-\delta f_{(\beta)\rightarrow (\alpha)}
\biggr].
\end{align}
Eqs.~(\ref{eq:perturbrho}) and~(\ref{eq:perturbv}) are two equations
which, when considered together with the perturbed Einstein equations,
form a complete set allowing us to follow the evolution of
cosmological perturbations.
Note that, in practice, we will rather use the re-scaled velocities $\varsigma_{(\alpha)}\equiv [\rho_{(\alpha)}+p_{(\alpha)}]v_{(\alpha)}^{\rm (gi)}$ for numerical integration.

\subsection{Perturbed conservation equations for scalar fields}
\label{subsec:perturbfield}

We now consider a collection of $N_\mathrm{field}$ scalar fields. As explained in
Sec.~\ref{subsec:scalar}, $N_\mathrm{field}$ scalar fields can always be seen as
$N_\mathrm{field}+1$ perfect fluids provided those ones interact in a specific
way. Of course, this ``technical trick'' remains true at the
perturbative level and one of the goals of this section is to
determine the form of the corresponding perturbed exchange vectors. As it was the
case for the background, this is achieved by requiring the
equations of motion to reduce to the equations of motion for perturbed
scalar fields, namely the perturbed Klein-Gordon equation. The
great advantage of working with $N_\mathrm{field}+1$ fluids (as opposed to $N_\mathrm{field}$ scalar
fields) is that, as it was the case for the background before, this
allows us to consistently implement the interaction between fields and ``real''
fluids at the perturbative level.

The perturbed kinetic fluids energy density, pressure and velocity are
given by the following expressions:
\begin{align}
\label{eq:perturbKfluid}
  \delta \rho_{K_a}^{\rm (gi)}&=\delta p_{K_a}^{\rm (gi)}
=\frac{\phi_a'}{a^2}\delta \phi_a^{\rm (gi)}{}'-
\frac{\phi_a'{}^2}{a^2}\Phi, \quad
v_{K_a}^{\rm (gi)}=-\frac{\delta \phi_a^{\rm (gi)}}{\phi_a'}, 
\end{align}
while the same quantities for the potential fluid can be expressed as
\begin{align}
  \label{eq:perturbVfluid}
\delta \rho_V^{\rm (gi)}&=-\delta p_V^{\rm (gi)}
=\sum_{a=1}^{N_\mathrm{field}}
V_{\phi_a}\delta\phi_a^{(\mathrm{gi})}.
\end{align}
Notice that,
given the form of its individual energy-momentum tensor, it is impossible to define a velocity $v_{V}^{\rm (gi)}$ for the potential fluid. However, as already mentioned after Eq.~(\ref{eq:rhopV}), since this quantity never appears in the equations of motion, this is actually not an issue. We also notice that the sound speed, defined by $c^2_{_\mathrm{S}}\equiv  p'/\rho'$, of the kinetic and potential fluids is equal to the equation of state parameter, $c_{_\mathrm{S}}^2=w$, and (anticipating a little bit over the 
following considerations) that those fluids have no intrinsic entropy
perturbations. 

Our next move is to consider 
Eq.~(\ref{eq:perturbrho}). We take the
expression of the energy density, pressure and velocity for the
kinetic fluids given by Eqs.~(\ref{eq:perturbKfluid}) and insert them
in 
Eq.~(\ref{eq:perturbrho}). This leads to the standard perturbed
Klein-Gordon equation
\begin{align}
\delta \phi_a^{\rm (gi)}{}''+2{\cal H}\delta \phi_a^{\rm (gi)}{}'
-\nabla^2\left[\delta \phi_a^{\rm (gi)}\right]
+a^2 \sum_{b=1}^{N_\mathrm{field}} V_{\phi_a\phi_b}\delta \phi_b^{\rm (gi)}{}
=4\phi_a'\Psi'-2a^2V_{\phi_a}\Psi,
\end{align}
provided the perturbed exchanges take the following form
\begin{align}
  \label{eq:perturbexchangeK1}
a\delta Q_{K_a\rightarrow V}&=-V_{\phi_a}\delta \phi_a^{\rm (gi)}{}'
+V_{\phi_a}\phi_a'\Psi-\sum_{b=1}^{N_\mathrm{field}} V_{\phi_a\phi_b}\phi_a'
\delta\phi_b^{\rm (gi)}, \\
\label{eq:perturbexchangeK2}
a\delta Q_{V\rightarrow K_a}&=0, \quad a\delta Q_{K_a\rightarrow K_b}=0.
\end{align}
If one does the same for the potential fluid, namely insert the
perturbed energy density and perturbed pressure given by
Eq.~(\ref{eq:perturbVfluid}) in 
Eq.~(\ref{eq:perturbrho})
with the
exchanges just established in Eqs.~(\ref{eq:perturbexchangeK1}),~
(\ref{eq:perturbexchangeK2}), then one verifies that the corresponding
equation is identically satisfied. The
choices~(\ref{eq:perturbexchangeK1}) and~(\ref{eq:perturbexchangeK2})
are therefore consistent. The next step is to proceed in a similar
fashion for the coefficients $\delta f_{(\alpha)\rightarrow
  (\beta)}$. Clearly, this has to be done by considering the momentum
conservation equation. Therefore, we insert the
expressions~(\ref{eq:perturbKfluid}) in 
Eq.~(\ref{eq:perturbv})
and 
find that it is automatically satisfied provided one takes
\begin{align}
\label{eq:dfscalar}
  \delta f_{K_a\rightarrow V}&=V_{\phi_a}\delta \phi_a^{\rm (gi)}-\phi_a'V_{\phi_a}
\frac{1}{\sum _{b=1}^{N_\mathrm{field}} \phi_b'{}^2}\sum _{c=1}^{N_\mathrm{field}}\phi_c'
\delta \phi_c^{\rm (gi)}, 
\quad
\delta f_{V\rightarrow K_a}=\delta f_{K_a\rightarrow K_b}=0.
\end{align}
The expression for the only non-vanishing coefficient, $\delta
f_{K_a\rightarrow V}$, is quite complicated but can be simplified if
one notices that, in the case under consideration in this section where one has
only scalar fields (and no ``real" fluids), the total velocity reads
\begin{align}
  v^{\rm (gi)}&=\sum _{(
    \alpha)}\frac{\rho_{(\alpha)}+p_{(\alpha)}}{\rho+p}v_{(\alpha)}^{\rm (gi)}
=\frac{1}{\sum _b\phi'_b{}^2/a^2}\sum_{a=1}^{N_\mathrm{field}}\frac{\phi_a'{}^2}{a^2}
\left[ -\frac{\delta \phi_a^{\rm (gi)}}{\phi_a'}\right].
\end{align}
We see that this reproduces exactly the expression appearing in
Eq.~(\ref{eq:dfscalar}). As a consequence, it follows that
\begin{align}
\label{eq: f with vtot}
\delta f_{K_a\rightarrow V}&=
-\phi_a'V_{\phi_a}v_{K_a}^{\rm (gi)}
+\phi_a'V_{\phi_a}v^{\rm (gi)}
=-aQ_{K_a\rightarrow V}\left[v^{\rm (gi)}-v_{K_a}^{\rm (gi)}\right]
\end{align}
It is important to keep in mind that, here, $v^{\rm (gi)}$
is the total velocity in the case where they are only
scalar fields. In a situation where they are scalar fields and fluids,
we have the same formula~\eqref{eq: f with vtot} but the expression of $v^{\rm (gi)}$ is
modified since the velocities of the fluids participate to the total
velocity.
Therefore, it is fair to acknowledge that another nonequivalent possibility would be to define the momentum exchange with Eq.~\eqref{eq:dfscalar}.
Moreover, let us notice that the exchanges between the fictitious kinetic and potential fluids cannot be written in the form used in Eq.~\eqref{eq:Qfluid}.

\subsection{Perturbed conservation equations in presence of scalar fields
  and fluids in interaction}
\label{subsec:pertfieldfluid}

We now consider the situation which is probably the most relevant for
the present article, namely the case where there are scalar fields and
``real'' fluids in interaction. As already mentioned, this interaction
is described covariantly and non-perturbatively by Eq.~(\ref{eq:defQperfectfluid}), modeling the
exchange between the ``fictitious'' fluids
representing the scalar fields and the ``real'' fluids, as well as the ``real'' fluids self-interactions. We have already shown that this
implies $Q_{(\alpha)\rightarrow (\beta)} =-\Gamma_{(\alpha)\rightarrow
  (\beta)}\rho_{(\alpha)}$ for the background, see Eq.~(\ref{eq:Qbackground}). Then, at the perturbed level, one has
\begin{align}
\delta Q_{(\alpha)\rightarrow (\beta)}^{\mu}=-\Gamma_{(\alpha)\rightarrow (\beta)}
\delta \rho_{(\alpha)}u^{\mu}-\Gamma_{(\alpha)\rightarrow (\beta)}
\rho_{(\alpha)}\delta u^{\mu}.
\end{align}
This implies that the time component of the perturbed exchange vector
can be expressed as
\begin{align}
\delta Q^0_{(\alpha)\rightarrow (\beta)}
=-\frac{1}{a}\Gamma_{(\alpha)\rightarrow (\beta)}\delta \rho_{(\alpha)}
+\Gamma _{(\alpha)\rightarrow (\beta)}\rho_{(\alpha)}\frac{\phi}{a},
\end{align}
and, comparing to Eq.~(\ref{eq:pertQzero}), this leads to
\begin{align}
  \label{eq:deltaQgi}
\delta Q_{(\alpha)\rightarrow (\beta)}^{(\mathrm{gi})}=-\Gamma_{(\alpha)\rightarrow (\beta)}
\delta \rho_{(\alpha)}^{(\mathrm{gi})}.
\end{align}
Then, the scalar $\delta f_{(\alpha)\rightarrow (\beta)}$ which is
associated with the spatial component of the exchange vector remains
to be determined. This last one can be written as
\begin{align}
\delta Q^i_{(\alpha)\rightarrow (\beta)}
&=-\Gamma_{(\alpha)\rightarrow (\beta)}\rho_{(\alpha)}\delta u^i_{(\alpha)}
=-\Gamma_{(\alpha)\rightarrow (\beta)}\rho_{(\alpha)}\delta u^i
\nonumber 
+\Gamma_{(\alpha)\rightarrow (\beta)}\rho_{(\alpha)}
\left[\delta u^i-\delta u^i_{(\alpha)}\right].
\end{align}
Comparing to Eq.~(\ref{eq:pertQi}), this implies that
$\delta f^i_{(\alpha)\rightarrow (\beta)}= \Gamma_{(\alpha)\rightarrow
  (\beta)}\rho_{(\alpha)} \left[\delta u^i-\delta
u^i_{(\alpha)}\right]$, that is to say
\begin{align}
  \label{eq:deltaf}
\delta f_{(\alpha)\rightarrow (\beta)}=
a\Gamma_{(\alpha)\rightarrow (\beta)}\rho_{(\alpha)}
\left[v^{\rm (gi)}-v_{(\alpha)}^{\rm (gi)}\right].
\end{align}
This completes our derivation of the exchange vector
describing the interaction between scalar field and ``real'' fluids.

\subsection{Conserved quantities for scalar fields and fluids}
\label{subsec:conserved}

\subsubsection{General case}
\label{subsubsec:generalconserved}

In this section, we turn to the definition of the so-called
``conserved quantities'' (a name which might not be totally appropriate in the present case since these quantities will not always
be ``conserved", or ``constant", in the presence of several scalar fields and fluids). These quantities play an important role in the theory of
cosmological perturbations for two reasons. Firstly, their behavior
is, at least in some regimes, quite simple which is of great help to
understand the behavior of the perturbations and to test and check
numerical calculations. Secondly, they appear in the definition of
non-adiabatic or entropy perturbations which correspond to typical signatures of multi-fluid systems.

We introduce two gauge-invariant ``conserved'' quantities for a given
fluid ``$(\alpha)$'', $\zeta_{(\alpha)}$ and ${\cal R}_{(\alpha)}$,
which are related to the individual energy density and velocity perturbations, respectively. Their definition reads~\cite{Bardeen:1980kt,Mukhanov:1990me,Malik:2002jb,Lyth:2003im,Malik:2004tf}
\begin{align}
\label{eq:defzetaR}
  \zeta_{(\alpha)}=-\Psi-{\cal H}
  \frac{\delta \rho_{(\alpha)}^{(\mathrm{gi})}}{\rho'_{(\alpha)}},
  \quad {\cal R}_{(\alpha)}=\Psi-{\cal H}v_{(\alpha)}^{(\mathrm{gi})}.
  \end{align}
One can also introduce the corresponding ``total'' quantities (as
opposed to individual) by means of the following expressions,
$\zeta=-\Psi-{\cal H}\delta \rho^{(\mathrm{gi})}/\rho'$ and ${\cal
  R}=\Psi-{\cal H}v^{(\mathrm{gi})}$, where we recall that $\delta
\rho^{(\mathrm{gi})}=\sum_{(\alpha)}\delta
\rho_{(\alpha)}^{(\mathrm{gi})}$,
$\rho'=\sum_{(\alpha)}\rho_{(\alpha)}'$ and $v^{(\mathrm{gi})}$ is the
total velocity defined in Eq.~(\ref{eq:totalvelocity}). It is then
easy to show that
\begin{align}
  \zeta=\sum_{(\alpha)}\frac{\rho'_{(\alpha)}}{\rho'}\zeta_{(\alpha)},
  \quad
      {\cal R}=\sum_{(\alpha)}\frac{\rho_{(\alpha)}+p_{(\alpha)}}{\rho+p}
      {\cal R}_{(\alpha)},
\end{align}
where $\rho$ and $p$ are the total energy density and pressure,
respectively. The quantities $\zeta$ and ${\cal R}$ are nothing but
the weighted sum of the individual $\zeta_{(\alpha)}$ and ${\cal R}_{(\alpha)}$. We notice, however, that the weight is not the same for $\zeta$ and ${\cal R}$ unless the individual fluids are separately conserved in which case $\rho'_{(\alpha)}/\rho'=[\rho_{(\alpha)}+p_{(\alpha)}]/(\rho+p)$. The quantities $\zeta$ and ${\cal R}$ correspond to the well-known curvature perturbations on constant energy density slices ($\zeta=-\psi|_{\delta \rho =0}$) and on co-moving slices ($\mathcal{R}=\psi|_{v+B=0}$), respectively.

Our next move is to derive the equation of motion for the individual
conserved quantities. Using the equations of motion established
before, straightforward but lengthy calculations lead to
\begin{align}
\label{eq:derzetaalpha}
  \zeta_{(\alpha)}'&=-\frac{1}{3{\cal H}}\nabla^2\left[\Psi-{\cal R}_{(\alpha)}
    \right]+\frac{3{\cal H}^2}{\rho'_{(\alpha)}}\delta p_\mathrm{nad}^{(\alpha)}
-\frac{a}{3{\cal H}\rho_{(\alpha)}'}\sum_{(\beta)}
  \left[Q_{(\alpha)\rightarrow (\beta)}-Q_{(\beta)\rightarrow (\alpha)}\right]
  \nabla ^2{\cal R}_{(\alpha)}
  \nonumber \\ &
  -\frac{a}{\rho_{(\alpha)}'}\left({\cal H}-\frac{{\cal H}'}{\cal H}\right)
\sum_{(\beta)}\left[Q_{(\alpha)\rightarrow (\beta)}-Q_{(\beta)\rightarrow (\alpha)}\right]
\left[\zeta_{(\alpha)}-\zeta\right]
\nonumber \\ &
-\frac{a{\cal H}}{\rho_{(\alpha)}'}
\sum_{(\beta)}
\left\{
\delta Q_{(\alpha)\rightarrow (\beta)}^{(\mathrm{gi})}
  -\delta Q_{(\beta)\rightarrow (\alpha)}^{(\mathrm{gi})}
-\frac{\delta \rho_{(\alpha)}}{\rho'_{(\alpha)}}
\left[Q'_{(\alpha)\rightarrow (\beta)}
  -Q'_{(\beta)\rightarrow (\alpha)}\right]\right\}.
  \end{align}
 Several comments are in order at this stage. Firstly, the quantity
$\delta p_\mathrm{nad}^{(\alpha)}$ is defined by $\delta
p_\mathrm{nad}^{(\alpha)} =\delta p_{(\alpha)}-c_{(\alpha)}^2\delta
\rho_{(\alpha)}$, where $c_{(\alpha)}$ is the sound speed for the
fluid ``$(\alpha)$'' [the sound speed has already been defined in the
  text, after Eq.~(\ref{eq:perturbVfluid})]. Physically, it represents
intrinsic entropy (or non adiabatic -nad-) perturbations. For a
perfect fluid with constant equation of state, such as the ``real''
and ``fictitious'' fluids considered before, this quantity is
vanishing. However, this is not the case for a scalar field as can be
checked by direct inspection. Secondly, the term between curled
brackets in the last line of the above equation can be seen as a kind
of ``intrinsic exchange perturbations''. For interactions characterized
by Eqs.~(\ref{eq:Qfluid}),~(\ref{eq:defQperfectfluid})
and~(\ref{eq:Qbackground}), it vanishes. For more complicated
interactions, such as the ones we have introduced between the
``fictitious'' fluids, it is not necessarily zero and can contribute
to the evolution of $\zeta_{(\alpha)}$. Thirdly, in absence of
exchanges between the fluids and in absence of intrinsic entropy
perturbations, we see that $\zeta_{(\alpha)}$ is a conserved quantity
on large scales. However, if interactions between the fluids are
present, $\zeta_{(\alpha)}$ can evolve even on large scales, thanks to
the term in the second line. Fourthly, it is also possible to
establish the equation satisfied by the ``total'' $\zeta$. Indeed,
$\zeta$ corresponds to a fluid which is conserved (since its energy
momentum tensor has to satisfy the Einstein equations, see above). As
a consequence, all terms related to exchange vectors have to
disappear. Moreover, for the same reason, $\rho'$ can be replaced with
$-3{\cal H}(\rho+p)$. It follows that $\zeta$ obeys the equation
\begin{align}
\label{eq:dertotzeta}
  \zeta'&=-\frac{1}{3{\cal H}}\nabla^2\left(\Psi-{\cal R}
    \right)-\frac{{\cal H}}{\rho+p}\delta p_\mathrm{nad},
  \end{align}
with $\delta p_\mathrm{nad}=\delta p-c_{_\mathrm{S}}^2\delta \rho$ and
$c_{_\mathrm{S}}^2=p'/\rho'$, these definitions involving the total physical quantities (namely total, background and perturbed, energy density and pressure). We see that $\zeta$ can evolve on large
scales in presence of non-adiabatic pressure.

Let us now present the equation of motion of the individual quantity
${\cal R}_{(\alpha)}$. Straightforward but quite lengthy manipulations lead to 
\begin{align}
\label{eq:derRalpha}
  {\cal R}_{(\alpha)}'&=\left({\cal H}-\frac{{\calH}'}{{\calH}}\right)
  \left[{\cal R}-{\cal
      R}_{(\alpha)}\right]
  -c_{(\alpha)}^2\frac{\rho_{(\alpha)}'}{\rho_{(\alpha)}+p_{(\alpha)}}\left[{\cal
    R}_{(\alpha)}+\zeta_\alpha\right]
\nonumber \\ &
  +\frac{a}{\rho_{(\alpha)}+p_{(\alpha)}}
  \sum_{(\beta)}\left[Q_{(\alpha)\rightarrow (\beta)}-Q_{(\beta)\rightarrow (\alpha)}\right]
\left[{\cal R}-{\cal
    R}_{(\alpha)}\right]
+\frac{{\cal H}}{\rho_{(\alpha)}+p_{(\alpha)}}\delta p_\mathrm{nad}^{(\alpha)}
\nonumber \\ &
-\frac{{\cal H}}{\rho_{(\alpha)}+p_{(\alpha)}}\sum_{(\beta)}
\left[\delta f_{(\alpha)\rightarrow (\beta)}
  -\delta f_{(\beta)\rightarrow (\alpha)}\right].
\end{align}
We remark that, contrary to the case of $\zeta_{(\alpha)}$, the scalar
$\delta f_{(\alpha)\rightarrow (\beta)}$ now participates to the
equation of motion of ${\cal R}_{(\alpha)}$. The same reasoning as the one 
used for $\zeta$ allows us to derive the equation of motion of ${\cal
  R}$: all terms depending on the exchange vectors should not be
present in that equation since this is an equation for a fluid (the
total fluid) which is conserved and $\rho'$ can be replaced by
$-3{\cal H}(\rho+p)$. One obtains that
\begin{align}
\label{eq:dertotR}
  {\cal R}'&=
  3{\cal H}c_{_\mathrm{S}}^2\left({\cal
    R}+\zeta\right)
+\frac{{\cal H}}{\rho+p}\delta p_\mathrm{nad}.
\end{align}
As it was already the case before, we notice that the presence of non-adiabatic pressure can cause the evolution of ${\cal R}$ on large scales. However, in the present case, one may also wonder about the impact of the first term in Eq.~(\ref{eq:dertotR}). At first sight, it could also be responsible for an evolution of ${\cal R}$ on large scales. However, one can show that this term can in fact be expressed in terms of the gradient of the Bardeen potential, see for instance Eq.~(\ref{eq:phizetaR}). Therefore, in fact, the presence of non-adiabatic pressure is the only cause for a potential evolution of ${\cal R}$ on large scales, as it was the case for $\zeta$.

\subsubsection{Definitions of non-adiabatic perturbations}
\label{subsubsec:defentropy}

We conclude this section~\ref{subsec:conserved} by introducing the definitions of non-adiabatic perturbations. These definitions are directly related to the individual conserved quantities discussed before. The
energy density entropy perturbations $S_{(\alpha)(\beta)}$ and the
velocity entropy perturbations $V_{(\alpha)(\beta)}$ are respectively defined
by
\begin{align}
\label{eq:defentropypert}
  S_{(\alpha)(\beta)}=3\left[\zeta_{(\alpha)}-\zeta_{(\beta)}\right], \quad
  V_{(\alpha)(\beta)}={\cal R}_{(\alpha)}-{\cal R}_{(\beta)}.
  \end{align}
Using Eq.~(\ref{eq:defzetaR}), one can also write
$S_{(\alpha)(\beta)}=-3{\cal H}[\delta
\rho_{(\alpha)}^{(\mathrm{gi})}/\rho_{(\alpha)}' -\delta
\rho_{(\beta)}^{(\mathrm{gi})}/\rho_{(\beta)}']$ which, if each fluid
is separately conserved, reduces to
\begin{align}
  \label{eq:naddeltarho}
  S_{(\alpha)(\beta)}=\frac{\delta \rho_{(\alpha)}^{(\mathrm{gi})}}{\rho_{(\alpha)}
    +p_{(\alpha)}}-\frac{\delta \rho_{(\beta)}^{(\mathrm{gi})}}{\rho_{(\beta)}
    +p_{(\beta)}}.
\end{align}
This matches the standard expression for energy density entropy
perturbations. In the same way, velocity entropy perturbations can be
re-expressed as
\begin{align}
  V_{(\alpha)(\beta)}=-{\cal H}\left[v_{(\alpha)}^{(\mathrm{gi})}
    -v_{(\beta)}^{(\mathrm{gi})}\right],
  \end{align}
and, as expected (and as the name indicates), is related to the difference in velocities of the two fluids.

\section{Reheating after multi-field inflation}
\label{sec:applic2field}

Having introduced the formalism describing a physical situation where
scalar fields and fluids are present and interacting, we are now in a position
where we can make use of this formalism to study reheating after multi-field inflation. The
equations derived in the previous sections are completely general and can, in principle, be applied to a situation with an arbitrary number of fields and fluids in interaction. In this section, however, we restrict ourselves to two-field inflationary
scenarios where each field can decay into two fluids with
constant but otherwise arbitrary equation of state. This does not restrict the generalities of the results established in this article since, as already mentioned, from the previous equations, it would be straightforward to apply the following considerations to a scenario with more fields and/or fluids. The main advantage of this assumption is that this will allow to write and work with concrete equations of motion and to discuss several physical questions in an especially convenient manner. Moreover, this  simple framework covers in fact a very large landscape of models.

\subsection{Equations of motion}
\label{subsec:eomtwofield}

In the following, the two fields will be denoted $\phi_\ell$ and
$\phi_\mathrm{h}$ and the two perfect fluids will be named ``fluid 1'' and ``fluid
2''. $\phi_\ell$ and $\phi_\mathrm{h}$ is the notation used for the
double inflation model where one field is said to be ``light'' and the
other ``heavy''. We will consider this model in great details in what
follows, see Sec.~\ref{sec:doubleinf}, but, at this stage, the potential of the model
$V(\phi_\ell,\phi_\mathrm{h})$ remains arbitrary and the indices
``$\ell$'' and ``h'' must only be viewed as a convenient way to
distinguish the two fields without carrying the meaning it will
acquire when, later, we come explicitly to double inflation. The model
we have just introduced is described by the following Lagrangian
\begin{align}
  {\cal L}_\mathrm{twofields-inf}=-\frac12 \partial^\mu \phi_\ell \partial_\mu
  \phi_\ell -\frac12 \partial^\mu \phi_\mathrm{h} \partial_\mu
  \phi_\mathrm{h} -V(\phi_\ell,\phi_\mathrm{h})
+{\cal L}_\mathrm{matter}+{\cal L}_\mathrm{int}.
  \end{align}
As explained before, the two fields responsible for inflation also
interact with other components of matter that, phenomenologically, we
represent by perfect fluids. These fluids are described by ${\cal L}_\mathrm{matter}$ and the  interaction between matter and the inflaton fields is supposed to be given
by the term ${\cal L}_\mathrm{int}$.  Although
evidently ever-present, it becomes relevant only at the end of inflation and
will account for the disintegration of the two inflaton fields
explaining the ``graceful exit'', \textit{i.e.} how the Universe smoothly evolves
from inflation to the standard Big Bang model. Following the formalism
used in this article, we will not describe the interaction between the
scalar fields and the fluids by specifying ${\cal L}_\mathrm{int}$ but
we will rather proceed as reviewed earlier, at the level of the non-conservation of the individual energy-momentum tensors.
Indeed, according to the previous
considerations, the two scalar fields are in fact equivalent to three
fluids, two kinetic ones and one potential one, with energy densities and
pressures given by
\begin{align}
\rho_{K_\ell}&=\frac{\phi_\ell^2{}'}{2a^2}=p_{K_\ell}, \quad
\rho_{K_\mathrm{h}}=\frac{\phi_\mathrm{h}^2{}'}{2a^2}=p_{K_\mathrm{h}}, \quad
\rho_V=V(\phi_\ell,\phi_\mathrm{h})=-p_V.
\end{align}
We know from Eqs.~(\ref{eq:perturbexchangeK1})
and~(\ref{eq:perturbexchangeK2}) that the exchanges between those
three fluids are given by
\begin{align}
aQ_{K_\ell\rightarrow V}&=-\phi_\ell'V_{\phi_\ell}, 
\quad aQ_{V\rightarrow K_\ell}=0, \quad
aQ_{K_\mathrm{h}\rightarrow V}=-\phi_\mathrm{h}'V_{\phi_\mathrm{h}}, 
\quad aQ_{V\rightarrow K_\mathrm{h}}=0,  
\end{align}
in order to recover the usual equations of motion for the
fields. Then, as already discussed at length before, the crucial ingredient
is that the interactions between scalar fields and fluids are obtained
by coupling ``real'' cosmological fluids to the ``fictitious'' kinetic ones only (and not to
the potential fluid). In practice, using Eq.~(\ref{eq:Qbackground}),
one takes
\begin{align}
Q_{K_\ell \rightarrow (1)}& =-\Gamma_{\ell 1}\rho_{K_\ell}, \quad
Q_{(1)\rightarrow K_\ell}=0, \quad Q_{K_\ell \rightarrow (2)}
=-\Gamma_{\ell 2}\rho_{K_\ell}, \quad Q_{(2)\rightarrow K_\ell}=0,
\nonumber \\
Q_{K_\mathrm{h} \rightarrow (1)} &=-\Gamma_{\mathrm{h} 1}\rho_{K_\mathrm{h}},
\quad Q_{(1) \rightarrow K_\mathrm{h}}=0, \quad
Q_{K_\mathrm{h} \rightarrow (2)} =-\Gamma_{\mathrm{h} 2}\rho_{K_\mathrm{h}},
\quad Q_{(2) \rightarrow K_\mathrm{h}}=0,
\end{align}
Finally, we make also the hypothesis that the decay products (the ``real'' fluids) can
interact among themselves. This will be described by
\begin{align}
Q_{(1)\rightarrow (2)}& =-\Gamma_{12}\rho_{1}, \quad
Q_{(2)\rightarrow (1)}=-\Gamma_{21}\rho_{2}.
\end{align}
To summarize, the parameters of the model will be the parameters
appearing in the potential $V(\phi_\ell,\phi_\mathrm{h})$ that, as
already mentioned, we do not specify, the equation of state
of the two perfect ``real'' fluids, $w_1$ and $w_2$, the parameters describing the
decay of the scalar fields into fluids one and two, $\Gamma_{\ell 1}$,
$\Gamma_{\ell 2}$, $\Gamma_{\mathrm{h}1}$, $\Gamma_{\mathrm{h}2}$ and
the possible interaction between the decay products, $\Gamma_{12}$,
$\Gamma_{21}$.

\subsubsection{Background equations of motion for two-field inflation}
\label{subsubsec:backgroundtwofields}

We now describe the equations that control the evolution of this
system at the background and perturbative levels.
At the background level, we have the Friedman equation
\begin{align}
H^2=\frac{1}{3\Mp^2}\left(
\rho_{K_\ell}+\rho_{K_\mathrm{h}}+\rho_V+\rho_1+\rho_2\right)
=\frac{1}{3\Mp^2}\left[\frac12\dot{\phi}^2_\ell+\frac12\dot{\phi}^2_\mathrm{h}
  +V\left(\phi_\mathrm{h},\phi_\ell\right)+\rho_1+\rho_2\right],
\end{align}
where $\rho_1$ and $\rho_2$ are the energy densities associated with
the fluids one and two, respectively. We also have the two Klein-Gordon
equations for the $\phi_\ell$ and $\phi_\mathrm{h}$ fields, namely
\begin{align}
\ddot{\phi}_\ell
+\left[3H+\frac{1}{2}\left(\Gamma_{\ell 1}+\Gamma_{\ell 2}\right)\right]
\dot \phi_\ell +\frac{\partial V}{\partial {\phi_\ell}} &=0, \quad
\ddot{\phi}_\mathrm{h}
+\left[3H+\frac{1}{2}\left(\Gamma_{\mathrm{h}1}+\Gamma_{\mathrm{h}2}\right)\right]
\dot \phi_\mathrm{h} + \frac{\partial V}{\partial {\phi_\mathrm{h}}}=0.
\end{align}
Finally, we have the two conservation equations for fluid one and fluid
two which can be written as
\begin{align}
\dot{\rho}_1+3H(1+w_1)\rho_1
&=\frac{1}{2}\Gamma_{\ell 1}\dot \phi_{\ell}^2
+\frac{1}{2}\Gamma_{\mathrm{h} 1}\dot \phi_{\mathrm{h}}^2
-\Gamma_{12}\rho_1+\Gamma_{21}\rho_2,
\\
\dot{\rho}_2+3H(1+w_2)\rho_2
&=\frac{1}{2}\Gamma_{\ell 2}\dot \phi_{\ell}^2
+\frac{1}{2}\Gamma_{\mathrm{h} 2}\dot \phi_{\mathrm{h}}^2
+\Gamma_{12}\rho_1-\Gamma_{21}\rho_2.
\end{align}
The previous formulas form a closed set of equations which, when solved, provides a complete solution for the background
behavior.

\subsubsection{Perturbed equations of motion for two-field inflation}
\label{subsubsec:perturbedtwofields}

We now turn our attention to the equations of motion for the perturbations. We have
seen that, at the perturbative level, the exchanges between the
inhomogeneous fluids are described by the two scalars $\delta
Q_{(\alpha)\rightarrow (\beta)}^{(\mathrm{gi})}$ and $\delta
f_{(\alpha)\rightarrow (\beta)}$. In the case considered here, upon
using Eqs.~(\ref{eq:deltaQgi}), these coefficients can be expressed as
\begin{align}
  \delta Q_{K_\ell\rightarrow (1)}^{(\mathrm{gi})}&
  =-\Gamma_{\ell1}\delta \rho_{K_\ell}^{(\mathrm{gi})}, \quad
  \delta Q_{K_\ell\rightarrow (2)}^{(\mathrm{gi})}=-\Gamma_{\ell2}
  \delta \rho_{K_\ell}^{(\mathrm{gi})},
  \nonumber \\
 \delta Q_{K_\mathrm{h}\rightarrow (1)}^{(\mathrm{gi})}&
  =-\Gamma_{\mathrm{h}1}\delta \rho_{K_\mathrm{h}}^{(\mathrm{gi})}, \quad
  \delta Q_{K_\mathrm{h}\rightarrow (2)}^{(\mathrm{gi})}=-\Gamma_{\mathrm{h}2}
  \delta \rho_{K_\mathrm{h}}^{(\mathrm{gi})},
  \nonumber \\
  \delta Q_{(1)\rightarrow (2)}^{(\mathrm{gi})}&=-\Gamma_{12}\delta \rho_1^{(\mathrm{gi})},
  \quad 
\delta Q_{(2)\rightarrow (1)}^{(\mathrm{gi})}=-\Gamma_{21}\delta \rho_2^{(\mathrm{gi})},
\end{align}
all the other quantities $\delta Q_{(\alpha)\rightarrow
  (\beta)}^{(\mathrm{gi})}$ being zero. Regarding the coefficients
related to momentum transfers, they are given by Eq.~(\ref{eq:deltaf})
and read
\begin{align}
  \label{eq:coefdf1}
\delta f_{K_\ell\rightarrow (1)}&=a\Gamma_{\ell 1}\frac{\phi_\ell'{}^2}{2a^2}
\left[v^{(\mathrm{gi})}
  -v_{K_\ell}^{(\mathrm{gi})}\right],
\quad \delta f_{K_\ell\rightarrow (2)}=a\Gamma_{\ell 2}\frac{\phi_\ell'{}^2}{2a^2}
\left[v^{(\mathrm{gi})}
-v_{K_\ell}^{(\mathrm{gi})}\right],
\\
\label{eq:coefdf2}
\delta f_{K_\mathrm{h}\rightarrow (1)}&=a\Gamma_{\mathrm{h} 1}
\frac{\phi_\mathrm{h}'{}^2}{2a^2}
\left[v^{(\mathrm{gi})}
  -v_{K_\mathrm{h}}^{(\mathrm{gi})}\right],
\quad \delta f_{K_\mathrm{h}\rightarrow (2)}=a\Gamma_{\mathrm{h} 2}
\frac{\phi_\mathrm{h}'{}^2}{2a^2}
\left[v^{(\mathrm{gi})}
  -v_{K_\mathrm{h}}^{(\mathrm{gi})}\right],\\
\label{eq:coefdf3}
\delta f_{(1)\rightarrow (2)}&= a\Gamma_{12}\rho_1\left[v^{(\mathrm{gi})}
-v_1^{(\mathrm{gi})}\right], \quad \delta f_{(2)\rightarrow (1)}= a\Gamma_{21}
\rho_2\left[v^{(\mathrm{gi})}-v_2^{(\mathrm{gi})}\right],
\end{align}
the other coefficients $\delta f_{(\alpha)\rightarrow (\beta)}$
vanishing. In the above expressions, $v^{(\mathrm{gi})}$ is the total
velocity defined by Eq.~(\ref{eq:totalvelocity}) which, in the present
context, reads
\begin{align}
\label{eq:totalv}
v^{(\mathrm{gi})}&=\frac{1}{\phi_\mathrm{h}'{}^2/a^2
  +\phi_\ell'{}^2/a^2+(1+w_1)\rho_1+(1+w_2)\rho_2}
\biggl[-\frac{\phi_\mathrm{h}'{}^2}{a^2}
  \frac{\delta \phi_\mathrm{h}^{(\mathrm{gi})}}{\phi_\mathrm{h}'}
-\frac{\phi_\ell'{}^2}{a^2}\frac{\delta \phi_\ell^{(\mathrm{gi})}}{\phi_\ell'}
\nonumber \\ &
+(1+w_1)\rho_1v_1^{(\mathrm{gi})}+(1+w_2)\rho_2 v_2^{(\mathrm{gi})}\biggr].
\end{align}
We see that quantities related to fluids and scalar fields all
participate to the expression of the total velocity.

We are now in a position where the equations of motion can be
derived. From the perturbed Einstein equations (the time-space
component), we get
\begin{align}
  \label{eq:perEinstein}
\frac{\dot{\Phi}}{H}&=-\Phi-\frac{1}{2\Mp^2H}\left[
-\dot{\phi}_\ell \delta \phi_\ell^{(\mathrm{gi})}
-\dot{\phi}_\mathrm{h} \delta \phi_\mathrm{h}^{(\mathrm{gi})}
+a\rho_1(1+w_1)v_1^{(\mathrm{gi})}+a\rho_2(1+w_2)v_2^{(\mathrm{gi})}\right].
\end{align}
Notice that, in the present context, the two Bardeen potentials are equal: $\Phi=\Psi$. From the time component of the conservation
equation~(\ref{eq:perturbrho})
for the
perturbed ``fictitious'' kinetic fluids, we obtain the perturbed
Klein-Gordon equations for the two fields in presence of
energy-momentum exchanges. They read
\begin{align}
  \label{eq:perKGl}
\ddot{\delta \phi}_\ell^{(\mathrm{gi})}
  &+\left[3H+\frac{1}{2}\left(\Gamma_{\ell 1}+\Gamma _{\ell 2}\right)
\right]
    \dot{\delta \phi}_{\ell}^{(\mathrm{gi})}
+\frac{\partial ^2V}{\partial \phi_\ell^2}
\delta \phi_\ell^{(\mathrm{gi})} +\frac{\partial ^2V}
{\partial \phi_\ell \partial \phi_\mathrm{h}}\delta \phi_\mathrm{h}^{(\mathrm{gi})}
+\frac{k^2}{a^2}\delta \phi_\ell^{(\mathrm{gi})}
+2\frac{\partial V}{\partial \phi_\ell} \Phi
\nonumber \\ &
+\frac{1}{2}\left(\Gamma_{\ell 1}+\Gamma_{\ell 2}\right)
\dot{\phi}_\ell \Phi-4\dot{\phi}_\ell\dot{\Phi}=0, 
\\
\label{eq:perKGh}
\ddot{\delta \phi}_\mathrm{h}^{(\mathrm{gi})}
&+\left[3H+\frac{1}{2}\left(\Gamma_{\mathrm{h} 1}+\Gamma _{\mathrm{h} 2}\right)
  \right]\dot{\delta \phi}_{\mathrm{h}}^{(\mathrm{gi})}
+\frac{\partial ^2V}{\partial \phi_\mathrm{h}^2}
\delta \phi_\mathrm{h}^{(\mathrm{gi})} +\frac{\partial ^2V}
{\partial \phi_{\mathrm{h}} \partial \phi_\ell}\delta \phi_\ell^{(\mathrm{gi})}
+\frac{k^2}{a^2}\delta \phi_\mathrm{h}^{(\mathrm{gi})}
+2\frac{\partial V}{\partial \phi_\mathrm{h}} \Phi
\nonumber \\ &
+\frac{1}{2}\left(\Gamma_{\mathrm{h} 1}+\Gamma_{\mathrm{h} 2}\right)
\dot{\phi}_\mathrm{h} \Phi-4\dot{\phi}_\mathrm{h}\Phi=0.
\end{align}
In a similar way, Eq.~(\ref{eq:perturbrho}),
written for the fluid one and two lead to the equations controlling
the evolution of the perturbed energy density for those fluids
\begin{align}
  \label{eq:perdrho1}
\dot{\delta \rho_1}^{(\mathrm{gi})}&+3H(1+w_1)\delta \rho_1^{(\mathrm{gi})}
-3(1+w_1)\rho_1\dot{\Phi}-a(1+w_1)\rho_1\frac{k^2}{a^2}v_1^{(\mathrm{gi})}
-\Gamma_{\ell 1}\left[\delta \rho_{K_\ell}^{(\mathrm{gi})}
  +\rho_{K_\ell}\Phi\right]
\nonumber \\ &
-\Gamma_{\mathrm{h} 1}\left[\delta \rho_{K_\mathrm{h}}^{(\mathrm{gi})}
+\rho_{K_\mathrm{h}}\Phi\right]
+\Gamma_{12}\left[\delta \rho_1^{(\mathrm{gi})}+\rho_1\Phi\right]
-\Gamma_{21}\left[\delta \rho_2^{(\mathrm{gi})}+\rho_2\Phi\right]
=0,
\\
\label{eq:perdrho2}
\dot{\delta \rho_2}^{(\mathrm{gi})}&+3H(1+w_2)\delta \rho_2^{(\mathrm{gi})}
-3(1+w_2)\rho_2\dot{\Phi}-a(1+w_2)\rho_2\frac{k^2}{a^2}v_2^{(\mathrm{gi})}
-\Gamma_{\ell 2}\left[\delta \rho_{K_\ell}^{(\mathrm{gi})}
  +\rho_{K_\ell}\Phi\right]
\nonumber \\ &
-\Gamma_{\mathrm{h} 2}\left[\delta \rho_{K_\mathrm{h}}^{(\mathrm{gi})}
+\rho_{K_\mathrm{h}}\Phi\right]
+\Gamma_{21}\left[\delta \rho_2^{(\mathrm{gi})}+\rho_2\Phi\right]
-\Gamma_{12}\left[\delta \rho_1^{(\mathrm{gi})}+\rho_1\Phi\right]
=0.
\end{align}
Notice that the quantity $\delta
\rho_{K_a}^{(\mathrm{gi})}+\rho_{K_a}\Phi$ that often appears in the
above formulas can also be re-written as $\delta
\rho_{K_a}^{(\mathrm{gi})}+\rho_{K_a}\Phi= \dot{\phi}_a\dot{\delta
  \phi_a}^{(\mathrm{gi})} -\dot{\phi}_a^2\Phi/2$. We also remark that
the time component of the conservation equation does not explicitly depend on
the coefficients $\delta f_{(\alpha)\rightarrow (\beta)}$. As a
consequence, if one only tracks perturbed energy densities (and/or
energy density non-adiabatic perturbations), one could be under the impression that they can be ignored and that the above equations are sufficient. However, this is not the case because the velocities affect the evolution of the energy densities and the velocities equation of motion, see for instance Eq.~(\ref{eq:perturbv}), do depend on the coefficients $\delta f_{(\alpha)\rightarrow (\beta)}$. Therefore, even if at the end one is only interested in the density perturbations, the momentum exchanges must be specified.

Finally, we now turn to the space component of the conservation
equations, see Eq.~(\ref{eq:perturbv}).
Let us first write the space component of
the conservation equation for the ``fictitious'' kinetic fluid
associated to, say, $\phi_\ell$ (the conclusion is the same for
$\phi_\mathrm{h}$). Using Eq.~(\ref{eq:perturbKfluid}), this gives
\begin{align}
-\frac{\phi_\ell''}{a^2}\delta \phi_\ell^{(\mathrm{gi})}-\frac{2}{a^2}{\cal H}
\phi_\ell'\delta \phi_\ell^{(\mathrm{gi})}&=a\left[Q_{K_\ell\rightarrow V}
+Q_{K_\ell\rightarrow (1)}+Q_{K_\ell\rightarrow (2)}\right]v^{(\mathrm{gi})}
+\delta f_{K_\ell\rightarrow V}
\nonumber \\ & +\delta f_{K_\ell\rightarrow (1)}
+\delta f_{K_\ell\rightarrow (2)}.
\end{align}
However, the time component of the background equation of motion gives
$\phi_{K_\ell}'\left(\phi_{K_\ell}'' +2{\cal
  H}\phi_{K_\ell}'\right)/a^2 =a\left[Q_{K_\ell\rightarrow
    V}+Q_{K_\ell\rightarrow (1)} +Q_{K_\ell\rightarrow (2)}\right]$.
Plugging this expression in the space component of the conservation
equation leads to
\begin{align}
\delta f_{K_\ell\rightarrow V}+\delta f_{K_\ell \rightarrow (1)}
+\delta f_{K_\ell \rightarrow (2)}
=
a\left[Q_{K_\ell \rightarrow V}+Q_{K_\ell\rightarrow (1)}
+Q_{K_\ell \rightarrow (2)}\right]\left[v_{K_\ell}^{(\mathrm{gi})}-v^{(\mathrm{gi})}\right],
\end{align}
which is automatically satisfied, thanks to
Eq.~(\ref{eq:coefdf1}). This result makes sense because, for scalar
fields, there is no other equation of motion than the Klein-Gordon
equation, which is of second order in time and which we have already derived before.

Let us now study the space component of the conservation equation for
the ``real" fluids. Making use of Eq.~(\ref{eq:perturbv}),
one arrives at
\begin{align}
  \label{eq:v1}
\dot{v}_1^{(\mathrm{gi})}&+\left(1-3w_1\right)Hv_1^{(\mathrm{gi})}
+\frac{\Phi}{a}+\frac{w_1}{1+w_1}\frac{1}{a}
\frac{\delta \rho_1^{(\mathrm{gi})}}{\rho_1}=-\frac{w_1}{(1+w_1)\rho_1}
Q_{(1)\rightarrow (2)}v_1^{(\mathrm{gi})}
\nonumber \\ &
+\frac{1}{\rho_1}Q_{K_\ell\rightarrow (1)}
\left[v_1^{(\mathrm{gi})}-\frac{v_{K_\ell}^{(\mathrm{gi})}}{1+w_1}\right]
+\frac{1}{\rho_1}Q_{K_\mathrm{h}\rightarrow (1)}
\left[v_1^{(\mathrm{gi})}-\frac{v_{K_\mathrm{h}}^{(\mathrm{gi})}}{1+w_1}\right]
\nonumber \\ &
+\frac{1}{\rho_1}Q_{(2)\rightarrow (1)}
\left[v_1^{(\mathrm{gi})}-\frac{v_2^{(\mathrm{gi})}}{1+w_1}\right],
\\
\label{eq:v2}
\dot{v}_2^{(\mathrm{gi})}&+\left(1-3w_2\right)Hv_2^{(\mathrm{gi})}
+\frac{\Phi}{a}+\frac{w_2}{1+w_2}\frac{1}{a}
\frac{\delta \rho_2^{(\mathrm{gi})}}{\rho_2}=-\frac{w_2}{(1+w_2)\rho_2}
Q_{(2)\rightarrow (1)}v_2^{(\mathrm{gi})}
\nonumber \\ &
+\frac{1}{\rho_2}Q_{K_\ell \rightarrow (2)}
\left[v_2^{(\mathrm{gi})}-\frac{v_{K_\ell}^{(\mathrm{gi})}}{1+w_2}\right]
+\frac{1}{\rho_2}Q_{K_\mathrm{h}\rightarrow (2)}
\left[v_2^{(\mathrm{gi})}-\frac{v_{K_\mathrm{h}}^{(\mathrm{gi})}}{1+w_2}\right]
\nonumber \\ &
+\frac{1}{\rho_2}Q_{(1)\rightarrow (2)}
\left[v_2^{(\mathrm{gi})}-\frac{v_1^{(\mathrm{gi})}}{1+w_2}\right].
\end{align}

Eqs.~(\ref{eq:perEinstein}), (\ref{eq:perKGl}), (\ref{eq:perKGh}),
(\ref{eq:perdrho1}), (\ref{eq:perdrho2}), (\ref{eq:v1}) and
(\ref{eq:v2}) represent a complete set allowing us to track the
perturbations during inflation and afterwards. We stress that these
equations are valid for any model of two-field inflation and, in this
sense, are quite general. To our knowledge, although implicitly
present in Ref.~\cite{Malik:2008im}, this is the first time that they
are written explicitly. Similar equations have been studied in
Ref.~\cite{Visinelli:2014qla} but for one scalar field and one fluid only. Maybe
the closest related work is Ref.~\cite{Choi:2008et} which investigates the
generation of entropy fluctuations after multi-field inflation (this
paper studies models where there is a large number of fields during
inflation). But the derivation of the conservation equations does not
follow from a systematic formalism as we have done and, moreover, the
space component of the conservation equations is not presented because
Ref.~\cite{Choi:2008et} does not study how velocity non-adiabatic perturbations are
produced (although, as we have discussed before, it is not possible to ignore them since they influence the evolution of the perturbed density contrasts). We notice that, when the comparison is possible, the
equations presented here are consistent with those of
Ref.~\cite{Choi:2008et}.

\subsection{Quantization of cosmological perturbations in multi-field inflation}
\label{subsec:quantization}

According to the theory of inflation, all the inhomogeneities (CMB anisotropies, large scale structures, \dots ) that we observe today originate 
from quantum fluctuations that were amplified by gravitational instability and stretched to astrophysical distances by the cosmic expansion during inflation~\cite{Mukhanov:1981xt,Hawking:1982cz,Guth:1982ec} (see also Refs.~\cite{Martin:2004um,Martin:2007bw,Martin:2015qta}).

In the context of single-field inflation, this means that the perturbed metric, $\delta \hat{g}_{\mu \nu}(\eta,{\bf x})$ [in practice, the Bardeen potential $\hat{\Phi}(\eta,{\bm x})]$ and the quantity representing matter, the scalar field fluctuations $\delta \hat{\phi}(\eta,{\bf x})$,
must be quantum operators. In that case, everything can be reduced to the study of a single variable, the so-called Mukhanov-variable~\cite{Mukhanov:1981xt,Kodama:1985bj,Mukhanov:1990me}\footnote{Here, we denote this variable by $q(\eta,{\bm x})$ and not by the more traditional symbol $v(\eta,{\bm x})$ in order to avoid a possible confusion with the velocities of the fluids.}, $\hat{q}(\eta,{\bm x})=z(\eta)\hat{\cal R}(\eta,{\bm x})$, where $z(\eta)=a(\eta)\sqrt{2\epsilon_1}\Mp$. Then, this quantity is expanded in terms of creation and annihilation operators,
\begin{align}
\label{eq:operatorexpansion}
    \hat{q}(\eta,{\bm x})=\frac{1}{(2\pi)^{3/2}}
    \int \dd {\bm k}
    \left[ q_{k}(\eta) \hat{c}_{\bm k}  +  q_k^*(\eta) \hat{c}_{-{\bm k}}^\dagger \right]
    e^{i{\bm k}\cdot {\bm x}},
\end{align}
with $ \left[\hat{c}_{\bm k}, \hat{c}^\dagger_{{\bm k}^\prime} \right]= \delta^{(3)}\left({\bm k}-{\bm k}^\prime \right)$.
We have also introduced the mode function $q_k(\eta)$ that controls the evolution of the operator $\hat{q}(\eta,{\bm x})$. Notice that, here, we denote the mode function with the same symbol as the corresponding quantum operator. No confusion can arise with the Fourier transform of the quantum operator (which also carries an index ${\bm k}$) since the mode function does not carry the hat symbol. In this framework, the perturbed Einstein equations discussed in the previous sections must be viewed as operator equations. However, upon inserting the expansion~(\ref{eq:operatorexpansion}) in these operator equations, one obtains ordinary differential equations for the mode functions $q_k(\eta)$. Of course, the differential equations for the mode functions have in fact the same form as the operator differential equations. The two-point correlation function of the Mukhanov-Sasaki variable leads to the definition of the power spectrum ${\cal P}_q$, namely
\begin{align}
    \left \langle \hat{q}^2(\eta,{\bm x})
    \right \rangle=\int_0^{+\infty}\frac{\dd k}{k} {\cal P}_q(k),
\end{align}
with 
\begin{align}
    {\cal P}_q=\frac{k^3}{2\pi^2}\left \vert 
    q_k(\eta)\right\vert ^2.
\end{align}

This procedure can be extended to more degrees of freedom, for instance in a multi-field inflationary setup, which is the case of interest in this paper. With $N_\mathrm{field}$ scalar fields labeled by $a$, there exists a set of $N_\mathrm{field}$ independent annihilation and creation operators that, however, need not be aligned with the scalar fields themselves, i.e. there exists a frame labeled by numbers $A$ in which $\left[\hat{c}_{\bm k}^A, \hat{c}^{\dagger B}_{{\bm k}^\prime} \right]= \delta^{AB} \delta^{(3)}\left({\bm k}-{\bm k}^\prime \right)$. Then, the multi-field Sasaki-Mukhanov variables $q^a(\eta,{\bm x})$ are promoted to quantum operators and we have the expansion~\cite{GrootNibbelink:2001qt,Weinberg:2008zzc,Achucarro:2010jv}:
\begin{align}
\hat{q}^a(\eta,{\bm x})=\frac{1}{(2\pi)^{3/2}}
    \int \dd {\bm k}\sum_{A=1}^{N_\mathrm{field}}
    \left[ q_{A,k}^a(\eta) \hat{c}_{A,{\bm k}}  +  q_{A,k}^{a*}(\eta) \hat{c}_{A,-{\bm k}}^\dagger \right]
    e^{i{\bm k}\cdot {\bm x}},
\end{align}
where we now have $N_\mathrm{field}{}^2$ mode functions $q^a_{A,{\bm k}}(\eta)$. The calculation of the two-point correlation function,
\begin{align}
    \left \langle \hat{q}^a(\eta,{\bm x})\hat{q}^b(\eta,{\bm x})
    \right \rangle=\int_0^{+\infty}\frac{\dd k}{k} {\cal P}_q^{ab}(k),
\end{align}
leads to a definition of the ``generalized" power spectrum ${\cal P}_q^{ab}$ which reads
\begin{align}
    {\cal P}_q^{ab}=\frac{k^3}{2\pi^2}\sum_{A=1}^{N_\mathrm{field}}
    q_{A,k}^a(\eta) \left[q_{A,k}^{b}(\eta)\right]^*.
\end{align}
Interestingly enough, we see that the two-point correlation function receives contributions from all independent modes labeled by ``$A$".

In this article, we deal with a situation which is slightly different from what was described before. The reason is of course that we have, at the same time, the presence of scalar fields and fluids and their associated fluctuations. Cosmological fluids can be quantized in a way that is very similar to what was described above, once their Mukhanov-Sasaki variables with canonical kinetic terms have been identified, see Refs.~\cite{Kodama:1985bj,Mukhanov:1990me} for details. As for a scalar field, in the case of a single fluid, the Mukhanov-Sasaki variable is defined from the conserved quantity, ${\cal R}$, by $q(\eta,{\bm x})=z(\eta){\cal R}(\eta,{\bm x})$, with $z(\eta)=a(\eta)\sqrt{2\epsilon_1}\Mp/c_\mathrm{S}$, see Ref.~\cite{Kodama:1985bj,Mukhanov:1990me}. Notice that the quantity $z(\eta)$ differs from the case of a scalar field since there is an additional factor $c_\mathrm{S}$ participating to its expression. In the present context, the situation is more complicated since we deal with several fluids. One should, therefore, introduce several Mukhanov-Sasaki variables, one for each fluid. To our knowledge, this question has not been (yet) studied in a comprehensive manner in the literature but it is clear that each degree of freedom is related to the individual ${\cal R}_{(\alpha)}$. We will come back to this problem in the section where we discuss the initial conditions for the perturbations, see Sec.~\ref{subsec:initial}. In any case, the individual ${\cal R}_{(\alpha)}$ become quantum operators $\hat{\cal R}_{(\alpha)}(\eta,{\bm x})$ that can be expanded in terms of creation and annihilation operators and that possess their own mode functions. 

As explained before, once the system has been quantized, the individual $\hat{\cal R}_{(\alpha)}$ become quantum operators with their associated mode functions.  Clearly, for consistency, this 
is also the case for any other quantity participating to the description of matter since they are all related to the $\hat{\cal R}_{(\alpha)}$. For instance, in the case of scalar fields, the field  fluctuations $\delta \hat{\phi}^a(\eta,{\bm x})$ can be expanded as
\begin{align}
\label{eq:expansiondphi}
\delta \hat{\phi}^a(\eta,{\bm x})=\frac{1}{(2\pi)^{3/2}}
    \int \dd {\bm k}\sum_{A=1}^{N_\mathrm{field}+N_\mathrm{fluid}}
    \left[ \delta \phi_{A,k}^a(\eta) \hat{c}_{A,{\bm k}}  +  \delta \phi _{A,k}^{a*}(\eta) \hat{c}_{A,-{\bm k}}^\dagger \right]
    e^{i{\bm k}\cdot {\bm x}},
\end{align}
where $\delta \phi^a_{A,k}(\eta)$ are the associated mode functions. In a similar way, for a hydrodynamical fluid, the perturbed energy density $\delta \hat{\rho}_{(\alpha)}$ and/or the perturbed velocity $\hat{v}_{(\alpha)}$, must also be viewed as quantum operators. For instance the perturbed energy density can be expressed as
\begin{align}
\label{eq:expansiondrho}
\delta \hat{\rho}_{(\alpha)}(\eta,{\bm x})=\frac{1}{(2\pi)^{3/2}}
    \int \dd {\bm k}\sum_{A=1}^{N_\mathrm{field}+N_\mathrm{fluid}}
    \left[ \delta \rho_{A,k}^{(\alpha)}(\eta) \hat{c}_{A,{\bm k}}  +  \delta \rho _{A,k}^{(\alpha)*}(\eta) \hat{c}_{A,-{\bm k}}^\dagger \right]
    e^{i{\bm k}\cdot {\bm x}},
\end{align}
where we have introduced the mode functions $\delta \rho_{A,k}^{(\alpha)}(\eta)$ that are obviously different from, say, the mode functions of the perturbed scalar field. However, it is crucial to remark that the creation and annihilation operators in Eqs.~(\ref{eq:expansiondphi}) and~(\ref{eq:expansiondrho}) are really the same quantities. In general, with $N_\mathrm{field}$ scalar fields and $N_\mathrm{fluid}$ cosmological fluids, the index $A$ runs from $1$ to $N_\mathrm{field}+N_\mathrm{fluid}$.
In our case of interest with $2$ scalar fields and $2$ cosmological fluids, the index $A$ runs on the $4$ independent oscillators.

The derivation of the differential equations obeyed by the mode functions of the system is straightforward. As already mentioned, Eqs.~(\ref{eq:perEinstein}), (\ref{eq:perKGl}), (\ref{eq:perKGh}),
(\ref{eq:perdrho1}), (\ref{eq:perdrho2}), (\ref{eq:v1}) and
(\ref{eq:v2}) must now be viewed as differential equations for quantum operators. Then, we can insert the canonical expansions of each operators and the same equations for the mode functions is obtained, however duplicated for each independent oscillator ``$A$". Concretely, this gives for the perturbed Einstein equations
\begin{align}
  \label{eq:perEinstein-A}
\frac{\dot{\Phi}_A}{H}&=-\Phi_A-\frac{1}{2\Mp^2H}\left[
-\dot{\phi}_\ell \delta \phi_{\ell,A}^{(\mathrm{gi})}
-\dot{\phi}_\mathrm{h} \delta \phi_{\mathrm{h},A}^{(\mathrm{gi})}
+a\rho_1(1+w_1)v_{1,A}^{(\mathrm{gi})}+a\rho_2(1+w_2)v_{2,A}^{(\mathrm{gi})}\right] \,.
\end{align}
Since, as already mentioned, $A$ runs on four independent oscillators, the above equation really means four independent equations. For the mode functions of the field fluctuations operators, one obtains
\begin{align}
  \label{eq:perKGl-A}
\ddot{\delta \phi}_{\ell,A}^{(\mathrm{gi})}
  &+\left[3H+\frac{1}{2}\left(\Gamma_{\ell 1}+\Gamma _{\ell 2}\right)
\right]
    \dot{\delta \phi}_{\ell,A}^{(\mathrm{gi})}
+\frac{\partial ^2V}{\partial \phi_\ell^2}
\delta \phi_{\ell,A}^{(\mathrm{gi})} +\frac{\partial ^2V}
{\partial \phi_\ell \partial \phi_\mathrm{h}}\delta \phi_{\mathrm{h},A}^{(\mathrm{gi})}
+\frac{k^2}{a^2}\delta \phi_{\ell,A}^{(\mathrm{gi})}
+2\frac{\partial V}{\partial \phi_\ell} \Phi_A
\nonumber \\ &
+\frac{1}{2}\left(\Gamma_{\ell 1}+\Gamma_{\ell 2}\right)
\dot{\phi}_\ell \Phi_A-4\dot{\phi}_\ell\dot{\Phi}_A=0 \,,
\\
\label{eq:perKGh-A}
\ddot{\delta \phi}_{\mathrm{h},A}^{(\mathrm{gi})}
&+\left[3H+\frac{1}{2}\left(\Gamma_{\mathrm{h} 1}+\Gamma _{\mathrm{h} 2}\right)
  \right]\dot{\delta \phi}_{\mathrm{h},A}^{(\mathrm{gi})}
+\frac{\partial ^2V}{\partial \phi_\mathrm{h}^2}
\delta \phi_{\mathrm{h},A}^{(\mathrm{gi})} +\frac{\partial ^2V}
{\partial \phi_{\mathrm{h}} \partial \phi_\ell}\delta \phi_{\ell,A}^{(\mathrm{gi})}
+\frac{k^2}{a^2}\delta \phi_{\mathrm{h},A}^{(\mathrm{gi})}
+2\frac{\partial V}{\partial \phi_\mathrm{h}} \Phi_A
\nonumber \\ &
+\frac{1}{2}\left(\Gamma_{\mathrm{h} 1}+\Gamma_{\mathrm{h} 2}\right)
\dot{\phi}_\mathrm{h} \Phi_A-4\dot{\phi}_\mathrm{h}\Phi_A=0 \,.
\end{align}
Finally, it remains the equations of motion controlling the behavior of the mode functions associated to the ``real" fluids present in the system. They read
\begin{align}
\label{eq:perdrho1-A}
&\dot{\delta \rho}_{1,A}^{(\mathrm{gi})}+3H(1+w_1)\delta \rho_{1,A}^{(\mathrm{gi})}
-3(1+w_1)\rho_1\dot{\Phi}_A-a(1+w_1)\rho_1\frac{k^2}{a^2}v_{1,A}^{(\mathrm{gi})}
\nonumber \\ &
-\Gamma_{\ell 1}\left[\delta \rho_{K_\ell,A}^{(\mathrm{gi})}
  +\rho_{K_\ell}\Phi_A \right]
-\Gamma_{\mathrm{h} 1}\left[\delta \rho_{K_\mathrm{h},A}^{(\mathrm{gi})}
+\rho_{K_\mathrm{h}}\Phi_A\right]
+\Gamma_{12}\left[\delta \rho_{1,A}^{(\mathrm{gi})}+\rho_1\Phi_A\right]
\nonumber \\ &
-\Gamma_{21}\left[\delta \rho_{2,A}^{(\mathrm{gi})}+\rho_2\Phi_A\right]
=0 \,, \\ 
\label{eq:perdrho2-A}
&\dot{\delta \rho}_{2,A}^{(\mathrm{gi})}+3H(1+w_2)\delta \rho_{2,A}^{(\mathrm{gi})}
-3(1+w_2)\rho_2\dot{\Phi}_A-a(1+w_2)\rho_2\frac{k^2}{a^2}v_{2,A}^{(\mathrm{gi})}
\nonumber \\ &
-\Gamma_{\ell 2}\left[\delta \rho_{K_\ell,A}^{(\mathrm{gi})}
  +\rho_{K_\ell}\Phi_A\right]
-\Gamma_{\mathrm{h} 2}\left[\delta \rho_{K_\mathrm{h},A}^{(\mathrm{gi})}
+\rho_{K_\mathrm{h}}\Phi_A\right]
+\Gamma_{21}\left[\delta \rho_{2,A}^{(\mathrm{gi})}+\rho_2\Phi_A\right]
\nonumber \\ &
-\Gamma_{12}\left[\delta \rho_{1,A}^{(\mathrm{gi})}+\rho_1\Phi_A\right]
=0 \,,
\end{align}
\begin{align}
  \label{eq:v1-A}
\dot{v}_{1,A}^{(\mathrm{gi})}&+\left(1-3w_1\right)Hv_{1,A}^{(\mathrm{gi})}
+\frac{\Phi_A}{a}+\frac{w_1}{1+w_1}\frac{1}{a}
\frac{\delta \rho_{1,A}^{(\mathrm{gi})}}{\rho_1}=-\frac{w_1}{(1+w_1)\rho_1}
Q_{(1)\rightarrow (2)}v_{1,A}^{(\mathrm{gi})}
\nonumber \\ &
+\frac{1}{\rho_1}Q_{K_\ell\rightarrow (1)}
\left[v_{1,A}^{(\mathrm{gi})}-\frac{v_{K_\ell,A}^{(\mathrm{gi})}}{1+w_1}\right]
+\frac{1}{\rho_1}Q_{K_\mathrm{h}\rightarrow (1)}
\left[v_{1,A}^{(\mathrm{gi})}-\frac{v_{K_\mathrm{h},A}^{(\mathrm{gi})}}{1+w_1}\right]
\nonumber \\ &
+\frac{1}{\rho_1}Q_{(2)\rightarrow (1)}
\left[v_{1,A}^{(\mathrm{gi})}-\frac{v_{2,A}^{(\mathrm{gi})}}{1+w_1}\right] \,,
\\
\label{eq:v2-A}
\dot{v}_{2,A}^{(\mathrm{gi})}&+\left(1-3w_2\right)Hv_{2,A}^{(\mathrm{gi})}
+\frac{\Phi_A}{a}+\frac{w_2}{1+w_2}\frac{1}{a}
\frac{\delta \rho_{2,A}^{(\mathrm{gi})}}{\rho_2}=-\frac{w_2}{(1+w_2)\rho_2}
Q_{(2)\rightarrow (1)}v_{2,A}^{(\mathrm{gi})}
\nonumber \\ &
+\frac{1}{\rho_2}Q_{K_\ell \rightarrow (2)}
\left[v_{2,A}^{(\mathrm{gi})}-\frac{v_{K_\ell,A}^{(\mathrm{gi})}}{1+w_2}\right]
+\frac{1}{\rho_2}Q_{K_\mathrm{h}\rightarrow (2)}
\left[v_{2,A}^{(\mathrm{gi})}-\frac{v_{K_\mathrm{h},A}^{(\mathrm{gi})}}{1+w_2}\right]
\nonumber \\ &
+\frac{1}{\rho_2}Q_{(1)\rightarrow (2)}
\left[v_{2,A}^{(\mathrm{gi})}-\frac{v_{1,A}^{(\mathrm{gi})}}{1+w_2}\right] \,.
\end{align}
The above differential equations are the equations to be integrated in order to follow the evolution of the system. In particular, once the mode functions are known, the two-point correlation functions of any combination of operators can be evaluated. However, in order to be able to carry out this task, we need to know the initial conditions for each quantity. We now turn to this question.

\subsection{Initial conditions for the perturbations}
\label{subsec:initial}

\subsubsection{Warm-up: the case of a single fluid}
\label{subsubsec:warmup}

One of the great advantage of the inflationary theory is its ability to suggest natural and well-defined initial conditions. These initial conditions can be introduced in several different ways. Here, as a warm-up, we discuss one method and show how it is related to the formalism introduced before in the simple (and standard) case where there is only one scalar field or one perfect fluid (namely, one degree of freedom). 

Let us start with the case of one scalar field. The equations for the conserved quantities $\zeta$ and ${\cal R}$ (here, obviously, the total $\zeta$ and ${\cal R}$ are the same as the individual conserved quantities since, in our example, there is only one degree of freedom) are given by Eq.~(\ref{eq:dertotzeta}) and Eq.~(\ref{eq:dertotR}). In the present context, the non-adiabatic pressure that appears in those equations is the intrinsic non-adiabatic pressure of a scalar field. It is given by
    \begin{align}
        \delta p_\mathrm{nad}=-2\Mp^2(1-\cs^2)\frac{k^2}{a^2}\Phi_{\bm k}.
    \end{align}
We see that it is non-vanishing unless $\cs=1$, namely a scalar field is a fluid with non-vanishing intrinsic non-adiabatic pressure, but it is proportional to $k^2/a^2$ and, therefore, becomes irrelevant on large scales. Then, using the above expression of $\delta p_\mathrm{nad}$ in Eq.~(\ref{eq:dertotR}) and the fact that the Bardeen potential can be written as
\begin{align}
\label{eq:phizetaR}
    \Phi_{\bm k}=-\frac32 \frac{a^2(\rho+p)}{k^2\Mp^2}(\zeta_{\bm k}+{\cal R}_{\bm k}),
    \end{align}
it is easy to establish that the quantity ${\cal R}_{\bm k}'$ can be re-expressed as ${\cal R}_{\bm k}'=3{\cal H}({\cal R}_{\bm k}+\zeta_{\bm k})$. Then, deriving this expression once, and using the $\delta p_\mathrm{nad}$-independent expression of $\zeta_{\bm k}'+{\cal R}_{\bm k}'$ obtained by summing up Eq.~(\ref{eq:dertotzeta}) and Eq.~(\ref{eq:dertotR}), one arrives at
\begin{align}
    {\cal R}_{\bm k}''+2\frac{z'}{z}{\cal R}_{\bm k}'+k^2{\cal R}_{\bm k}=0,
    \end{align}
where we recall that $z\equiv a\Mp \sqrt{2\epsilon_1}$. Finally, introducing the Mukhanov-Sasaki variable, see above, 
$q=z{\cal R}$, we obtain the following equation of a parametric oscillator
\begin{align}
\label{eq:SM-eq-single-field}
    q_{\bm k}''+\left(k^2-\frac{z''}{z}\right)q_{\bm k}=0.
\end{align}
At the beginning of inflation, the physical wavelengths of Fourier modes of astrophysical interest today were much smaller than the Hubble radius, meaning that $k^2\gg z''/z$. As a consequence, in this regime, the solution of Eq. ~(\ref{eq:SM-eq-single-field}) reads
\begin{align}
\label{eq:Qsf}
    q_{\bm k}(\eta)=A_{\bm k}e^{ik(\eta-\eta_\uini)}+B_{\bm k}e^{-ik(\eta-\eta_\uini)},
\end{align}
where $A_{\bm k}$ and $B_{\bm k}$ are integration constants. Then, quantizing the fluctuations and assuming that the initial state is the vacuum, adiabatic or Wentzel Kramer Brillouin (WKB), Bunch-Davies state, it follows that $A_{\bm k}=(2k)^{-1/2}$ and $B_{\bm k}=0$. 

The above reasoning can easily be repeated if one now assumes that there is only one perfect fluid in the Universe. The main difference with the scalar field calculation presented above is that we now have $\delta p_\mathrm{nad}=0$. Let us restart from Eq.~(\ref{eq:dertotR}) and take  
the derivative of this equation and, then, replace $\zeta'_{\bm k}$ using Eq.~(\ref{eq:dertotzeta}). One arrives at 
\begin{align}
    {\cal R}_{\bm k}''=\frac{{\cal H}'}{{\cal H}}{\cal R}_{\bm k}'+\frac{(\cs^2)'}{\cs^2}{\cal R}_{\bm k}'+3{\cal H}\cs^2{\cal R}_{\bm k}'
    -\cs^2k^2{\cal R}_{\bm k}+\cs^2k^2\Phi_{\bm k}.
\end{align}
The last term in the above equation can be expressed in terms of $\zeta_{\bm k}+{\cal R}_{\bm k}$ using Eq.~(\ref{eq:phizetaR}) 
and, thanks to Eq.~(\ref{eq:dertotR}), $\zeta_{\bm k}+{\cal R}_{\bm k}$ is proportional to ${\cal R}_{\bm k}'$. As a consequence, one obtains a closed equation for ${\cal R}_{\bm k}$ which reads
\begin{align}
    {\cal R}_{\bm k}''=\left[\frac{{\cal H}'}{\cal H}+\frac{(\cs^2)'}{\cs^2}+3{\cal H}\cs^2-{\cal H}\epsilon_1\right]{\cal R}_{\bm k}'-\cs^2k^2{\cal R}_{\bm k}=-2\frac{z'}{z}
    {\cal R}_{\bm k}'-\cs^2k^2{\cal R}_{\bm k},
\end{align}
where $z=a\Mp\sqrt{2\epsilon_1}/\cs$. Notice that $z$ is not well-defined for a pressure-less fluid. Then, as we did for the case of a scalar field, we define the Mukhanov-Sasaki variable of the fluid, $q$, by $q=z{\cal R}$ and it follows that
\begin{align}
    q_{\bm k}''+\left(\cs^2k^2-\frac{z''}{z}\right)q_{\bm k}=0.
\end{align}
This equation is very similar to Eq.~(\ref{eq:Qsf}), the only difference being the appearance of the sound velocity in the gradient term and in the definition of $z(\eta)$ (which, if $\cs$ is constant, cancels out in the term $z''/z$). If the Fourier mode under consideration is such that its wavelength is much smaller than the Hubble horizon initially, then the solution of the above equation reads  $q_{\bm k}(\eta)=A_{\bm k}e^{i\cs k(\eta-\eta_\uini)}+B_{\bm k}e^{-i\cs k(\eta-\eta_\uini)}$. Quantizing the hydrodynamical fluctuations and assuming that they are initially placed in the vacuum state leads to the following initial conditions: $A_{\bm k}=(2\cs k)^{-1/2}$ and $B_{\bm k}=0$. 

Endowed with the initial conditions for the quantity $q_{\bm k}(\eta)$ determined above, one can infer the initial conditions of any other variable. However, this is not straightforward and, as a preparation to the multi-fluid case, we explain how it can be done in the single fluid case. Let us first assume that there is only one scalar field.
We have two methods to derive the initial conditions of the Bardeen potential. Using Eq.~(\ref{eq:phizetaR}) and the fact that ${\cal R}_{\bm k}+\zeta_{\bm k}={\cal R}_{\bm k}'/(3{\cal H})$, one can easily establish that (notice that this equation is exact)
\begin{align}
    \Phi_{\bm k}=-\frac{{\cal H}\epsilon_1}{k^2}\left(\frac{q_{\bm k}}{z}\right)',
\end{align}
or
\begin{align}
\label{eq:BardeenQ}
    a\Phi_{\bm k}=-\frac{1}{2k^2}\frac{\cal H}{\Mp}\sqrt{2\epsilon_1}\left(q_{\bm k}'-\frac{z'}{z}q_{\bm k}\right),
\end{align}
from which one can write
\begin{align}
\label{eq:Bardeeninisf}
    \Phi_{\bm k}=-\frac{i}{(2k)^{3/2}}\frac{\cal H}{\Mp}\sqrt{2\epsilon_1}
    \left(1-\frac{z'}{ikz}\right)
     e^{ik(\eta-\eta_\uini)}\simeq -\frac{i}{(2k)^{3/2}}\frac{\cal H}{\Mp}\sqrt{2\epsilon_1}
    e^{ik(\eta-\eta_\uini)},
\end{align}
the second term between the parenthesis in the above equation being negligible in the small-scale limit. Therefore, we have that $\Phi_{\bm k}\sim (2k)^{-3/2}$. Then, since $a\delta \phi_{\bm k}^{(\mathrm{gi})}=q_{\bm k}-z\Phi_{\bm k}$, one has $\delta \phi_{\bm k}^{(\mathrm{gi})}\sim (2k)^{-1/2}$, the second term proportional to the Bardeen potential giving a sub-dominant contribution. 

A second way to derive those results, which was also the method used in Ref.~\cite{Visinelli:2014qla}, consists in the following. The single-field version of the Einstein equation, see Eq.~(\ref{eq:perEinstein-A}), can be written as
\begin{align}
\label{eq:perteinsteinsinglesf}
    \Phi_{\bm k}'=-{\cal H} \Phi_{\bm k}+\frac{\phi'}{2\Mp^2}\delta \phi_{\bm k}^{(\mathrm{gi})}.
\end{align}
We are interested in the sub-Hubble behavior of $\Phi_{\bm k}$ and $\delta \phi_{\bm k}^{(\mathrm{gi})}$. In this regime, given the solution found above for $q_{\bm k}$, one can write $\Phi_{\bm k}={\cal A}_\Phi e^{ik (\eta-\eta_\uini)}$ and $\delta \phi_{\bm k}^{(\mathrm{gi})}={\cal A}_{\delta \phi}e^{ik(\eta-\eta_\uini)}$, where ${\cal A}_\Phi$ and ${\cal A}_{\delta \phi}$ are slowly varying overall amplitudes. Inserting these expressions into Eq.~(\ref{eq:perteinsteinsinglesf}) and using the definition of $q_{\bm k}$, $q_{\bm k}=a\delta \phi_{\bm k}^{(\mathrm{gi})}+z\Phi_{\bm k}$, in order to express ${\cal A}_{\delta \phi}$ in terms of $q_{\bm k}$ and ${\cal A}_\Phi$, one obtains
\begin{align}
    a{\cal A}'_\Phi+ik\left[1+\frac{{\cal H}(1+\epsilon_1)}{ik}\right]
    a{\cal A}_\Phi=\frac12\frac{\cal H}{\Mp}\sqrt{2\epsilon_1}q_{\bm k}.
\end{align}
In this equation, the derivative ${\cal A}_\Phi'$ can be neglected since ${\cal A}_\Phi$ is changing very slowly; moreover, given that $k/{\cal H}\gg 1$ on sub-Hubble scales, one arrives at exactly Eq.~(\ref{eq:Bardeeninisf}). Therefore, this also completely determines the initial conditions of $\Phi_{\bm k}$ in terms of those of $q_{\bm k}$ and gives results similar to the first method. As a consequence, as announced before, once the initial conditions for the Mukhanov-Sasaki variable are known, the initial conditions for any other relevant quantities can be automatically inferred.

The same reasoning can also be applied to the situation where there is only one perfect fluid in the Universe. In that case, an exact result is 
\begin{align}
    \Phi_{\bm k}=-\frac{{\cal H}\epsilon_1}{\cs ^2k^2}\left(\frac{q_{\bm k}}{z}\right)',
\end{align}
or
\begin{align}
\label{eq:BardeenQfluid}
    a\Phi_{\bm k}=-\frac{\cs }{2\cs^2k^2}\frac{\cal H}{\Mp}\sqrt{2\epsilon_1}\left(q_{\bm k}'-\frac{z'}{z}q_{\bm k}\right).
\end{align}
Then, we introduce the same WKB ansatz already discussed before with one important difference though. Instead of oscillatory terms $\propto e^{ik\eta}$, one needs to take 
into account the fact that, in a fluid, fluctuations propagate with a speed which not the speed of light. As a consequence, the relevant quantities characterizing the fluid will be written as a slow-varying amplitude times $e^{i\cs k(\eta-\eta_\uini)}$. Then, repeating the previous considerations leads to the following equation
\begin{align}
\label{eq:Bardeenfluidic}
    \Phi_{\bm k}=-\frac{i\cs}{(2\cs k)^{3/2}}\frac{\cal H}{\Mp}\sqrt{2\epsilon_1}
    \left(1-\frac{z'}{i\cs kz}\right)
     e^{i\cs k(\eta-\eta_\uini)}\simeq 
  -\frac{i\cs }{(2\cs k)^{3/2}}\frac{\cal H}{\Mp}\sqrt{2\epsilon_1}
     e^{i\cs k(\eta-\eta_\uini)},
\end{align}
and, therefore, that $\Phi_{\bm k}\propto (2\cs k)^{-3/2}$. Notice, as expected,  that the above equation is exactly Eq.~(\ref{eq:Bardeeninisf}) if one takes $\cs=1$.

The second method can also be used to check the validity of the result. In case of one perfect fluid, Eq.~(\ref{eq:perEinstein-A}) reduces to
\begin{align}
\label{eq:perteinsteinsingle}
    \Phi_{\bm k}'=-{\cal H}\Phi_{\bm k}-\frac{a^2}{2\Mp^2}\rho(1+w)v^{(\mathrm{gi})},
\end{align}
and using similar considerations as the ones presented before in the case of a single scalar field, we obtain 
\begin{align}
    a{\cal A}'_\Phi+i\cs k\left[1+\frac{{\cal H}(1+\epsilon_1)}{i\cs k}\right]
    a{\cal A}_\Phi=\frac{\cs}{2}\frac{\cal H}{\Mp}\sqrt{2\epsilon_1}q_{\bm k}.
\end{align}
that is to say the same formula as for a scalar field except for the presence of the sound velocity. In the large scale limit, one obtains exactly Eq.~(\ref{eq:Bardeenfluidic}) for the Bardeen potential and we conclude that $\Phi_{\bm k}\propto (2\cs k)^{-3/2}$.

For the velocity, we can use the definition of $q_{\bm k}$, namely ${\cal H}{\cal A}_v={\cal A}_\Phi-q_{\bm k}/z$ and, therefore, ${\cal A}_v\propto (2\cs k)^{-1/2}$. In particular, the contribution coming from the Bardeen potential is sub-dominant as expected.

Finally, the initial conditions for the density contrast remains to be discussed. In the single fluid case, the perturbed energy density conservation equation, see Eq.~(\ref{eq:perdrho1-A}), reads 
\begin{equation}
    \delta \rho'+3{\cal H}(1+w)\delta \rho-3(1+w)\rho\Phi'
    -(1+w)\rho k^2 v=0,
\end{equation}
yielding
\begin{align}
    {\cal A}_{\delta \rho}'+ik\cs \left[1
    +\frac{3{\cal H}(1+w)}{ik\cs}\right]{\cal A}_{\delta \rho}=3(1+w)\rho\left({\cal A}_\Phi'+ik\cs {\cal A}_\Phi\right)+(1+w)\rho k^2{\cal A}_v.
\end{align}
In the right hand side, the dominant term is the term proportional to ${\cal A}_v$ which scales as $\propto k^{3/2}$ since the term proportional to ${\cal A}_\Phi$ ``only" scales $\propto k^{-1/2}$. As a consequence, one can write
\begin{align}
\label{eq:Adrho}
    {\cal A}_{\delta \rho}=\frac{(1+w)\rho}{i\cs^2}\cs k {\cal A}_v,
\end{align}
and deduce that ${\cal A}_{\delta \rho}\propto k^{1/2}$. It is interesting to test the consistency of this result with the other conservation equation which, in the single fluid case, reads [see Eq.~(\ref{eq:v1-A})]
\begin{align}
    v'+3{\cal H}(1-3w)v+\Phi+\frac{w}{1+w}\frac{\delta \rho}{\rho}=0.
\end{align}
Inserting the WKB ansatz in this equation, one obtains
\begin{align}
    {\cal A}_v'+i\cs k\left[1+\frac{3{\cal H}(1-3w)}{i\cs k}
    \right]{\cal A}_v=-{\cal A}_\Phi-\frac{w}{(1+w)\rho}{\cal A}_{\delta \rho}.
\end{align}
In the right hand side the term proportional to ${\cal A}_\Phi$ is subdominant since it scales $\propto k^{-3/2}$ while the term proportional to ${\cal A}_{\delta \rho}$ is proportional to $k^{-1/2}$. As a consequence, one obtains
\begin{align}
    {\cal A}_{\delta \rho}=-\frac{(1+w)\rho}{w}i\cs k{\cal A}_v,
\end{align}
which is exactly Eq.~(\ref{eq:Adrho}) since $w=\cs^2$ for a perfect fluid with constant equation of state. We conclude that this is entirely consistent with the results obtained from the energy density perturbation conservation equation and that the relevant initial conditions have now been completely specified.

\subsubsection{The many-fluid case}
\label{subsubsec:manyic}

In the previous section, we have shown how to connect the formalism presented in Sec.~\ref{sec:pert} to the approach utilizing the Mukhanov-Sasaki variable. In principle, the generalization to the many-fluid case is straightforward.

The first step consists in defining a Mukhanov-Sasaki variable for each fluid present in the system. For scalar fields, this is known to be $q_a=a\delta \phi_a^{(\mathrm{gi})}+z_a\Phi$ with $z_a=a\phi'_a/{\cal H}$, see Ref.~\cite{Gordon:2000hv}. In the case of a single scalar field, we recall that $q=a\delta \phi^{(\mathrm{gi})}+z\Phi$, where, as already introduced before, $z=a\sqrt{2\epsilon_1}\Mp=a\phi'/{\cal H}$. In some sense, $z_a$ is a generalization of the second manner of writing the $z$ variable (namely a generalization of $a\phi'/{\cal H}$ and ``not" of $a\Mp\sqrt{2\epsilon_1}$). 

For perfect fluids, to our knowledge, the question has not been studied as thoroughly as in the case of scalar fields.
We recall that, for a single fluid $q=z{\cal R}=z\left[\Psi-{\cal H}v_{(\mathrm{gi})}\right]$, with $z=a\sqrt{2\epsilon_1}\Mp/c_{{_\mathrm S}}$. The question is then to define a variable $q_{(\alpha)}$ and a quantity $z_{(\alpha)}$ for each fluid. One possibility for $z_{(\alpha)}$ would be 
$z_{(\alpha)}=a\sqrt{2\epsilon_1}\Mp/c_{(\alpha)}$. However, there is also another possibility which seems closer to what is done in the case of scalar fields. Indeed the one-fluid definition of $z$ can also be written as $z=a^2/\cal H \sqrt{(\rho+p)/(p'/\rho')}$, where, here, $\rho$ and $p$ are the total energy density and pressure, respectively, which (obviously!) are also the energy density and the pressure of the fluid under consideration since we assumed there is only one degree of freedom. In the multi-fluid situation, this suggests the introduction of the 
quantity $z_{(\alpha)}$ defined by
\begin{align}
    z_{(\alpha)}=\frac{a^2}{\cal H}
   \sqrt{\frac{\rho_{(\alpha)}+p_{(\alpha)}}{p'_{(\alpha)}/\rho'_{(\alpha)}}}.
\end{align}
It is important to notice that $z_{(\alpha)}$ can no longer be expressed in terms of $\epsilon_1$ because $\epsilon_1$ is now determined by the total energy density and pressure. Then, one can define $q_{(\alpha)}$ by $q_{(\alpha)}=z_{(\alpha)}{\cal R}_{(\alpha)}=z_{(\alpha)}\left[\Psi-{\cal H}v_{(\alpha)}^{(\mathrm{gi})}\right]$. 

Having defined the generalized Mukhanov-Sasaki variables for each fluid of the system, the next step consists in establishing from the general equations of Sec.~\ref{sec:pert}, the equations satisfied by the $q_a$ and the $q_{(\alpha)}$. These equations will obviously be coupled. Finally, one needs to take the large scale limit, $k/{\cal H}\rightarrow \infty$, in order to guess the initial conditions for the $q_a$ and the $q_{(\alpha)}$. Technically, this is clearly a complicated task.

However, there exists a route which is equivalent and much easier. First of all, we can remark that, initially, the couplings between the scalar fields and the fluids can be neglected.
Technically, this is due to the fact that the physical momenta go to infinity in this regime and, therefore, become the leading contribution in the equation of motion for the Mukhanov-Sasaki variable. As a consequence, the $N_\mathrm{field}$ scalar fields and the $N_\mathrm{fluid}$ fluids can be treated separately. The initial conditions for a collection of $N_\mathrm{field}$ perturbed fields have been studied in great details in the literature and it is standard to show that one has~\cite{GrootNibbelink:2001qt,Weinberg:2008zzc,Achucarro:2010jv} 
\begin{align}
\label{eq:iniqscalarfield}
    q_{a,k}^A(\eta) \rightarrow \frac{1}{\sqrt{2k}}
    e^{ik(\eta-\eta_\uini)}\delta _a^A.
\end{align}
In this expression, it is worth noticing the 
presence of $\delta_a^A$ which means that, initially, the canonical variables are not ``mixed". Only the time evolution of those variables, in presence of an interaction between them, will be able to mix them.

The treatment of a collection of $N_\mathrm{fluid}$ fluids is less standard. As already mentioned, in principle, one should establish the equations of motion of the variables $q_{(\alpha)}$. However, in the large scale limit, we expect the Bardeen potential not to play an important role. As a consequence, one can introduce a simplified version of $q_{(\alpha)}$, $\frak{q} _{(\alpha)}$, defined by $\frak{q}_{(\alpha)}=-z_{(\alpha)} \calH v_{(\alpha)}^{(\mathrm{gi})}$. Assuming that each fluid has a constant equation of state, which is the case of interest in this paper, we find that $\frak{q}_{(\alpha)}$ obeys the equation
\begin{equation}
   \frak{q}_{(\alpha)}^{\prime\prime}+\frak{q}_{(\alpha)}\left[w_\alpha k^2 - \frac{z_{(\alpha)}^{\prime\prime}}{z_{(\alpha)}} + \calH^2  \epsilon_1\epsilon_2  \right]=0. 
\end{equation}
This equation is sufficient to fix the initial conditions in the sub-Hubble regime and we take
\begin{align}
\label{eq:iniMSfluid}
    \frak{q}_{(\alpha),k}^A\rightarrow \frac{1}{\sqrt{2 c_{(\alpha)}k}}e^{ic_{(\alpha)}k(\eta-\eta_\uini)}\delta_{(\alpha)}^A,
\end{align}
where, as it was the case for the scalar fields, we have introduced ``non-mixing" initial conditions, see the presence of $\delta _{(\alpha)}^A$. When the Bardeen potential is sub-dominant on sub-Hubble scales, one expects $\frak{q}_{(\alpha),k}^A\simeq q_{(\alpha),k}^A$.

Here, a comment is in order. We see that Eq.~(\ref{eq:iniMSfluid}) is ill-defined when $c_{(\alpha)}=0$, namely in the case where the fluid under consideration is pressure-less, which may be relevant in the present context. In that case, however, there is no oscillatory modes anymore and, moreover, the physical wavelengths of the Fourier amplitudes are never inside the sonic horizon (which simply vanishes), a necessary criterion to be able to single out well-defined initial conditions. Carrying out the quantization of such a system seems therefore difficult. On the other hand, we need to choose some initial conditions in order to perform numerical calculations. Here, we will simply take a small but non-vanishing pressure for the matter fluid: $c_\mathrm{m}=0.01$. Actually, having $c_\mathrm{m}\neq 0$ is physically well-justified since, in a realistic microscopic description, we expect non-relativistic matter to be made of particles moving at small, but, crucially, non-vanishing, velocities. Although it may be interesting to discuss this question more deeply, at the practical level, we have noticed that the numerical behavior of the system does not depend on those initial conditions. Therefore the physical conclusions obtained in this paper will not be sensitive to this issue.

Having chosen the initial conditions for the ``canonical variables" $q_a$ and $q_{(\alpha)}$, we now discuss how the initial conditions for the other relevant quantities can be determined. This is done with the method presented in Sec.~\ref{subsubsec:warmup}.  Notice, however, that the scalar field fluctuations and perfect fluid fluctuations propagate with different speeds. The different perfect fluid perturbations also propagate with different speeds since, \textit{a priori}, the fluids do not have the same equation of state. To deal with this issue, we write $\Phi^A={\cal A}_\Phi^Ae^{ic_Ak(\eta-\eta_\uini)}$, with $c_A=(1,1,c_1,c_2)$. In the same manner, we write the two perturbed scalar fields as $\delta \phi_{\ell, \mathrm{h}}^A={\cal A}_{\ell,\mathrm{h}}^Ae^{ic_Ak(\eta-\eta_\uini)}$, the two perturbed energy densities as $\delta \rho_{1,2}^A={\cal A}_{1,2}^Ae^{ic_Ak(\eta-\eta_\uini)}$ and, finally, the two velocities as $v_{1,2,A}^{(\mathrm{gi})}={\cal A}_{v_1,v_2}^Ae^{ic_Ak(\eta-\eta_\uini)}$. It is interesting to see that the scalar field fluctuations which, \textit{a priori}, propagate $\propto e^{ik(\eta-\eta_\uini)}$ can also acquire modes $\propto e^{ic_1k(\eta-\eta_\uini)}$ and $\propto e^{ic_2k(\eta-\eta_\uini)}$ (the same remark could be done for perfect fluids perturbations). Whether we initially populate the mode, say, $\Phi^1  = {\cal A}_\Phi^1 e^{ic_1k(\eta-\eta_\uini)}$ is precisely the choice of the initial conditions. In a standard situation, one could also 
decide to populate this type of mode but, without interactions between scalar fields and perfect fluids, this is not physically very relevant and, in any case, does not correspond to the usual choice of initial conditions. In the present context, however, this discussion is much more relevant since the interactions are able to sustain such a mode.

The next step consists in inserting the above expressions of $\Phi^A$, $\delta \phi_{\ell,\mathrm{h}}^A$, $\delta \rho_{1,2}^A$ and $v_{1,2,A}^{\mathrm{(gi)}}$ in Eqs.~(\ref{eq:perEinstein-A}), (\ref{eq:perKGl-A}), (\ref{eq:perKGh-A}), (\ref{eq:perdrho1-A}), (\ref{eq:perdrho2-A}), (\ref{eq:v1-A}) and~(\ref{eq:v2-A}), and neglect the interactions terms. As in Sec.~\ref{subsubsec:warmup}, we consider that the amplitudes ${\cal A}_\Phi^A$, ${\cal A}_{\ell,\mathrm{h}}^A$, ${\cal A}_{1,2}^A$ and ${\cal A}_{v_1,v_2}^A$ are slowly varying functions of time. Moreover, we use the initial conditions obtained from the quantum-mechanical considerations presented earlier in this section. In particular, the requirement~(\ref{eq:iniqscalarfield}) implies that
\begin{align}
\label{eq:inisf}
    {\cal A}_\ell^A=\frac{K_\ell}{a}\delta_{\ell}^A-\frac{z_\ell}{a}{\cal A}_\Phi^A, \quad {\cal A}_\mathrm{h}^A=\frac{K_\mathrm{h}}{a}\delta_{\mathrm{h}}^A-\frac{z_\mathrm{h}}{a}{\cal A}_\Phi^A,
\end{align}
and we also have
\begin{align}
\label{eq:inifluid}
    {\cal A}_{v_1}^A=-\frac{K_1}{{\cal H}z_1}\delta_1^A
    +\frac{1}{\cal H}{\cal A}_{\Phi,1}^A, 
    \quad 
    {\cal A}_{v_2}^A=-\frac{K_2}{{\cal H}z_2}\delta_2^A
    +\frac{1}{\cal H}{\cal A}_{\Phi,2}^A,
\end{align}
where the precise definitions of the background functions $z_\ell$, $z_\mathrm{h}$, $z_1$ and $z_2$ have been introduced before. In fact, at this stage, all we need to know is that these quantities are background quantities and we will see that they do not play a crucial role in determining the initial conditions. The quantities $K_\ell$, $K_\mathrm{h}$, $K_1$ and $K_2$, on the contrary, determine the scaling of the initial conditions and are fixed by the quantum-mechanical considerations described above, namely $K_\ell=K_\mathrm{h}=(2k)^{-1/2}$ and $K_{1,2}=(2c_{1,2}k)^{-1/2}$. 

Let us start with the equation for $\Phi^A$. Substituting the previous WKB ansatz in Eq.~(\ref{eq:perEinstein-A}), one arrives at 
\begin{align}
&{\cal A}_\Phi^A{}'+ic_Ak\left\{1+\frac{\cal H}{ic_Ak}
+\frac{1}{2\Mp^2ic_Ak}\left[\frac{\Phi_\ell'z_\ell}{a}
+\frac{\Phi_\mathrm{h}'z_\mathrm{h}}{a}
+\frac{a^2\rho_1(1+w_1)}{\cal H}
+\frac{a^2\rho_2(1+w_2)}{\cal H}\right]\right\}
{\cal A}_\Phi^A
\nonumber \\ &
=\frac{1}{2\Mp^2}\left[\frac{\phi_\ell'}{a}K_\ell \delta _\ell^A+\frac{\phi'_\mathrm{h}}{a}K_\mathrm{h}\delta_\mathrm{h}^A+\frac{a^2\rho_1(1+w_1)}{{\cal H}z_1}K_1\delta_1^A
+\frac{a^2\rho_2(1+w_2)}{{\cal H}z_2}K_2\delta_2^A
\right],
\end{align}
from which one obtains
\begin{align}
   \label{eq:bardeenquantumini}
    {\cal A}_\Phi^A\simeq \frac{1}{2\Mp^2ic_Ak}\left[\frac{\phi_\ell'}{a}K_\ell \delta _\ell^A+\frac{\phi'_\mathrm{h}}{a}K_\mathrm{h}\delta_\mathrm{h}^A+\frac{a^2\rho_1(1+w_1)}{{\cal H}z_1}K_1\delta_1^A
+\frac{a^2\rho_2(1+w_2)}{{\cal H}z_2}K_2\delta_2^A
\right].
\end{align}

We see that each component of ${\cal A}_\Phi^A$ is determined by the corresponding fluid, that is to say the component ${\cal A}_\Phi^\ell$ is determined by $K_\ell$, ${\cal A}_\Phi^\mathrm{h}$ is determined by $K_\mathrm{h}$ and so on. As for the single fluid case, one has ${\cal A}_\Phi^A\propto k^{-3/2}$. Then, using Eqs.~(\ref{eq:inisf}) and~(\ref{eq:inifluid}), one deduces that, at leading order, ${\cal A}_{\ell}^A\propto k^{-1/2}\delta _\ell^A$, ${\cal A}_\mathrm{h}^A\propto k^{-1/2}\delta _\mathrm{h}^A$, ${\cal A}_{v_1}^A\propto k^{-1/2}\delta_1^A$ and ${\cal A}_{v_2}^A\propto k^{-1/2}\delta_2^A$. At this stage, a comment is in order about the off-diagonal terms of the components ${\cal A}_\ell^A$, ${\cal A}_\mathrm{h}^A$, ${\cal A}_{v_1}^A$ and ${\cal A}_{v_2}^A$. According to the previous considerations, these quantities are next-to-leading order. For instance, according to Eq.~(\ref{eq:inisf}), if $A\neq \ell$, one has ${\cal A}_\ell^A\propto-z_\ell{\cal A}_\Phi^A/a$, which means a scaling $\propto k^{-3/2}$ instead of $\propto k^{-1/2}$ for the diagonal component ($A=\ell$).
However, one has to remember that these conclusions are based on the quantization of the variable $\frak{q}_{(\alpha)}$ introduced before, which is equivalent to $q_{(\alpha)}$ only if the Bardeen contribution is neglected. As a consequence, rigorously, the present considerations do not allow us to derive the sub-leading contributions exactly. In the following, for convenience, we will just assume that the off-diagonal terms initially vanish. Again, at the practical level, this point does not play an important role in the following since the evolution of the system is largely independent of those initial conditions.

Finally, the initial conditions for the energy density perturbations remain to be established. Using Eqs. (\ref{eq:perdrho1-A}) and~(\ref{eq:perdrho2-A}), it is easy to show that
\begin{align}
    {\cal A}_1^A{}'+ikc_A\left[1+\frac{3{\cal H}(1+w_1)}{ic_Ak}\right]{\cal A}_1^A
    =3(1+w_1)\rho_1\left({\cal A}_\Phi^A{}'+ic_Ak {\cal A}_\Phi^A\right)+(1+w_1)\rho_1k^2 {\cal A}_{v_1}^A,
\end{align}
and a similar equation for ${\cal A}_2^A$. It follows that
\begin{align}
    {\cal A}_1^A\simeq \frac{(1+w_1)\rho_1}{ic_A^2}c_Ak{\cal A}_{v_1}^A\simeq -\frac{(1+w_1)\rho_1}{ic_A^2{\cal H}z_1}c_Ak K_1 \delta _{1}^A,
\end{align}
and, again, a similar expression for ${\cal A}_2^A$. This gives the following scaling ${\cal A}_{1,2}\propto k^{1/2}\delta _{1,2}^A$, at leading order. As in the single fluid case, one can check that the above expression can also be recovered from (or is consistent with) the conservation equations involving the velocity of the fluids. Again, for the off-diagonal terms, which are next-to-leading order, we assume that they initially vanish. The previous comments on this assumption are also valid in the present case.

\subsection{Numerical Codes}
\label{subsec:numerics}

It is clear that Eqs.~(\ref{eq:perEinstein}), (\ref{eq:perKGl}), (\ref{eq:perKGh}), (\ref{eq:perdrho1}), (\ref{eq:perdrho2}), (\ref{eq:v1}) and (\ref{eq:v2}) are too complicated to permit the obtention of analytical solutions. We have therefore integrated them numerically and, in this section, we briefly present the methods that we have used to find numerical solutions.
Actually, we have written two independent codes to follow the evolution of the background and linear fluctuations  in these kinds of setups, in order to be able to cross-check the numerical results.
One code is using \texttt{Python3} and the other is using \texttt{Fortran}.

The Python code uses the \texttt{LSODA} method from the routine \texttt{solve\_ivp} of the package \texttt{scipy.integrate}, in order to numerically evolve the two-fields-two-fluids system of $8$ coupled background functions $\left(a,H,\phi_\ell,\phi_\mathrm{h},\pi_\ell,\pi_\mathrm{h},\rho_1,\rho_2\right)\,,$ where $\pi_{\ell,\mathrm{h}}$ are the conjugate momenta of the two scalar fields, and $72=9 \times 2 \times 4$ coupled linear perturbations $\mathrm{Re}/\mathrm{Im}\left(\Phi^A,\delta\phi_\ell^A,\delta\phi_\mathrm{h}^A,\delta \pi_\ell^A, \delta \pi_\mathrm{h}^A,\delta\rho_1^A,\delta\rho_2^A,v_1^A,v_2^A\right)_{A=\ell,\mathrm{h},1,2}\,.$ 
The time variable for the integration is chosen to be the number of $e$-folds $N$, which constrains the value of the scale factor $a$ at each time step.
Moreover, the Hubble parameter $H$ is evolved with the dynamical Friedmann equation $\dot{H}=-(\rho+p)/(2\Mp^2)$ with $\rho,p$ the total energy density and pressure,
while the other Friedmann equation is used as a diagnosis of the accuracy (energy conservation) of the numerical computation: $\mathcal{E}=\left|1-3 H^2 \Mp^2/\rho\right|$ provides a dimensionless quantity measuring the numerical error.
During inflation, $\mathcal{E}$ is of order $10^{-11}$ in our fiducial two-field model of Sec.~\ref{sec:doubleinf}, showing that the numerical implementation is evidently very accurate, even though it grows quickly up to $\lesssim 1 \%$ after the oscillations of the lightest field at the end of inflation.
Indeed, to follow the dynamics of the fluids (background and linear fluctuations) after the end of reheating, we abruptly drop the scalar fields in the numerical evolution in order for the code not to try to resolve the tiny oscillations that must remain even though they constitute a negligible amount of the total energy density, and we believe that this violent procedure (the derivative of $H$ is formally infinite at this point) is at the origin of the quite important decrease in the numerical accuracy.
Although more clever solutions might be possible in order to keep a good numerical accuracy during radiation and matter domination, this does not play an important role since we can still compute accurately each quantity at the end of reheating, a few $e$-folds after the end of inflation, which is the main goal of this article, and only numerical predictions well within the radiation era (or even in the matter era) should be taken more cautiously.

Our Fortran code integrates the same equations (and, therefore, calculate the behavior of the same variables) with a time parameter which is also taken to be the number of e-folds. The method of integration is a fifth-order Runge-Kutta method with adaptive step size control~\cite{Press1996}. As in the Python code, when it becomes difficult to follow the evolution of the inflaton fields after reheating, their 
contribution is automatically put to zero when the corresponding energy density becomes smaller than a chosen threshold. This is done at the background and perturbed level.

As mentioned before, we have cross-checked our results and no significant difference between the results of the two codes has been found.

\section{Application: double inflation}
\label{sec:doubleinf}

\subsection{Description of the model}
\label{subsec:descriptiondouble}

In this section, we apply the formalism studied before to a specific model of multi-field inflation. In order to remain as simple as possible, we consider that the phase of inflation is well-described by
the ``double inflation'' model~\cite{Polarski:1992dq,Peter:1994dx,Langlois:1999dw,Tsujikawa:2002qx} characterized by the following
Lagrangian
\begin{align}
\label{eq:doublelagrangian}
  {\cal L}_\mathrm{dbl-inf}=-\frac12 \partial^\mu \phi_\ell \partial_\mu
  \phi_\ell -\frac12 \partial^\mu \phi_\mathrm{h} \partial_\mu
  \phi_\mathrm{h} -\frac12
  m_\ell^2\phi_\ell^2-\frac12m_\mathrm{h}^2\phi_\mathrm{h}^2
  +{\cal L}_\mathrm{matter}+{\cal L}_\mathrm{int}.
  \end{align}
In this model there are two fields, $\phi_\ell$ and $\phi_\mathrm{h}$
and the main difference with the general case of two-field inflation considered before is that, now, the potential is separable
$V(\phi_\ell,\phi_\mathrm{h})=V_\ell(\phi_\ell)+V_\mathrm{h}(\phi_\mathrm{h})$,
each individual term being simply the potential for a massive
field. The mass of $\phi_\ell $ is $m_\ell$ and that of
$\phi_\mathrm{h}$ is $m_\mathrm{h}$ with 
\begin{equation}
    R \equiv  \frac{m_\mathrm{h}}{m_\ell} > 1 \,,
\end{equation}
which explains why one field is called ``light'', $\phi_\ell$,
and the other ``heavy'', $\phi_\mathrm{h}$. As explained before, the
two fields responsible for inflation also interact with other
components of matter represented here by ${\cal L}_\mathrm{matter}$. In the present context, we interpret ${\cal L}_\mathrm{matter}$ as the Lagrangian describing the decay products of the inflaton fields, the interaction between the inflaton fields and those decay products being given by ${\cal L}_\mathrm{int}$. \textit{A priori}, a complete model requires the microscopic description of the decay products (especially if a perturbative calculation is carried out), namely an explicit form for ${\cal L}_\mathrm{matter}$ in terms of other fundamental fields. 

In this article, however, we follow a different route and use a phenomenological description in which the decay products are modeled by perfect
fluids, see Secs.~\ref{sec:background}, \ref{sec:pert}. In this section, we will consider two fluids with constant
equations of state: the first fluid will be radiation
with $w_\gamma=1/3$ and the second fluid will be pressure-less with
vanishing equation of state $w_\mathrm{m}=0$. Moreover, as already discussed before, we assume that the 
light field can only disintegrate into radiation and the heavy field into pressure-less matter. Finally, we also make the hypothesis that the
decay products cannot interact among themselves. As described before, the two scalar fields are in fact equivalent to three
fluids, two ``kinetic fluids" and one ``potential fluid" with energy densities and
pressures given by
\begin{align}
\rho_{K_\ell}&=\frac{\phi_\ell^2{}'}{2a^2}=p_{K_\ell}, \quad
\rho_{K_\mathrm{h}}=\frac{\phi_\mathrm{h}^2{}'}{2a^2}=p_{K_\mathrm{h}}, \quad
\rho_V=V=-p_V.
\end{align}
We know from the previous considerations, see Sec.~\ref{sec:background}, that the exchanges between those
three fluids are given by
\begin{align}
aQ_{K_\ell\rightarrow V}&=-\phi_\ell'V_{\phi_\ell}, 
\quad aQ_{V\rightarrow K_\ell}=0, \quad
aQ_{K_\mathrm{h}\rightarrow V}=-\phi_\mathrm{h}'V_{\phi_\mathrm{h}}, 
\quad aQ_{V\rightarrow K_\mathrm{h}}=0,  
\end{align}
in order to recover the usual equations of motion for the
fields. The above assumptions concerning the interaction between fields and fluids, using
Eq.~(\ref{eq:Qbackground}), are equivalent to 
\begin{align}
Q_{K_\ell \rightarrow \gamma}& =-\Gamma_{\ell \gamma}\rho_{K_\ell}, \quad
Q_{\gamma \rightarrow K_\ell}=0, \quad
Q_{K_\mathrm{h} \rightarrow \mathrm{m}} =-\Gamma_{\mathrm{h} \mathrm{m}}\rho_{K_\mathrm{h}},
\quad Q_{\mathrm{m} \rightarrow K_\mathrm{h}}=0,
\end{align}
all other, \textit{a priori} possible, terms vanishing. To summarize, the model is characterized by four parameters, $m_\ell$,
$m_\mathrm{h}$ (or $m_\ell$ and $R$), $\Gamma_{\ell \gamma}$ and
$\Gamma_{\mathrm{h}\mathrm{m}}$.

\subsection{Background Evolution}
\label{subsec:backdouble}

\subsubsection{Generalities}
\label{subsubsec:generalities}

The equations of motion controlling the evolution of the background
fields $\phi_\ell(t)$, $\phi_\mathrm{h}(t)$ and of the fluid energy
densities, $\rho_\gamma(t)$, $\rho_\mathrm{m}(t)$ are the Friedman
equation
\begin{align}
  H^2=\frac{1}{3\Mp^2}\left[\frac{\dot \phi_\ell^2}{2}
    +\frac{\dot \phi_\mathrm{h}^2}{2}+V_\ell(\phi_\ell)+V_\mathrm{h}(\phi_\mathrm{h})
    +\rho_\gamma+\rho_\mathrm{m}\right],
\end{align}
the two Klein-Gordon equations for the light and heavy fields, namely
\begin{align}
\label{eq:KGdecaylh}
\ddot{\phi}_\ell
+\left(3H+\frac{1}{2}\Gamma_{\ell \gamma}\right)
\dot \phi_\ell +m_\ell^2\phi_\ell &=0, \quad
\ddot{\phi}_\mathrm{h}
+\left(3H+\frac{1}{2}\Gamma_{\mathrm{h} \mathrm{m}}\right)
\dot \phi_\mathrm{h} +m_\mathrm{h}^2\phi_\mathrm{h} =0, 
\end{align}
and the two conservation equations for the radiation and pressure-less
fluids
\begin{align}
\label{eq:rhogammam}
\dot{\rho}_\gamma+4H\rho_\gamma
&=\frac{1}{2}\Gamma_{\ell \gamma}\dot \phi_{\ell}^2,
\quad
\dot{\rho}_\mathrm{m}+3H\rho_\mathrm{m}
=\frac{1}{2}\Gamma_{\mathrm{h} \mathrm{m}}\dot \phi_{\mathrm{h}}^2.
\end{align}
We now describe the different epochs of evolution of the
scenario. Initially, the terms $\Gamma_{\ell\gamma}$ and
$\Gamma_\mathrm{hm}$ are chosen to be negligible,
$\Gamma_{\ell\gamma}\ll H$ and $\Gamma_\mathrm{hm}\ll H$. This implies
that any pre-existing amount of radiation and/or matter (if any) plays no role during the initial
phase of the model.  In that case, we are left with two standard
Klein-Gordon equations, $\ddot{\phi}_{\ell,
  \mathrm{h}}+3H\dot{\phi}_{\ell,\mathrm{h}}
+m_{\ell,\mathrm{h}}^2\phi_{\ell,\mathrm{h}}=0$ and a simplified
Friedman equation, $H^2=(\dot{\phi}_\ell^2/2+\dot{\phi}_\mathrm{h}^2/2
+V_\ell+V_\mathrm{h})/(3\Mp^2)$. Nevertheless, no exact, analytical,
solution, to this system of equations is known and we will have to rely on
numerical calculations, see below. However, the exact numerical solutions can be
well-understood by means of the slow-roll approximation. We therefore
introduce the (hierarchy of) Hubble-flow parameters $\epsilon_n$
defined by~\cite{Leach:2002ar}
\begin{align}
  \epsilon_{n+1}=\frac{{\rm d}\ln \vert \epsilon_n\vert }{{\rm d}N},
\end{align}
with $\epsilon_0=H_\uini/H$. In particular, the first Hubble-flow parameter,
$\epsilon_1=-\dot{H}/H^2=1-\ddot{a}/(aH^2)$ indicates whether inflation
occurs since $\epsilon_1<1$ is equivalent to $\ddot{a}>0$. In the
present case, one has
\begin{align}
  \label{eq:eps1}
  \epsilon_1=\frac{1}{2H^2\Mp^2}\left(\dot{\phi}_\ell^2
  +\dot{\phi}_\mathrm{h}^2\right)=3\frac{\dot{\phi}_\ell^2/2
    +\dot{\phi}_\mathrm{h}^2/2}{\dot{\phi}_\ell^2/2+\dot{\phi}_\mathrm{h}^2/2
    +V(\phi_\ell,\phi_\mathrm{h})}.
  \end{align}
Therefore, in order to have inflation, one needs
$\dot{\phi}_\ell^2/(H^2\Mp^2)\ll 1$ and
$\dot{\phi}_\mathrm{h}^2/(H^2\Mp^2)\ll 1$ and, as usual, this
corresponds to a situation where the fields have sub-dominant kinetic
energy compared to their potential energy. This also implies that the
slow-roll Friedman equation can be approximated as $H^2\sim
(m^2_\ell\phi_\ell^2+m_\mathrm{h}^2\phi_\mathrm{h}^2)/(6 \Mp^2)$.

The validity of the slow-roll approximation also depends on field
acceleration, a piece of information which is encoded in the second
Hubble-flow parameter given by
\begin{align}
  \epsilon_2=2\epsilon_1+\frac{2}{H}
  \frac{\ddot{\phi}_\ell\dot{\phi}_\ell+\ddot{\phi}_\mathrm{h}\dot{\phi}
    _\mathrm{h}}{\dot{\phi}_\ell^2
    +\dot{\phi}_\mathrm{h}^2}.
  \end{align}
Interestingly enough, and contrary to the single-field case,
$\epsilon_2\ll 1$ does not necessarily imply that
$\ddot{\phi}_\mathrm{h}/(H\dot{\phi}_\mathrm{h})\ll 1$ and
$\ddot{\phi}_\ell/(H\dot{\phi}_\ell)\ll 1$ separately. In the
following, we will nevertheless assume that this is true, in which
case the two Klein-Gordon equations can be approximated as
$3H\dot{\phi}_{\ell, \mathrm{h}}+m_{\ell, \mathrm{h}}^2\phi_{\ell,
  \mathrm{h}}\sim 0$.

The previous considerations also allow us to re-write the first
Hubble flow parameter as
\begin{align}
  \label{eq:eps1field}
  \epsilon_1\sim 2\Mp^2\frac{\phi_\ell^2+R^4\phi_\mathrm{h}^2}{(\phi_\ell^2
    +R^2\phi_\mathrm{h}^2)^2}.
  \end{align}
As a consistency check, one verifies that, if the heavy field
dominates, $m_\mathrm{h}\phi_\mathrm{h}\gg m_\ell \phi_\ell$, or
$R\phi_\mathrm{h}\gg \phi_\ell$, then the
expression~(\ref{eq:eps1field}) reduces to $\epsilon_1\simeq
2\Mp^2/\phi_\mathrm{h}^2$ which is, as expected, the single-field
expression of the first Hubble flow parameter. If, on the contrary,
the light field dominates, $\phi_\ell\gg R\phi_\mathrm{h}$, then
$\epsilon_1\simeq 2\Mp^2/\phi_\ell^2$ and, again, one recovers the
single-field result (even though strictly speaking this rather requires the more constraining inequality $\phi_\ell \gg R^2\phi_\mathrm{h}$).

\subsubsection{Slow-roll evolution of the background} 
\label{subsubsec:paramsr}

As already mentioned, if the slow-roll approximation is satisfied, then the equations of
motion can be analytically integrated. Let us briefly (since this solution
is standard in the literature) recall how this is derived, mainly
in order to clarify the role of the integration constants. Summing up the two slow-roll Klein-Gordon equations~(\ref{eq:KGdecaylh}) (where, in accordance with the previous discussion, we do not - yet - take into account the decay terms) and using the Friedmann equation, one arrives at
\begin{align}
      3H^2\left(\phi_\ell\frac{{\rm d}\phi_\ell}{{\rm d}N}+\phi_\mathrm{h}
      \frac{{\rm d}\phi_\mathrm{h}}{{\rm d}N}\right)=-m_\ell^2\phi_\ell^2
      -m_\mathrm{h}^2\phi_\mathrm{h}^2=-6\Mp^2H^2,
\end{align}
which results in ${\rm d}(\phi_\ell^2/\Mp^2+\phi_\mathrm{h}^2/\Mp^2)/{\rm d}s=4$.
Here, the variable $s$ is such that ${\rm d}s/{\rm d}N=-1$ which means that $s=-\ln(a/a_\mathrm{p})=-N+N_\mathrm{p}$, $a_\mathrm{p}$ being the scale factor at a particular time ``$t_\mathrm{p}$" that we do not need to specify for the moment. 
Then, if one chooses the parameterization,  $\phi_\ell=r(s) \mathrm{cos}[\theta(s)]$, $\phi_\mathrm{h}=r(s) \mathrm{sin}[\theta(s)]$, the above equation becomes a differential equation for $r(s)$ only, which can easily be solved. One finds $r(s)=\sqrt{r_0^2 + 4 \Mp^2 (s-s_0)}$ where $r_0$ and $s_0$ are constants of integration. Moreover, using the slow-roll equation for the light field, one can show that the angle $\theta(s)$ obeys
\begin{align}
\frac{2 \Mp^2}{r^2(s)} (R^2-1) \frac{{\rm d}s}{{\rm d} \theta} = \frac{1+ R^2 \tan^2 \theta}{\tan \theta},
\end{align}
which can also be solved exactly, leading to 
\begin{align}
    s=s_0 + \frac{r_0^2}{4 \Mp^2}  \left[ - 1 + \frac{\mathrm{cos}^2\theta_0}{\mathrm{cos}^2\theta} \left( \frac{\mathrm{tan}\theta}{\mathrm{tan}\theta_0} \right)^\frac{2}{R^2-1} \right]\, ,
    \end{align}
where $\theta_0=\theta(s_0)$. As a consequence, the solutions 
for the light and heavy fields can be expressed as
\begin{align}
\label{eq:fieldsr}
    \phi_\ell =2 \Mp C^{1/2} \left(\tan \theta\right)^{1/(R^2-1)}, 
    \quad \phi_\mathrm{h}=2 \Mp C^{1/2} \left(\tan \theta\right)^{R^2/(R^2-1)},
\end{align}
    where $C$ is a constant given by
    \begin{align}
        C=\frac{r_0^2}{4\Mp^2} \frac{\cos^2 \theta_0}{(\tan \theta_0)^{2/(R^2-1)}}.
    \end{align}
At this stage, without loss of generality, we can use our freedom to choose the peculiar time $N_\mathrm{p}$ (corresponding to $t_\mathrm{p}$ introduced above) in order to simplify the above expressions. A convenient choice is $N_\mathrm{p}=N_0+r_0^2/(4\Mp^2)$ which implies that $s_0=r_0^2/(4\Mp^2)$. Then, it immediately follows that the 
expression for $s$ simplifies to 
\begin{align}
\label{eq:sfunctiontheta}
    s=C \frac{(\tan \theta)^{2/(R^2-1)}}{\cos ^2\theta}.
\end{align}
It is also interesting to express the quantity $C$ in a way which is more convenient, in particular when numerical and analytical estimates are compared. Using the above expression of $C$, it is easy 
to establish that
\begin{align}
\label{eq:expressionC}
    C=N_\mathrm{tot} \frac{\cos ^2 \theta_\uini}{(\tan \theta _\uini)^{2/(R^2-1)}}\left[1
    -\frac{\cos ^2 \theta_\uini}{\cos ^2\theta_\mathrm{end}}
    \left(\frac{\tan \theta_\mathrm{end}}{\tan \theta _\mathrm{ini}}\right)^{2/(R^2-1)}\right]^{-1},
    \end{align}
where $\theta_\uini$ and $\theta_\mathrm{end}$ are the value of $\theta $ at the beginning and at the end of inflation, respectively. The quantity $N_\mathrm{tot}=N_\mathrm{end}-N_\uini$ is the total number of e-folds during inflation. At this point, one needs to elaborate a little bit on what we exactly mean by the end of inflation. Of course, the end of inflation is defined by $\epsilon_1=1$. However, the previous considerations are based on the slow-roll approximation which, \textit{a priori}, ceases to be valid when the heavy field starts oscillating (in this regime, its kinetic energy equals its potential energy and the slow-roll approximation breaks down). However, when it is the case, the energy density of the heavy field becomes negligible and the Universe is dominated by the light field which is still slow-rolling. As a consequence, the slow-roll
expression for $\phi_\ell$ and for the Hubble parameter $H$ found above can be extended in this regime. As mentioned before, in this case, $\epsilon_1\simeq 2\Mp^2/\phi_\ell^2$ and the end 
of inflation is given by $\tan \theta_\mathrm{end}\simeq (2C)^{(1-R^2)/2}$, which implies $s_\mathrm{end}=1/2$ and, therefore, $N_\mathrm{p}=N_\mathrm{end}+1/2$. Another reasoning is, however, possible. We can also say that the end of inflation cannot be predicted accurately within the slow-roll approximation but that it certainly happens when $\theta_\mathrm{end}\ll 1$. As a consequence, given Eq.~(\ref{eq:sfunctiontheta}), $s_\mathrm{end}=-N_\mathrm{end}+N_\mathrm{p}\simeq 0$, namely $N_\mathrm{p}=N_\mathrm{end}$. We conclude that analytical approximations lead to $N_\mathrm{p}=N_\mathrm{end}+{\cal O}(1)$ but that the extra factor of order one cannot be unambiguously determined in this framework. For simplicity, in the following, we will ignore it and take $N_\mathrm{p}\simeq N_\mathrm{end}$, which, with the help of Eq.~(\ref{eq:expressionC}), results in 
$C=N_\mathrm{tot} \cos ^2 \theta_\uini/(\tan \theta _\uini)^{2/(R^2-1)}$. Let us also notice that this last expression can be used to estimate the total number of e-folds in terms of the initial values of the fields. Straightforward manipulations lead to
\begin{align}
\label{eq:Ntot}
    N_\mathrm{tot}
    &=\frac{\phi_\ell^2\vert_\uini}{4\Mp^2}
    \left[1+\left(\frac{\phi_\mathrm{h}\vert_\uini}{\phi_\ell\vert_\uini}\right)^{2}\right].
\end{align}

Finally, with the choice $N_\mathrm{p}=N_0+r_0^2/(4\Mp^2)$, one has $r(s)=2\Mp \sqrt{s}$ and the Hubble parameter can be written as 
  \begin{align}
      \label{eq:Hubblegene}
      H^2(s)=\frac23sm_\ell^2
      \left[1+(R^2-1)\sin^2\theta\right],
\end{align}
an expression which is a good approximation of the exact Hubble parameter during the whole inflation duration even when the heavy field is no longer slow-rolling (as explained above).

It is also interesting to relate the variable $\theta$ to the cosmic time. Using the chain equation
\begin{align}
\label{eq: dt to dtheta}
    \dd t=\frac{\dd N}{H}=\frac{\dd N}{\dd s}
    \frac{\dd s}{\dd \theta}\frac{1}{H}\dd \theta
    =-\frac{\dd s}{\dd \theta}\frac{1}{H}\dd \theta,
    \end{align}
the cosmic time can be expressed by mean of the following integral
\begin{align}
\frac{m_\ell}{\sqrt{6}}(t-t_\mathrm{p})
=-\frac{C^{1/2}}{R^2-1}
\int _{\theta_\mathrm{p}}^\theta 
\frac{\sqrt{1+R^2 \tan^2 \theta}}{\sin \theta \cos \theta}
(\tan \theta)^{\frac{1}{R^2-1}}\dd \theta,
\end{align}
which can be explicitly calculated. As a consequence, the relation between $\theta $ and $t$ can be expressed as
\begin{align}
\label{eq:thypergeo}
\frac{m_\ell}{\sqrt{6}}(t-t_\mathrm{p})
&=-C^{1/2}\left(\tan \theta \right)^{\frac{1}{R^2-1}}
{}_2F_1\left[-\frac{1}{2},\frac{1}{2(R^2-1)},1+\frac{1}{2(R^2-1)}; -R^2\tan ^2 \theta\right]
\nonumber \\ &
+C^{1/2}
\left(\tan \theta_\mathrm{p} \right)^{\frac{1}{R^2-1}}
{}_2F_1\left[-\frac{1}{2},\frac{1}{2(R^2-1)},1+\frac{1}{2(R^2-1)}; -R^2\tan ^2 \theta_\mathrm{p}\right],
\end{align}
where ${}_2F_{1}(.)$ is an hypergeometric function~\cite{Gradshteyn:1965aa,Abramovitz:1970aa}. In this expression, one is of course free to choose the peculiar time as one wishes.

\begin{figure}
    \centering    
    \includegraphics[width=0.75\linewidth]{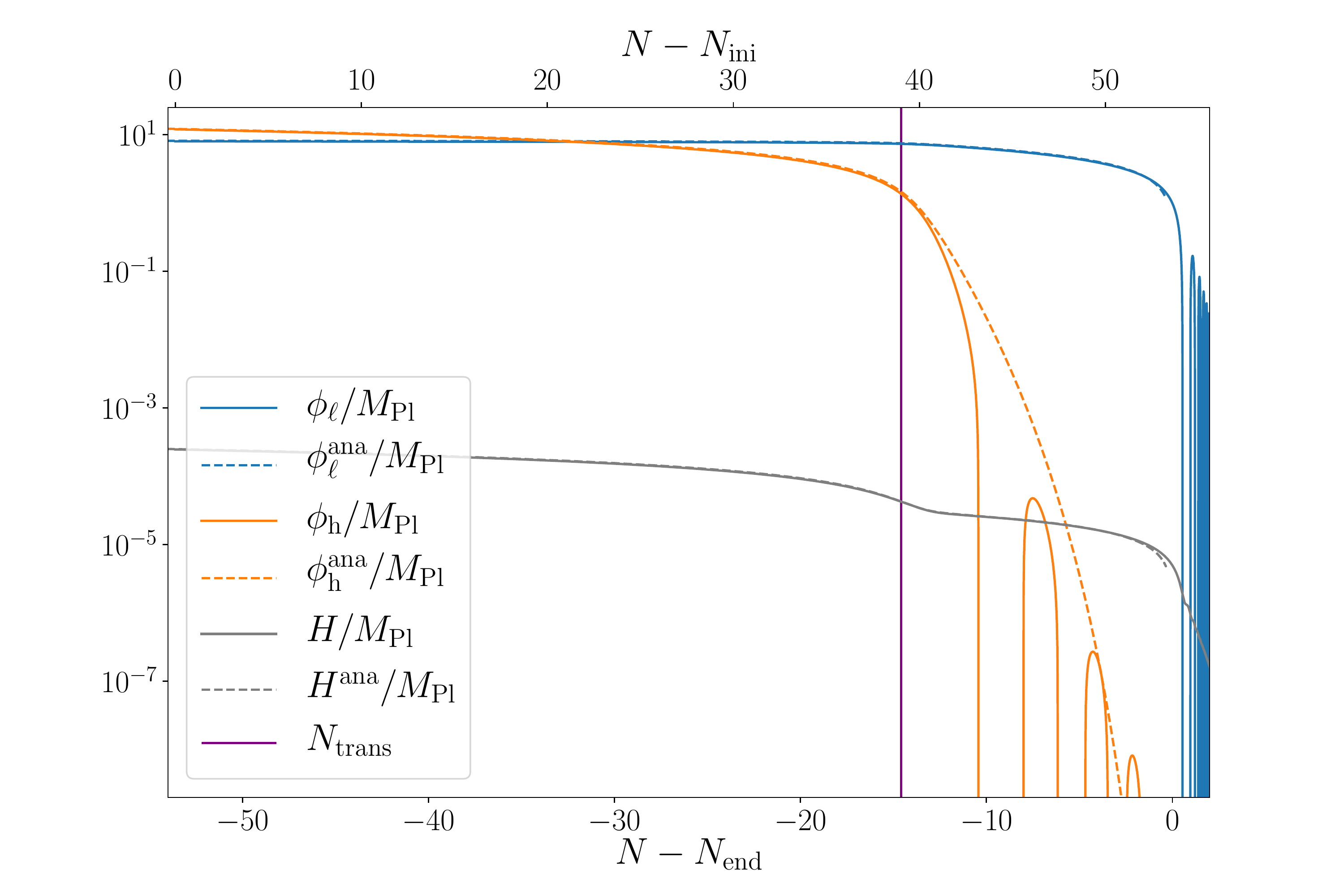}
    \caption{Light and heavy scalar fields and Hubble parameter during inflation, in Planck mass units. Exact numerical solutions (solid lines, respectively blue, orange and gray) and approximate, slow-roll, analytical solutions ``$\mathrm{ana}$" (dashed lines, respectively blue, orange and gray) given by Eqs.~\eqref{eq:fieldsr}--\eqref{eq:Hubblegene} are both displayed. The vertical purple line indicates the transition between the regime dominated by the heavy field and the regime dominated by the light field.}
    \label{fig:exactfield}
\end{figure}

Let us now study the exact solutions of the Friedmann and Klein-Gordon equations and how the approximations developed above perform. In the following, unless specified otherwise, we will always work with the ``fiducial" parameters $m_\ell=10^{-5}\Mp$, $R=5$, $\Gamma_{\ell \gamma}=10^{-6}\Mp$ and $\Gamma_{\mathrm{hm}}=10^{-5}\Mp$.
The initial conditions are chosen so that $\phi_\ell\vert_\uini=8\Mp $, $\phi_\mathrm{h}\vert_\uini=12\Mp$ and with a negligible amount of radiation and pressure-less matter (we come back to this question below).
The velocities of the scalar fields are fixed according to the slow-roll approximation of their Klein-Gordon equation. This implies the 
following initial value for the Hubble parameter, $H_\uini = 2.5\times 10^{-4}\Mp$. The corresponding exact, numerical, solutions for $\phi_\ell(N)$ (solid blue line), $\phi_\mathrm{h}(N)$ (solid orange line) and the Hubble parameter $H(N)$ (solid gray line) during inflation are presented in Fig.~\ref{fig:exactfield}. We have also plotted the slow-roll approximated evolution of the fields and of the Hubble parameter, Eqs.~\eqref{eq:fieldsr}-\eqref{eq:Hubblegene}, see the dashed blue, orange and gray lines. In brief, the approximation evidently appears to be very good in the slow-roll regime. \textit{A priori}, this regime is left when the fields start oscillating and, as a result, we see that the solid and dashed lines separate. Note, however, as already mentioned above, that even when the heavy field has started its oscillations, the slow-roll solutions for the light field and the Hubble parameter are still valid approximations because the universe becomes quickly dominated by the light field which is still slowly rolling. 

Let us now describe Fig.~\ref{fig:exactfield} in more quantitative terms. One first sees that the heavy field has initially a vacuum expectation value higher than that of the light field and, therefore, starts ``higher" in the potential. This regime can be understood by means of Eqs.~(\ref{eq:fieldsr}) and~(\ref{eq:Hubblegene}). One can indeed assume that, initially, the heavy field completely dominates the energy budget of the universe. This is a good approximation as can be checked if, for instance, one uses it to calculate the initial value of the Hubble parameter. This gives $H_\uini/\Mp\simeq m_\mathrm{h}\phi_\mathrm{h}\vert_\uini/(\sqrt{6}\Mp)$ leading, with the fiducial parameters of Fig.~\ref{fig:exactfield}, to $H_\uini\simeq 2.5 \times 10^{-4}\Mp$, a value in perfect agreement with the exact, numerical, value given above. In the heavy phase, when $\phi_\mathrm{h}\gg \phi_\ell$, $\cos ^2\theta $ is a small quantity (for the initial conditions studied here this is admittedly not very accurate since one ``only" has $\cos ^2\theta _\uini=4/13\simeq 0.31$ but it could be smaller if the difference between $\phi_\mathrm{h}\vert_\uini$ and $\phi_\ell\vert_\uini$ were chosen to be larger) and 
\begin{align}
    s\simeq C \left(\cos ^2 \theta\right)^{-R^2/(R^2-1)}
    \left(1-\frac{\cos^2 \theta }{R^2-1}+\cdots \right)
    \simeq \frac{C}{\cos ^2 \theta},
\end{align}
where (in the last expression only) we have also assumed $R\gg 1$. As a consequence, $\theta\simeq \arccos[(C/s)^{1/2}]$ and one obtains
\begin{align}
\label{eq:philini}
\phi_\ell=\Mp \sqrt{4s}\cos[\theta(s)]\simeq 2\Mp C^{1/2}\simeq \phi_\ell\vert_\uini,
\end{align}
which is a constant, in agreement with what we observe in Fig.~\ref{fig:exactfield} during the domination of the heavy field. Notice that this approximation also leads to a simplified expression for the constant $C$, namely 
\begin{align}
\label{eq:defc}
C\simeq \frac{\phi^2_\ell\vert_\uini}{4 \Mp^2}.
\end{align}
Regarding the heavy field, in the regime discussed here, its slow-roll trajectory can be approximated by
$\phi_\mathrm{h}=\Mp \sqrt{4s}\sin[\theta(s)]\simeq  \Mp \sqrt{4(s-C)}$, or
\begin{align}
\label{eq:phihsr}
\phi_\mathrm{h}^2(N)=-\phi_\ell^2\vert_\uini-4\Mp^2(N-N_\uend)=\phi_\mathrm{h}\vert_\uini^2-4\Mp^2(N-N_\uini),
\end{align}
where, in the last expression, we have made use of Eq.~(\ref{eq:Ntot}). This last expression corresponds to the single-field trajectory for a large field model (with a quadratic potential) which, of course, makes sense since, in this regime, the heavy field completely dominates the energy budget of the Universe. It is also easy to calculate the evolution of the Hubble parameter, and the result reads
\begin{align}
\label{eq:Hubblesr}
    H^2\simeq -\frac16 m_\mathrm{h}^2\frac{\phi_\ell^2\vert_\uini}{\Mp^2}
    -\frac23m_\mathrm{h}^2=H^2\vert_\uini-\frac{2}{3}
    m_\mathrm{h}^2\left(N-N_\uini \right),
\end{align}
where, again, we have used Eq.~(\ref{eq:Ntot}). As expected, one also recovers the single-field expression of the Hubble parameter. 

Finally, making use of Eq.~(\ref{eq:Ntot}) yet another time, with the fiducial parameters chosen before, one has $N_\mathrm{tot}\simeq 52$ to be compared with the exact, numerical, value $N_\mathrm{tot}=53.6$.

The regime described above lasts until the energy density of the heavy field (which decreases with time) equals that of the light field (which stays approximately constant). This transition occurs when $m_\mathrm{h}\phi_\mathrm{h}=m_\ell \phi_\ell$ or $\theta_\mathrm{trans}=\arctan(1/R)$. Let us notice that, at 
the transition $\cos \theta_\mathrm{trans}=R/\sqrt{R^2+1}$ and, in the limit $R\gg 1$, we see that $\cos \theta_\mathrm{trans}$ is not small (but rather of order one) and, therefore, one expects to observe deviations from the above trajectories already at the transition (or even before). Using Eq.~(\ref{eq:sfunctiontheta}), which is valid even if $\cos \theta $ is not small, one can evaluate the time of transition quite precisely. One obtains
\begin{align}
\label{eq:strans}
    s_\mathrm{trans}=-N_\mathrm{trans}+N_\uend=
    CR^{2R^2/(1-R^2)}(1+R^2)\simeq 
    \frac{\phi_\ell^2\vert_\uini}{4\Mp^2}R^{2R^2/(1-R^2)}(1+R^2),
\end{align}
where we have used the simplified expression of $C$, see Eq.~(\ref{eq:defc}). For the fiducial parameters used before, one finds $N_\mathrm{trans}-N_\uend\simeq -14.6$, in good agreement with the exact, numerical, value $N_\mathrm{trans}-N_\uend = -14.49$.
Notice that, in Fig.~\ref{fig:exactfield}, the crossing between the blue and orange lines corresponds to the time at which the vacuum expectation value of the heavy field becomes smaller than the expectation value of the light field.
This time has clearly nothing to do with the transition mentioned before and is not associated with any change in the physical properties of the system.
In fact, the passage from the phase dominated by the heavy field to the phase dominated by the light field manifests itself by the small dropout in the evolution of the Hubble parameter that can be seen in Fig.~\ref{fig:exactfield} and which occurs around $N_\mathrm{trans}$.
At this time, using Eq.~(\ref{eq:Hubblegene}), the Hubble parameter is given by
\begin{align}
\label{eq:Htrans}
    H_\mathrm{trans}^2&=\frac23s_\mathrm{trans}m_\ell^2
    \left[1+(R^2-1)\sin^2\theta_\mathrm{trans}\right]
    =\frac43 s_\mathrm{trans}m_\ell^2\frac{R^2}{R^2+1}
    \\
        &\simeq 
    \frac13\frac{\phi_\ell^2\vert_\uini}{\Mp^2}
    m_\mathrm{h}^2R^{2R^2/(1-R^2)},
    \end{align}
that is to say $H_\mathrm{trans}\simeq 4.3 \times 10^{-5}\Mp$ to be compared with the value observed in Fig.~\ref{fig:exactfield}, namely $4.27 \times 10^{-5} \Mp$.

After the transition, the contribution from the heavy field becomes negligible and the universe is dominated by the light field which is no longer frozen (as can be checked in Fig.~\ref{fig:exactfield}) and starts to move. In this 
regime $\theta \ll 1$ [we have seen before that, already at the transition, $\theta _\mathrm{trans}=\arctan(1/R)\ll 1$] and, as a consequence, we can write
\begin{align}
    s\simeq C \, \theta ^{2/(R^2-1)}\left[1+\frac{3R^2-1}{3(R^2-1)}\theta^2
    +\cdots \right]\simeq C\, \theta ^{2/R^2}\left(1+\theta^2
    +\cdots\right),
\end{align}
where, in the last equality, we have used $R\gg 1$. This implies that $\theta \simeq (s/C)^{R^2/2}$ and this phase corresponds to values of $s/C$ such that $s/C\ll 1$. Then, using the slow-roll trajectory~(\ref{eq:fieldsr}) where we neglect the cosine term given that $\theta \ll 1$, one obtains
\begin{align}
    \phi_\ell^2\simeq  4\Mp^2(N_\mathrm{end}-N).
    \end{align}
In a similar way, using  Eq.~(\ref{eq:Hubblegene}), one finds that
    \begin{align}
    \label{eq:hubblelightphase}
        H^2\simeq \frac23 m_\ell^2 s=\frac23m_\ell^2 (N_\mathrm{end}-N).
        \end{align}
Notice that, at the end of inflation, by definition, $s=0$ and, therefore, one finds $\phi_\ell \vert_\uend=H_\uend=0$. It is easy to see that this a consequence of the fact that we chose $N_\mathrm{p}=N_\mathrm{end}$. If, instead, as discussed above, we had chosen $N_\mathrm{p}=N_\mathrm{end}+1/2$ then we would 
have found $H^2=2m_\ell^2(N_\mathrm{end}-N+1/2)/3$ and, therefore, $H_\mathrm{end}^2=m_\ell^2/3$. This is completely consistent with the fact that, during the epoch dominated by the light field, $\epsilon_1\simeq 2\Mp^2/\phi_\ell^2$ implying that $\phi_\ell\vert_\uend \simeq \sqrt{2}\Mp$ and $H_\uend\simeq m_\ell/\sqrt{3}\simeq 5.8 \times 10^{-6} \Mp$, to be compared with the exact numerical value $H_\uend = 4.97 \times 10^{-6} \Mp$.

\subsubsection{Matter and radiation energy densities during slow-roll}
\label{subsubsec:mgammasr}

\begin{figure}
    \centering    
    \includegraphics[width=0.75\linewidth]{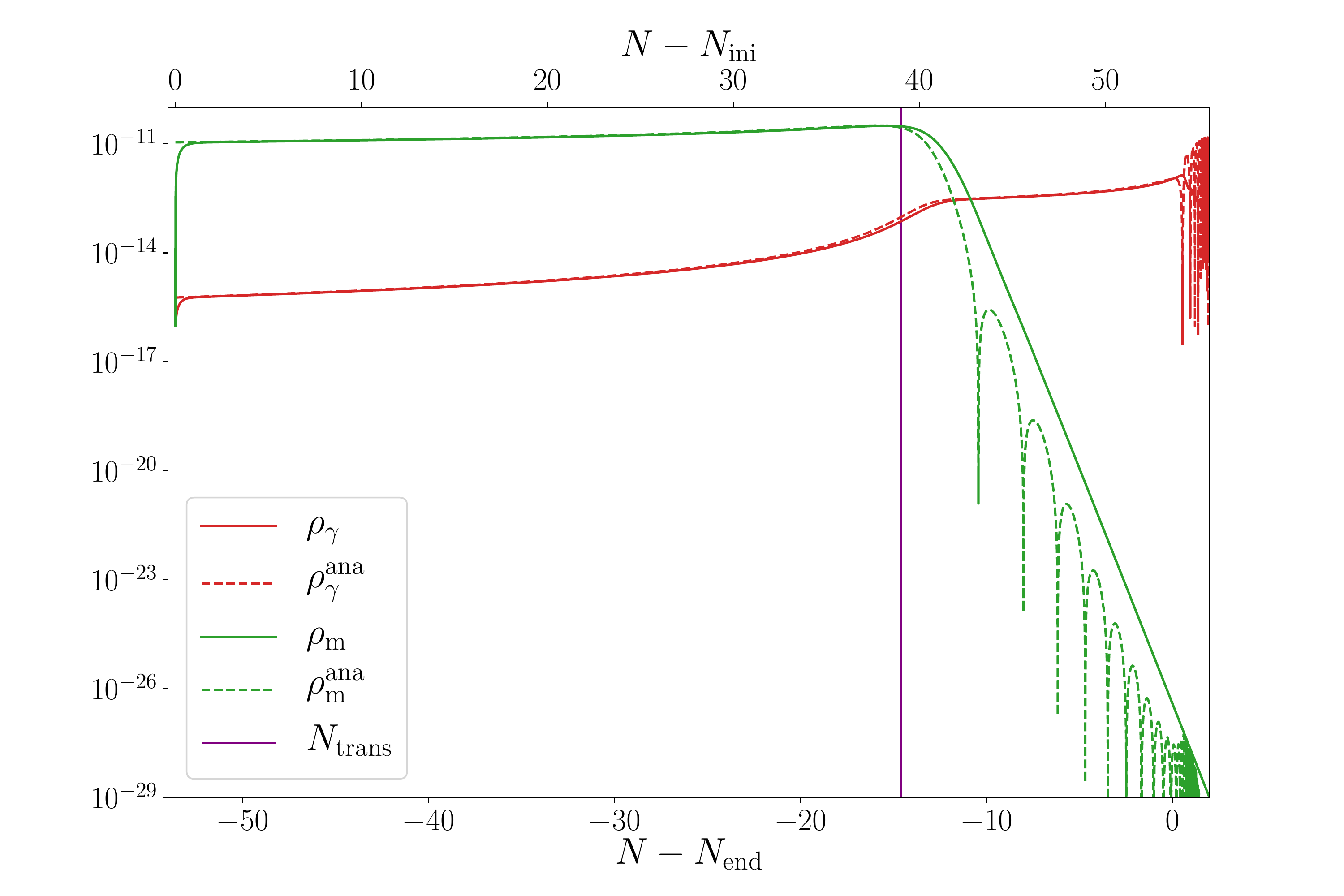}
    \caption{Behavior of the fluids' energy densities during multi-field inflation. Exact numerical results (solid lines) as well as the analytical approximations ``$\mathrm{ana}$" (dashed lines) given by Eqs.~\eqref{eq:mthermo}--\eqref{eq:radthermo}, are represented both for radiation (red) and matter (green).
    The agreement is nearly perfect until the time when the scalar field to which the fluid is coupled, leaves the slow-roll regime and begins to oscillate.
    The vertical purple line indicates the transition between the regime dominated by the heavy field and the regime dominated by the light field.}
    \label{fig:thermo}
\end{figure}

Let us now study how matter and radiation behave in the regime described above. The conservation equation for matter~(\ref{eq:rhogammam}) can be re-written
\begin{align}
    \frac{\dd}{\dd t}\left(a^3\rho_\mathrm{m}\right)=\frac{\Gamma_\mathrm{hm}}{2}a^3\dot{\phi}_\mathrm{h}^2,
\end{align}
and the solution reads
\begin{align}
\label{eq:solthermo}
    a^3(t)\rho_\mathrm{m}(t)-a^3(t_\uini)\rho_\mathrm{m}(t_\uini)
    =\frac{\Gamma_\mathrm{hm}}{2}\int _{t_\uini}^t a^3(\tau) \dot{\phi}^2_\mathrm{h}(\tau) \dd \tau. 
\end{align}
A rough approximation consists in assuming that, since the heavy field is in slow-roll, the quantity $\dot{\phi}_\mathrm{h}^2$ can be taken outside the integral. Then, using an exponential (de Sitter) scale factor, we arrive at
\begin{align}
\label{eq:mthermo}
    \rho_\mathrm{m}(t)\simeq
    \left(\frac{a_\uini}{a}\right)^3\left[\rho_\mathrm{m}(t_\uini)
    -\frac{\Gamma_\mathrm{hm}}{6H}\dot{\phi}_\mathrm{h}^2\right]
    +\frac{\Gamma_\mathrm{hm}}{6H}\dot{\phi}_\mathrm{h}^2
    \longrightarrow
    \frac{\Gamma_\mathrm{hm}}{6H}\dot{\phi}_\mathrm{h}^2\simeq 
    \frac{\Gamma_\mathrm{hm}m_\mathrm{h}^4}{54 H^3}\phi_\mathrm{h}^2(t).
\end{align}
This expression is valid as long as the heavy field is in slow-roll that is to say, roughly speaking, until the transition time. Similar considerations lead to a solution for the radiation energy density, namely
\begin{align}
\label{eq:radthermo}
    \rho_\gamma(t)\simeq 
    \left(\frac{a_\uini}{a}\right)^4\left[\rho_\gamma(t_\uini)
    -\frac{\Gamma_{\ell\gamma}}{8H}\dot{\phi}_\ell^2\right]
    +\frac{\Gamma_{\ell \gamma}}{8H}\dot{\phi}_\ell^2
    \longrightarrow
    \frac{\Gamma_\mathrm{\ell \gamma}}{8H}\dot{\phi}_\ell^2\simeq 
    \frac{\Gamma_\mathrm{\ell \gamma}m_\ell^4}{72 H^3}\phi_\ell^2(t).
\end{align}
Since the light field is in slow-roll until the end of inflation, this solution is valid until that time. The two solution~(\ref{eq:mthermo}) and~(\ref{eq:radthermo}) are represented in Fig.~\ref{fig:thermo} and compared to the correspond exact, numerical, solutions. Evidently, they match very well the numerical results.

\subsubsection{Decays of inflaton fields}
\label{subsubsec:decay}

Let us now discuss the decays of the heavy and light fields.
The decay of the heavy field happens first. Predicting, with good accuracy, the time at which it happens turns out to be a central issue in this article since we will see in the following sections that, in fact, this quantity determines the level of non-adiabatic perturbations in the post-inflationary universe. Usually, the criterion that allows us to identify when the decay starts is $H\simeq {\cal O}(1)\Gamma _{\mathrm{hm}}$ or, dimensionally, $t_\mathrm{decay,h}\simeq {\cal O}(1)/\Gamma_\mathrm{hm}$. This criterion can be quite accurate especially in the context of single-field models where $H$ evolves rapidly towards the end of inflation. However, it is interesting to notice that this is no longer true in the present context, a fact that, to our knowledge, has not been fully appreciated in the existing literature. The reason for this difficulty is that the Hubble parameter changes very slowly in this regime (recall that $H$ still obeys the slow-roll approximation since the light field is now dominant). As a consequence, a substantial error in the determination of the decay time can easily be made resulting in a large error in
the predicted level of non-adiabatic perturbations in the post-inflationary Universe [and in the value of 
$\rho_\mathrm{h}(t)$ at the end of inflation]. Therefore, it is important to track the behavior of the background during the heavy field decay with some accuracy. We now turn to this question. 

In order to further motivate the need for an accurate (hence, unfortunately, more complicated) treatment of the question, which is given in Appendix~\ref{app:decay}, we start with a treatment which is both intuitive and simple. As argued in this section, it will turn out that this approach is not precise enough in order to reach our goal, namely calculate analytically, with good accuracy, the non-adiabatic perturbations after inflation. 

Soon after the transition between the phase dominated by the heavy field and the phase dominated by the light field, the heavy field oscillates and decays. In order to study the detailed behavior of $\phi_\mathrm{h}(t)$, one writes $\phi_\mathrm{h}(t)=(a/a_\mathrm{p})^{-3/2}\phi_{\mathrm{h},\mathrm{p}} e^{-\Gamma_{\mathrm{hm}}(t-t_\mathrm{p})/4}g_\mathrm{h}(t)$, where we recall that $t_\mathrm{p}$ is a particular time and $g_\mathrm{h}(t)$ is a function defined by the previous equation. Let us also emphasize here that this particular time needs not to be chosen as before. In the present context, it should be thought as the time from which the above writing of $\phi_\mathrm{h}(t)$ is relevant, see below for more discussions about this point, which turns out to be quite important. Then, using Eq.~(\ref{eq:KGdecaylh}) the function $g_\mathrm{h}(t)$ obeys the equation
\begin{align}
\label{eq:g(t)text}
    \ddot{g}_\mathrm{h}(t)+m_\mathrm{h}^2\left(1-\frac32\epsilon_1 \frac{H^2}{m_\mathrm{h}^2}
    -\frac94 \frac{H^2}{m_\mathrm{h}^2}-\frac{1}{16}\frac{\Gamma_\mathrm{hm}^2}{m_\mathrm{h}^2}-\frac34 \frac{H}{m_\mathrm{h}}\frac{\Gamma_{\mathrm{hm}}}{m_\mathrm{h}}\right)g_\mathrm{h}(t)=0.
\end{align}
This is the equation of a parametric oscillator with its time-dependent frequency given by
\begin{align}
\label{eq:defw2text}
    \omega^2_\mathrm{h}(t)=m_\mathrm{h}^2-\frac32\epsilon_1 H^2
    -\frac94 H^2-\frac{1}{16}\Gamma_\mathrm{hm}^2-\frac34 H\Gamma_{\mathrm{hm}}.\
\end{align}
Unfortunately, the exact time dependence of $\omega ^2_\mathrm{h}(t)$ is too complicated to permit an explicit integration of Eq.~(\ref{eq:g(t)text}). A plot of $\omega^2_\mathrm{h}$ is given in Appendix~\ref{app:decay}, see Fig.~\ref{fig:omegas}. One of the main feature of this plot is to show that $\omega_\mathrm{h}^2$ is negative during the heavy field slow-roll regime. In fact, it remains negative until a time $t_\mathrm{osc}$ such that $\omega_\mathrm{h}^2(t_\mathrm{osc})=0$ (which is, therefore, a turning point) after which $\omega^2_\mathrm{h}>0$ and it is only in this regime that $\phi_\mathrm{h}(t)$ starts oscillating. It is also important to notice that the transition 
time lies in the region where $\omega_\mathrm{h}^2<0$, that is to say $t_\mathrm{trans}<t_\mathrm{osc}$.

During the decay of the heavy field, the background is still in slow-roll and, therefore, the Hubble parameter evolves slowly. As a first approximation, we can thus assume that $\omega ^2_\mathrm{h}$ remains constant in this regime. Then, it is straightforward to write the solution of Eq.~(\ref{eq:g(t)text}) which reads $g_\mathrm{h}(t)=\cos\left[\omega_\mathrm{h}(t-t_\mathrm{p})\right]+B\sin\left[\omega_\mathrm{h}(t-t_\mathrm{p})\right]$, where $B$ is an integration constant. Notice that $g_\mathrm{h}(t_\mathrm{p})=1$ (as appropriate given the definition of $g_\mathrm{h}$) and $B=\dot{g}_\mathrm{h}(t_\mathrm{p})/\omega_\mathrm{h}$. From the above considerations, it is clear that $t_\mathrm{p}$ must be chosen such that $t_\mathrm{p}>t_\mathrm{osc}$ (and, as a consequence, $t_\mathrm{p}$ cannot be chosen to be $t_\mathrm{trans}$). For times $t\gg t_\mathrm{p}$, $\phi_\mathrm{h}(t)$ has undergone several oscillations and we can calculate the time average of the solution (denoted by $\langle \cdot\rangle$ in the following). This leads to
\begin{align}
\label{eq:meanphi2}
    \left\langle \phi_\mathrm{h}^2\right \rangle = \phi^2_\mathrm{p}\frac12 \left(1+B^2
    \right)\left(\frac{a_\mathrm{p}}{a}\right)^{3}
    e^{-\Gamma_\mathrm{hm}(t-t_\mathrm{p})/2},
\end{align}
and 
\begin{align}
\label{eq:meandotphi2}
    \left\langle \dot{\phi}_\mathrm{h}^2\right \rangle 
    =m_\mathrm{h}^2\phi^2_\mathrm{p}\frac12 \left(1+B^2
    \right)\left(\frac{a_\mathrm{p}}{a}\right)^{3}
    e^{-\Gamma_\mathrm{hm}(t-t_\mathrm{p})/2}.
\end{align}
We see that the two above equations imply equipartition between kinetic and potential energy. Notice that the term $\propto \epsilon_1 H^2/m_\mathrm{h}^2$ has been neglected in order to be consistent with the assumption that the Hubble parameter is constant. This gives an expression for the heavy field energy density, $\rho_\mathrm{h}\simeq m_\mathrm{h}^2\left\langle \phi_\mathrm{h}^2\right \rangle $.

The expression of the field derivative~(\ref{eq:meandotphi2}) could also be used to calculate the evolution of the matter energy density, see Eq.~(\ref{eq:solthermo}). Indeed, contrary to the slow-roll case, the evolution of the heavy field is no longer slow and, as a consequence, the term $\dot{\phi}_\mathrm{h}^2$ cannot be taken outside the integral. However, using the above analytical formula~(\ref{eq:meandotphi2}), it can be calculated explicitly. Indeed, inserting Eq.~(\ref{eq:meandotphi2}) in Eq.~(\ref{eq:solthermo}), one arrives at
\begin{align}
\label{eq:rhomnaive}
\rho_\mathrm{m}(t)=m_\mathrm{h}^2\phi_\mathrm{p}^2\frac12 (1+B^2)
    \left(\frac{a_\mathrm{p}}{a}\right)^3
    \left[1-e^{-\Gamma_\mathrm{hm}(t-t_\mathrm{p})/2}\right] + \rho_\mathrm{m}^\mathrm{p} \left(\frac{a_\mathrm{p}}{a}\right)^3 \, . 
\end{align}
Pushing the solution~(\ref{eq:mthermo}) up to $t_\mathrm{p}$, we have $\rho_\mathrm{m}^\mathrm{p} \sim \Gamma_\mathrm{hm} m_\mathrm{h}^4 \phi_\mathrm{p}^2 /(54 H^3)$, hence the second term is negligible as soon as $1- e^{- \Gamma_\mathrm{hm} (t-t_p)/2} >  \Gamma_\mathrm{hm} m_\mathrm{h}^2 /\left[27(1+B^2) H^3\right]$ which is very quickly true, roughly $\Gamma_\mathrm{hm}(t-t_\mathrm{p}) > 0.05$. Therefore, the second term can be neglected and, following the considerations presented above, we have  thus obtained simple expressions for $\rho_\mathrm{h}(t)$ and $\rho_\mathrm{m}(t)$.

Then, the time of decay of the heavy field, $t_\mathrm{decay,h}$ is defined by the condition $\rho_\mathrm{h}(t_\mathrm{decay,h})=\rho_\mathrm{m}(t_\mathrm{decay,h})$ and, using the above expressions, we arrive at
\begin{align}
\label{eq:tdecayln2}
    t_\mathrm{decay,h}-t_\mathrm{p}&=\frac{2}{\Gamma_\mathrm{hm}}\ln 2.
\end{align}
We see it has the form predicted above (namely, inversely proportional to the decay rate) with the factor ${\cal O}(1)$ being simply $2\ln 2\simeq 1.39$.

Finally, if one wants to express the time of decay not in terms of cosmic time but in terms of the number of e-folds, one can use the fact that the oscillatory behavior of the heavy field takes place during the phase dominated by the light field during which $H^2=2m_\ell^2(N_\mathrm{end}-N)/3$, see Eq.~(\ref{eq:hubblelightphase}). As a consequence, given that $\dd N=H\dd t$, one finds 
\begin{align}
\label{eq:linktN}
    t-t_\mathrm{p}=-\frac{\sqrt{6}}{m_\ell}\left[\left(N_\mathrm{end}-N\right)^{1/2}-\left(N_\mathrm{end}-N_\mathrm{p}\right)^{1/2}\right],
\end{align}
which allows us to relate $t$ and $N$. This relation is a simplified version of the exact~(\ref{eq:thypergeo}). It follows that the number of efolds at which the heavy field decay occurs would be given by the expression
\begin{align}
\label{eq:Ndecay}
    N_\uend-N_\mathrm{decay,h}=\left[\left(N_\uend-N_\mathrm{p}\right)^{1/2}-\frac{m_\ell}{\Gamma_\mathrm{hm}}\frac{2}{\sqrt{6}}\ln 2\right]^2.
\end{align}

\begin{figure}
    \centering    
    \includegraphics[width=0.75\linewidth]{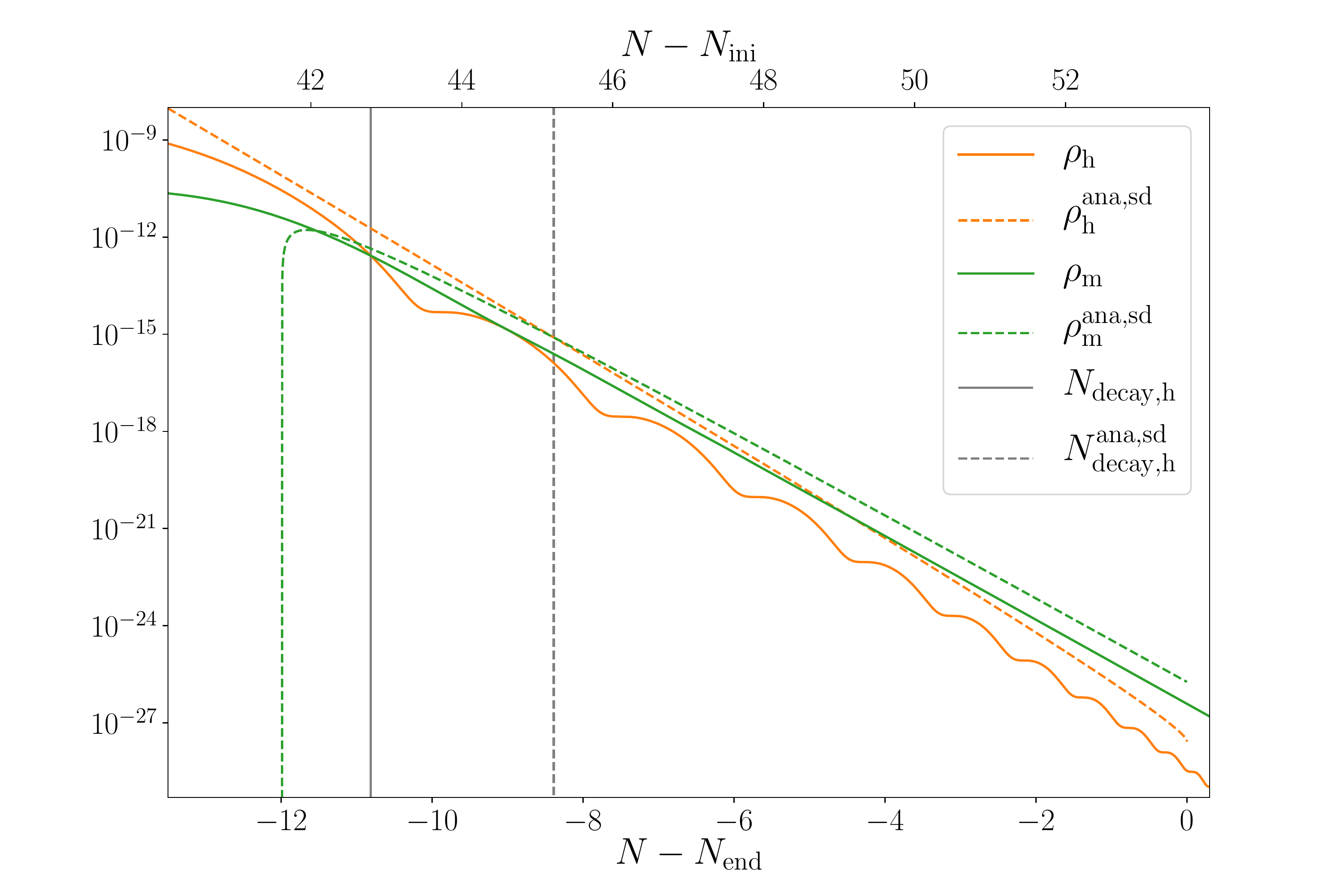}
    \caption{Behavior of $\rho_\mathrm{h}$ and $\rho_\mathrm{m}$ during the decay of the heavy field. Here we show both their exact numerical values (respectively the orange and green solid lines), and their analytical formulas inferred from the simple decay approximation presented in this section, ``$\mathrm{ana,sd}$" (respectively the orange and green dashed lines) as given by Eqs.~\eqref{eq:meanphi2}--\eqref{eq:meandotphi2} and Eq.~\eqref{eq:rhomnaive}.
    Moreover the time of decay of the heavy field, defined as the time when $\rho_\mathrm{h}=\rho_\mathrm{m}$ is shown with the vertical gray lines both from the numerical approach (solid line, $N_\mathrm{decay,h}=-10.81$) and the simple decay analytical approximation (dashed line, $N_\mathrm{decay,h}^\mathrm{ana,sd}=-8.39$).
    Clearly, this simple formalism is not sufficient to encapsulate the physics of the decay of the heavy field: a more rigorous approach is proposed in the Appendix~\ref{app:decay}, see in particular Fig.~\ref{fig: rho decay WKB}.
    }
    \label{fig:rhonaive}
\end{figure}

At this stage, we would need to choose the particular time $t_\mathrm{p}$. As already mentioned, this cannot be $t_\mathrm{trans}$ because it lies in the region where $\omega_\mathrm{h}^2<0$. It cannot be $t_\mathrm{osc}$ either since $\omega(t_\mathrm{osc})=0$ and the solution for $g_\mathrm{h}(t)$ in terms of trigonometric functions presented before becomes ill-defined. In fact, the trigonometric solution for $g_\mathrm{h}(t)$ introduced before is nothing but the WKB solution of Eq.~(\ref{eq:g(t)text}) in the regime where  $\omega_\mathrm{h}^2$ is constant and positive. The WKB approximation becomes valid if $\dot{\omega}_\mathrm{h}/(2\omega_\mathrm{h}^2)<1$. Therefore, a natural definition of $t_\mathrm{p}$ would be the time such that $\dot{\omega}_\mathrm{h}=2{\cal C}\omega_\mathrm{h}^2$ where ${\cal C}\leq 1$ is a constant which quantifies how restrictive the WKB criterion is chosen (namely, for instance, ${\cal C}=1$ or $0.1$, \dots). This time will be noted $t_\mathrm{p}=t_\mathrm{wkb}$ in the following. In the vicinity of the turning point (since we expect $t_\mathrm{p}=t_\mathrm{wkb}$ to be close to $t_\mathrm{osc}$) a good approximation for the time-dependent frequency is $\omega_\mathrm{h}^2\simeq m_\mathrm{h}^2-9H^2/4$, which implies that $\dot \omega=9 \epsilon_1 H^3/(4\omega)$. As a consequence $t_\mathrm{wkb}$ (or rather $\theta _\mathrm{wkb}$) is a solution of the following equation
\begin{align}
    \frac{H^2}{m_\mathrm{h}^2}=\left[\frac94+\left(
    \frac{9\epsilon_1}{8{\cal C}}\right)^{2/3}\right]^{-1},
\end{align}
where $H$ is given by Eq.~(\ref{eq:Hubblegene}) [with $s$  defined by Eq.~(\ref{eq:sfunctiontheta})] and $\epsilon_1$ by Eq.~(\ref{eq:eps1srappendix}).
These considerations allow us to calculate the quantities appearing in the WKB solution for $g_\mathrm{h}(t)$ (namely, $\omega_\mathrm{wkb}$, $B$, \dots)  and to plot $\rho_\mathrm{h}=m_\mathrm{h}^2\left\langle \phi_\mathrm{h}^2\right \rangle $, see Eq.~(\ref{eq:meanphi2}) and $\rho_\mathrm{m}$, see Eq.~(\ref{eq:rhomnaive}). They are represented and compared to the exact, numerical, solutions in Fig.~\ref{fig:rhonaive}. This lead to $N_\mathrm{decay,h}\simeq -8.39$ to be compared with the exact result $N_\mathrm{decay,h}\simeq -10.81$. 

Unfortunately, we notice that this result is not precise enough to predict with good accuracy the behaviors of $\rho_\mathrm{h}(t)$ and $\rho_\mathrm{m}(t)$ and, therefore, the final level of non-adiabatic perturbations after the end of inflation. Therefore, if one wants an analytical result for $N_\mathrm{decay,h}$, one needs better approximations for the behavior of $\rho_\mathrm{h}(t)$ and $\rho_\mathrm{m}(t)$ around the time of decay of the heavy field. Such a calculation (which is more involved at the technical level) is carried out in great details in  Appendix~\ref{app:decay}.

Finally, as can also be observed in Fig.~\ref{fig:exactfield}, the light field starts to oscillate and decays when $H\simeq \Gamma_{\ell \gamma}$, that is to say
\begin{align}
 s_{\ell,\mathrm{decay}}=   -N_{\ell,\mathrm{decay}}+N_\mathrm{end}\simeq \frac32{\cal O}(1)
\frac{\Gamma_{\ell \gamma}^2}{m_\ell^2}.
\end{align}
Numerically, for the parameters considered here, this gives $N_{\ell,\mathrm{decay}}\simeq N_\mathrm{end}$, namely around the end of inflation in agreement with Fig.~\ref{fig:exactfield}. This estimate is better than the estimate of the heavy field decay time because, around the end of inflation, the Hubble parameter evolves much more abruptly than when the heavy field starts oscillating. This also leads to an estimate of the reheating temperature of this model. Given that, just after the end of inflation, the universe is dominated by radiation, one can write $3\Mp^2H^2_{\ell,\mathrm{decay}}\simeq 3\Mp^2\Gamma_{\ell \gamma}^2=\pi^2g_*T_\mathrm{rh}^4/30$ or
\begin{align}
    g_*^{1/4}T_\mathrm{rh}\simeq \left(\frac{90}{\pi^2}\right)^{1/4}\sqrt{\Gamma_{\ell \gamma }\Mp},
\end{align}
where $g_*$ is the number of relativistic degrees of freedom just after the fields decay. With our fiducial parameters, one finds $g_*^{1/4}T_\mathrm{rh}\simeq 0.0023 \Mp\simeq 5.53\times 10^{15}\, \mbox{GeV}$.

\begin{figure}
    \centering
    \begin{subfigure}{0.48\textwidth}
        \centering
        \includegraphics[width=1.\linewidth]{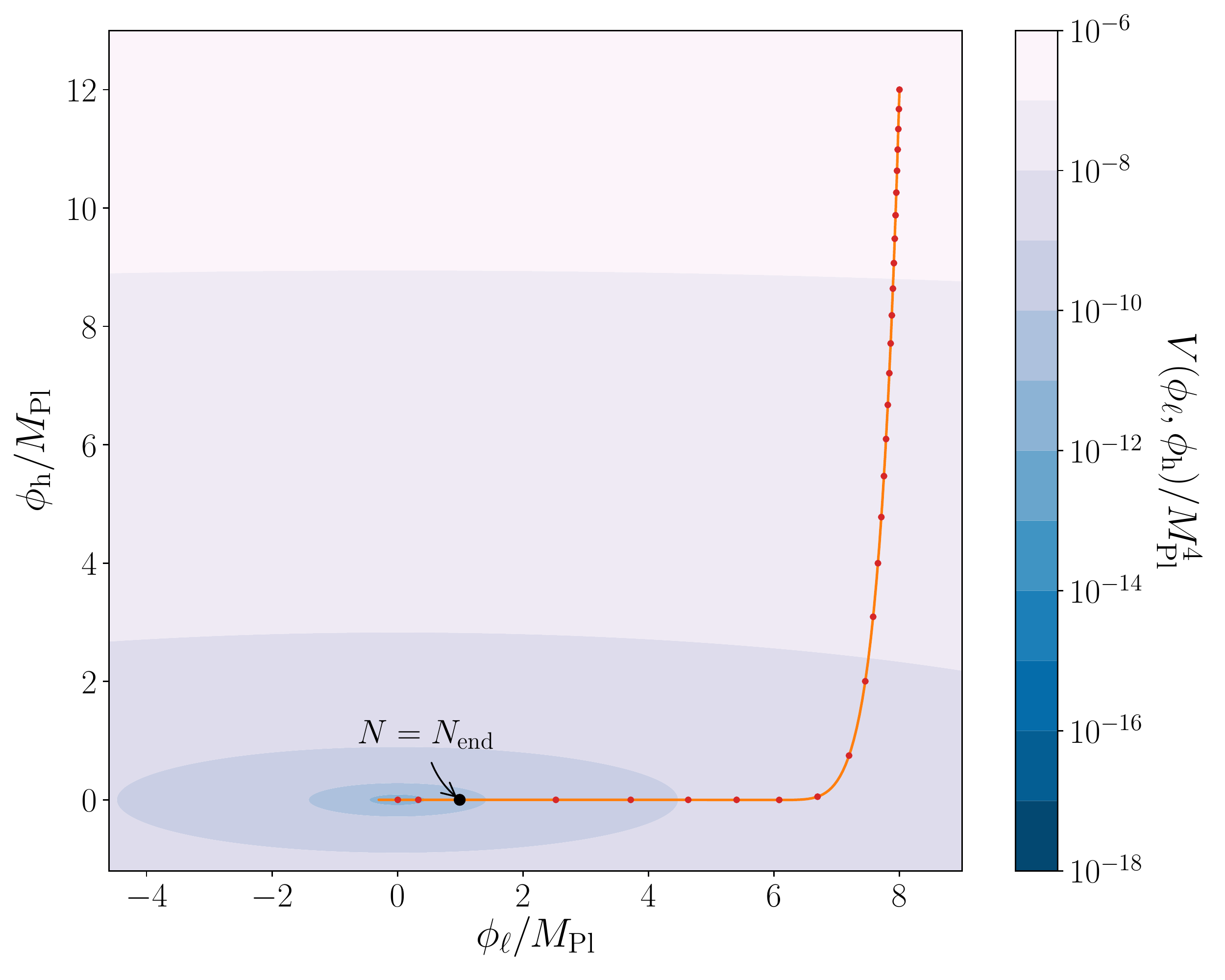}
        \caption{Linear scaling for the whole trajectory: the field-space turn is clearly visible.}
    \end{subfigure}%
    \hfill
    \begin{subfigure}{0.48\textwidth}
        \centering
        \includegraphics[width=1.\linewidth]{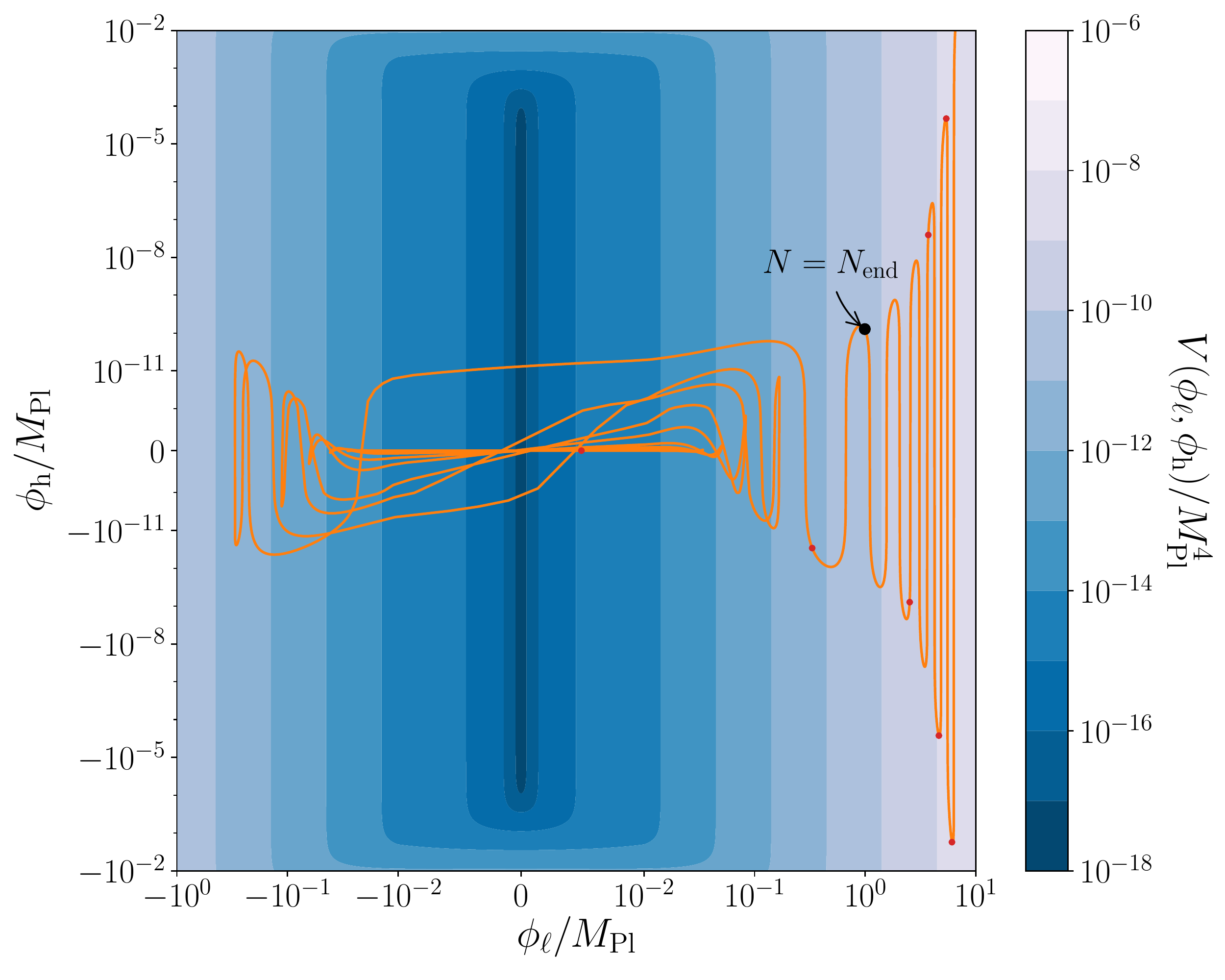}
        \caption{Logarithmic scaling for the last $e$-folds of inflation: the fields' oscillations are resolved.}
    \end{subfigure}
    \caption{Field-space $\left(\phi_\ell,\phi_\mathrm{h}\right)$ background trajectory from the numerical evolution (orange line) superimposed on the two-field potential contour levels.
    The instantaneous direction of the trajectory is aligned with the potential gradient and is therefore effectively single-field like, except around the turn which is a truly multi-field feature.
    Red dots along the trajectory are equally spaced by $\Delta N=2$ $e$-folds of expansion and the black dot denotes the end of inflation.}
    \label{fig:2dfield}
\end{figure}

\subsubsection{Evolution in field space}
\label{subsubsec:fieldspace}

In Fig.~\ref{fig:2dfield}, we have again represented the trajectory of the 
system but, now, in the two-dimensional space $\left(\phi_\ell, \phi_\mathrm{h}\right)$ superimposed on the potential contour levels. On the left panel, we have focused on the slow-roll regime and the two phases, dominated respectively by the heavy and light fields, are clearly visible. The sharp turn separating these two epochs is also easy to identify. On the right panel, we have zoomed in on the vicinity of the minimum of the potential in order to see the oscillations of the fields occurring after the end of inflation. In those two panels, the ``non-isotropic" character of the contour levels is of course a consequence of the fact that the two fields do not have the same mass.

\begin{figure}
    \centering
    \begin{subfigure}{0.48\textwidth}
        \centering
        \includegraphics[width=1\linewidth]{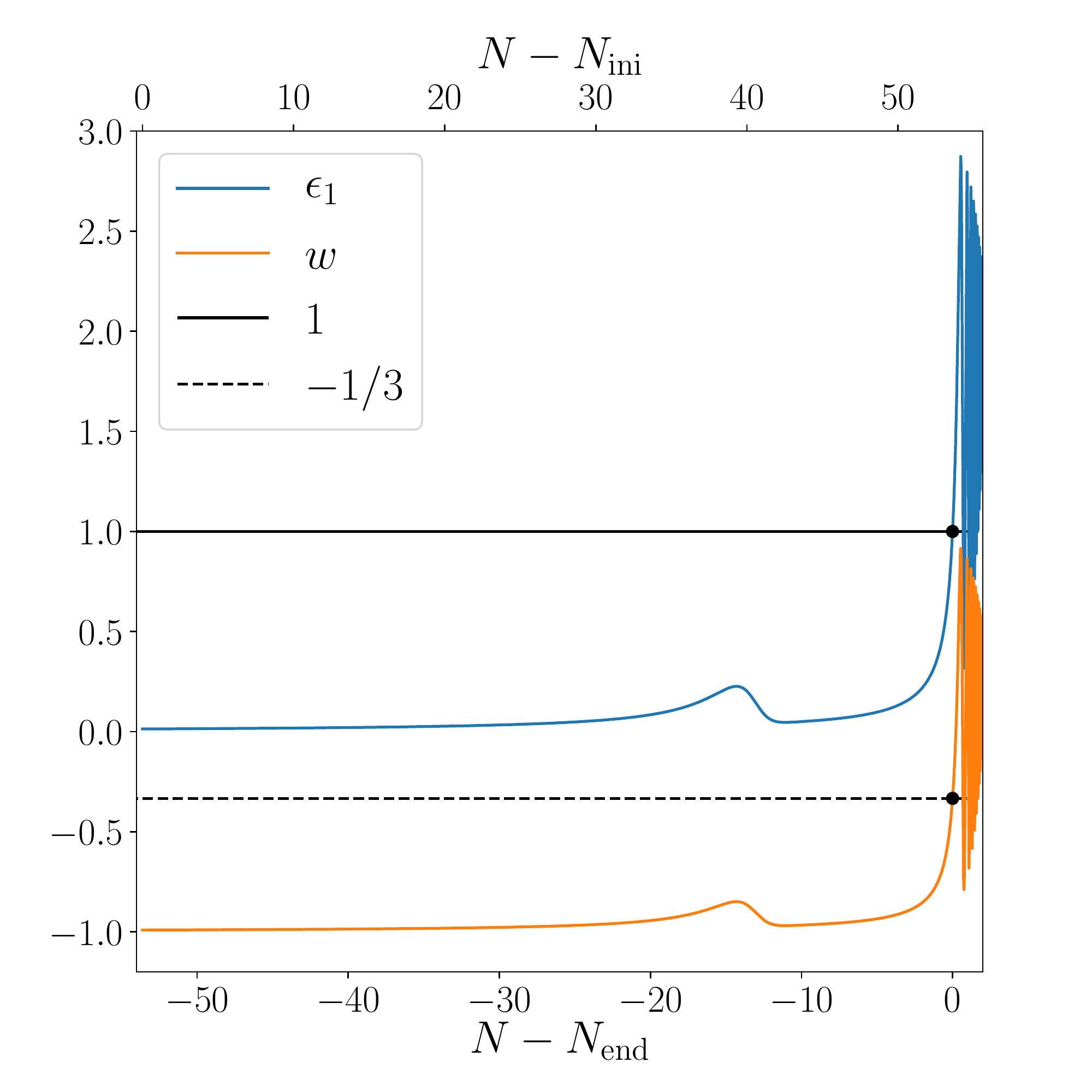}
        \caption{Set of parameters presented in the main \linebreak text and used in all other figures.}
    \end{subfigure}
    \hfill
    \begin{subfigure}{0.48\textwidth}
        \centering
        \includegraphics[width=1\linewidth]{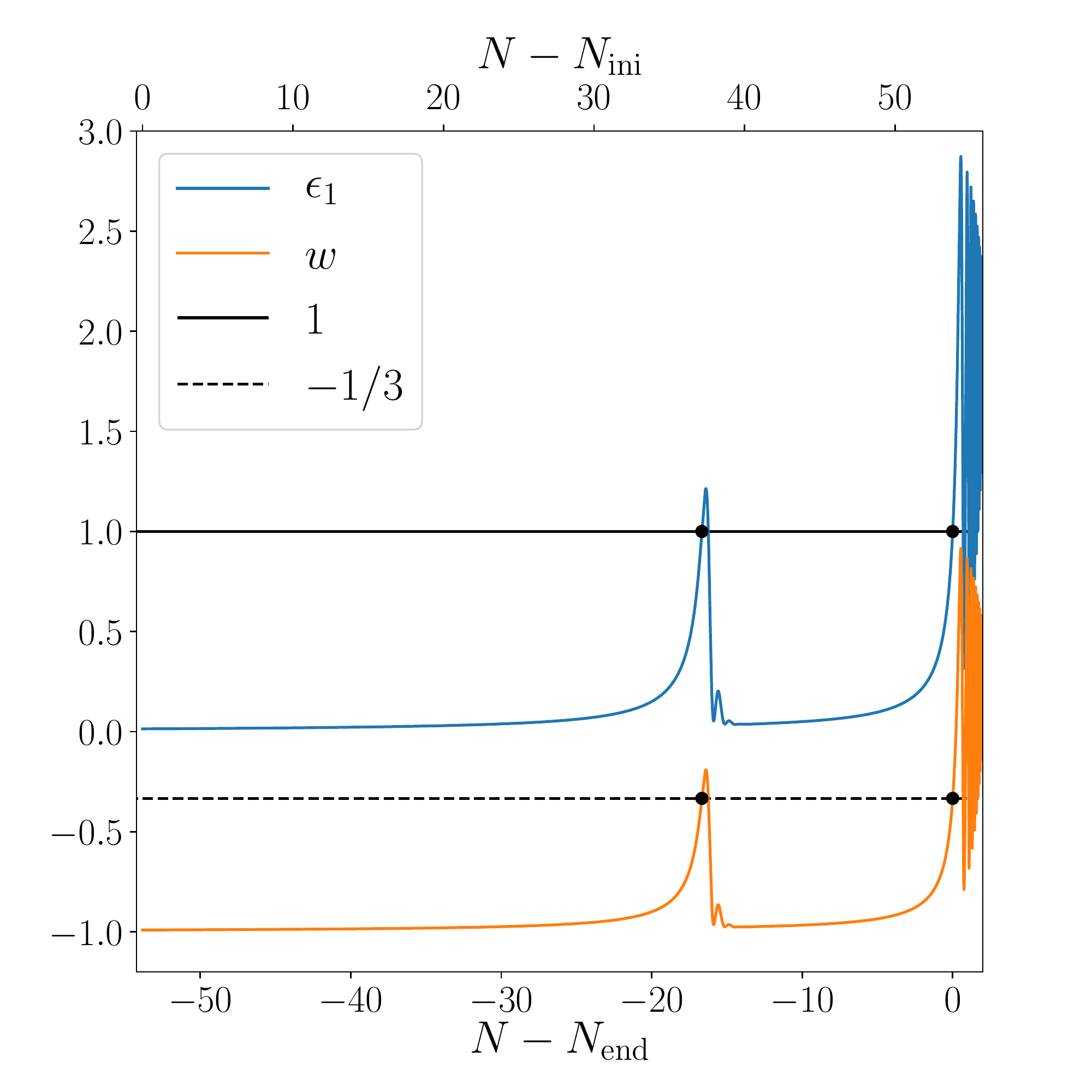}
        \caption{Same set of parameters, except for the ratio of masses taken to be $R=15$ here.}
    \end{subfigure}
    \caption{First Hubble-flow parameter $\epsilon_1=-\dot{H}/H^2$ and total equation of state of the universe $w=p/\rho$, for two sets of parameters differing only by the value of $R$. The solid (respectively dashed) black line represents the maximal value for $\epsilon_1$ (respectively $w$) in order for inflation to proceed: $1$ (respectively $-1/3$). Black dots represent times at which inflation is stopped, either temporarily or permanently. For $R=5$ (left panel), inflation slows down but does not stop when the heavy field reaches the minimum of its potential around $N-N_\mathrm{end} \simeq -14.6$, while for $R=15$ (right panel) its oscillations are so strong that inflation is transiently violated.
    In this work, we choose to focus on a region of parameter space such that inflation is not transiently violated, and keep $R=5$ in our fiducial model.
    At the end of inflation, the fast oscillations of the light field at the bottom of its quadratic potential around $N \simeq N_\mathrm{end}$, are responsible for an equation of state $w$ which averages to zero over a few oscillations, the Universe thus behaving, on average and on large scales, effectively as if it was dominated by a pressure-less fluid.}
    \label{fig:eps1}
\end{figure}

In Fig.~\ref{fig:eps1}, we have represented the evolution of the first Hubble-flow parameter (blue line) and the equation-of-state parameter, namely the pressure to energy density ratio (orange line), during inflation. The left panel corresponds to the fiducial parameters already used in Fig.~\ref{fig:exactfield} (and used throughout this article).
We see that it represents a situation where inflation is never interrupted. The time at which the heavy field (or, rather, the energy density associated to this field) becomes subdominant manifests itself as a small bump around $N-N_\mathrm{end} \simeq -14.6$ (already identified as a small dropout in the evolution of the Hubble parameter in Fig.~\ref{fig:exactfield}) but we notice that $\epsilon_1$ never reaches one (horizontal solid black line) and $w$ is never greater than $-1/3$ (horizontal dashed black line).
The end of inflation occurs when the light field starts to oscillate and is represented by the black dot.
The right panel represents the evolution of the same quantities with the same initial conditions for the fields and the same values of the parameters except that, now, $R=15$.
In that case, the physical situation is very different from the one depicted on the left panel. In particular, around $N-N_\mathrm{end} \simeq -16.4$ (``first" black dot), we notice that inflation temporarily comes to an end at the transition between the heavy and light phases.
Indeed, $\epsilon_1$ becomes larger than one and $w$ becomes larger than $-1/3$.
This can be understood as follows. At the transition, $R\phi_\mathrm{h}=\phi_\ell$ and, therefore, from Eq.~\eqref{eq:eps1field}, $\epsilon_1\simeq \Mp^2(1+R^2)/(2\phi_\ell^2)$. Using the crude approximation that the light field stays constant during the heavy phase, this can also be written as $\epsilon_1\simeq \Mp^2(1+R^2)/(2\phi_\ell\vert_\uini^2)$. As a consequence, inflation stops at the transition only if $R\gtrsim  \sqrt{2} \phi_\ell\vert_\uini/\Mp$, which gives in our case $R\gtrsim 11.3$ (exploring the parameter space numerically, we rather find this threshold value to be $R\gtrsim 12.9$). Therefore, the left panel of Fig.~\ref{fig:eps1} ($R=5$) corresponds to a case where inflation does not stop since $R$ is slightly below the threshold. On the contrary, the situation on the right panel corresponds to $R=15$ and one expects a transient violation of inflation which is what is observed.

We see that, despite its \textit{a priori} simplicity, the solutions for the fields in the double inflation model possess a rich variety of different behaviors. In order to simplify the discussion, in the following, we will always assume that inflation never stops before both fields have reached the bottom of their potential. In practice, as already mentioned, we will always work with the fiducial parameters corresponding to Fig.~\ref{fig:exactfield}.

\subsubsection{Evolution of the fractional energy densities}
\label{subsubsec:ratio}

\begin{figure}
    \centering
    \includegraphics[width=0.75\linewidth]{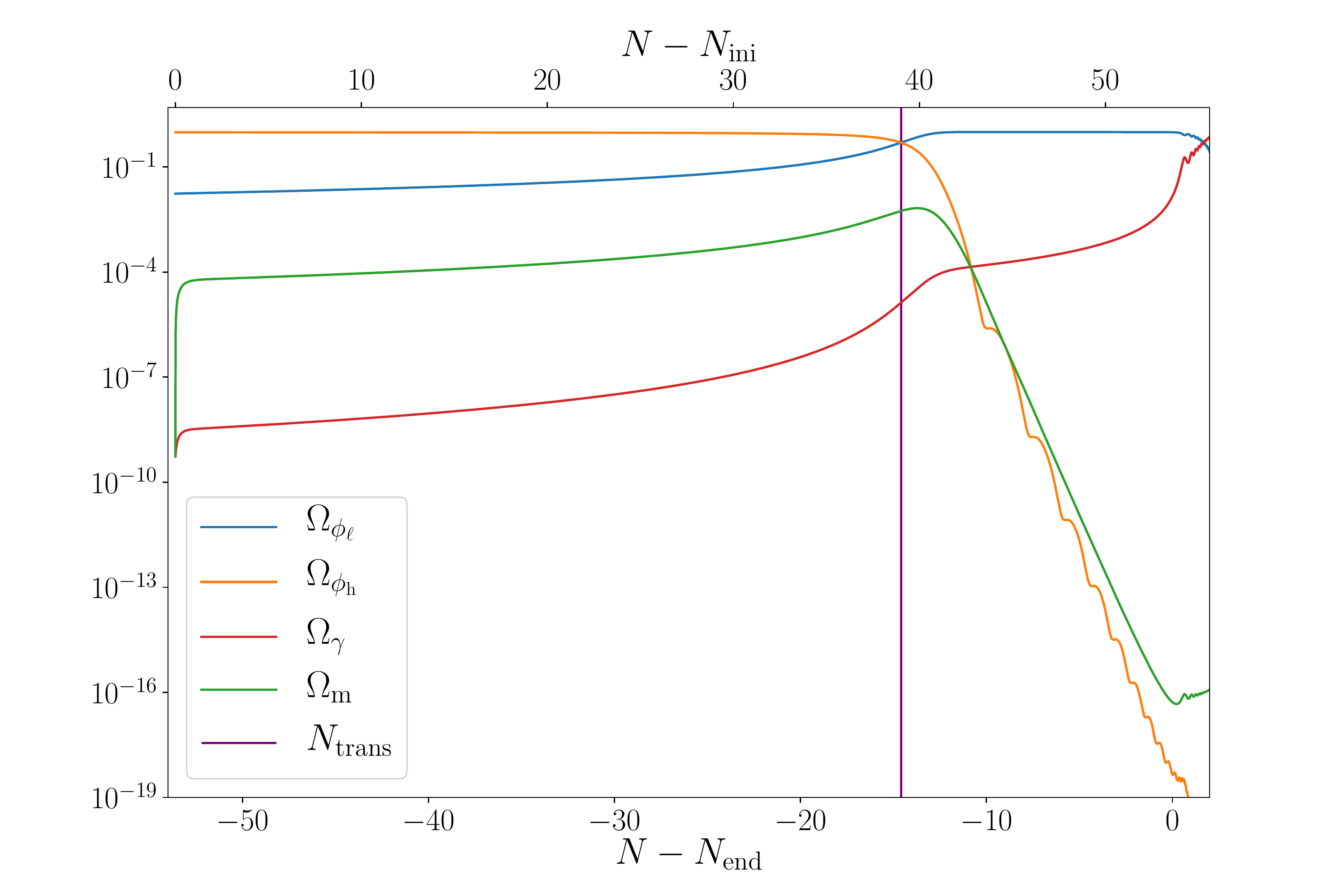}
    \caption{Numerical evolution of the energy budget in the universe during inflation.
    The four fundamental components of matter are represented: the light scalar field (blue line) and the heavy one (orange line), as well as the radiation fluid (red line) and the pressure-less, matter one (green line).
    The scalar field that is driving inflation is the one that is dominating the energy budget (by definition), and the contributions from the cosmological fluids are overall negligible but not astonishingly small.
    The vertical purple line indicates the transition between the regime dominated by the heavy field and the regime dominated by the light field, a time around which the energy from the heavy field is transferred to the matter fluid.
    Inflation stops at $N=N_\mathrm{end}$ after a total of $53.61$ $e$-folds of expansion from the initial time of the simulation, and the energy from the light scalar fields completes its transition to the radiation fluid, thus setting the stage for the radiation-dominated era.
    }
    \label{fig:omegainf}
\end{figure}

Yet another way to understand the evolution of the system is to plot the quantities $\Omega_i\equiv \rho_i/\rho_\mathrm{tot}$ where $\rho_i$ is the energy density of the fluid ``$i$" and $\rho_\mathrm{tot}$ is the total energy density. By definition, one has $\sum_i\Omega _i=1$.
The corresponding plot is presented in Fig.~\ref{fig:omegainf} where $\Omega_{\phi_\ell}$ corresponds to the solid blue line, $\Omega_{\phi_\mathrm{h}}$ to the solid orange line, $\Omega _\gamma $ (radiation) to the solid red line and, finally, $\Omega_\mathrm{m}$ (pressure-less matter) to the solid green line. Let us emphasize that 
it is possible to plot $\Omega_{\phi_\ell}$ and $\Omega_{\phi_\mathrm{h}}$ because, in the particular case of double inflation, it is possible to define separately a fluid associated with the heavy and light fields since the potential is separable. As already mentioned when Fig.~\ref{fig:exactfield} was described, at the beginning of inflation, the energy density associated to the heavy field dominates the energy budget of the universe. At $N_\mathrm{trans}-N_\mathrm{end} \simeq -14.6$, estimated in Eq.~(\ref{eq:strans}), the contribution of the light field takes over and the blue and orange lines intersect. At this time, the decay of the heavy field is apparent as the quantity $\Omega _{\phi_\mathrm{h}}$ sharply drops out (while oscillating). Then, the phase dominated by the light field starts until inflation comes to an end. During inflation, we also remark that the contributions originating from radiation and/or pressure-less matter remain negligible except, of course, when the end of the inflationary phase is approached.

\begin{figure}
    \centering
    \includegraphics[width=0.75\linewidth]{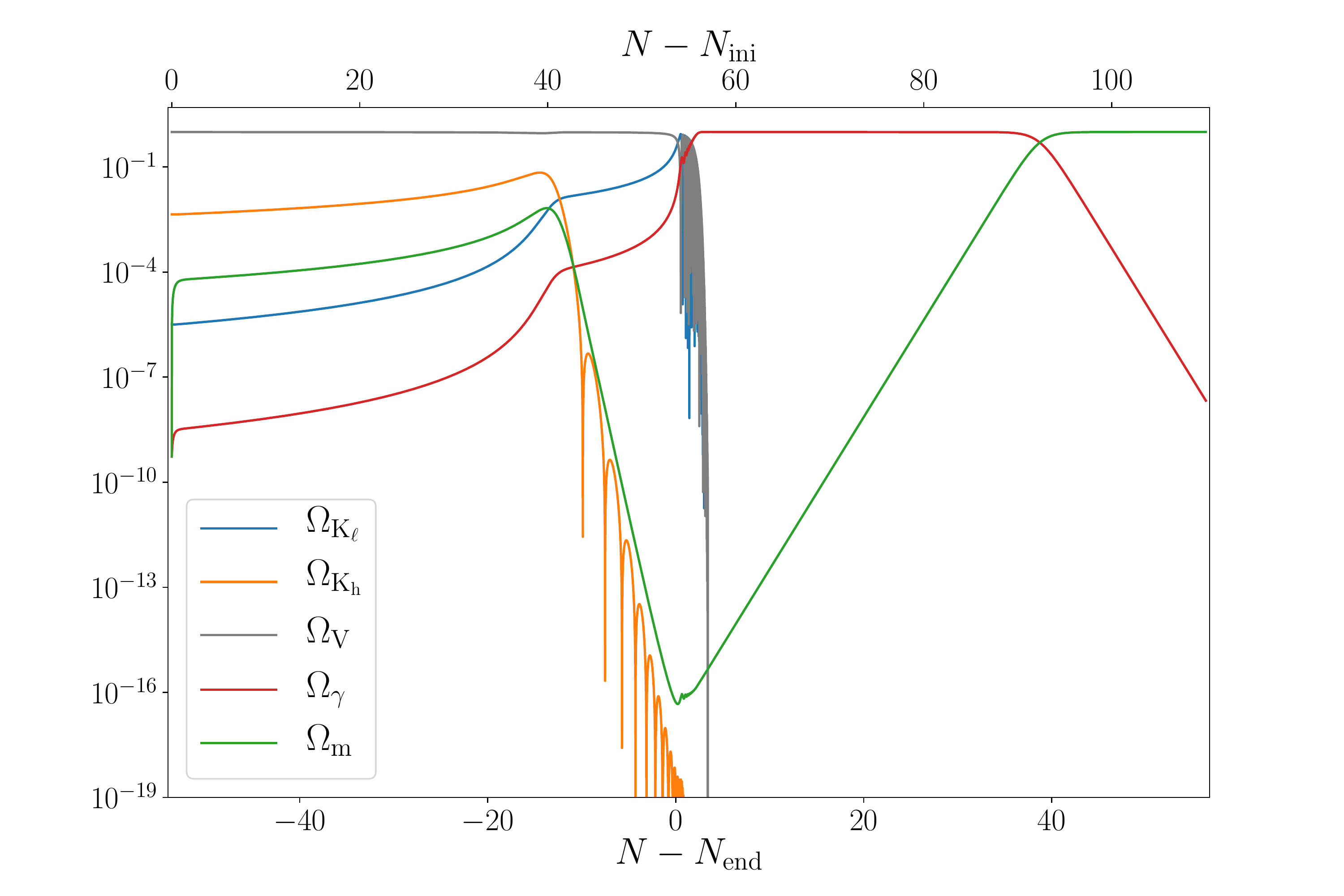}
    \caption{Energy budget $\Omega_{(\alpha)}=\rho_{(\alpha)}/\rho$ between the ``fictitious" and ``real" fluids in the universe during inflation, reheating, radiation domination and matter domination.
    During inflation, the potential fluid is always dominating, and the second most important contribution comes from the kinetic fluid corresponding to the scalar field that is driving inflation at that time.
    As already mentioned, the contributions from the cosmological fluids are then negligible but not extremely small.
    Clearly, when the heavy field reaches the minimum of its potential, it begins to oscillate and redshift, and completes its transfer of energy to the matter fluid (however the latter quickly begins to decay a bit less rapidly than $a^{-3}$ because it is then only sustained by the exponentially decreasing kinetic energy of the heavy field).
    At the end of inflation, both the kinetic energy of the light scalar field and the potential energy oscillate with opposite phases but equal amplitudes, and the universe therefore behaves on average and on large scales as in a matter-dominated epoch.
    This large kinetic energy of the light scalar field is efficiently transferred to the radiation fluid, whose contribution rapidly grows during reheating.
    Around $3$ $e$-folds after the end of inflation, reheating is complete and it is the onset of the radiation-dominated era, which lasts until radiation has so much redshifted (as $a^{-4}$) that the small remaining quantity of matter at the end of reheating (that redshifts slower as $a^{-3}$) eventually dominates, thus setting the stage for the matter-dominated era.}
    \label{fig:postinflation}
\end{figure}

Having described how inflation proceeds, let us now extend our
analysis to the post-inflationary universe. In
Fig.~\ref{fig:postinflation}, we have plotted the evolution of the
fictitious ``kinetic fluids" associated to the light field,
$\Omega_{\mathrm{K}_\ell}$ (solid blue line), to the heavy field,
$\Omega_{\mathrm{K}_\mathrm{h}}$ (solid orange line) and to the
``potential fluid", $\Omega_\mathrm{V}$ (solid gray line). We have
also represented the contribution of the ``real fluids", namely
radiation (solid red line) and pressure-less matter (solid green
line).  As expected, during inflation, we see that the ``potential
fluid" dominates over all the other components. At
$N=N_\mathrm{decay,h}$, the heavy field decays and one sees
$\Omega_{\mathrm{K}_\mathrm{h}}$ sharply dropping as
$\rho_{\mathrm{K}_\mathrm{h}}\propto
a^{-3}e^{-\Gamma_\mathrm{hm}t/2}$, see the considerations around
Eqs.~(\ref{eq:g(t)text}) and~(\ref{eq:defw2text}). Starting from the
same time, the pressure-less matter that has been produced by the
heavy field decay behaves as $\Omega_\mathrm{m}\propto a^{-3}$ (the
orange curve envelope has, initially, namely before the exponential
term in $\rho_{\mathrm{K}_\mathrm{h}}$ takes over, the same slope as
the solid green line). At the end of inflation, all fluids associated
to the scalar fields quickly become subdominant and radiation becomes
the main component in the universe, $\Omega _\gamma \simeq 1$. Then,
during the subsequent radiation-dominated epoch, pressure-less matter
is subdominant but its relative contribution grows as
$\Omega_\mathrm{m}\propto a$ given that $\rho_\gamma\propto 1/a^4$ and
$\rho_\mathrm{m}\propto 1/a^3$. Eventually, around
$N-N_\mathrm{end}\simeq 40$ $e$-folds after the end of inflation,
pressure-less matter becomes dominant and the matter-dominated era
starts. Equality occurs at the temperature
\begin{align}
T_\mathrm{eq}\simeq 2.43 \times 10^{18} \left(\frac{90}{\pi^2}\right)^{1/2}
\left(\frac{\Gamma_{\ell \gamma}}{\Mp}\right)^{1/2}
\Omega_\mathrm{m}\vert_\mathrm{rh} \, \mbox{GeV}.
\end{align}
This gives $T_\mathrm{eq}\simeq 0.73 \, \mbox{GeV}$, where 
we have used $\Omega_\mathrm{m}\vert_\mathrm{rh}\simeq 10^{-16}$ as can be checked in Fig.~\ref{fig:postinflation}. Notice that 
we do not use an analytical approximation for the quantity $\Omega_\mathrm{m}\vert _\mathrm{rh}$ due to the remarks made before, namely the fact that the decay of the heavy field is difficult to predict with good accuracy in the approximate framework used here (although it could be attempted using considerations presented in App.~\ref{app:decay}). The temperature obtained before is a temperature higher than the BBN scale. This means that the parameters chosen before are not very realistic and, in order to be realistic, one should change them, for instance lower the value of $\Gamma_{\ell \gamma }$. However, in the following, we will nevertheless continue to work with the fiducial parameters for various reasons. First, considering smaller values of the decay rates can introduce severe numerical problems and, second, since we have analytical estimates that are well-verified in the regime where numerical calculations are available, it would be easy to derive useful predictions in situations of physical interest simply by using these estimates. Here, our main goal is not to build a fully realistic multi-field scenario but to investigate and test the accuracy of the tools that have been developed to study these models.

\begin{figure}
    \centering
    \includegraphics[width=0.85\linewidth]{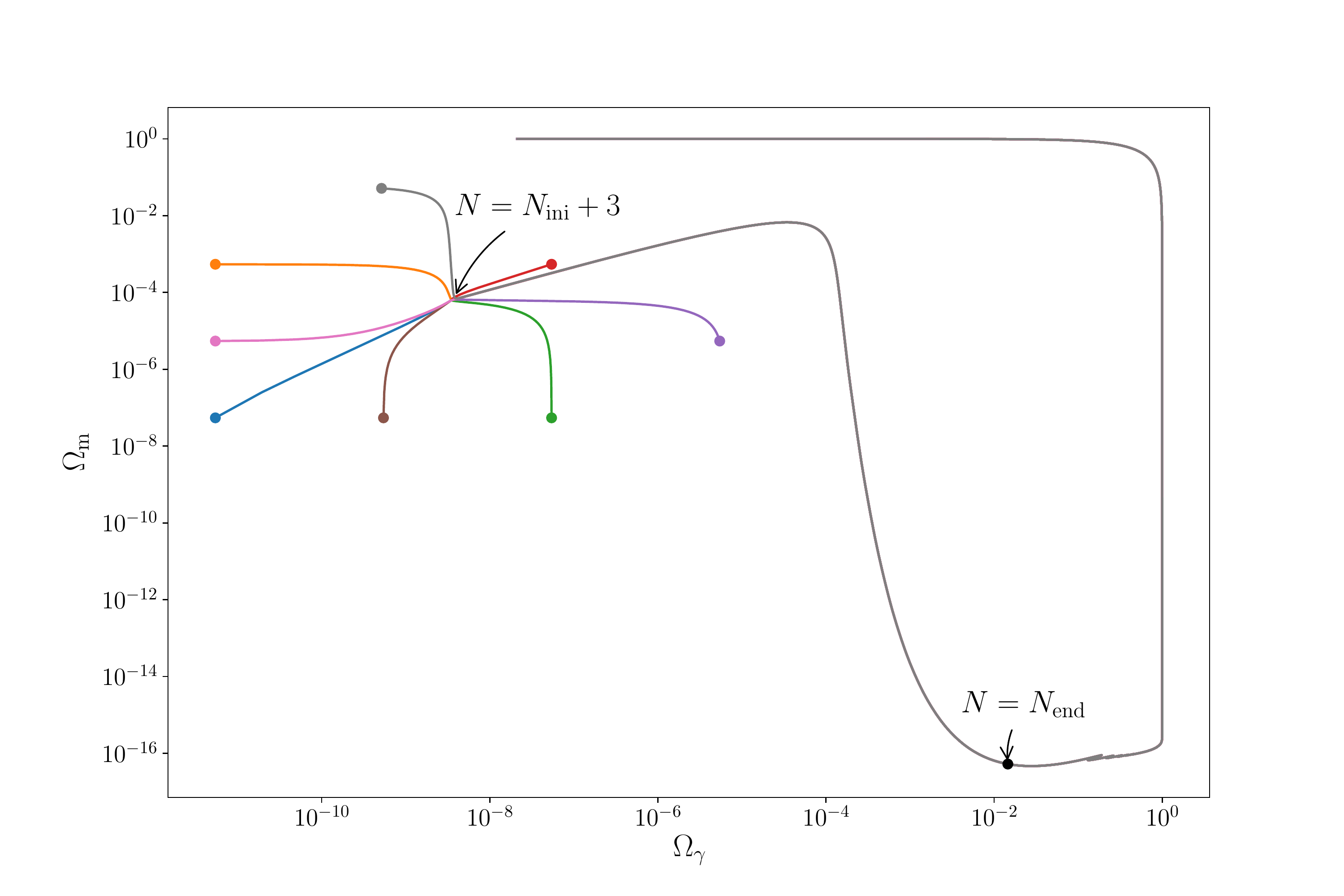}
    \caption{This figure shows the independence of the dynamics of the cosmological fluids on their initial conditions, as found with numerical simulations.
    Different colors correspond to different initial conditions for $\rho_\gamma$ and $\rho_\mathrm{m}$ and the colorful dots indicate these initial conditions.
    During inflation, the energy densities of the fluids clearly reach an attractor dynamics along which they are negligible, independently on their initial conditions, as can be understood with the investigation of Eqs.~\eqref{eq:mthermo}--\eqref{eq:radthermo}.
    The first turn in this $\left(\Omega_\gamma,\Omega_\mathrm{m}\right)$ two-dimensional space corresponds to the decay of the heavy scalar field, during which the contribution of the matter fluid is rapidly decaying, and the second turn coincides with the end of inflation, denoted by the black dot.
    Afterward, the radiation fluid dominates the energy budget until the time of equality, from which the matter fluid is dominating.}
    \label{fig:attractor_fluids}
\end{figure}

The last figure of this section is Fig.~\ref{fig:attractor_fluids} in which we have represented 
the ``trajectories" of the system in the two-dimensional space $(\Omega_\gamma,\Omega_\mathrm{m})$ for different initial conditions. The main conclusion is that the late time behavior of $\Omega_\gamma$ and $\Omega_\mathrm{m}$ is not sensitive to the choice of these initial conditions. Indeed, after $\simeq 3 \, \mbox{e-folds}$, the system joins an attractor as can be seen in Fig.~\ref{fig:attractor_fluids}. Therefore, we have established that our results do not depend on what we assume about the initial energy densities stored in radiation and pressure-less matter.

\subsection{Solutions for the perturbations}
\label{subsec:pertdouble}

\subsubsection{Adiabatic and non-adiabatic modes during inflation}
\label{subsubsec:adiaandnonadia}

\begin{figure}
    \centering
    \includegraphics[width=0.75\linewidth]{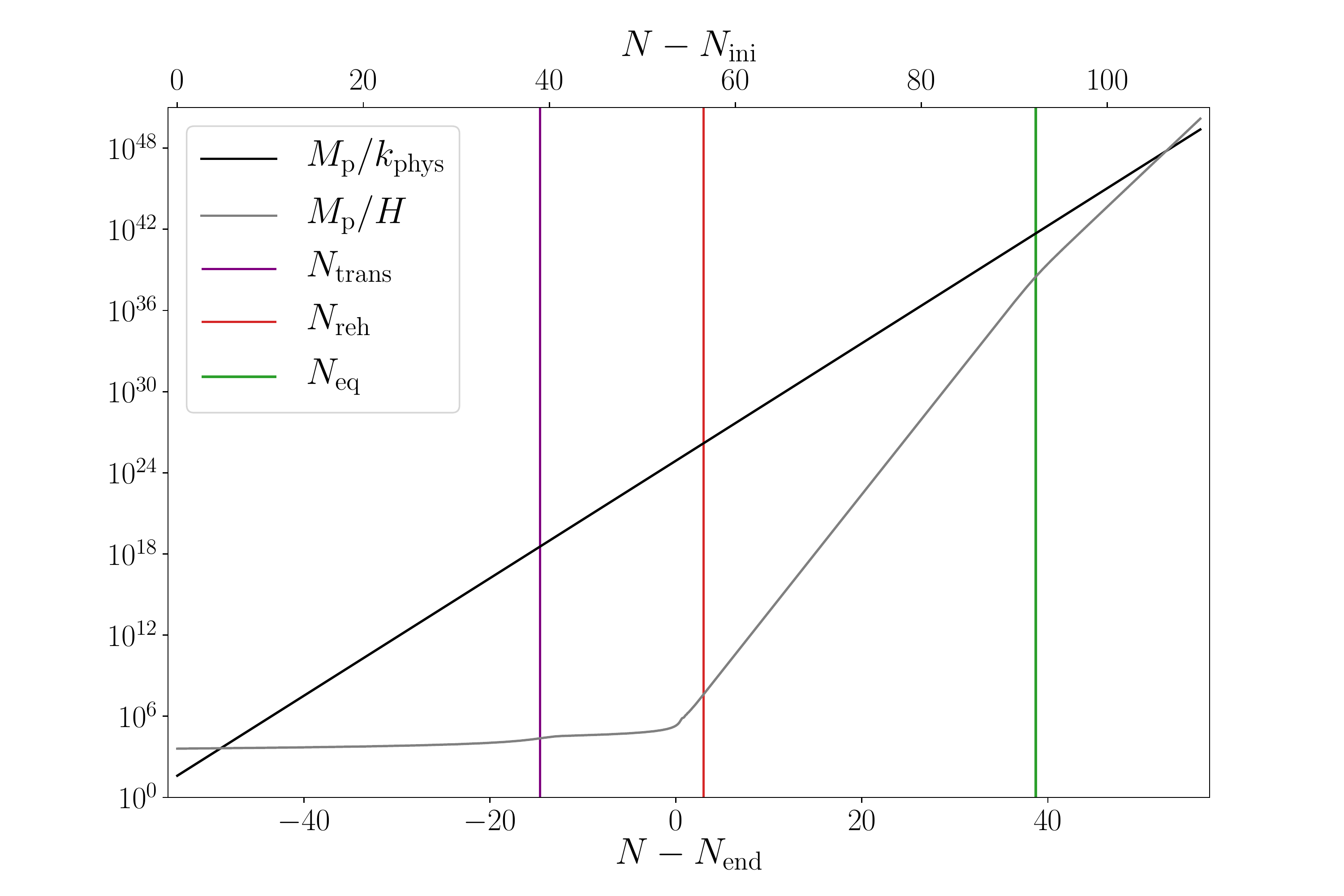}
    \caption{
    Cosmic history for the representative physical scale $k_\mathrm{phys}^{-1}$ (black line) corresponding to the mode $k_\mathrm{phys}=k/a$ with $k=100 \times a_\mathrm{ini}H_\mathrm{ini} \simeq 0.025 M_\mathrm{Pl}$.
    This physical wavelength is stretched from sub-Hubble scales where it behaves as in the Bunch-Davies vacuum, to super-horizon ones (the transition happens when the physical scale $k_\mathrm{phys}^{-1}=H^{-1}$, where $H^{-1}$ is the Hubble radius and is represented in gray), where crucially non-trivial dynamics may happen due to the multi-species nature of the model at hand.
    Then it re-enters the horizon of our observable Universe during the matter era.
    Particular times of interest are represented with vertical lines, such as the transition between the two regimes of inflation (purple line), the end of reheating and beginning of the radiation-dominated era (red line), and the time of equality from which matter is dominating (green line).
    }
    \label{fig: mode history}
\end{figure}

In this section, we investigate the solutions for the perturbations during the phase of inflation.
Here, and for the remainder of this work, we focus on a single wave-number $k$, chosen such that $k=100 \times a_\mathrm{ini}H_\mathrm{ini} \simeq 0.025 M_\mathrm{Pl}$, which therefore features approximately $4.6$ $e$-folds of sub-Hubble evolution before crossing the Hubble radius.
Actually, our numerical simulation enables to follow its evolution until it re-enters the Hubble radius in the matter-dominated era (though remember that the radiation-matter equality is not realistic in our fiducial model).
The cosmic history of this mode, stretched by inflation to super-horizon scales and then re-entering our observable Universe, is represented in Fig.~\ref{fig: mode history}.

We neglect, at least for the moment, the contributions of matter and radiation. As a consequence, the perturbed quantities that remain to be studied are the Bardeen potential and the two (heavy and light) field fluctuations. The equations of motion of these quantities are given by Eqs.~(\ref{eq:perEinstein-A}), (\ref{eq:perKGl-A})  and~(\ref{eq:perKGh-A}). During inflation, on large scales, we can also use the slow-roll approximation which transforms the equations of motion, which are second order differential equations, into first order differential equations. These considerations lead to simplified versions of Eqs.~(\ref{eq:perEinstein-A}), (\ref{eq:perKGl-A}) and~(\ref{eq:perKGh-A}) which, on large scales, can be expressed as
\begin{align}
\label{eq:eqBardeensr}
    \Phi_A&=\frac{1}{2\Mp^2H}\left(\dot{\phi}_{\ell}\delta \phi_{\ell,A}
    +\dot{\phi}_\mathrm{h}\delta \phi_{\mathrm{h},A}\right),
    \\
    \label{eq:eqdphiellsr}
     3H\dot{\delta \phi}_{\ell,A}+\frac{\partial^2 V_{\ell}}{\partial \phi_{\ell}^2}\delta \phi_{\ell,A}+2\frac{\partial V_{\ell}}{\partial \phi_{\ell}}\Phi_A&=0,
     \\
     \label{eq:eqdphihsr}
    3H\dot{\delta \phi}_{\mathrm{h},A}+\frac{\partial^2 V_{\mathrm{h}}}{\partial \phi_{\mathrm{h}}^2}\delta \phi_{\mathrm{h},A}+2\frac{\partial V_{\mathrm{h}}}{\partial \phi_{\mathrm{h}}}\Phi_A&=0.
\end{align}
Notice that, for the moment, we have not used the specific 
form of the potential (namely the fact that it is quadratic). The only property utilized above is that the potential is separable.

Then, it turns out that this system of equations can be solved exactly. The solution for the Bardeen potential 
reads~\cite{Polarski:1992dq}
\begin{align}
  \Phi_A &=-C_{1,A} \frac{\dot H}{H^2}-H\frac{{\rm d}}{{\rm d}t}
  \left(\frac{d_{\ell,A} V_\ell+d_{\mathrm{h},A}V_\mathrm{h}}{V}\right)
\\ 
\label{eq:solbardeensr}
& =C_{1,A}\epsilon_1+\frac{(d_{\ell,A}-d_{\mathrm{h},A})}
  {3(V_\ell+V_\mathrm{h})^2}
  \left[\left(\frac{\partial V_\ell}{\partial \phi_\ell}\right)^2
  V_\mathrm{h}
    -\left(\frac{\partial V_\mathrm{h}}{\partial \phi_\mathrm{h}}\right)^2V_\ell
    \right]  ,
  \end{align}
where we recall that $\epsilon_1$ is the first Hubble flow parameter, while the solutions for the field fluctuations take the form
\begin{align}
\label{eq:soldphilsr}
    \frac{\delta \phi_{\ell,A}}{\dot{\phi}_\ell}&=
    \frac{C_{1,A}}{H}-2H(d_{\mathrm{h},A}-d_{\ell,A})
    \frac{V_\mathrm{h}}{V_\ell+V_\mathrm{h}}, 
    \\
    \label{eq:soldphihsr}
    \frac{\delta \phi_{\mathrm{h},A}}{\dot{\phi}_\mathrm{h}}&=
    \frac{C_{1,A}}{H}+2H(d_{\mathrm{h},A}
    -d_{\ell,A})\frac{V_\ell}{V_\ell+V_\mathrm{h}}.
\end{align}
In these expressions, $C_{1,A}$, $d_{\mathrm{h},A}$ and $d_{\ell,A}$ are integration constants.

Following Ref.~\cite{Choi:2008et}, it is interesting to split the above solutions in adiabatic and non-adiabatic modes. Concretely, for the adiabatic components, we define
\begin{align}
\label{eq:adiasolfield}
  \Phi^\mathrm{ad}_A&= C_{1,A}\epsilon_1,
  \quad
  \delta \phi_{\ell,A} ^\mathrm{ad}=\frac{C_{1,A}}{H}\dot{\phi}_\ell,
  \quad
  \delta \phi_{\mathrm{h},A} ^\mathrm{ad}=\frac{C_{1,A}}{H}\dot{\phi}_\mathrm{h}.
\end{align}
The justification for these definitions is as follows. One can use these solutions to calculate the corresponding curvature perturbations. This means that we consider the standard expression for the curvature perturbation in terms of the Bardeen potential and the field fluctuations but when the concrete forms of $\Phi_A$ and $\delta \phi_{\mathrm{h},\ell,A}$ are used in this expression, only the adiabatic modes are taken into account. Explicitly, we have
\begin{align}
\label{eq:zetaadia}
  \zeta^\mathrm{ad}_{\ell,A}=-\Phi^\mathrm{ad}_A
  -H\frac{\delta \rho_{\ell,A}^\mathrm{ad}}{\dot{\rho}_\ell}
  \simeq -C_{1,A},
\end{align}
where $\delta \rho_{\ell,A}^\mathrm{ad}=\dot{\phi}_\ell\delta \dot{\phi}_{\ell,A}^\mathrm{ad}
  -\dot{\phi}_\ell^2\Phi^\mathrm{ad}_A+(\partial V_\ell/\partial \phi_\ell)
  \delta \phi_{\ell,A}^\mathrm{ad}$. Of course, one also has
$\zeta_{\mathrm{h},A}^\mathrm{ad}\simeq -C_{1,A}$ since the same calculation holds for
the heavy field. This implies that
$\zeta_{\mathrm{h},A}^\mathrm{ad}-\zeta_{\ell,A}^\mathrm{ad}=0$, which, therefore, indeed, corresponds to an adiabatic perturbation, justifying the split (and the notation) introduced before.

In a similar way, the non-adiabatic components of the full solution can be defined by the following formulas
\begin{align}
\Phi^\mathrm{nad}_A&=\frac{(d_{\ell,A}-d_{\mathrm{h},A})}
  {3(V_\ell+V_\mathrm{h})^2}
  \left[\left(\frac{\partial V_\ell}{\partial \phi_\ell}\right)^2
  V_\mathrm{h}
    -\left(\frac{\partial V_\mathrm{h}}{\partial \phi_\mathrm{h}}\right)^2V_\ell
    \right],
 \\
   \delta \phi_{\ell,A} ^\mathrm{nad}&=-2H\dot{\phi}_\ell(d_{\mathrm{h},A}-d_{\ell,A})
  \frac{V_\mathrm{h}}{V_\ell+V_\mathrm{h}},
  \\
\delta \phi_{\mathrm{h},A} ^\mathrm{nad}&=-2H\dot{\phi}_\mathrm{h}(d_{\ell,A}-d_{\mathrm{h},A})\frac{V_\ell}{V_\ell+V_\mathrm{h}}.
\end{align}
Using these results, one can establish the corresponding expressions for curvature perturbations. For the light field, one arrives at  
\begin{align}
\label{eq:zeta_hlA^nad1}
  \zeta_{\ell,A}^\mathrm{nad}&=-\frac{2}{9(V_\ell+V_\mathrm{h})^2}
  (d_{\mathrm{h},A}-d_{\ell,A})
  \left[\left(\frac{\partial V_\mathrm{h}}{\partial \phi_\mathrm{h}}\right)^2
    V_\ell-\left(\frac{\partial V_\ell}{\partial \phi_\ell}\right)^2
    V_\mathrm{h}\right]
\nonumber \\ &
  +2H^2(d_{\mathrm{h},A}-d_{\ell,A})\frac{V_\mathrm{h}}{V_\ell+V_\mathrm{h}},
\end{align}
and a similar equation for the quantity $\zeta_{\mathrm{h},A}^\mathrm{nad}$, where the indices $\ell$ and $\mathrm{h}$ are permuted. It follows that $\zeta_{\mathrm{h},A}^\mathrm{nad}-\zeta_{\ell,A}^\mathrm{nad}\neq 0$ which justifies why these branches of the full solutions are called non-adiabatic. In fact, using the above expressions, it is straightforward to show that
\begin{align}
\label{eq:entropylh}
  S_{\ell \mathrm{h},A}=3\left(\zeta_{\ell,A}^\mathrm{nad}-
  \zeta_{\mathrm{h},A}^\mathrm{nad}\right)
=6C_{3,A}H^2,
\end{align}
where we have defined the constant $C_{3,A}$ by $C_{3,A}=d_{\mathrm{h},A}-d_{\ell,A}$. We emphasize again that the above considerations are valid for any potential provided this one is separable.

\subsubsection{Fixing the initial conditions}
\label{subsubsec:icdouble}

\begin{figure}
    \centering
    \begin{subfigure}{\textwidth}
        \centering
        \includegraphics[width=1.\linewidth]{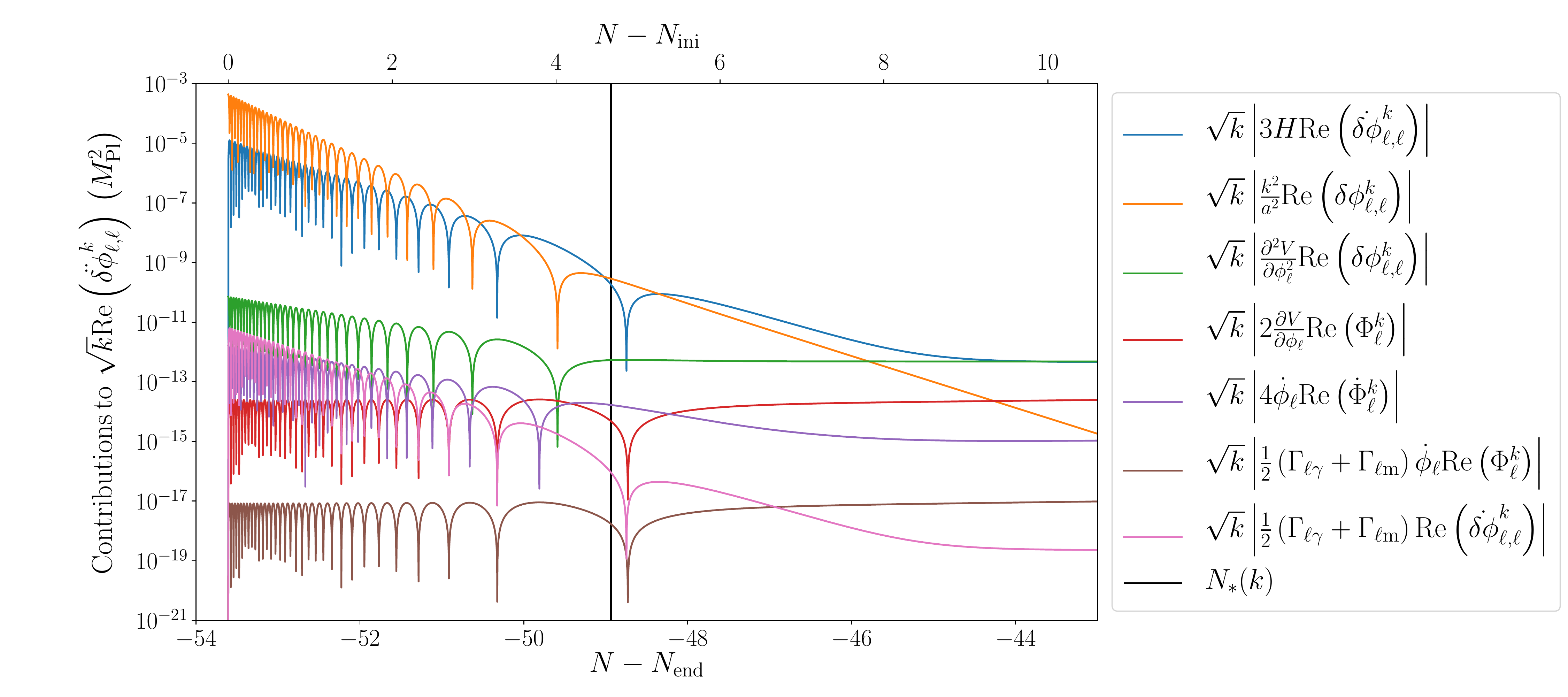}
        \caption{Light scalar field. The Bardeen contributions (red, purple and brown lines) remain sub-dominant for the times represented here.
        This is because the light field is not driving inflation and therefore the component of the Bardeen along the oscillator $A=\ell$, $\Phi_\ell$, is negligible until the transition time $N_\mathrm{trans}-N_\mathrm{end} = -14.6 $ which happens dozens of $e$-folds after Hubble crossing.}
    \end{subfigure}%
    \\
        \centering
    \begin{subfigure}{\textwidth}
        \centering
        \includegraphics[width=1.\linewidth]{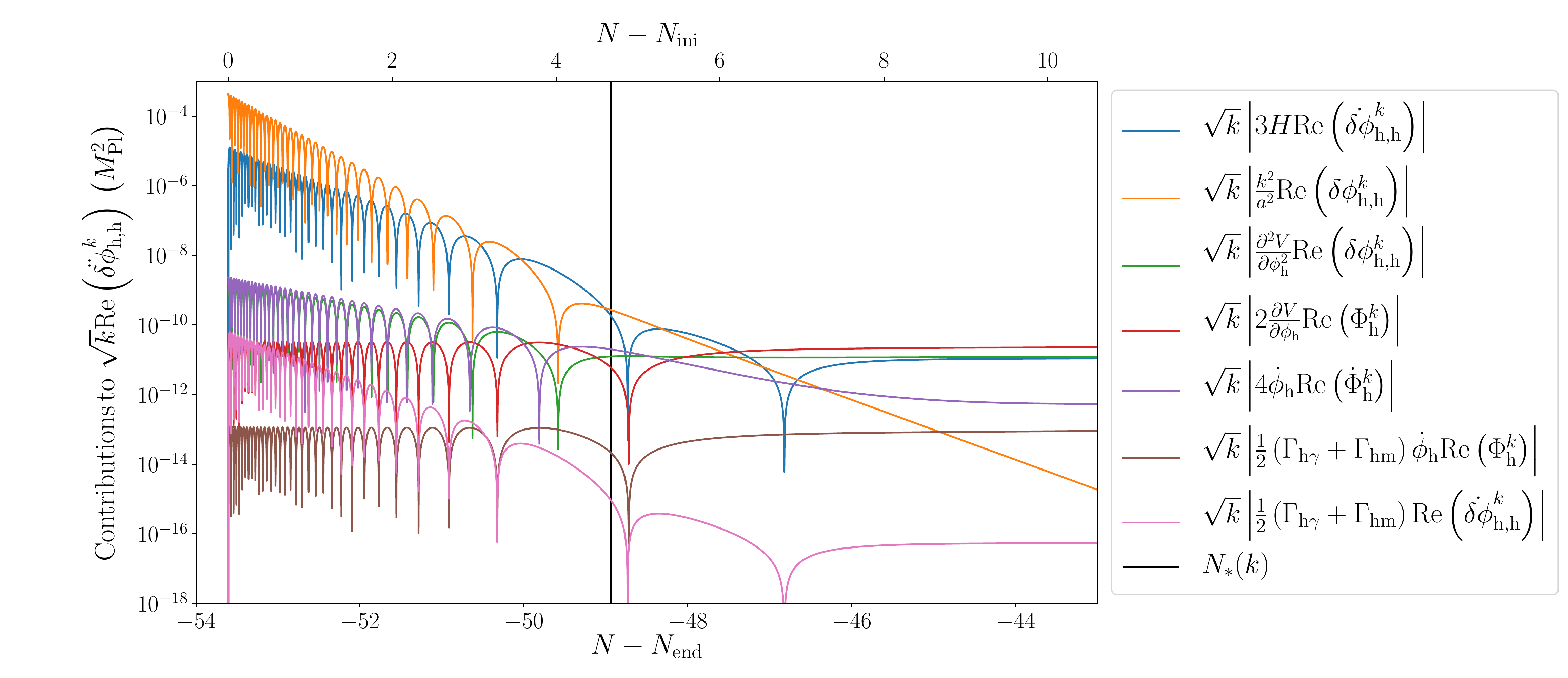}
        \caption{Heavy scalar field. Because the heavy scalar field is driving inflation, the Bardeen contributions can matter.
        On sub-Hubble scales, both the Bardeen contributions $\propto m_2^2$ and to $\dot{\phi}_\mathrm{h}$ (red and purple lines) are small but not completely negligible towards Hubble crossing: 
        neglecting them is a source of small inaccuracy in this treatment.
        On super-Hubble scales, the Bardeen contribution $\propto m_2^2$ (red line) becomes the dominant one and can not be neglected at all.
        }
    \end{subfigure}%
    \caption{Real parts (the imaginary ones are similar) of the contributions to the equations of motion of the perturbations of the light and heavy scalar fields, for the modes that are initially aligned with the corresponding oscillator: $\ddot{\delta\phi}_{\ell,\ell}^k$ (upper panel) and $\ddot{\delta\phi}_{\mathrm{h},\mathrm{h}}^k$ (lower panel).
    The Hubble-crossing time for the mode $k$ of interest is represented by the vertical black line. Before and around Hubble crossing, the two dominant terms are the Hubble friction $\propto H$ (blue line) and the quantum oscillations $\propto k^2/a^2$ (orange line). The mass term $\propto m_2^2$ (green line) comes as a third, small contribution around and just after Hubble crossing. The decay terms (brown and pink lines) play a negligible role in this regime.
    For the Bardeen contributions, see the corresponding captions.}
    \label{fig:perturbations-eom-l}
\end{figure}

\begin{figure}
    \centering
    \begin{subfigure}{0.5\textwidth}
        \centering
        \includegraphics[width=1.05\linewidth]{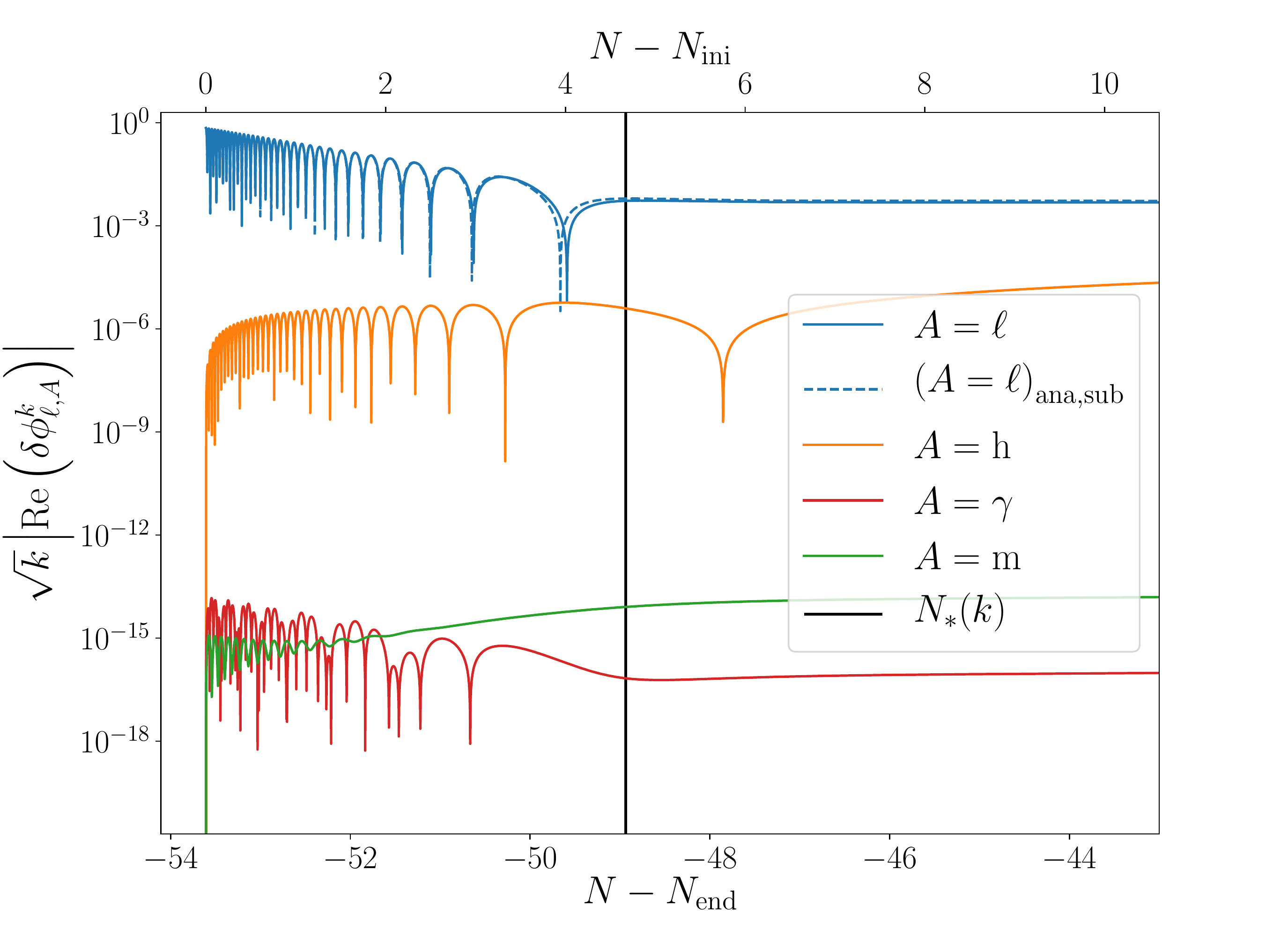}
        \caption{Light scalar field.}
    \end{subfigure}%
    \hfill
    \begin{subfigure}{0.5\textwidth}
        \centering
        \includegraphics[width=1.05\linewidth]{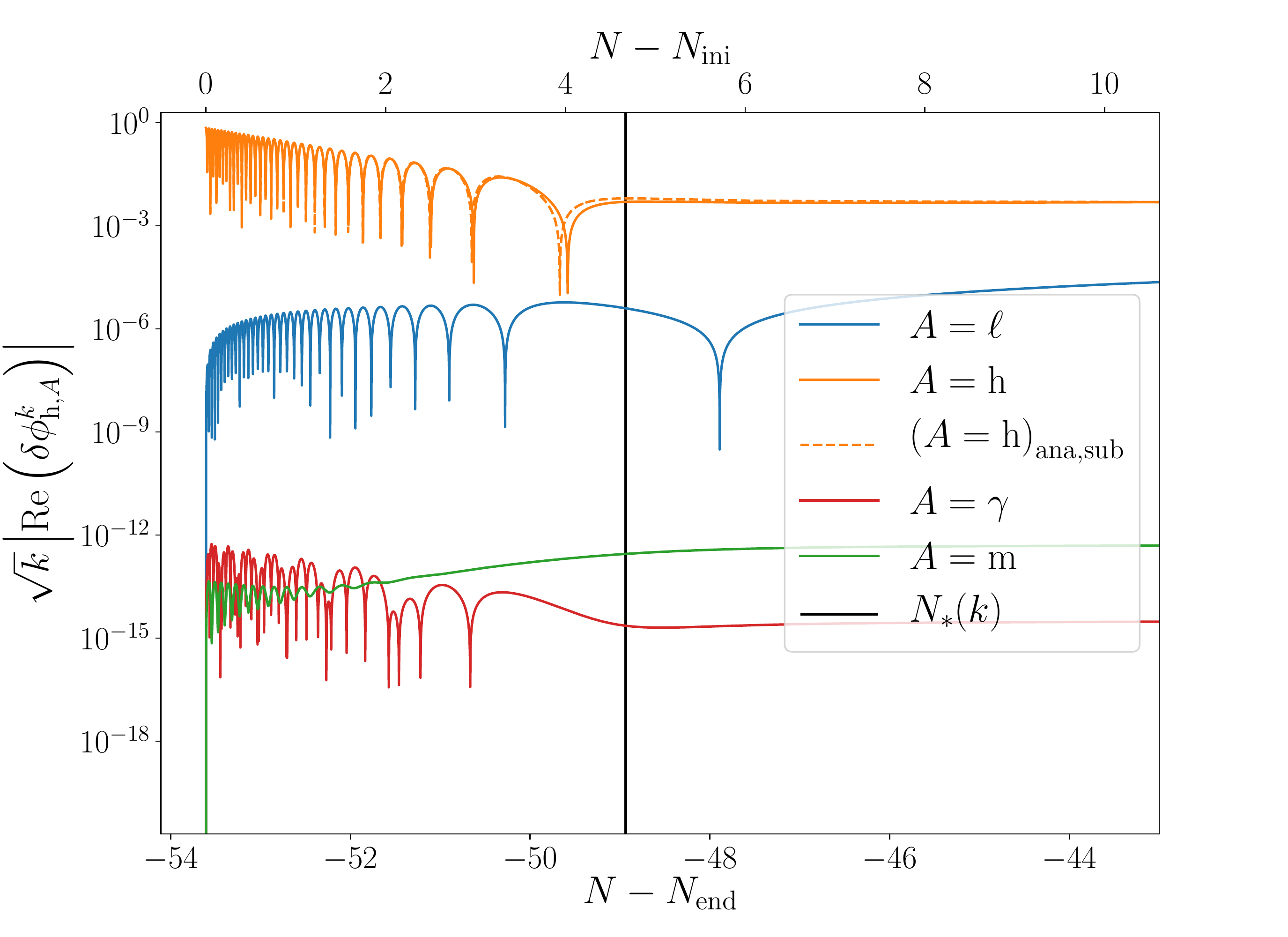}
        \caption{Heavy scalar field.}
    \end{subfigure}
    \caption{Real parts (the imaginary ones are similar) of the fields' perturbations $\delta\phi_{\ell,A}$ (upper panel) and $\delta\phi_{\mathrm{h},A}$ (lower panel), including the exact numerical solutions for the different components $A=\ell,\mathrm{h},\gamma,\mathrm{m}$ (respectively blue, orange, red and green solid lines), and the sub-Hubble, slow-roll analytic approximations ``$\mathrm{ana,sub}$" for the diagonal mode functions $\delta\phi_{\ell,\ell}$ and $\delta\phi_{\mathrm{h},\mathrm{h}}$ (respectively blue and orange dashed lines).
    The analytical approximations are extremely accurate, even though slightly better for the light field than for the heavy one (note that the non-diagonal mode functions such as $\delta\phi_{\ell,\mathrm{h}}$ and $\delta\phi_{\mathrm{h},\ell}$ are neglected in this treatment, which is indeed justified). 
    The vertical black line corresponds to the time of horizon crossing for the mode $k$ represented in these plots.}
    \label{fig:perturbations-Hankel-subHubble}
\end{figure}

\begin{figure}
    \centering
    \begin{subfigure}{0.5\textwidth}
        \centering
        \includegraphics[width=1.05\linewidth]{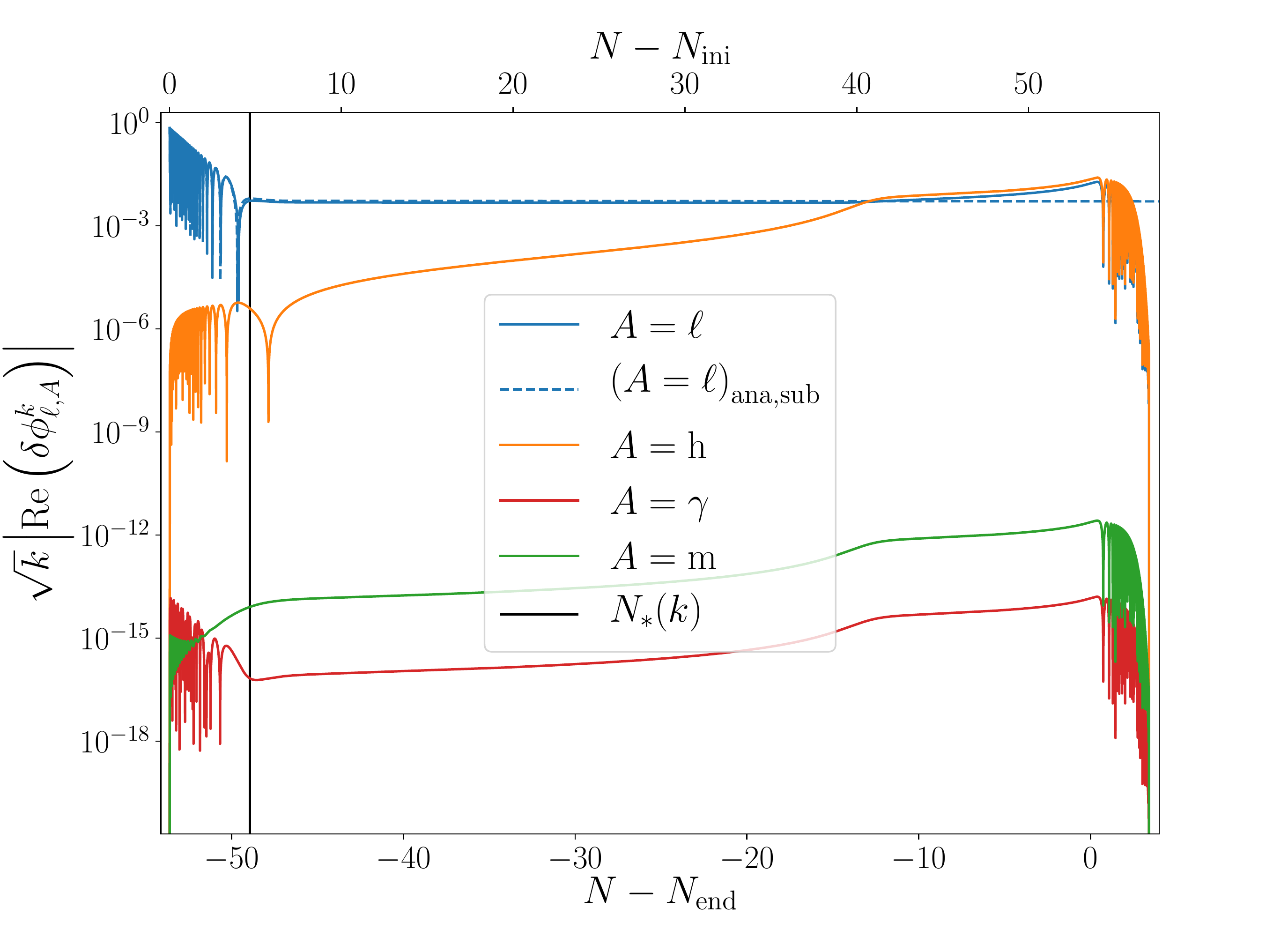}
        \caption{Light scalar field.}
    \end{subfigure}%
    \hfill
    \begin{subfigure}{0.5\textwidth}
        \centering
        \includegraphics[width=1.05\linewidth]{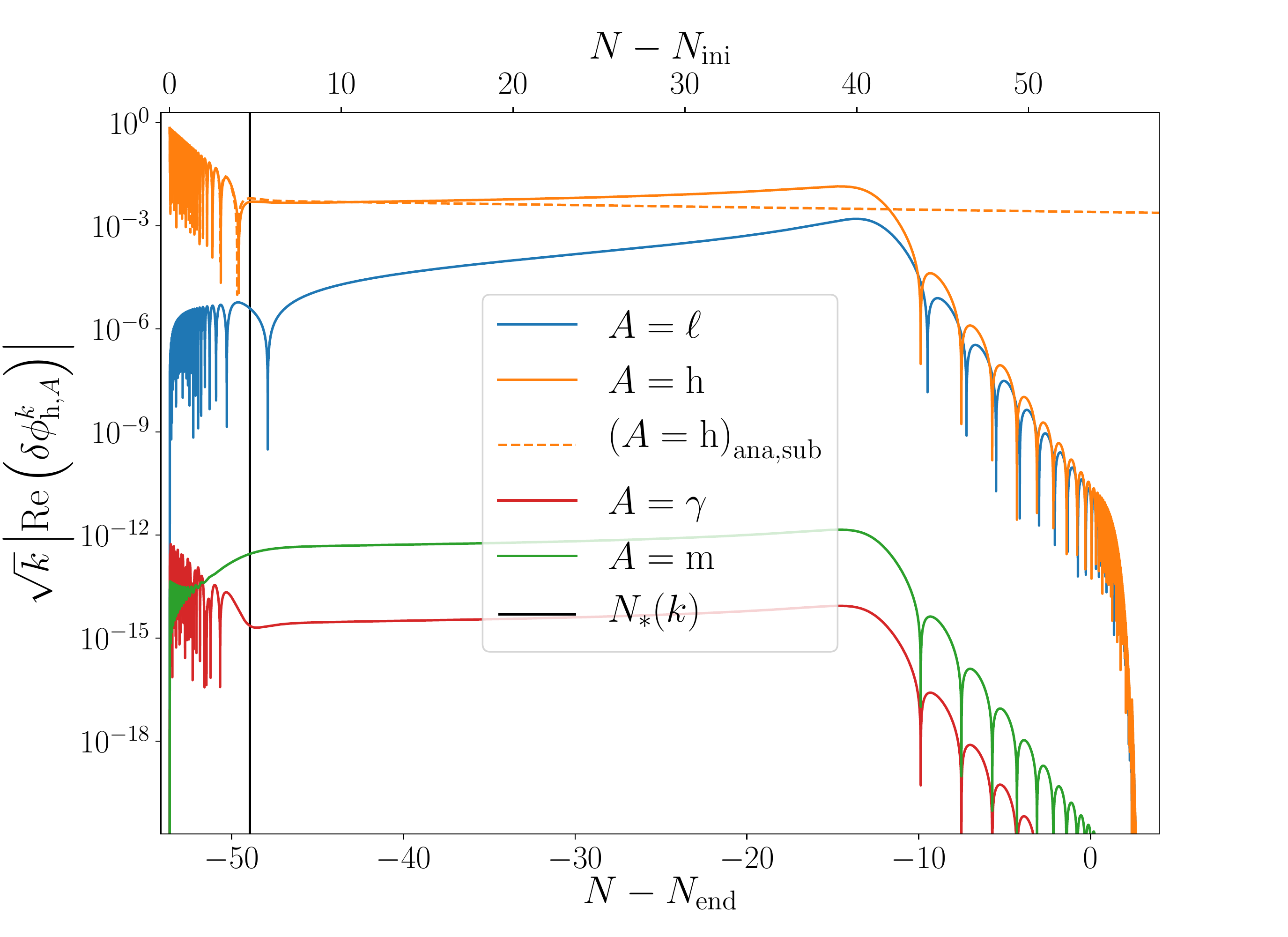}
        \caption{Heavy scalar field.}
    \end{subfigure}
    \caption{Same figures than the one represented in Fig.~\ref{fig:perturbations-Hankel-subHubble}, but unzoomed until the end of inflation (imaginary parts are similar).
    Clearly, the sub-Hubble approximations become inaccurate after horizon crossing, for mainly two reasons.
    First, the diagonal mode functions eventually become inaccurate.
    For the heavy field that is first driving inflation, we have understood that the Bardeen contribution, neglected in the above treatment, is actually not negligible and therefore results in the departure observed in the right panel between the orange solid and dashed lines representing $\delta\phi_{\mathrm{h},\mathrm{h}}$.
    The same phenomenon happens on the left panels for $\delta\phi_{\ell,\ell}$ (departure between the blue solid and dashed lines) after the light field begins to drive inflation.
    Secondly, the non-diagonal mode functions cease to be negligible, in particular the mode function $\delta\phi_{\ell,\mathrm{h}}$ (orange line in the left panel) receives an important contribution from the heavy field during the transition between the two regimes of inflation.
    This is also true, although in a less striking manner, for the other non-diagonal function $\delta\phi_{\mathrm{h},\ell}$ (blue line in the right panel).
    It is crucial to understand this super-horizon interaction in order to model correctly the physics of non-adiabatic perturbations in multi-field inflation.
    Clearly, it is not encapsulated by the approximations made in this section, and it is the aim of Sec.~\ref{subsubsec:largedouble} to address this question.
    Note that the oscillators $A=\gamma,\mathrm{m}$ of the scalar fields remain negligible throughout the whole inflationary epoch.
    }
    \label{fig:perturbations-Hankel-superHubble}
\end{figure}

To go further and have a complete knowledge of the large scales solutions~(\ref{eq:solbardeensr}), (\ref{eq:soldphilsr}) and~(\ref{eq:soldphihsr}), we must fix the constants $C_{1,A}$ and $C_{3,A}$. Inverting Eqs.~(\ref{eq:soldphilsr}) and~(\ref{eq:soldphihsr}), one obtains
\begin{align}
\label{eq:c1a}
    C_{1,A}&=-\frac{1}{\Mp^2}\left(\frac{V_\ell}{\partial V_\ell/\partial \phi_\ell}\delta \phi_{\ell,A}
    +\frac{V_\mathrm{h}}{\partial V_\mathrm{h}/\partial \phi_\mathrm{h}}\delta \phi_{\mathrm{h},A}\right),
    \\
    \label{eq:c3a}
    C_{3,A}&=\frac32\left(\frac{\delta \phi_{\ell,A}}{\partial V_\ell/\partial \phi_\ell}-\frac{\delta \phi_{\mathrm{h},A}}{\partial V_\mathrm{h}/\partial \phi_\mathrm{h}}\right),
\end{align}
and, therefore, it is sufficient to evaluate the time-dependent quantities that appear in the above formulas at a specific 
time to find the two constants $C_{1,A}$ and $C_{3,A}$ once and for all. This can be achieved with the help of the following considerations. The exact equations for field fluctuations are Eqs.~(\ref{eq:perKGl-A})  and~(\ref{eq:perKGh-A}). By plotting all the terms appearing in those equations, see Fig.~\ref{fig:perturbations-eom-l}, one can see that it is a good approximation to assume that, on sub-Hubble scales, both field fluctuations behave as
\begin{align}
  \label{eq:dphiapprox}
  \delta \phi_{\ell, \mathrm{h},A}''+2{\cal H}\delta \phi_{\ell, \mathrm{h},A}'
  +\left[k^2+m_{\ell,\mathrm{h}}^2 a^2(\eta)\right] \delta \phi_{\ell, \mathrm{h},A}\simeq 0 \,,
\end{align}
where all terms proportional to the Bardeen and to the decay constants $\Gamma_{\mathrm{hm}}$, $\Gamma_{\ell \gamma}$ were neglected. In fact, and this is going to play an important role in the following, one can show that the approximation discussed above should be valid not only on sub-Hubble scales, but also up to a time slightly after Hubble radius crossing. This is because, until that time, the terms that we have neglected in Eqs.~(\ref{eq:perKGl-A})  and~(\ref{eq:perKGh-A}) remain sub-dominant\footnote{
Note from Fig.~\ref{fig:perturbations-eom-l} that this approximation seems better verified for the perturbations of the light scalar field than the ones of the heavy one.
And indeed one can see in Figs.~\ref{fig:perturbations-Hankel-subHubble} and \ref{fig:perturbations-Hankel-superHubble} a better accuracy of the analytical approximations for the light fluctuations than for the heavy ones for times after Hubble crossing.
}.

Eq.~(\ref{eq:dphiapprox}) can be
rewritten if a way which is more suitable to discuss its solutions, namely 
\begin{align}
\label{eq:fieldfluctuations}
  f_{\ell,\mathrm{h},A}''+\left[k^2-\frac{a''}{a}+m_{\ell,\mathrm{h}}^2a^2(\eta)\right]f_{\ell, \mathrm{h},A}=0,
\end{align}
where one has defined $\delta \phi_{\ell,\mathrm{h},A}(\eta)\equiv f_{\ell,\mathrm{h},A}(\eta)/a(\eta)$. As expected, we obtain an equation which is in fact similar to the equation of a test massive field in an expanding space-time. Eq~(\ref{eq:fieldfluctuations}) should be solved with the following, Bunch-Davies, initial conditions
\begin{align}
\label{eq:inif}
f_{\ell,A}(\eta_\uini)&=\frac{1}{\sqrt{2k}}\delta_{\ell,A},
  \quad
f_{\mathrm{h},A}(\eta_\uini)= \frac{1}{\sqrt{2k}}\delta_{\mathrm{h},A},
  \\
  \label{eq:iniderf}
f_{\ell,A}'(\eta_\uini) &=
  -i\sqrt{\frac{k}{2}}\delta_{\ell,A},
  \quad
f_{\mathrm{h},A}'(\eta_\uini) 
= -i\sqrt{\frac{k}{2}}\delta_{\mathrm{h},A},
 \end{align}
with, in principle, $k\eta_\uini \rightarrow -\infty$. These conditions are such that, initially, each mode function labeled by $A$ only possesses a non-vanishing component along that direction. Of course, this is equivalent to the initial conditions already discussed in Eq.~(\ref{eq:iniqscalarfield}) and in the text after Eq.~(\ref{eq:bardeenquantumini}).

Let us now discuss the solutions of Eq.~(\ref{eq:fieldfluctuations}). If the physical wavelength of the Fourier mode under consideration is much smaller than the Hubble radius, namely $k\eta \ll -1$, then the two terms $a''/a$ and $m_{\ell,\mathrm{h}}^2a^2(\eta)$ are negligible. In this regime, one does not need to know the form of the scale factor explicitly in order to solve the equation. The corresponding solution, satisfying the initial conditions~(\ref{eq:inif}) and~(\ref{eq:iniderf}) reads
\begin{align}
\label{eq:fsmall}
    f_{\ell, \mathrm{h},A}(\eta)=\frac{1}{\sqrt{2k}}
    e^{-ik(\eta-\eta_\uini)}\delta_{\ell, \mathrm{h},A},
\end{align}
where the symbol $\delta_{\ell,\mathrm{h},A}$ means either $\delta_{\ell,A}$ or $\delta_{\mathrm{h},A}$ depending of whether the mode functions $f_{\ell,A}(\eta)$ or $f_{\mathrm{h},A}(\eta)$ are considered. Again, equivalent considerations have already been presented around Eq.~(\ref{eq:iniqscalarfield}).

As the Universe expands, the physical wavelength of the Fourier mode increases and the above solution ceases to be valid: around Hubble horizon crossing (in principle, before since we deal with an equation valid on small scales only), namely $k/(a_*H_*)= 1$, the two terms mentioned before are indeed no longer negligible. In that case, in order to solve Eq.~(\ref{eq:fieldfluctuations}) in this regime, the form of the scale factor is needed. Since the background is undergoing inflation, this can be obtained by means of the slow-roll approximation. At first order, $a(\eta)\propto a_*(\eta/\eta_*)^{-1-\epsilon_{1*}}$, where $\epsilon_{1*}$ is the first slow-roll parameter (evaluated at Hubble radius crossing), which can also be expanded as $a(\eta)\simeq -(H_*\eta)^{-1}(1+\epsilon_{1*}-\epsilon_{1*}\ln \eta/\eta_*+\dots )$. This approximation for the scale factor is expected to be valid only around Hubble radius crossing. Notice that, in the regime of validity of Eq.~(\ref{eq:fsmall}), the background should also be inflationary and, as a consequence, the slow-roll approximation should also be satisfied. However, if the number of e-folds separating this epoch to the epoch of Hubble radius crossing is too large, the assumption that $\epsilon_1$ is always constant becomes questionable. In other words, one has two epochs where slow-roll is valid, each one having its own slow-roll parameter, $\epsilon_{1\uini}$ and $\epsilon_{1*}$. Within each regimes, the slow-roll parameter can indeed be considered constant but these constants are \textit{a priori} not the same in different epochs. With the scale factor given above, Eq~(\ref{eq:fieldfluctuations}) takes the form
\begin{align}
\label{eq:massivefieldsr}
  f_{\ell,\mathrm{h},A}''+\left[k^2-\frac{1}{\eta^2}\left(2+3\epsilon_{1*}
  -\frac{m_{\ell,\mathrm{h}}^2}{H_*^2}\right)
  \right]f_{\ell,\mathrm{h},A}=0.
  \end{align}
Notice that, in the present context, it is justified to go beyond a simple de Sitter approximation (namely $\epsilon_{1*}=0$). Indeed, it is well-known that this one is not very accurate for a
background which resembles Large Field Models (LFI)~\cite{Martin:2013tda}. It would
certainly be better in a small field/Starobinsky (SFI/HI, see Ref.~\cite{Martin:2013tda}) context where, to a very good accuracy, the Hubble parameter remains constant. It is also interesting to remark that, in order to have a better (more accurate) solution, what is needed is not to consider the impact of the terms neglected in Eqs.~(\ref{eq:perKGl-A})  and~(\ref{eq:perKGh-A}), since, anyway, we showed before that they remain negligible in the range of e-folds we are interested in. Rather a better description of
the evolution of the background is what matters in order to improve the description of the system. Then, Eq~(\ref{eq:massivefieldsr}) can be solved in terms of Bessel function
\begin{align}
\label{eq:flarge}
  f_{\ell,\mathrm{h},A}(\eta)=(-k\eta)^{1/2}
\left[A_{\ell,\mathrm{h},A}\mathrm{H}_{\nu_{\ell,\mathrm{h}}}^{(1)}(-k\eta)
+B_{\ell,\mathrm{h},A}\mathrm{H}_{\nu_{\ell,\mathrm{h}}}^{(2)}(-k\eta)\right],
\end{align}
where $\nu_{\ell,\mathrm{h}}=3/2+\epsilon_{1*}-m_{\ell,\mathrm{h}}^2/(3H_*^2)$ and $\mathrm{H}_\nu^{(1,2)}$ are Hankel functions of the first and second kind. The quantities $A_{\ell,\mathrm{h},A}$ and $B_{\ell,\mathrm{h},A}$ are integration constants.

The next step consists in making the junction between the solutions~(\ref{eq:fsmall}) and~(\ref{eq:flarge}) at a time $\eta_\mathrm{j}$, where ``$\mathrm{j}$" stands for junction, which is such that $\eta_\uini<\eta_\mathrm{j}<\eta_*$ (that is to say before Hubble radius crossing). Requiring the continuity of the mode function and of its derivative, one finds 
\begin{align}
    A_{\ell, \mathrm{h},A} &=\frac{\pi z_\mathrm{j}^{1/2}}{4i}\frac{1}{\sqrt{2k}}
    e^{iz_\mathrm{j}+ik\eta_\uini}
    \left[\left(-\frac{1}{2z_\mathrm{j}}+\frac{\nu_{\ell,\mathrm{h}}}{z_\mathrm{j}}
    +i\right)\mathrm{H}_{\nu_{\ell,\mathrm{h}}}^{(2)}(z_\mathrm{j})
    -\mathrm{H}_{\nu_{\ell,\mathrm{h}}-1}^{(2)}(z_\mathrm{j})\right]\delta_{\ell,\mathrm{h},A},
\\
B_{\ell,\mathrm{h},A} &=-\frac{\pi z_\mathrm{j}^{1/2}}{4i}\frac{1}{\sqrt{2k}}
    e^{iz_\mathrm{j}+ik\eta_\uini}
    \left[\left(-\frac{1}{2z_\mathrm{j}}+\frac{\nu_{\ell,\mathrm{h}}}{z_\mathrm{j}}
    +i\right)\mathrm{H}_{\nu_{\ell,\mathrm{h}}}^{(1)}(z_\mathrm{j})
    -\mathrm{H}_{\nu_{\ell,\mathrm{h}}-1}^{(1)}(z_\mathrm{j})\right]\delta_{\ell,\mathrm{h},A},
\end{align}
where $z_\mathrm{j}\equiv -k\eta_\mathrm{j}$. Since the matching time is chosen to be before Hubble radius crossing, one has $z_\mathrm{j}\gg 1$ and the large argument limit of the Hankel functions can be used. As a result, one obtains
\begin{align}
\label{eq:coefA}
    A_{\ell,\mathrm{h},A}&\simeq \frac{1}{2}\sqrt{\frac{\pi}{k}}
    e^{ik\eta_\uini+i\pi \nu_{\ell,\mathrm{h}}/2+i\pi/4}
    \left[1+\mathcal{O}\left(\frac{1}{z_\mathrm{j}}\right)\right]\delta_{\ell,\mathrm{h},A},
    \\
\label{eq:coefB}
    B_{\ell,\mathrm{h},A}&\simeq \mathcal{O}\left(\frac{1}{z_\mathrm{j}^2}\right)
    \delta_{\ell,\mathrm{h},A}.
\end{align}
It can be noticed that, in the limit $z_\mathrm{j}\rightarrow \infty $, $B_{\ell,\mathrm{h},A}\rightarrow 0$. This case is formally identical to the situation where the solution~(\ref{eq:flarge}) is assumed to be valid all the way from $\eta_\uini$ to Hubble radius crossing and where the constants $A_{\ell,\mathrm{h},A}$ and $B_{\ell,\mathrm{h},A}$ are obtained by requiring the function~(\ref{eq:flarge}) to tend towards~(\ref{eq:fsmall}). 

Using Eqs.~(\ref{eq:coefA}) and~(\ref{eq:coefB}), the solution around Hubble radius crossing 
for the field fluctuation can be written as
\begin{align}
\label{eq:dphiHubble}
  a_\mathrm{ini}^{3/2}\delta \phi_{\ell,\mathrm{h},A}(\eta)
  &\simeq \frac{a_\mathrm{ini}}{a(\eta)}(-k\eta)^{1/2}
  \frac{1}{\sqrt{2k/a_\mathrm{ini}}}\sqrt{\frac{\pi}{2}}
  e^{i\pi(\nu+1/2)/2}
  e^{ik\eta_\mathrm{ini}}\mathrm{H}_\nu^{(1)}(-k\eta)
  \delta_{\ell,\mathrm{h},A}.
  \end{align}
Again, as explained above, although this expression is formally equivalent to the solution that one would obtain assuming that~(\ref{eq:flarge}) is valid from $\eta_\uini$ to Hubble radius crossing (since, formally, we took the limit $z_\mathrm{j}\rightarrow \infty$), we emphasize that we will consider that Eq.~(\ref{eq:dphiHubble}) describes the behavior of $\delta \phi_{\ell,\mathrm{h},A}(\eta)$ only in the vicinity of Hubble radius crossing. This will allow us to consider that $\epsilon_{1*}$ and $\epsilon_{1\uini}$ are not necessary equal, which leads to more accurate expressions.

The above solution is expressed in terms of conformal time. In order to make the connection with numerical calculations, it is necessary to express it in terms of the number of e-folds. From the exact expression
\begin{align}
    \dd N=-{\cal H}\dd \left(\frac{1+\epsilon_1}{\cal H}
    \right)+\frac{{\cal H}}{a}\dd \left(\frac{\epsilon_1}{H}\right),
\end{align}
one obtains, at first order in slow-roll, $\dd N\simeq -(1+\epsilon_1){\cal H}\dd(1/{\cal H})$ which can be integrated to ${\cal H}={\cal H}_\mathrm{p}e^{(N-N_\mathrm{p})/(1+\epsilon_{1\mathrm{p}})}$ where ``$\mathrm{p}$" just denotes a peculiar time. Integrating this relation leads to $\eta=-(1+\epsilon_{1\mathrm{p}})/{\cal H}_\mathrm{p}e^{-(N-N_\mathrm{p})/(1+\epsilon_{1\mathrm{p}})}$ where we have chosen an integration constant to match the standard de Sitter limit. In the following, we will choose the initial time to be the peculiar time for the first regime [namely the regime where the solution is given by Eq.~(\ref{eq:fsmall})], meaning
\begin{align}
   k\eta &=-(1+\epsilon_{1\uini})\frac{k}{a_\mathrm{ini}H_\mathrm{ini}}
  e^{-(N-N_\mathrm{ini})/(1+\epsilon_{1\uini})},
\end{align}
and the Hubble radius crossing time for the second regime [namely the regime where the solution is given by Eq.~(\ref{eq:flarge})] implying
\begin{align}
   k\eta &=-(1+\epsilon_{1*})\, 
  e^{-(N-N_*)/(1+\epsilon_{1*})},
\end{align}
where we have used $k/(a_*H_*)=1$.

In Fig.~\ref{fig:perturbations-Hankel-subHubble}, we have compared the exact, numerical, real parts of $\delta \phi_{\ell, \mathrm{h},A}$ with the analytical solution~(\ref{eq:dphiHubble}) from the initial time $\eta_\uini$ to a time slightly after Hubble radius crossing (indicated by the black solid vertical line).
Evidently, the two solutions match very well. It is especially interesting to remark that the analytical solution is still a very good approximation even a few e-folds after Hubble radius crossing even if the equation used is, in principle, only valid before Hubble radius crossing. In Fig.~\ref{fig:perturbations-Hankel-superHubble}, we have represented the same quantities but zoomed out.
It shows the limitations of the previous remark: one indeed notices that the agreement between the small-scale solution and the exact one after Hubble radius crossing is only valid for a few e-folds. Then, as expected, the two functions split.
Moreover, note that the analytical approximation neglects the mixing of perturbations: indeed $\delta \phi_{\ell,A} \propto \delta_{\ell, A}$ and $\delta \phi_{\mathrm{h},A} \propto \delta \phi_{\mathrm{h},A}$.

Given the previous considerations, our final move will be to use the solution (\ref{eq:dphiHubble}) evaluated at a time slightly after Hubble radius crossing; in this regime we can consider that $-k\eta \ll 1 $ and one can take the small argument limit in the Hankel function. This leads to the following expressions for the field fluctuations:
\begin{align}
    a_\mathrm{ini}^{3/2} \delta \phi_{\ell,\mathrm{h},A} (N)&=H_\mathrm{ini}^{-1/2} \frac{H_*}{H_\mathrm{ini}}\left( \frac{k}{a_\mathrm{ini} H_\mathrm{ini}}\right)^{-3/2} \exp \left[ - \frac{m_{\ell,\mathrm{h}}^2}{3 H_*^2} (N-N_*) \right] \mathcal{F}(\nu_{\ell,\mathrm{h}}) \delta_{\ell,\mathrm{h},A} \,, 
    \end{align}
with
    \begin{align}
\mathcal{F}(\nu) &= \frac{2^{\nu-1}}{\sqrt{\pi}} \Gamma(\nu) (1+\epsilon_*)^{1/2-\nu} e^{i \left[ \frac{\pi}{2} \left( \nu - 1/2 \right) + k\eta_\mathrm{ini} \right]} \,.
\end{align}
We see that the mode functions still possess a mild time dependence because of the exponential term [$m_{\ell,\mathrm{h}}^2/(3H_*^2)$ is a small factor]. This dependence vanishes in the limit of mass-less fields. This makes sense because, in this limit, on large scales, $f_{\ell,\mathrm{h},A}(\eta)\propto a(\eta)$ and, as a consequence, $\delta \phi_{\ell, \mathrm{h},A}(\eta)
=f_{\ell,\mathrm{h},A}(\eta)/a(\eta)$ is frozen.

Finally, in order to be in a position where the constants $C_{1,A}$ and $C_{3,A}$ can be calculated
according to Eqs.~(\ref{eq:c1a}) and~(\ref{eq:c3a}), we also need to evaluate the quantities $V_{\ell,\mathrm{h}}$ and $\partial V_{\ell,\mathrm{h}}/\partial \phi_{\ell,\mathrm{h}}$. Since we need to calculate those quantities just after Hubble horizon crossing and since this happens in the epoch dominated by the heavy field, $\phi_\ell$ and $\phi_\mathrm{h}$ (and $H$ as well as the first slow-roll parameter) can be replaced with their slow-roll trajectory during the epoch dominated by the heavy field.
First, using Eqs.~(\ref{eq:c1a}) and~(\ref{eq:c3a}), one arrives at
\begin{align}
    C_{1,A} (N_\mathrm{m})= &- \frac{H_\uini^{-1/2}}{2\Mp^2}\frac{H_*}{H_\mathrm{ini}}\left( \frac{k}{a_\mathrm{ini} H_\mathrm{ini}}\right)^{-3/2}  
    \Biggl[ \phi_\ell(N_\mathrm{m}) e^{-\frac{m_{\ell}^2}{3 H_*^2} (N_\mathrm{m}-N_*)} \mathcal{F}(\nu_{\ell})  \delta_{\ell,A} 
    \nonumber \\ &
    +
    \phi_\mathrm{h}(N_\mathrm{m}) e^{-\frac{m_{\mathrm{h}}^2}{3 H_*^2} (N_\mathrm{m}-N_*)} \mathcal{F}(\nu_{\mathrm{h}})  \delta_{\mathrm{h},A}  \Biggr] \, , \\
    C_{3,A}(N_\mathrm{m}) = &\frac{3 H_\uini^{-1/2}}{2}\frac{H_*}{H_\mathrm{ini}}\left( \frac{k}{a_\mathrm{ini} H_\mathrm{ini}}\right)^{-3/2}  \Biggl[\frac{\mathcal{F}(\nu_{\ell})  \delta_{\ell,A}}{m_\ell^2  \phi_\ell(N_\mathrm{m})} e^{-\frac{m_{\ell}^2}{3 H_*^2} (N_\mathrm{m}-N_*)}  
    \nonumber \\ & 
    - \frac{\mathcal{F}(\nu_{\mathrm{h}})  \delta_{\mathrm{h},A}}{m_\mathrm{h}^2  \phi_\mathrm{h}(N_\mathrm{m})} e^{-\frac{m_{\mathrm{h}}^2}{3 H_*^2} (N_\mathrm{m}-N_*)}   \Biggr] \, . 
\end{align}
In the above expressions, $N_\mathrm{m}$ should be understood as the e-fold number at which the coefficient $C_{1,A}$ is calculated.
As already discussed before, $N_\mathrm{m}$ should be chosen such that it corresponds to a time slightly after Hubble crossing time. 
In the following, we use the fiducial parameters already considered before, namely $m_\ell=10^{-5}\Mp$ and $m_\mathrm{h}=5\times 10^{-5}\Mp$ with the initial conditions $\phi_{\ell}\vert_\uini=8\Mp$, $\phi_\mathrm{h}\vert_\uini=12\Mp$. We recall that 
this implies that $H_\uini\simeq 2.5\times 10^{-4}\Mp$ and $\epsilon_{1\uini}\simeq 0.01388$.
We take the e-fold number at which we evaluate the constants $C_{1,A}$ and $C_{3,A}$ to be $3$ e-folds after the Hubble radius crossing, namely $N_\mathrm{m}-N_\uini\sim 7.67$ given that $N_*-N_\uini\sim 4.67$.
Then, we can make use of the slow-roll approximation to evaluate the other quantities.
We take $\phi_\ell(N_\mathrm{m}) \simeq \phi_\ell\vert_\uini=8\Mp$, see Eq.~(\ref{eq:philini}) and $\phi_\mathrm{h}(N_\mathrm{m})=\sqrt{\phi_\mathrm{h}^2\vert_\uini-4\Mp^2(N_\mathrm{m}-N_\uini)}\simeq 10.64\Mp$, see Eq.~(\ref{eq:phihsr}).
Given that Hubble radius crossing occurs during the phase dominated by the heavy, we can consider that $\epsilon_1=2\Mp^2/\phi_\mathrm{h}^2$, see the arguments presented after Eq.~(\ref{eq:eps1field}), implying $\epsilon_{1*}\simeq 0.0159$.
We also have, see Eq.~(\ref{eq:Hubblesr}), $H^2=H_\uini^2-2m_\mathrm{h}^2(N-N_\uini)/3$, which means $H_*\simeq 0.00228\Mp$. From these numerical values, it follows that $\nu_\ell\simeq 1.515$ and $\nu_\mathrm{h}\simeq 1.500$.
Using these numbers, we analytically find the values of $C_{1,A}$ and $C_{3,A}$ collected in the column  ``Analytic" of Table~\ref{table:C1A-and-C3A}, which should be compared to the exact, numerical values in column  ``Numeric"; the relative errors between those estimates are summarized in the last column ``Relative error".

\begin{table}
\large
\renewcommand{\arraystretch}{1.5} 
\begin{tabular}{| m{4cm} | m{2.7cm} | m{2.7cm} | m{2.7cm} |}
    \hline
    Quantity / Value & Numeric & Analytic & Relative error \\
    \hline
    \hline
    $\Rea C_{1,\ell}$  & $-0.122$ & $-0.124$ & $1.4 \%$ \\
    \hline
    $\Im C_{1,\ell}$  & $-0.112$ & $-0.113$ & $ 0.9 \%$\\
    \hline
    $\Rea C_{1,\mathrm{h}}$  & $-0.159$ & $-0.159$ & $ < 0.1\%$\\
    \hline
    $\Im C_{1,\mathrm{h}}$ & $-0.157$ & $-0.138$  & $12.1 \%$\\
    \hline
    \hline
    $\Rea C_{3,\ell} \Mp^2 \times 10^{-7} $ & $5.77$ & $5.79$ &  $0.3 \%$\\
    \hline
    $\Im C_{3,\ell} \Mp^2\times 10^{-7} $  & $5.26$ & $5.27$ &  $0.1    \%$\\
    \hline
    $\Rea C_{3,\mathrm{h}} \Mp^2\times 10^{-7} $ & $-0.158$  & $-0.169$  & $ 7.2 \%$\\
    \hline
    $\Im C_{3,\mathrm{h}} \Mp^2\times 10^{-7}  $  & $-0.142$ & $-0.146$ & $ 3.1 \%$\\
    \hline
\end{tabular}
\caption{Analytical predictions versus exact numerical results for the constants $C_{1,3,A}$ evaluated at the matching time $N_\mathrm{m}=N_*(k)+3$, where $N_*(k)$ corresponds to the time of Hubble crossing for the mode $k$ under investigation. These values are only very mildly dependent on the exact choice of $N_\mathrm{m}$. The relative error = $|$analytic$-$numeric$|$/$|$numeric$|$ is also displayed.}
\label{table:C1A-and-C3A}
\end{table}

Therefore, we have obtained accurate (except, maybe, for the imaginary part of $C_{1,\mathrm{h}}$), analytical, expressions for the constants $C_{1,A}$ and $C_{3,A}$ which allow us to follow the perturbations during inflation on large scales.
In particular, note the interesting ratio between the constants $C_{1,A}$ and $C_{3,A}$:
\begin{equation}
\label{eq:relC1C3}
    \frac{C_{3,\ell}}{C_{1,\ell}}= - \left.\frac{3 \Mp^2}{2 V_\ell}\right|_{N_\mathrm{m}}, \quad \frac{C_{3,\mathrm{h}}}{C_{1,\mathrm{h}}} =  \left.\frac{3 \Mp^2}{2 V_\mathrm{h}}\right|_{N_\mathrm{m}} \,.
\end{equation}

Finally, let us mention that other expressions of the two constants $C_1$ and $C_3$ (the ``standard" expressions that have been used in the literature so far) will be obtained in Sec.~\ref{sec:comparison}.

\subsubsection{Large scale solutions during inflation}
\label{subsubsec:largedouble}

Having determined the constants $C_{1,A}$ and $C_{3,A}$, we now have a complete knowledge of the solutions~(\ref{eq:solbardeensr}), (\ref{eq:soldphilsr}) and~(\ref{eq:soldphihsr}). This also gives the adiabatic part of curvature perturbations, see Eq.~(\ref{eq:zetaadia}) and, moreover, using the relations between $C_{1,A}$ and $C_{3,A}$, see Eq.~(\ref{eq:relC1C3}), the non-adiabatic parts of curvature perturbations can be re-written as
\begin{align}
\label{eq:zetaellell}
    \zeta_{\ell,\ell}^\mathrm{nad}&=- C_{1,\ell} \frac{V_\mathrm{h}}{V_\ell|_{N_\mathrm{m}}} \left[ 1 - \frac{2 V_\ell}{9 V} \frac{m_\ell^2}{H^2} (R^2-1) \right] \,, \\
    \label{eq:zetaellh}
    \zeta_{\ell,\mathrm{h}}^\mathrm{nad}&= C_{1,\mathrm{h}} \frac{V_\mathrm{h}}{V_\mathrm{h}|_{N_\mathrm{m}}} \left[ 1 - \frac{2 V_\ell}{9 V} \frac{m_\ell^2}{H^2} (R^2-1) \right] \,, \\
    \label{eq:zetahh}
    \zeta_{\mathrm{h},\mathrm{h}}^\mathrm{nad}&=- C_{1,\mathrm{h}} \frac{V_\ell}{V_\mathrm{h}|_{N_\mathrm{m}}} \left[ 1 + \frac{2 V_\mathrm{h}}{9 V} \frac{m_\ell^2}{H^2} (R^2-1) \right] \,, \\
    \label{eq:zetahell}
    \zeta_{\mathrm{h},\ell}^\mathrm{nad}&= C_{1,\ell} \frac{V_\ell}{V_\ell|_{N_\mathrm{m}}} \left[ 1 + \frac{2 V_\mathrm{h}}{9 V} \frac{m_\ell^2}{H^2} (R^2-1) \right] \,.
\end{align}

In Fig.~\ref{fig: zeta_ell,A}, we have represented the exact (numerical) real parts of $\zeta_{\ell,A}$ together with their (super-Hubble) analytical expression, $\zeta_{\ell,A}=\zeta_{\ell,A}^\mathrm{ad}+\zeta_{\ell,A}^\mathrm{nad}=-C_{1,A}+\zeta_{\ell,A}^\mathrm{nad}$, where we have used Eq.~(\ref{eq:zetaadia}) and where $\zeta_{\ell,A}^\mathrm{nad}$ are given by Eqs.~(\ref{eq:zetaellell}) and~(\ref{eq:zetaellh}). We see that, after Hubble radius crossing, the analytical and numerical curves match very well, thus confirming that our analytical approximations are very good. In Fig.~\ref{fig: zeta_ell,A ad-nad}, we have again represented the analytical $\zeta_{\ell,\ell}$ and $\zeta_{\ell,\mathrm{h}}$ but we have split these quantities into their adiabatic and non-adiabatic components. We see that $\zeta_{\ell,\ell}$ is initially dominated by its non adiabatic part, $\zeta_{\ell, \ell}\simeq \zeta_{\ell,\ell}^\mathrm{nad}$, which makes sense since, prior to the heavy field decay, one has $V_\mathrm{h}\gg V_\ell\vert_{N_\mathrm{m}}$, see Eq.~(\ref{eq:zetaellell}). After the decay, the non-adiabatic part strongly decreases and the adiabatic part takes over, $\zeta_{\ell,\ell}\simeq \zeta_{\ell,\ell}^\mathrm{ad}$. For $\zeta_{\ell,\mathrm{h}}$, initially, there is no clear hierarchy between the adiabatic and non-adiabatic contributions [see Eq.~(\ref{eq:zetaellh}) where the overall amplitude of $\zeta_{\ell,\mathrm{h}}^\mathrm{nad}$ is proportional to $V_\mathrm{h}/V_\mathrm{h}\vert _{N_\mathrm{m}}$] but, after the decay of the heavy field, the adiabatic component largely dominates and $\zeta_{\ell,\mathrm{h}}\simeq \zeta_{\ell,\mathrm{h}}^\mathrm{ad}$.

\begin{figure}
    \centering
    \begin{subfigure}{\textwidth}
        \centering
        \includegraphics[width=0.75\linewidth]{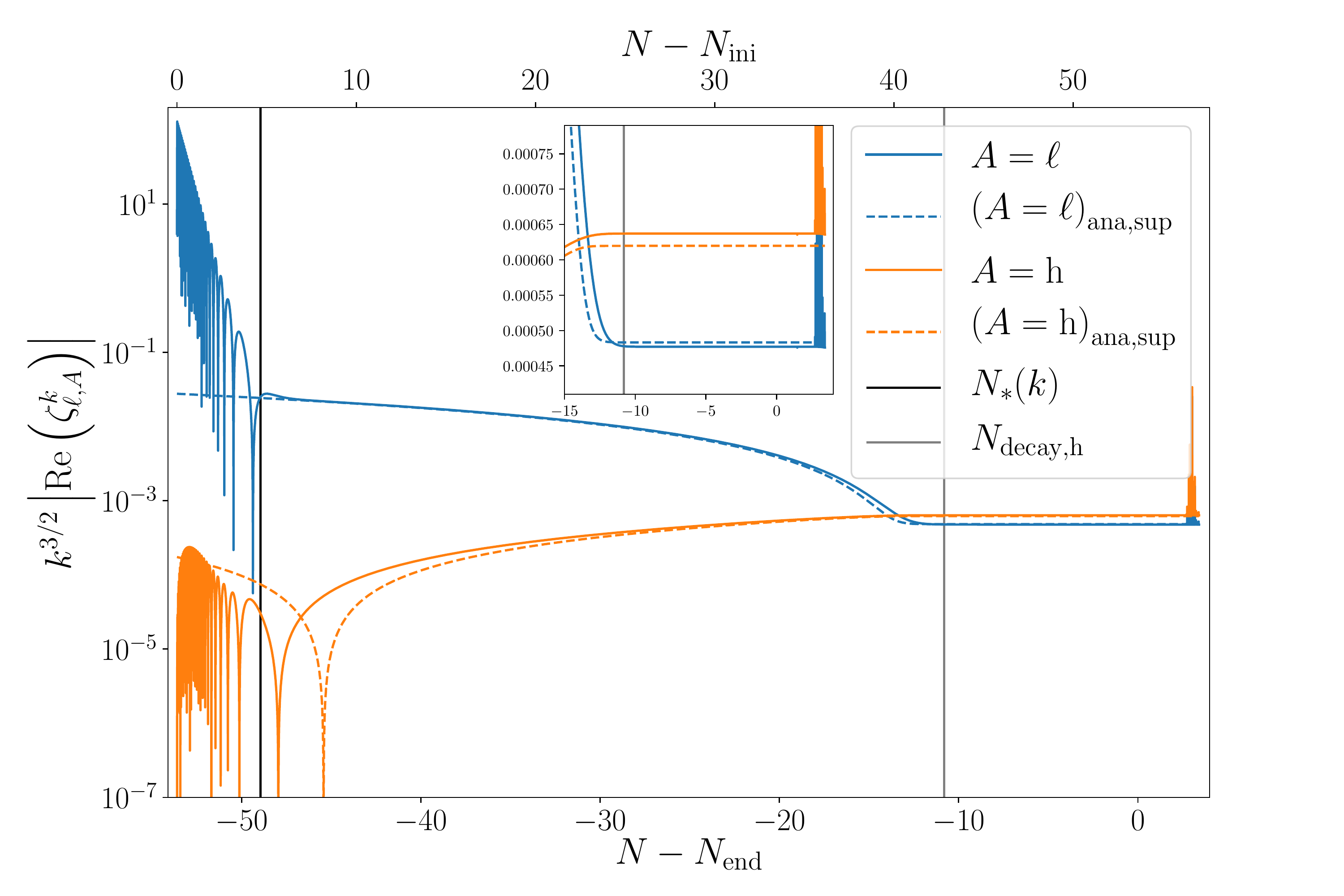}
        \caption{Clearly, both the diagonal mode function $\zeta_{\ell,\ell}$ and the non-diagonal one $\zeta_{\ell,\mathrm{h}}$ are now accurately described on super-Hubble scales by the formulas inferred from Eqs.~\eqref{eq:soldphilsr}--\eqref{eq:soldphihsr}. (there is of course a residual error, see the inset, that will be quantified in the final result).}
        \label{fig: zeta_ell,A}
    \end{subfigure}%
    \\
        \centering
    \begin{subfigure}{\textwidth}
        \centering
        \includegraphics[width=0.75\linewidth]{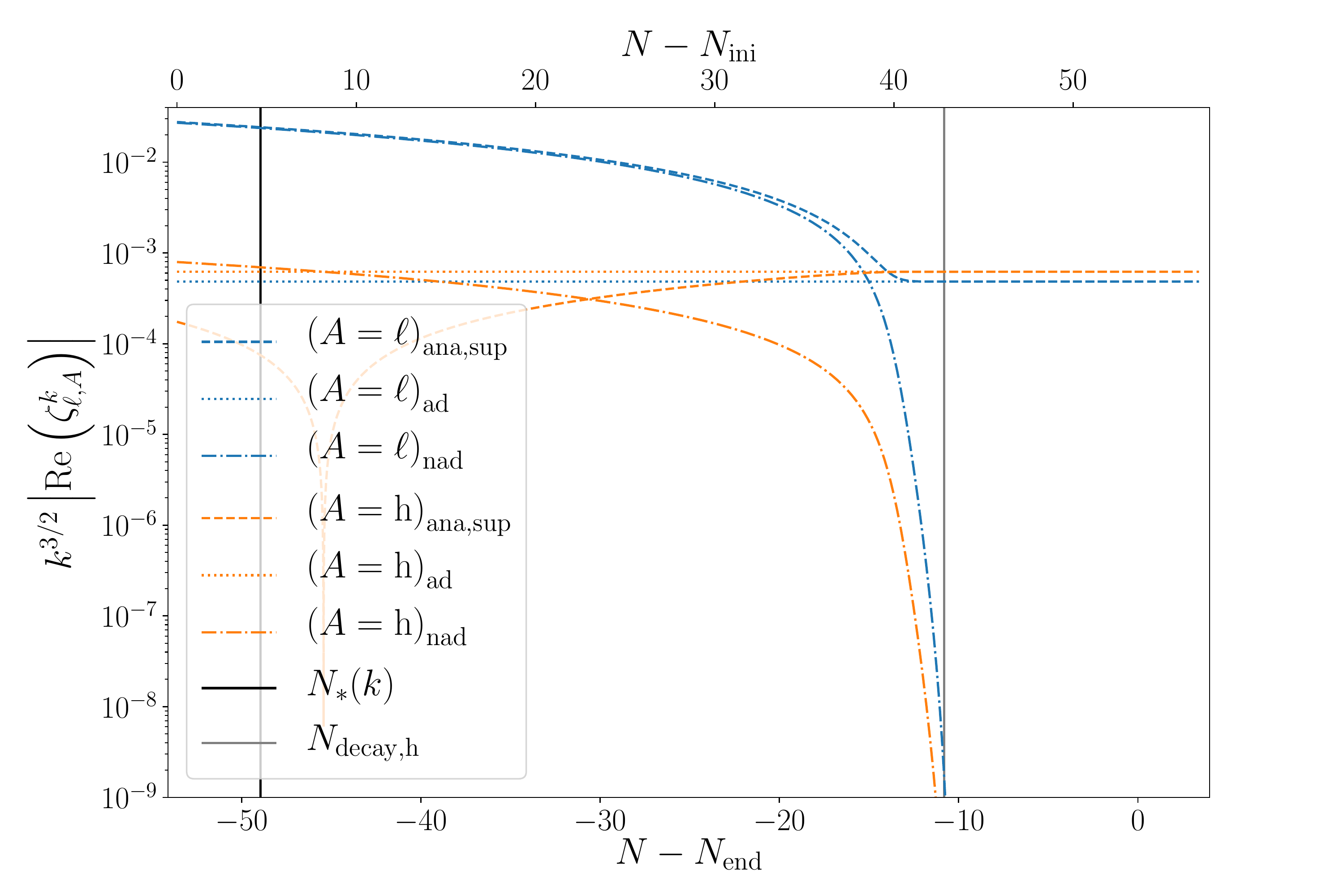}
        \caption{
        The non-adiabatic components of the light field's curvature perturbations are non-negligible only until the decay of the heavy field, after which adiabatic components largely dominate.
        }
        \label{fig: zeta_ell,A ad-nad}
    \end{subfigure}%
    \caption{Real parts (imaginary ones are similar) of the light field's curvature perturbations $\zeta_{\ell,\ell}$ and $\zeta_{\ell,\mathrm{h}}$ during inflation (respectively blue and orange solid lines in the upper panel), and comparison with analytical approximations on super-Hubble scales ``$\mathrm{ana,sup}$" (respectively blue and orange dashed lines in both panels) including their decomposition into their adiabatic parts $\zeta_{\ell,A}^\mathrm{ad}$ (respectively blue and orange dotted lines in lower panel)  defined in Eq.~\eqref{eq:zetaadia} and non-adiabatic ones $\zeta_{\ell,A}^\mathrm{nad}$  (respectively blue and orange dotted-dashed lines in lower panel) defined in Eq.~\eqref{eq:zeta_hlA^nad1}.
    The vertical black line corresponds to the time of horizon crossing for the mode $k$ of interest, and the vertical gray one to the time of decay of the heavy field, after which the light field drives inflation.
    }
    \label{fig: zeta_ell,A tot}
\end{figure}

\begin{figure}
    \centering
    \begin{subfigure}{\textwidth}
        \centering
        \includegraphics[width=0.75\linewidth]{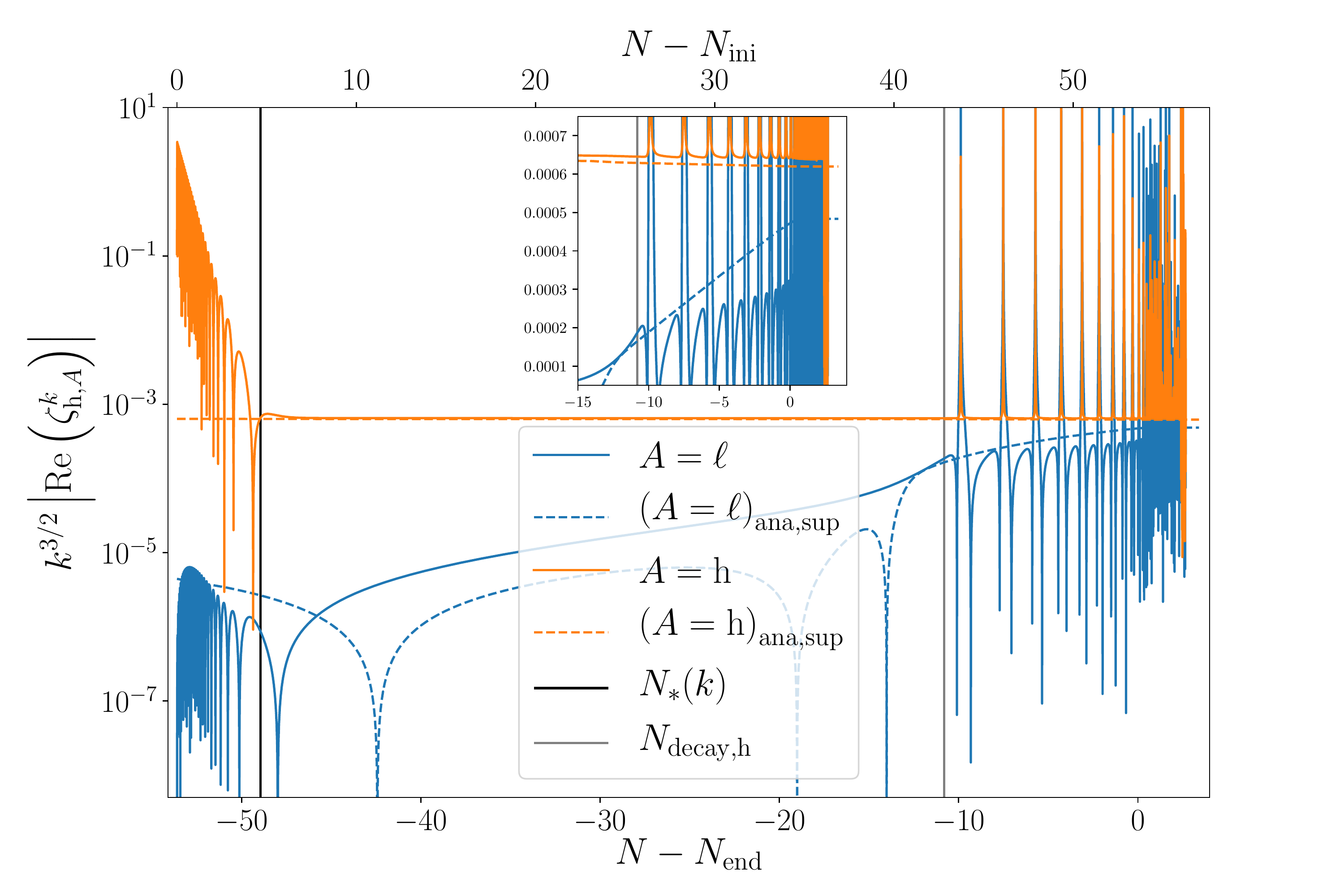}
        \caption{Both the non-diagonal mode function $\zeta_{\mathrm{h},\ell}$ and the diagonal one $\zeta_{\mathrm{h},\mathrm{h}}$ are now accurately described (there is of course a residual error, see the inset, that will be quantified in the final result).
        The sharp oscillations correspond to a numerical artifact (see how the analytical approximation follows accurately the envelop of these oscillations).}
        \label{fig: zeta_h,A}
    \end{subfigure}%
    \\
        \centering
    \begin{subfigure}{\textwidth}
        \centering
        \includegraphics[width=0.75\linewidth]{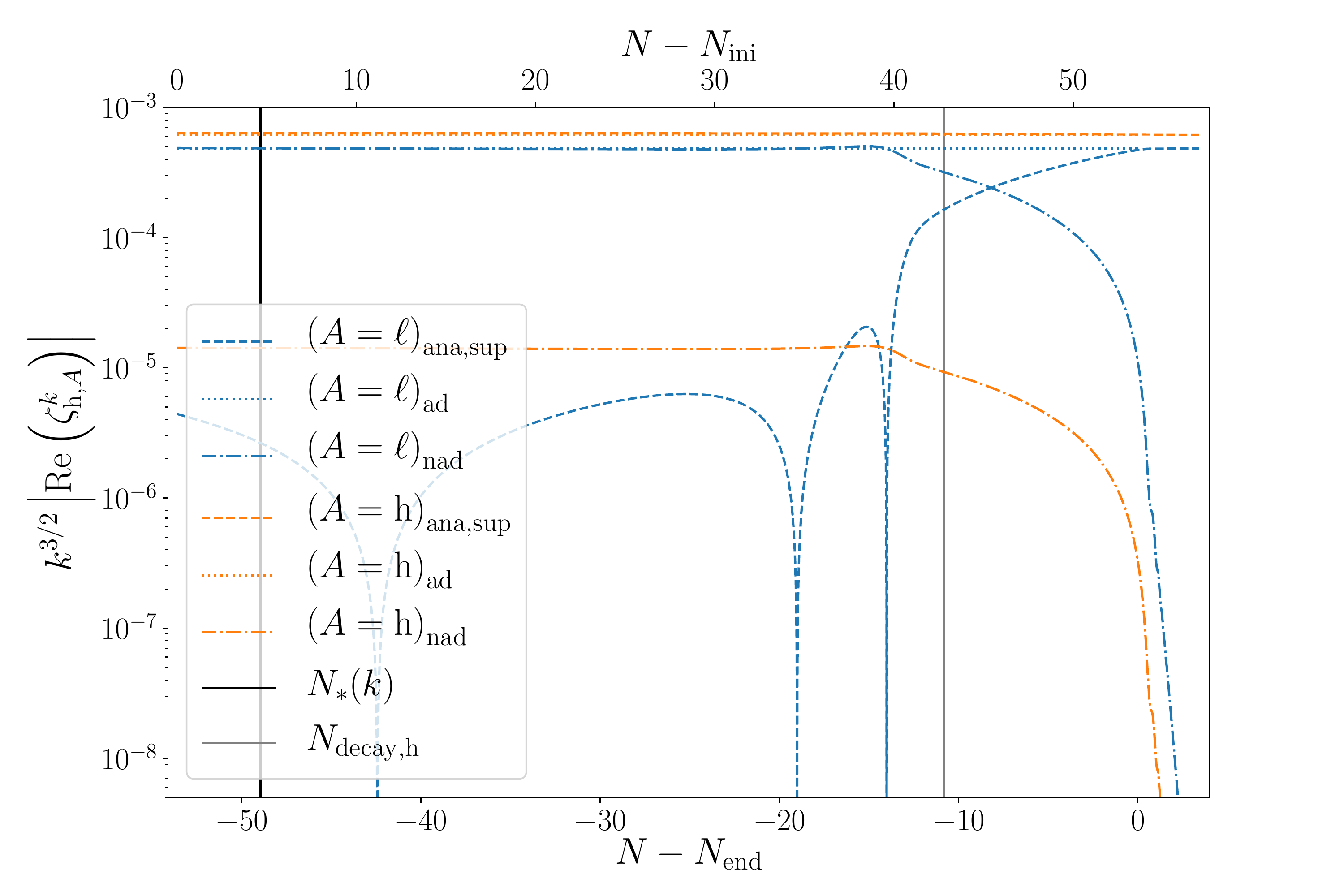}
        \caption{The non-adiabatic component plays an important role for the curvature perturbation $\zeta_{\mathrm{h},\ell}$ even at the decay of the heavy field. This will be the main source of final non-adiabatic perturbations.}
        \label{fig: zeta_h,A ad-nad}
    \end{subfigure}%
    \caption{Real parts (imaginary ones are similar) of the heavy field's curvature perturbations $\zeta_{\mathrm{h},\ell}$ and $\zeta_{\mathrm{h},\mathrm{h}}$ during inflation (respectively blue and orange solid lines in the upper panel), and comparison with analytical approximations on super-Hubble scales ``$\mathrm{ana,sup}$" (respectively blue and orange dashed lines in both panels) including their decomposition into their adiabatic parts $\zeta_{\mathrm{h},A}^\mathrm{ad}$ (respectively blue and orange dotted lines)  defined in a way similar to Eq.~\eqref{eq:zetaadia} and non-adiabatic ones $\zeta_{\mathrm{h},A}^\mathrm{nad}$  (respectively blue and orange dotted-dashed lines) defined in a way similar to Eq.~\eqref{eq:zeta_hlA^nad1}.
    }
    \label{fig: zeta_h,A tot}
\end{figure}

In Fig.~\ref{fig: zeta_h,A}, we have plotted the numerical and analytical (super-Hubble) real parts of $\zeta_{\mathrm{h},A}$. We see that, after Hubble radius crossing, both types of solution match well, the agreement being much better for the component $\zeta_{\mathrm{h},\mathrm{h}}$ than for $\zeta_{\mathrm{h},\ell}$. After the decay of the heavy field, both exact components start to oscillate while the analytical $\zeta_{\mathrm{h},\mathrm{h}}$ and $\zeta_{\mathrm{h},\ell}$ continue to monotonically evolve. We think that those oscillations are numerical artifacts: during the oscillations of the heavy field, there are exact cancellations of $\delta \rho_\mathrm{h}/\rho_\mathrm{h}$ that are not exactly reproduced at the numerical level and, therefore, result in these oscillations.
In Fig.~\ref{fig: zeta_h,A ad-nad}, we have represented the real parts of the analytical components $\zeta_{\mathrm{h},\ell}$ and $\zeta_{\mathrm{h},\mathrm{h}}$ as well as their adiabatic and non-adiabatic contributions.
Before the heavy field's decay, $\zeta_{\mathrm{h},\ell}^\mathrm{ad}$ and $\zeta_{\mathrm{h},\ell}^\mathrm{nad}$ are constant and almost equal.
After the decay, the non-adiabatic part starts to decrease and we are left with  $\zeta_{\mathrm{h},\ell}\sim \zeta_{\mathrm{h},\ell}^\mathrm{ad}$. For $\zeta_{\mathrm{h},\mathrm{h}}$, the non-adiabatic part is always subdominant (and is constant before the decay and then decreases) and, as a consequence, we always have $\zeta_{\mathrm{h},\mathrm{h}}\sim \zeta_{\mathrm{h},\mathrm{h}}^\mathrm{ad}$.

\begin{figure}
    \centering
    \includegraphics[width=0.75\linewidth]{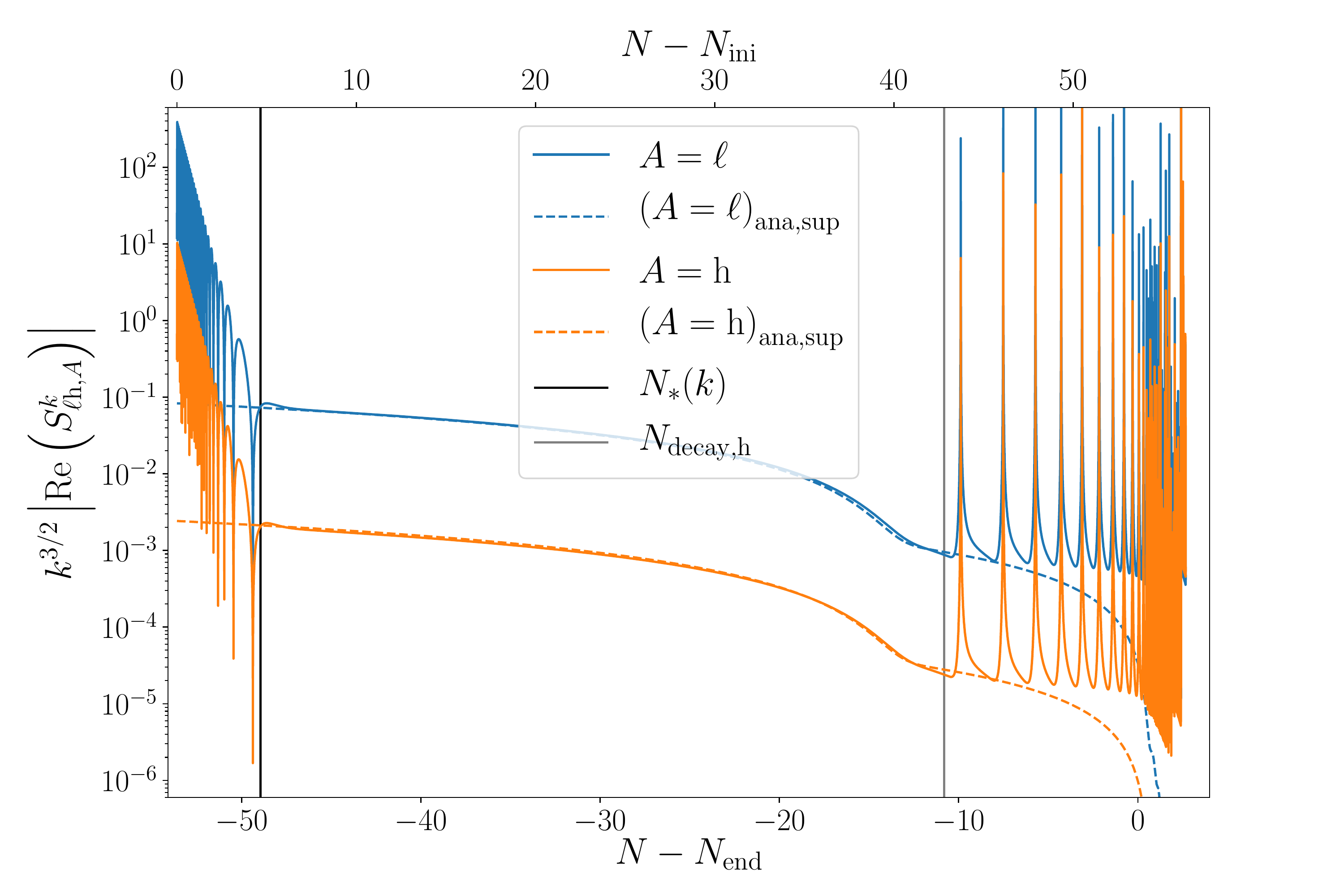}
    \caption{
    Real parts (the imaginary ones are similar) of the non-adiabatic fluctuations $S_{\ell\mathrm{h},\ell}$ and $S_{\ell\mathrm{h},\mathrm{h}}$, including the exact numerical solutions (respectively blue and orange solid lines) and their analytical approximations  on super-Hubble scales ``$\mathrm{ana,sup}$" (respectively blue and orange dashed lines).
    The approximations are again very accurate, at least until the time of decay for the heavy field from which any way the non-adiabatic component is transferred to the  ``real" cosmological fluids as we will see in the following.
    }
    \label{fig:Slh}
\end{figure}

Finally, in Fig.~\ref{fig:Slh}, we have represented the exact solutions for the real parts of the non-adiabatic perturbations during inflation, $S_{\ell \mathrm{h},A}$, compared to their analytical approximations given by Eq.~(\ref{eq:entropylh}). Again, after Hubble radius crossing, we observe a very good agreement between the two. After the heavy field's decay, the numerical solutions start to oscillate for the same reason as given above.

We conclude this section by stating that, during inflation and on super-Hubble scales, we have a good analytical control on the behavior of the various individual curvature perturbation components.

\subsubsection{Matching to the radiation-dominated era}
\label{subsubsec:matchingdouble}

\begin{figure}
    \centering
    \begin{subfigure}{\textwidth}
        \centering
        \includegraphics[width=0.75\linewidth]{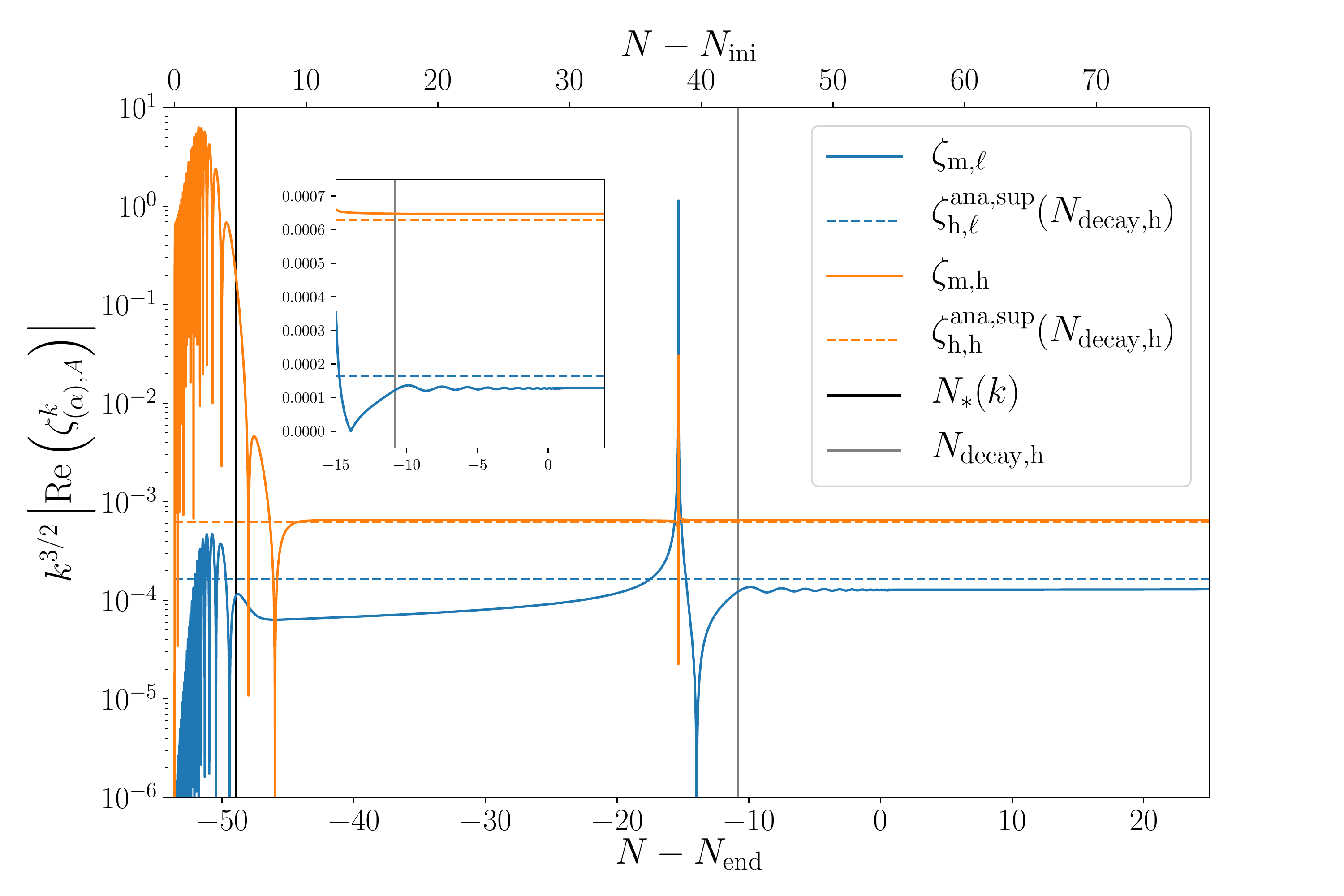}
        \caption{Matter fluid}
        \label{fig:zetam}
    \end{subfigure}%
    \\
        \centering
    \begin{subfigure}{\textwidth}
        \centering
        \includegraphics[width=0.75\linewidth]{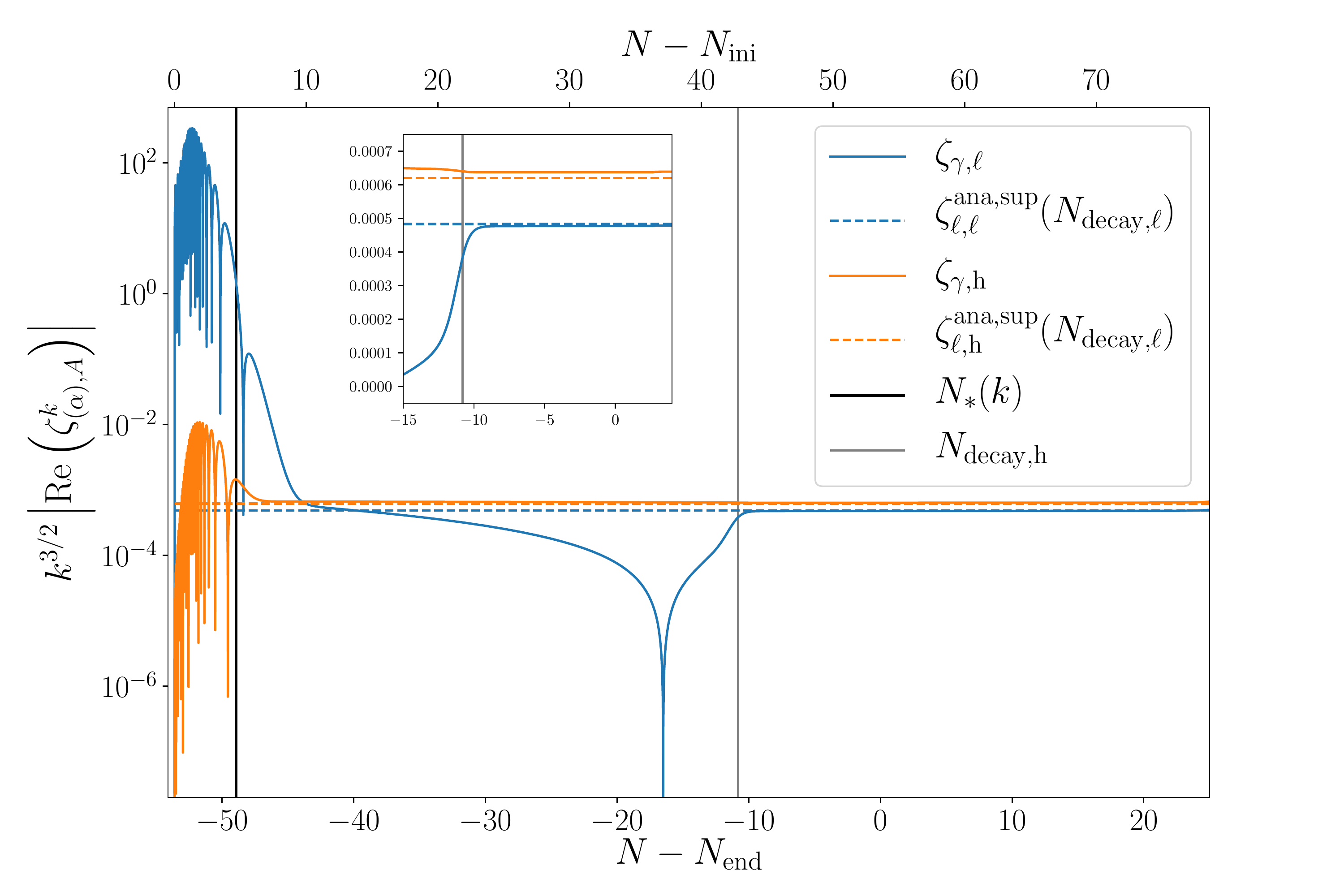}
        \caption{Radiation fluid}
        \label{fig:zetagamma}
    \end{subfigure}%
    \caption{
     Real parts (the imaginary ones are similar) of the fluids' curvature fluctuations $\zeta_{(\alpha),A}$ and their analytical approximations valid after the decay of the corresponding scalar field, as given by Eq.~\eqref{eq:continuityzeta}.
    The asymptotic solutions for the fluid's curvature perturbations in the radiation era are very accurate again (there is of course a residual error, see the inset, that will be quantified in the final result). Upper panel: $\zeta_{\mathrm{m},\ell}$ and  $\zeta_{\mathrm{m},\mathrm{h}}$ and their analytical approximations $\zeta_{\mathrm{h},A}^\mathrm{ana,sup}\left(N_\mathrm{decay,h}\right)$  (respectively blue and orange dashed lines). Lower panel: $\zeta_{\gamma,\ell}$ and  $\zeta_{\gamma,\mathrm{h}}$ (respectively blue and orange solid lines) and their analytical approximations $\zeta_{\ell,A}^\mathrm{ana,sup}\left(N_\mathrm{decay,\ell}\right)$ (respectively blue and orange dashed lines).
    }
    \label{fig:zetam-zetagamma}
\end{figure}

We are now in a position where we can turn to one of the main questions studied in this paper, namely the calculation of non-adiabatic perturbations in the post-inflationary Universe. Clearly this requires another matching, this time between  quantities defined in the inflationary phase and quantities defined in the radiation-dominated era. Since our assumption is that the heavy field decays in pressure-less matter and the light field in radiation, the fundamental property used in order to relate inflationary and post-inflationary quantities will be the continuity of the corresponding (individual) curvature perturbations at the time of decay.
This fact simply follows from the continuity of the various physical quantities through the decay of the scalar fields in perfect fluids; for example $\rho(t_\mathrm{decay,h})=\rho_\mathrm{h}(t_\mathrm{decay,h}^-)+\rho_\mathrm{other}(t_\mathrm{decay,h}^-)=\rho_\mathrm{m}(t_\mathrm{decay,h}^+)+\rho_\mathrm{other}(t_\mathrm{decay,h}^+)$ implies that $\rho_\mathrm{h}(t_\mathrm{decay,h}^-)=\rho_\mathrm{m}(t_\mathrm{decay,h}^+)$, and similar relations hold for the first derivative of the energy densities, as well as for the energy density and velocity perturbations.
Concretely, this means that, after the decays,
\begin{align}
\label{eq:continuityzeta}
    \zeta_{\mathrm{m},A}=\zeta_{\mathrm{h},A}(t_\mathrm{decay, \mathrm{h}}), \quad \zeta_{\gamma,A}=\zeta_{\ell,A}(t_\mathrm{decay, \ell}),
\end{align}
and, using Eq.~(\ref{eq:defentropypert}), it follows that
\begin{align}
\label{eq:SmgammaA}
    S_{\mathrm{m}\gamma,A}=3\left[\zeta_{\mathrm{h},A}(t_\mathrm{decay, \mathrm{h}})-\zeta_{\ell,A}(t_\mathrm{decay, \ell})\right].
\end{align}
In Fig.~\ref{fig:zetam-zetagamma}, we have represented the evolution, during and after inflation, of matter and radiation curvature perturbations in order to check the conditions~(\ref{eq:continuityzeta}). In Fig.~\ref{fig:zetam}, we have represented the real parts of $\zeta_{\mathrm{m},\ell}$ and $\zeta_{\mathrm{m},\mathrm{h}}$ as functions of the number of e-folds during inflation. The two horizontal dotted blue and orange lines corresponds to the values of $\zeta_{\mathrm{m},\ell}$ and $\zeta_{\mathrm{m},\mathrm{h}}$ (respectively) at the time of the heavy field decay. We see that, after this time (and, for $\zeta_{\mathrm{m,h}}$, it is even already true since just after Hubble radius crossing), we indeed have $\zeta_{\mathrm{m},\ell}(N)\rightarrow \zeta_{\mathrm{h},\ell}^\mathrm{ana,sup}(N_\mathrm{decay,h})$ and $\zeta_{\mathrm{m},\mathrm{h}}(N)\rightarrow \zeta_{\mathrm{h},\mathrm{h}}^\mathrm{ana,sup}(N_\mathrm{decay,h})$. We also notice that the agreement is better for the second limit than for the first one. In Fig.~\ref{fig:zetagamma}, we have plotted the real parts of $\zeta_{\gamma ,\ell}$ and $\zeta_{\gamma, \mathrm{h}}$ and studied how they tend towards $\zeta_{\ell, \ell}^\mathrm{ana,sup}(N_\mathrm{decay,\ell})$ and $\zeta_{\ell, \mathrm{h}}^\mathrm{ana,sup}(N_\mathrm{decay,\ell})$, respectively. As can be seen in the figures (and in the insets), these limits are well-verified.
We conclude that Fig.~\ref{fig:zetam-zetagamma} validate Eqs.~(\ref{eq:continuityzeta}).

\begin{figure}
    \centering
    \includegraphics[width=0.75\linewidth]{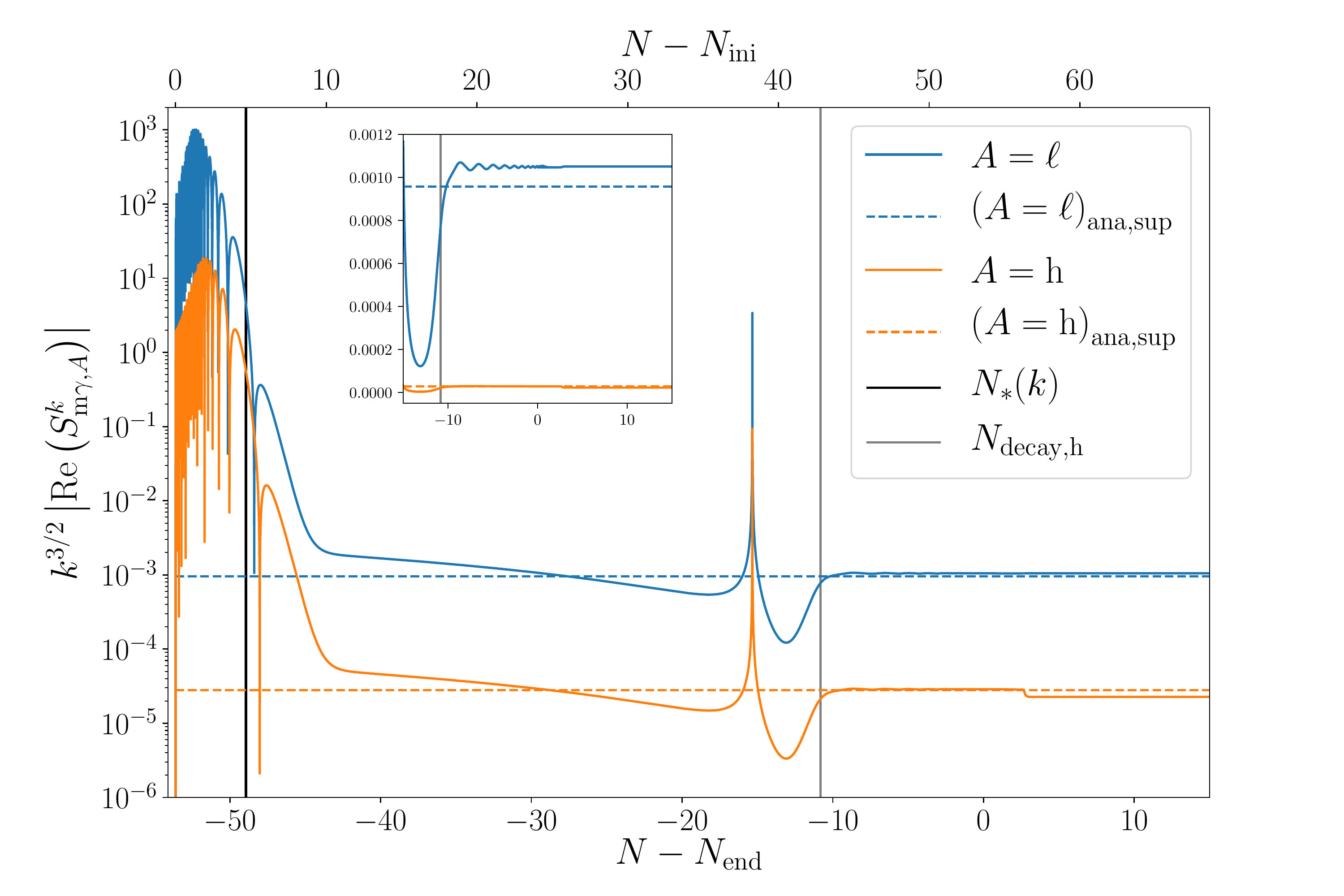}
    \caption{
    Real parts (the imaginary ones are similar) of the non-adiabatic fluid's fluctuations $S_{\mathrm{m}\gamma,\ell}$ and  $S_{\mathrm{m}\gamma,\mathrm{h}}$ (respectively blue and orange solid lines) and their analytical approximations valid after the decay of the heavy and respectively light fields (respectively blue and orange dashed lines).
    The asymptotic solutions for the fluid's non-adiabatic perturbations in the radiation era are very accurate again, and the residual errors of $\sim 5 \%$ for the component $A=\ell$ and visible in the inset, correspond to the ones of the final result.
    Note that the small drop around $N-N_\mathrm{end}=2.7$ (only visible by eye in a logarithmic scale and for the smallest component $A=\mathrm{h}$) is a numerical artifact due to the fact that the heavy field is abandoned from the numerical evolution at that time because its tiny oscillations are computationally expensive to follow (we have checked that by changing the time at which the heavy field is abandoned, the position and the height of the drop changes accordingly, but always remains negligible).
    }
    \label{fig:Smr}
\end{figure}

In order to calculate the inflationary curvature perturbations, $\zeta_{\mathrm{h},A}$ and $\zeta_{\ell,A}$, we make use of Eqs.~(\ref{eq:zetaellell}), (\ref{eq:zetaellh}), (\ref{eq:zetahh}) and~(\ref{eq:zetahell}). A first remark is that the quantities $\zeta_{\ell, \ell}^\mathrm{nad}(t_{\mathrm{decay},\ell})$ and $\zeta_{\ell, \mathrm{h}}^\mathrm{nad}(t_{\mathrm{decay},\ell})$ are proportional to $V_\mathrm{h}\vert _{N_{\mathrm{decay},\ell}}/V_\ell\vert_{N_\mathrm{m}}$
and $V_\mathrm{h}\vert _{N_{\mathrm{decay},\ell}}/V_\mathrm{h}\vert_{N_\mathrm{m}}$, respectively, see Eqs.~(\ref{eq:zetaellell}), and~(\ref{eq:zetaellh}), which implies that they are in fact sub-dominant quantities since, by the time of the light field decay, the contribution of the heavy field is highly suppressed. This can be easily checked by inspection of Figs.~\ref{fig: zeta_ell,A tot} and~\ref{fig: zeta_h,A tot}. Indeed, if one compares Fig.~\ref{fig: zeta_ell,A} and~\ref{fig: zeta_h,A} (in particular the insets), we notice that the quantities $\zeta_{\ell,\ell}$, $\zeta_{\ell, \mathrm{h}}$, $\zeta_{\mathrm{h},\ell}$ and $\zeta_{\mathrm{h},\mathrm{h}}$ are, very roughly speaking, all of the same order of magnitude at the end of inflation.
However, what matters are in fact the non-adiabatic components since, by definition, the adiabatic ones cancel out in the expression of $S_{\mathrm{m}\gamma}$; and if one now looks at Figs.~\ref{fig: zeta_ell,A ad-nad} and~\ref{fig: zeta_h,A ad-nad}, we notice that there is a clear hierarchy between the quantities $\zeta_{\ell,\ell}^\mathrm{nad}$, $\zeta_{\ell, \mathrm{h}}^\mathrm{nad}$ on one hand and $\zeta_{\mathrm{h},\ell}^\mathrm{nad}$, $\zeta_{\mathrm{h},\mathrm{h}}^\mathrm{nad}$ one the other hand, namely $\zeta_{\mathrm{h},\ell}^\mathrm{nad}, \zeta_{\mathrm{h},\mathrm{h}}^\mathrm{nad}\gg \zeta_{\ell,\ell}^\mathrm{nad}, \zeta_{\ell, \mathrm{h}}^\mathrm{nad}$. As a consequence, 
$\zeta_{\ell,\ell}^\mathrm{nad}, \zeta_{\ell, \mathrm{h}}^\mathrm{nad}$ can be neglected in Eq.~(\ref{eq:SmgammaA}) which, therefore, reduces to the following expression
\begin{equation}
       S_{\mathrm{m}\gamma,A}\simeq 3\delta_{A,\ell} \zeta_{\mathrm{h},\ell}^\mathrm{nad}\left(N_{\mathrm{decay},\mathrm{h}}\right) + 3\delta_{A,\mathrm{h}} \zeta_{\mathrm{h},\mathrm{h}}^\mathrm{nad}\left(N_{\mathrm{decay},\mathrm{h}}\right) \,.
\end{equation}
We see that the non-adiabatic perturbations in the post-inflationary Universe is in fact completely determined by what happens at the time of decay of the heavy field.

Then, in order to calculate $\zeta_{\mathrm{h},\ell}^\mathrm{nad}$ and $\zeta_{\mathrm{h},\mathrm{h}}^\mathrm{nad}$, we use Eqs.~(\ref{eq:zetahh}) and~(\ref{eq:zetahell}). Since the decay of the heavy field happens during the phase of inflation dominated by the light field, the slow-roll approximation for the background can be used, which leads to the following results:  $\phi_\ell^2 \simeq 4 \Mp^2 (N_\mathrm{end}- N)$ and $V_\mathrm{h} / V \ll 1$. As a consequence, from Eqs.~(\ref{eq:zetahh}) and~(\ref{eq:zetahell}), one obtains
\begin{align}
\label{eq:Smgammaell}
    S_{\mathrm{m}\gamma,\ell} &\simeq C_{1 \ell} \frac{12 \Mp^2}{\phi_{\ell,\mathrm{ini}}^2} \left(N_\mathrm{end}- N_{\mathrm{decay},\mathrm{h}} \right) \,, \\
\label{eq:Smgammah}
    S_{\mathrm{m}\gamma,\mathrm{h}} &\simeq - C_{1 \mathrm{h}} \frac{1}{R^2} \frac{12 \Mp^2}{\phi_{\mathrm{h}}^2(N_\mathrm{m}) } \left(N_\mathrm{end}- N_{\mathrm{decay},\mathrm{h}} \right) \,.
\end{align}
In Fig.~\ref{fig:Smr}, we have represented the real part of $S_{\mathrm{m}\gamma,A}$ as a function of the number of e-folds compared to the numerical estimates given by Eqs.~(\ref{eq:Smgammaell}) and~(\ref{eq:Smgammah}). We see that, after the decay of the heavy field, the numerical exact curves freeze out to a value which match well the analytical approximations.

Then, we calculate the non-adiabatic power spectrum of $S_{\mathrm{m}\gamma}$, $P_{S_{\mathrm{m}\gamma}}$, which involves the different components of $S_{\mathrm{m}\gamma,A}$. It can be expressed as
\begin{equation}
\label{eq:psnad}
    P_{S_{\mathrm{m}\gamma}} \propto |S_{\mathrm{m}\gamma,\ell}|^2 + |S_{\mathrm{m}\gamma,\mathrm{h}}|^2 \simeq |S_{\mathrm{m}\gamma,\ell}|^2 \left\{ 1 + \left| \frac{C_{1,\mathrm{h}}}{C_{1,\ell}} \right|^2 \frac{1}{R^4} \left[ \frac{\phi_{\ell,\mathrm{ini}}^2}{\phi_\mathrm{h}(N_\mathrm{m})}  \right]^4 \right\} \,.
\end{equation}

\begin{table}
\renewcommand{\arraystretch}{1.5} 
\begin{tabular}{| m{2.3cm} || m{1.4cm} || m{1.5cm} | m{.9cm} || m{1.5cm} | m{.9cm} || m{1.5cm} | m{.9cm} |}
    \hline
    Quantity \linebreak / Value & Numeric & Analytic $\big(N_\mathrm{decay,h}\big)$ & error & Analytic $\big(N_\mathrm{decay,h}^{_\mathrm{WKB}}\big)$ & error & Analytic $\big(N_\mathrm{decay,h}^\mathrm{ana,sd}\big)$ & error \\
    \hline
    \hline
    $\Rea S_{\mathrm{m}\gamma, \ell}$ & $-0.269$ & $-0.251$ & $6.5\%$ & $-0.241$ & $10.5\%$ & $-0.195$ & $27.4\%$ \\
    \hline
    $\Im S_{\mathrm{m}\gamma, \ell}$ & $-0.245$ & $-0.229$ & $6.6\%$ & $-0.219$ & $10.6\%$ & $-0.177$  & $27.5\%$ \\
    \hline
    $\Rea S_{\mathrm{m}\gamma, \mathrm{h}} \times 10^{3}$ & $7.35$  & $7.29 $ & $0.9\%$ & $6.98$ & $5.1\%$ & $5.66 $ & $23.1\%$  \\ 
    \hline
    $\Im S_{\mathrm{m}\gamma, \mathrm{h}} \times 10^{3} $ & $ 6.60 $  & $6.33 $ & $4.2\%$ & $6.06 $ & $8.3\%$ & $4.91$ & $25.7\%$  \\
    \hline
    \hline
    $P_{S_{\mathrm{m}\gamma}} $ & $ 0.132 $  & $0.116$ & $12.7\%$ & $0.106 $ & $20.0\%$ & $0.070$ & $47.4\%$  \\
    \hline
\end{tabular}
\caption{Numerical results versus analytical approximations following different treatments to find the time of decay of the heavy field, for the non-adiabatic fluctuations $S_{\mathrm{m}\gamma, A}$ and its power spectrum $P_{S_{\mathrm{m}\gamma}} = \left|S_{\mathrm{m}\gamma, \ell}\right|^2 + \left|S_{\mathrm{m}\gamma, \mathrm{h}}\right|^2 $ at the end of reheating and beginning of the radiation-domination era.
The analytical approximations crucially depend on the value of the time of the decay for the heavy field, so three possibilities are shown: $N_\mathrm{decay,h}=-10.81$ for the value from the numerical evolution of the background, $N_\mathrm{decay,h}^{_\mathrm{WKB}}=-10.35$ for the analytical treatment with the WKB approximation presented in the App.~\ref{app:decay} (see Fig.~\ref{fig: rho decay WKB}) and $N_\mathrm{decay,h}^\mathrm{ana,sd}=-8.39$ with the simple analytical treatment that was shown in Sec.~\ref{subsubsec:decay} not to be very precise (see Fig.~\ref{fig:rhonaive}).}
\label{table:Smr}
\end{table}

From Eqs.~(\ref{eq:Smgammaell}), (\ref{eq:Smgammah}) and~(\ref{eq:psnad}), we see that the crucial quantity that needs to be calculated in order to evaluate the amplitude of the non-adiabatic perturbations is $N_{\mathrm{decay},\mathrm{h}}$ or $N_\uend-N_{\mathrm{decay},\mathrm{h}}$. As discussed in Sec.~\ref{subsubsec:decay}, the decay time of the heavy field is the time before which the amount of cold dark matter should be negligible, and after which the heavy field should be negligible. Therefore, in Sec.~\ref{subsubsec:decay}, this time was taken to be $\rho_\mathrm{h}(N_{\mathrm{decay},\mathrm{h}}) = \rho_\mathrm{m}(N_{\mathrm{decay},\mathrm{h}})$. In this paper, given the importance of calculating $N_{\mathrm{decay},\mathrm{h}}$, we have developed different methods to estimate it. First, there are of course numerical calculations leading to a semi-analytical determination of $P_{S_{\mathrm{m}\gamma}}$, namely $P_{S_{\mathrm{m}\gamma}}$ is found analytically according to Eq.~(\ref{eq:psnad}) but, in this equation, $N_{\mathrm{decay},\mathrm{h}}$ is evaluated numerically. A second way is to use the simple considerations presented in Sec.~\ref{subsubsec:decay}, which leads to Eq.~(\ref{eq:tdecayln2}). Finally, we have also worked out a more precise approach in Appendix~\ref{app:decay}, based on the WKB approach, which leads to a third estimate of the non-adiabatic perturbations. These results are summarized in Table.~\ref{table:Smr}. We see that the first method leads to an error of $\sim 10\%$, the WKB-based one to an error of $\sim 20\% $ and, finally, the estimate of Eq.~(\ref{eq:tdecayln2}) to a $\sim 50\%$ error. In the case of the first method, since $N_{\mathrm{decay},\mathrm{h}}$ is calculated exactly, the $10\%$ error originates from the condition~(\ref{eq:continuityzeta}). Since it is difficult to see how other analytical approximations could be designed, it represents, to some extent, the incompressible error of any analytical approach.
Note that these numbers apply to the set of parameters ($R=5,\Gamma_\mathrm{hm}=10^{-5} \Mp$) that was studied in detail in this section. In Sec.~\ref{sec:comparison}, we explore other values of these parameters for which the analytical approach may be less precise, in which case the numerical resolution appears to be the only reliable method to accurately predict the amount of non-adiabatic perturbations at the end of inflation.

\section{Comparison with the standard approach}
\label{sec:comparison}

In this section, we compare our (numerical and analytical) calculations of the non-adiabatic perturbations to the calculations of Refs.~\cite{Polarski:1992dq,Peter:1994dx} and Ref.~\cite{Langlois:1999dw}, which also derive an analytical formula for $P_{S_{\mathrm{m}\gamma}}$ in the double inflation scenario.

Before carrying out this comparison, we first recall, very briefly, how the results of Refs.~\cite{Polarski:1992dq,Peter:1994dx} and Ref.~\cite{Langlois:1999dw} are obtained. Since the details of reheating and of the decay process are neglected in those references, each fluid is always individually conserved and the amplitude of the non-adiabatic modes can be defined by Eq.~(\ref{eq:naddeltarho}), namely
\begin{align}
    S_{\mathrm{m}\gamma}=\frac{\delta \rho_\mathrm{m}}{\rho_\mathrm{m}}
    -\frac{3}{4}\frac{\delta \rho_\gamma}{\rho_\gamma}
    \simeq \frac{\delta \rho_\mathrm{m}}{\rho_\mathrm{m}},
\end{align}
where the last expression is again the fact, discussed in the previous section, that the amplitude of the non-adiabatic perturbation is dominated by what happens at the heavy field decay. The question is then to relate $\delta \rho _\mathrm{m}/\rho_\mathrm{m}$ to $\delta {\rho_\mathrm{h}}/\rho_\mathrm{h}$. In the literature, $\delta \phi_\mathrm{h}/\phi_\mathrm{h}$ is taken to
be constant during the oscillations of the heavy field (because $\delta \phi_\mathrm{h}$ and $\phi_\mathrm{h}$ obey the same equation of motion in this regime) which covers the phase between the heavy field decay and the end of inflation. At the end of inflation, one can thus write
\begin{align}
\label{eq:matchinghm}
  \frac{\delta \rho_\mathrm{m}}{\rho_\mathrm{m}}\sim \frac{\delta \rho_\mathrm{h}}{\rho_\mathrm{h}} \sim
  2 \frac{\delta \phi_\mathrm{h}}{\phi_\mathrm{h}},
\end{align}
which comes from the continuity of energy density and
$\rho_\mathrm{h}\sim m_\mathrm{h}^2\langle \phi_\mathrm{h}\rangle
^2/2$.

Then, our next goal is to make the connection between the previous considerations and the super Hubble slow-roll solution given by Eq.~(\ref{eq:soldphihsr}), which can be written as 
\begin{align}
    \delta \phi_\mathrm{h}=\frac{C_1}{H}\dot{\phi}_\mathrm{h}+2HC_3\frac{V_\ell}{V_\ell+V_\mathrm{h}}\dot{\phi}_\mathrm{h}=-2C_1\frac{\Mp^2}{\phi_\mathrm{h}}\frac{V_\mathrm{h}}{V_\mathrm{h}+V_\ell}-\frac{2}{3}
    C_3\frac{\partial V_\mathrm{h}}{\partial \phi_\mathrm{h}}\frac{V_\ell}{V_\mathrm{h}+V_\ell}.
\end{align}
From this expression, we conclude that, if the light field dominates ($V_\ell \gg V_\mathrm{h}$), then the first branch of the solution $\propto C_1$ can be ignored and one has 
\begin{align}
    \frac{\delta \phi_\mathrm{h}}{\phi_\mathrm{h}}
    \simeq -\frac23 C_3 m_\mathrm{h}^2.
\end{align}
As a consequence, the quantity $\delta \phi_\mathrm{h}/\phi_\mathrm{h}$ also remains constant once the light field has started to dominate, which occurs slightly before the decay of the heavy field (if one can really define a time of decay for the heavy field, see the discussion in the previous sections). In some sense, these considerations allow us to fix the value of the constant $\delta \phi_\mathrm{h}/\phi_\mathrm{h}$ at the end of inflation in 
Eq.~(\ref{eq:matchinghm}) and one can write
\begin{align}
  S_{\mathrm{m}\gamma}\simeq\frac{\delta \rho_\mathrm{m}}{\rho_\mathrm{m}}\simeq  -\frac43 m_\mathrm{h}^2C_3.
  \end{align}
One recovers the fact that non-adiabatic perturbations solely depend on the constant $C_3$.

Then, the constant $C_3$ is evaluated with the help of Eq.~(\ref{eq:c3a}). In this formula, the quantities $\delta \phi_\ell$ and $\delta \phi_\mathrm{h}$ are written
\begin{align}
    \delta \phi_\ell =\frac{H_*}{\sqrt{2k^3}}e_\ell, 
    \quad \delta \phi_\mathrm{h} =\frac{H_*}{\sqrt{2k^3}}e_\mathrm{h},
\end{align}
where we recall that $H_*$ denotes the value of the Hubble parameter at horizon crossing and $e_i$ ($i=\ell,\mathrm{h}$) are classical random Gaussian processes with $\langle e_i({\bm k})\rangle =0$ and $\langle e_i({\bm k})e_j^*({\bm k}')\rangle=\delta_{ij}\delta({\bm k}-{\bm k}')$. It follows that Eq.~(\ref{eq:c3a}) can be re-written as
\begin{align}
    S_{\mathrm{m}\gamma}=\sqrt{\frac{2}{k^3}}H_*
    \left(\frac{e_\mathrm{h}}{\phi_\mathrm{h}}-R^2\frac{e_\ell}{\phi_\ell}\right).
\end{align}
As a consequence, using Eqs.~(\ref{eq:fieldsr}), (\ref{eq:sfunctiontheta}) and~(\ref{eq:Hubblegene}), one arrives at
\begin{align}
\label{eq: Smgamma Langlois}
    P_{S_{\mathrm{m}\gamma}}=\left\vert S_{\mathrm{m}\gamma}\right\vert^2
    =\frac{1}{3k^3}\frac{m_\ell^2}{\Mp^2}
    \left[1+(R^2-1)\sin ^2 \theta_*\right]
    \left(\frac{1}{\sin ^2\theta_*}+\frac{R^4}{\cos^2 \theta_*}\right),
\end{align}
an equation that should be compared to Eq.~(\ref{eq:psnad}). Numerically, this equation gives (we use $\theta_*=0.962$ from $N_*=4.67$ determined numerically and $-N_*+N_\mathrm{end}\simeq C/\mathrm{cos}^2\theta_*$) $P_{S_{\mathrm{m}\gamma}}=0.072$, to be compared with the numerical result $P_{S_{\mathrm{m}\gamma}}=0.132$ and the analytical approximations presented in this work: the relative error found with Eq.~\eqref{eq: Smgamma Langlois} is $45.5 \%$, which is comparable to our worst analytical estimate with $N_\mathrm{decay,h}^\mathrm{ana,sd}$, see Table.~\ref{table:Smr}. 

Therefore, we conclude that modeling the reheating mechanism is crucial in order to get an accurate prediction for non-adiabatic perturbations after the end of multi-field inflation, in the post-inflationary Universe. Ignoring the details of reheating can indeed lead to an error as big as $\sim 50\%$ (or worse for some other values of the parameters, see below)

It is also interesting to compare the amplitude of adiabatic and non-adiabatic perturbations. In the radiation-dominated era, we have $\zeta_\gamma \simeq \zeta_\ell^\mathrm{ad}=-C_1$, see Eq.~(\ref{eq:zetaadia}). Using Eq.~(\ref{eq:c1a}), one has
\begin{align}
    \zeta_\gamma=\frac{1}{2\Mp^2}\frac{H_*}{\sqrt{2k^3}}
    \left(\phi_\mathrm{h}e_\mathrm{h}+\phi_\ell e_\ell\right),
\end{align}
from which we deduce
\begin{align}
    P_{\zeta_\gamma}=\frac{C^2}{3k^3}\frac{m_\ell^2}{\Mp^2}
    \left[1+(R^2-1)\sin ^2 \theta_*\right]
    \frac{(\tan \theta _*)^{4/(R^2-1)}}{\cos ^4\theta_*}.
\end{align}
It follows, using the expression~(\ref{eq:defc}) of the constant $C$, that the non-adiabatic to adiabatic power spectrum ratio can be expressed as
\begin{align}
\label{eq:ratiostandard}
    \frac{P_{S_{\mathrm{m}\gamma}}}{P_{\zeta_\gamma}}=
    \frac{16\Mp^4}{\phi_\ell^4\vert_\uini}
    \left(\tan \theta_*\right)^{4/(1-R^2)}
    \left(\frac{\cos ^4\theta_*}{\sin ^2\theta_*}+R^4\cos ^2 \theta_*\right).
\end{align}
For our fiducial parameters used above (leading to $\theta_*\simeq 0.962$), one finds $P_{S_{\mathrm{m}\gamma}}/P_{\zeta_\gamma}\simeq 0.75$. This should be compared to our results, $P_{S_{\mathrm{m}\gamma}}/P_{\zeta_\gamma}\simeq 1.65$ numerically and $P_{S_{\mathrm{m}\gamma}}/P_{\zeta_\gamma}\simeq (1.60,1.46,0.96)$ analytically with the three possible choices of $N_\mathrm{decay,h}$ as explained in Sec.~\ref{subsubsec:matchingdouble}. Notice that the precision on the non-adiabatic power spectrum to adiabatic power spectrum ratio seems better than the numbers quoted in Tables.~\ref{table:C1A-and-C3A} and~\ref{table:Smr} (for instance, it is $3\%$ with an exact $N_\mathrm{decay,h}$). This is due to the fact that both quantities are underestimated and, as a consequence, the error in their ratio is smaller than the error in each of them.

\begin{figure}
    \centering
    \begin{subfigure}{\textwidth}
        \centering
        \includegraphics[width=0.8\linewidth]{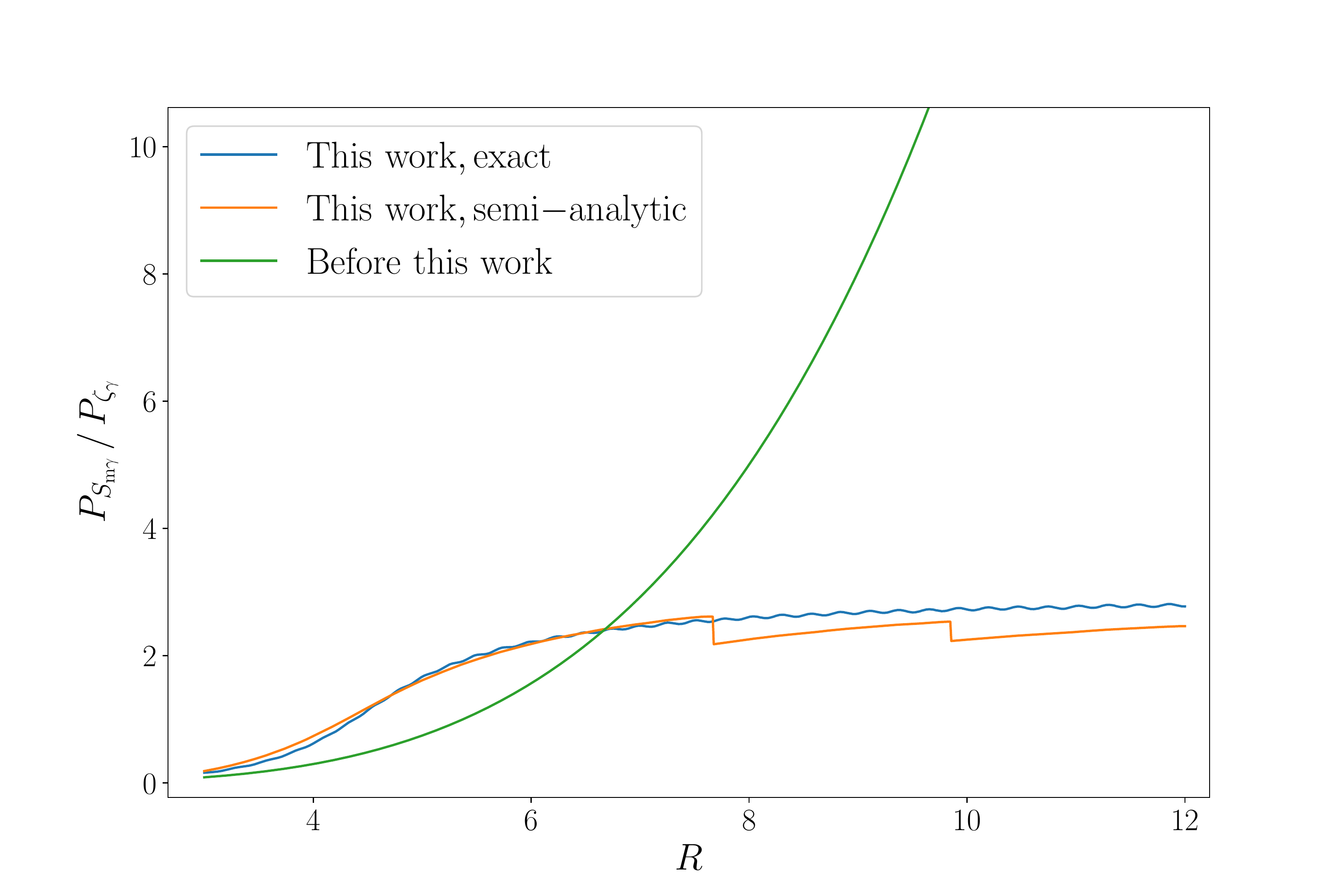}
        \caption{Ratio $P_{S_{\mathrm{m}\gamma}}/P_{\zeta_\gamma}$ as a function of $R$}
    \end{subfigure}%
    \\
        \centering
    \begin{subfigure}{\textwidth}
        \centering
        \includegraphics[width=0.8\linewidth]{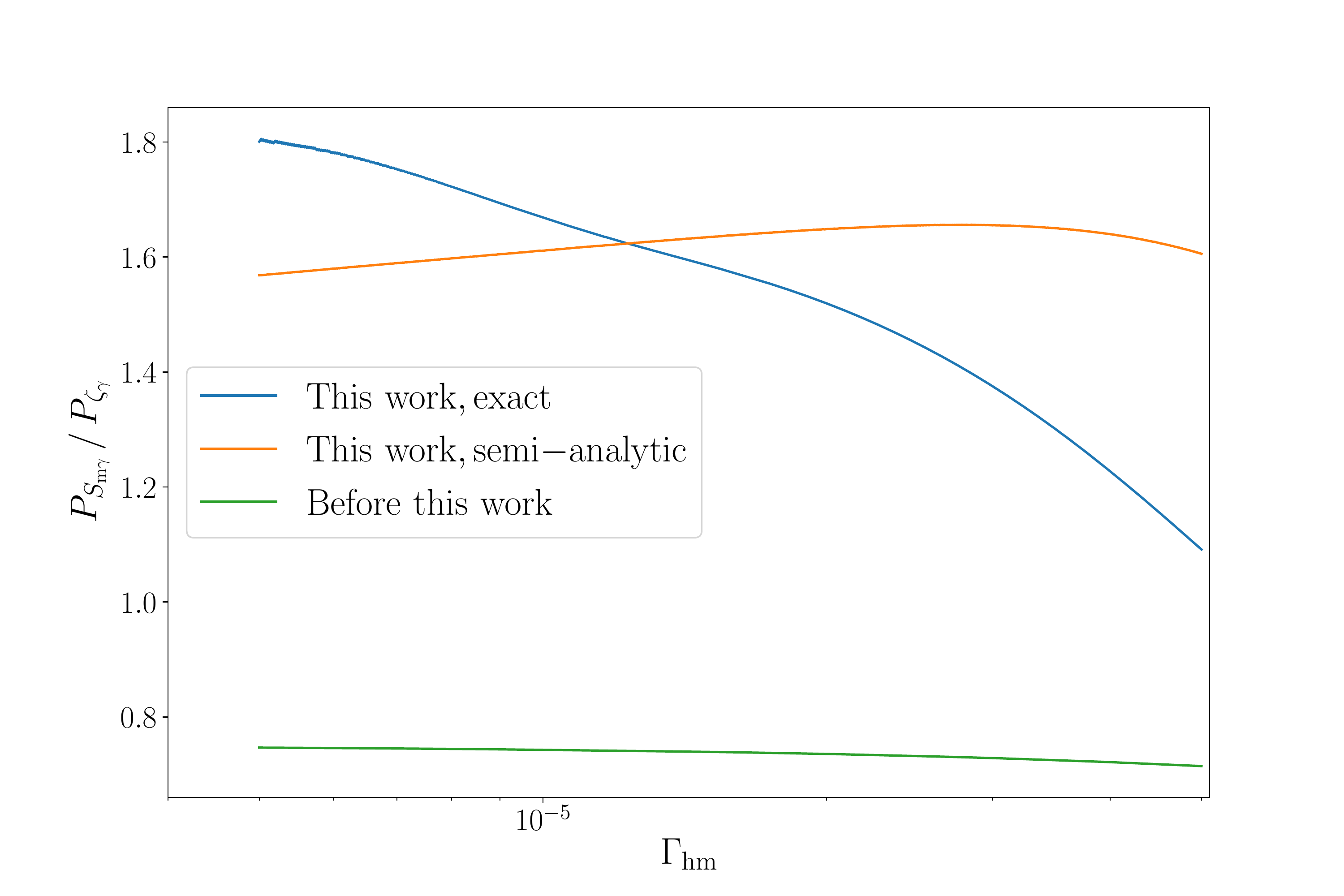}
        \caption{Ratio $P_{S_{\mathrm{m}\gamma}}/P_{\zeta_\gamma}$ as a function of $\Gamma_\mathrm{hm}$ (in $\Mp$ units)}
    \end{subfigure}%
    \caption{Ratio $P_{S_{\mathrm{m}\gamma}}/P_{\zeta_\gamma}$
    as a function of the parameter $R$, panel (a), and as a function of the parameter $\Gamma_\mathrm{hm}$ (in $\Mp$ units), panel (b). The blue line represents the exact result, obtained by means of numerical calculations. The orange line corresponds to the analytical approach developed in this paper with $N_\mathrm{decay,h}$ determined numerically. Finally, the green line corresponds to the standard formulation, see Eq.~(\ref{eq:ratiostandard}).}
    \label{fig:varying-params}
\end{figure}

In Fig.~\ref{fig:varying-params}, we have represented the ratio $P_{S_{\mathrm{m}\gamma}}/P_{\zeta_\gamma}$ versus $R$, for a fixed $\Gamma_{\mathrm{hm}}$ (taken at its fiducial value $\Gamma_{\mathrm{hm}} = 1.0 \times 10^{-5} \Mp$ ) and versus $\Gamma_{\mathrm{hm}}$ for a fixed $R$ (also taken at its fiducial value $R=5$) for three cases: the blue curve corresponds to the exact (numerical) result, the orange curve corresponds to the semi-analytical result obtained in this paper and the green curve is obtained from Eq.~(\ref{eq:ratiostandard}). In panel (a), we see that our semi-analytical result reproduces well the exact result while the standard formulation poorly performs, especially for $R\gtrsim 8$. The two drops that can be seen on the orange curve are due to the way $N_\mathrm{decay,h}$ is calculated. 
$N_\mathrm{decay,h}$ is obtained by determining when $\rho_\mathrm{h}$ and $\rho_\mathrm{m}$ intersects. Since these two quantities oscillate in this regime, the intersection can occur around one specific oscillation and then, suddenly, when the value of $R$ changes, jumps to another oscillation resulting in the behavior observed in Fig.~\ref{fig:varying-params}.
In panel (b), we notice that the agreement between the exact and analytical results (blue and orange lines) is still quite good but, admittedly, less good than in panel (a). On the other hand, the standard formula (green line) still poorly performs.
Notice that the latter is almost independent of $\Gamma_{\mathrm{hm}}$, a slight dependence being nevertheless present through the value of $\theta_*$. We also remark that the agreement between the numerical and analytical results deteriorates when  $\Gamma_{\mathrm{hm}}$ increases compared to our fiducial value $\Gamma_{\mathrm{hm}}=1.0 \times 10^{-5} \Mp$.
However, when this is the case, one enters a regime where, in the Klein-Gordon equation, the contribution of the term proportional to $\Gamma_{\mathrm{hm}}$ is no longer negligible compared to $H_\uini$ at the onset of inflation.
This means that we are in a situation which is reminiscent of warm inflation where the considerations presented in this work are no longer valid.
Moreover, although the departure is less visible, it seems also that the analytical approximation is less accurate when $\Gamma_{\mathrm{hm}}$ decreases too much.
But again, this can be understood: if $\Gamma_{\mathrm{hm}}$ is very small, then the instantaneous decay approximation for the heavy field is even less justified, and even though one can still define the time $N_\mathrm{decay,h}$ as the time when $\rho_\mathrm{h}=\rho_\mathrm{m}$, it becomes less relevant to use it as a particularly interesting time.
This shows once more that, in the general case, the system should be solved numerically in order to derive reliable predictions for the cosmological observables.

\begin{figure}
    \centering
    \begin{subfigure}{0.5\textwidth}
        \centering
        \includegraphics[width=1.\linewidth]{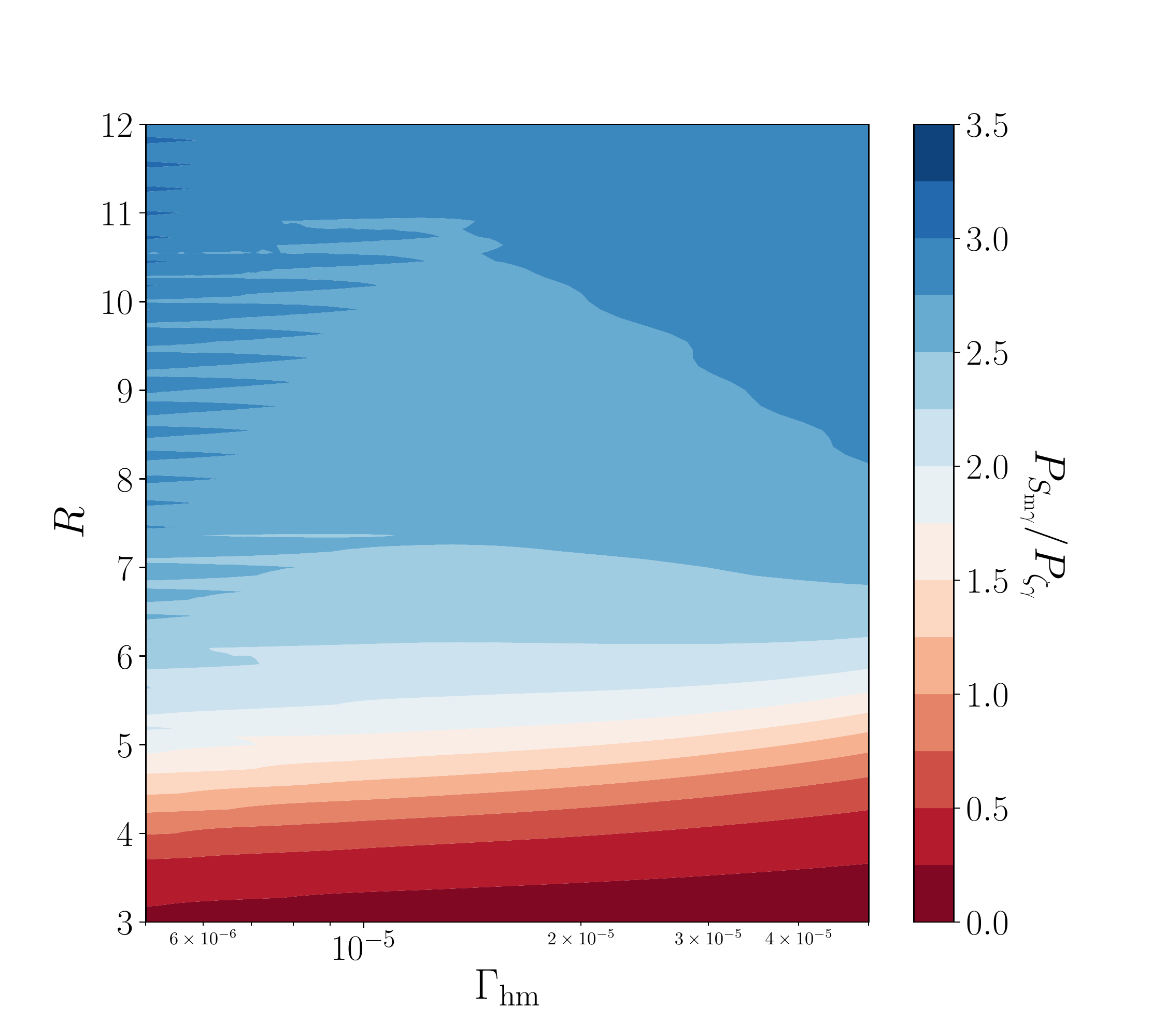}
        \caption{Full numerical result.}
    \end{subfigure}%
    \hfill
    \begin{subfigure}{0.5\textwidth}
        \centering
        \includegraphics[width=1.\linewidth]{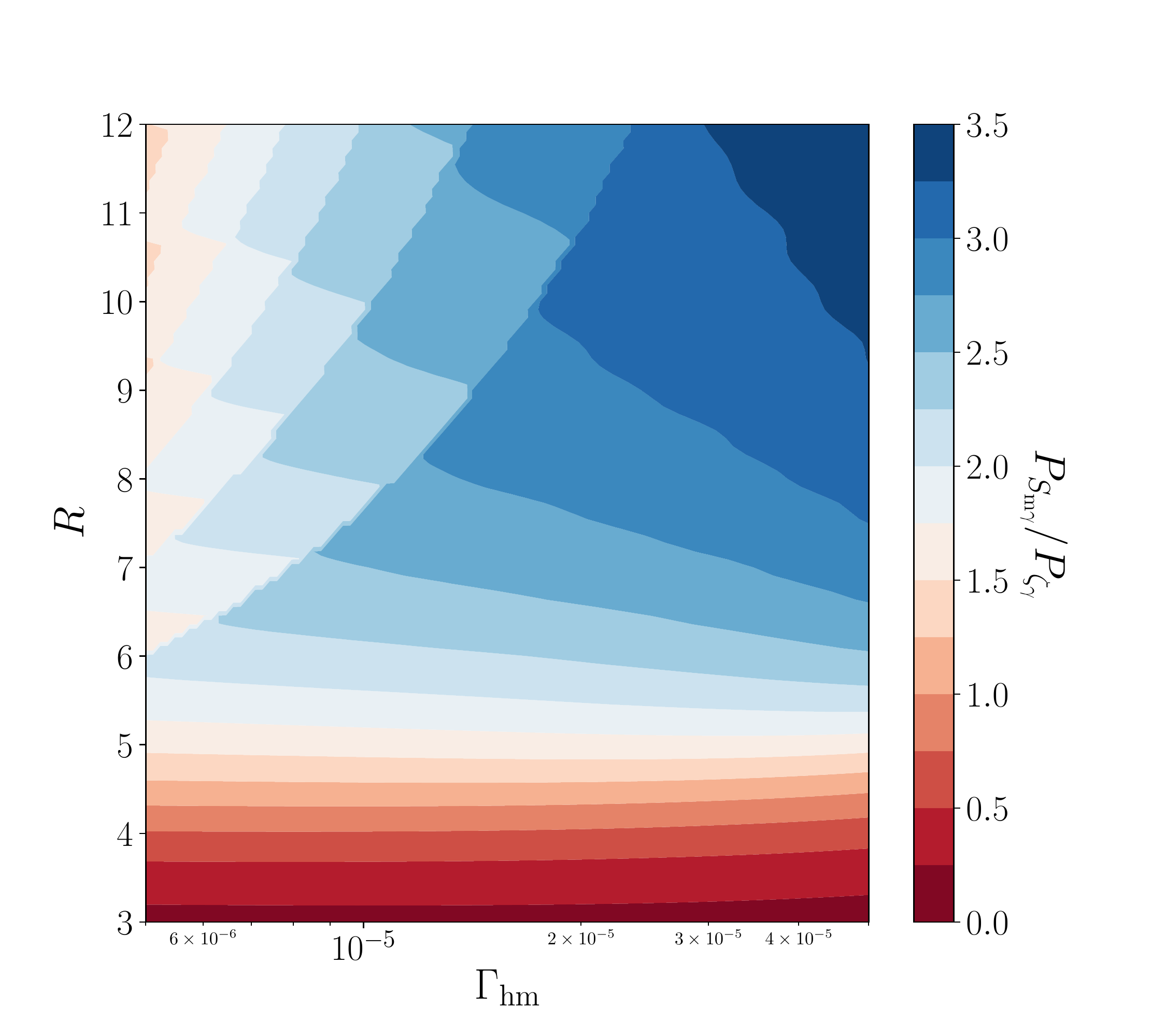}
        \caption{Analytical (approximated) result.}
    \end{subfigure}
    \caption{Contour levels of the ratio $P_{S_{\mathrm{m}\gamma}}/P_{\zeta_\gamma}$ as a function of $R$ and $\Gamma_\mathrm{hm}$ (in $\Mp$ units). The figure in the left panel is obtained by numerical calculations and is the results of $10000$ simulations (namely $100$ values for $R$ and $100$ values for $\Gamma_\mathrm{hm}$). The right panel is obtained by means of the analytical approximations discussed in this work.}
    \label{fig:varying-params2d}
\end{figure}

In Fig.~\ref{fig:varying-params2d}, we present a comparison of the ratios  $P_{S_{\mathrm{m}\gamma}}/P_{\zeta_\gamma}$, as a function of the parameters $R$ and $\Gamma_\mathrm{hm}$, obtained numerically and analytically. This confirms that our analytical approximations correctly reproduce the exact results, especially for $R\lesssim 7$. For larger values of $R$, the agreement is less good and we observe a complicated structure in parameter space. As already mentioned above, this structure is due to difficulties in handling $N_\mathrm{decay,h}$. This structure is the two-dimensional manifestation of the drops observed in the orange line in  Fig.~\ref{fig:varying-params}, panel (a). Fig.~\ref{fig:varying-params2d} also confirms that the dependence of $P_{S_{\mathrm{m}\gamma}}/P_{\zeta_\gamma}$ is stronger in the $R$-direction than in the $\Gamma_\mathrm{hm}$, a property which can also be noticed 
in Fig.~\ref{fig:varying-params}. However, numerical calculations indicate that this dependence is less flat than suggested by analytical approximations.

We conclude by stressing again that a detailed modeling of the reheating epoch can affect the predictions of multi-field scenarios and in a regime in which non-adiabatic perturbations represent a non-negligible fraction of the adiabatic ones.

\section{Conclusions}
\label{sec:conclusions}

In multi-field inflation, because curvature perturbations are not necessarily conserved on large scales, one expects the predictions of a model to depend on what happens during reheating, and the main goal of this article was to investigate whether this effect can indeed be relevant. Moreover, beyond being a question of principle, this issue is also important, and timely, because future, high-accuracy, experiments could  maybe, open up a new window on the micro-physics of inflation. 

In order to carry out the aforementioned task at the practical level, we have introduced a detailed description of reheating which explicitly takes into account the interactions between the inflaton fields and their decay products. As an illustration of a property that can be affected by the physical processes at play during reheating, we have chosen to consider the amplitude of non-adiabatic perturbations in the post-inflationary Universe. 

We have then developed new numerical codes, and new analytical techniques, allowing us to track the behavior of the perturbations throughout reheating. Concretely, these tools were used in the context of a specific model, namely double inflation.

The main result of this study is that, indeed, in the context of multi-field inflation, the predictions can be quite sensitive to the details of what is going on during reheating. In the case studied here, we have estimated that different modelings of the reheating phase can result in predictions that differ by almost $\sim 50\%$. Of course, this number is to be taken with a grain of salt given that it was obtained for a particular model and for a particular observable. However, we think it illustrates well the main conclusion of this article. Another conclusion is that it is possible to develop analytical techniques which can reasonably approximate the exact calculations, at the $\sim 10\%$ level. However, for a more realistic and, therefore, more complicated model (with, for instance, potential and kinetic interactions between the inflaton fields, more complicated potentials, more decay channels \dots), it is very likely that no analytical approach will be able to reach a good level of accuracy. In this respect, the codes that we have developed for this paper should be very useful for future investigations in multi-field inflation.

As a final remark, let us notice that it would be interesting to test the robustness of the above mentioned lessons by exploring whether other effects in double inflation, new models and/or new observables are also dependent on the details of reheating at the level found before. For instance, one possibility would be that the inflaton fields decay in other scalar fields rather than in perfect fluids. For example, the light field could decay into the field $\chi$ with $-{\cal L}_\mathrm{matter}=(\partial \chi)^2/2+V(\chi)$ and $V(\chi)\propto \mathrm{exp}\left[-\sqrt{3(1+w)}\chi/\Mp\right]$, where $w$ is a free parameter~\cite{Gonzalez:2018jax}. This potential is used in the model of power-law inflation. An important feature of this model is that the equation of state of $\chi$ is constant and given by $w$ (hence the name). In the context of power-law inflation, $w$ is chosen sufficiently close to minus one which makes the potential very flat as required for an inflationary model. However, if $w=1/3$, and, therefore, if the potential is rather steep, then when the inflaton fields have decayed away and the energy budget of the Universe is dominated by the field $\chi$, a radiation-dominated epoch starts. In other words, a field $\chi$ with the above potential provides an example of a microscopic description of radiation. Then, the interaction between $\chi$ and the inflaton field could typically be taken as $-{\cal L}_\mathrm{int}=\sigma \phi_\ell \chi^2$, where $\sigma$ is a constant of dimension one. This Lagrangian describes the decay of one ``$\phi_\ell $- particle" into two ``$\chi$-particles". This represents an interesting example where the Lagrangian~(\ref{eq:doublelagrangian}) is entirely explicit. Then, the calculations of the present article could be carried out again with this new model in order to see if similar conclusions are reached. We hope to return to these questions soon.

\begin{acknowledgments}
We would like to thank David Kaiser for insightful discussions after the publication of the first version of this manuscript.
\end{acknowledgments}

\appendix

\section{Background evolution at heavy field decay}
\label{app:decay}

In this appendix, we consider again the problem of modeling the behavior of the background around the decay time of the heavy field. This question was already treated in Sec.~\ref{subsubsec:decay} and our aim in this appendix is to improve this treatment. Here, we study in detail the behavior of the system when the heavy field leaves the slow-roll regime and starts to oscillate and to decay. As argued before, see Sec.~\ref{subsubsec:matchingdouble}, this is important because tracking what happens to the background in this regime with good accuracy turns out to be crucial in order to predict the level of non-adiabatic perturbations that remains after the end of inflation.

\subsection{Time-dependent frequency}
\label{subapp:w}

\begin{figure}
\begin{center}
        \includegraphics[width=0.75\linewidth]{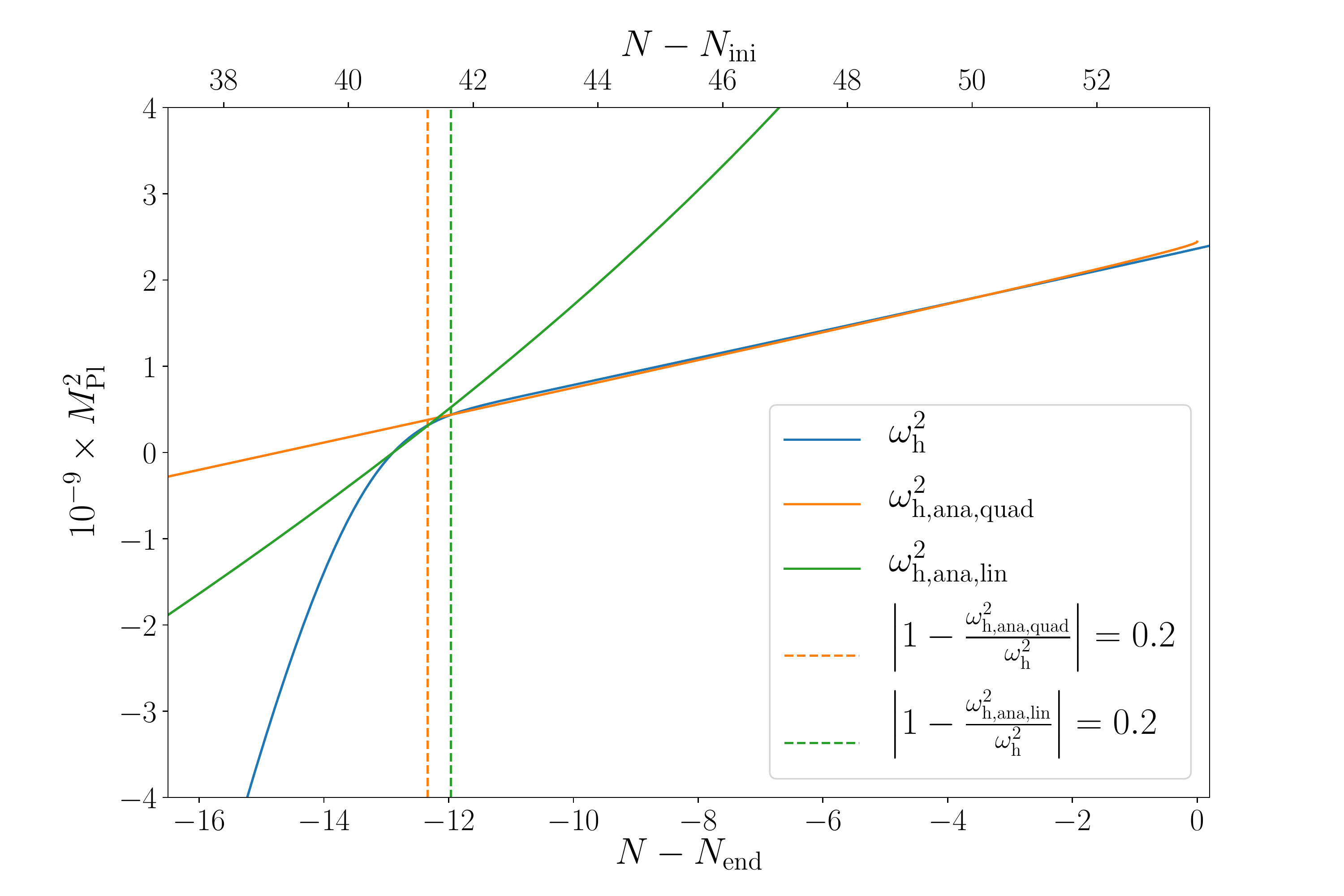}
    \caption{Squared pulsation $\omega_\mathrm{h}^2$, defined in Eq.~(\ref{eq:defwh}), versus number of e-folds: exact numerical result (blue line), quadratic approximation $\omega_\mathrm{h,ana,quad}^2$  (orange line) given by Eq.~\eqref{eq:quadraticw2} and that becomes roughly valid after the time represented by the orange vertical dashed line, and linear approximation $\omega_\mathrm{h,ana,lin}^2$ (green line) given by Eq.~ \eqref{eq:linearw2} and that is roughly valid until the time represented by the green vertical dashed line.
    Both approximations are faithful in the time domain between the two vertical dashed lines, which also coincides roughly with the time at which the WKB solution becomes correct (see the next figure).}
    \label{fig:omegas}
\end{center}
\end{figure}

Since we know that the heavy field decays and oscillates, as it was already done in Sec.~\ref{subsubsec:decay}, it is especially convenient to write this quantity as
\begin{align}
\label{eq:defg}
\phi_\mathrm{h}(t)=\phi_{\mathrm{h},\mathrm{p}}\left(\frac{a_\mathrm{p}}{a}\right)^{3/2} e^{-\Gamma_{\mathrm{hm}}(t-t_\mathrm{p})/4}\, g_\mathrm{h}(t), 
\end{align}
with ``$\mathrm{p}$" meaning evaluated at a particular time. The quantity $\phi_\mathrm{h,p}$ is the vacuum expectation value of the heavy field at this particular time. This particular time will be specified soon. Eq.~(\ref{eq:defg}) defines the new function $g_\mathrm{h}(t)$. Then, using Eq.~(\ref{eq:KGdecaylh}), it is straightforward to show that $g_\mathrm{h}(t)$ obeys the equation [see also Eqs.~(\ref{eq:g(t)text}) and~(\ref{eq:defw2text}) where these quantities were already discussed in the main text], 
\begin{align}
\label{eq:g(t)}
    \ddot{g}_\mathrm{h}(t)+\omega^2_\mathrm{h}(t)g_\mathrm{h}(t)=0,
\end{align}
with
\begin{align}
\label{eq:defwh}
\omega^2_\mathrm{h}(t)=m_\mathrm{h}^2\left(1-\frac32\epsilon_1 \frac{H^2}{m_\mathrm{h}^2}
    -\frac94 \frac{H^2}{m_\mathrm{h}^2}-\frac{1}{16}\frac{\Gamma_\mathrm{hm}^2}{m_\mathrm{h}^2}-\frac34 \frac{H}{m_\mathrm{h}}\frac{\Gamma_{\mathrm{hm}}}{m_\mathrm{h}}\right).
\end{align}
This effective frequency is represented in Fig.~\ref{fig:omegas} (blue solid line). Analytically, viewed as a function of cosmic time, the explicit form of $\omega^2_\mathrm{h}(t)$ is complicated and, as a consequence, Eq.~(\ref{eq:g(t)}) cannot be integrated exactly. We see that, at the beginning of inflation, $\omega^2_\mathrm{h}(t)$ is negative, vanishes at a time that we denote in the following by $t_\mathrm{osc}$, which is a turning point, and then becomes positive which entails the presence of oscillations in the solution.

The position of the turning point can be determined with the help of the following considerations. In the neighborhood of the turning point, one can use an approximation for $\omega_\mathrm{h}(t)$ and write $\omega^2_\mathrm{h}\simeq m_\mathrm{h}^2-9H^2/4$. The turning point is therefore located at a time when $H_\mathrm{osc}\simeq 2m_\mathrm{h}/3$ and using the expression~(\ref{eq:Hubblegene}), it occurs at a value of $\tan \theta_\mathrm{osc}$ which is a solution to the following equation
\begin{align}
\label{eq:eqosc}
    \left(\tan \theta_\mathrm{osc}\right)^{2/(R^2-1)}\left(1+R^2\tan^2 \theta_\mathrm{osc}\right)=\frac{2R^2}{3C}.
\end{align}
This equation is a polynomial equation in $\tan ^2\theta_\mathrm{osc}$ which, unfortunately, cannot be solved exactly. However, one can use Newton's method to find an approximate solution. Starting from the initial guess $\tan \theta_\mathrm{osc}^2\sim R^{-2}$, the first iteration leads to
\begin{equation}
\label{eq:soloscapprox}
    \tan ^2\theta_\mathrm{osc}\simeq \frac{3 - R^2 + 2R^2(R^2-1)R^{2/(R^2-1)}/(3C) }{R^2(1+R^2)} \,.
\end{equation}
The exact (numerical) solution to Eq.~(\ref{eq:eqosc}), for the fiducial parameters, is $\tan ^2\theta_\mathrm{osc}\simeq 0.0101$ to be compared with the solution~(\ref{eq:soloscapprox}), $\tan ^2\theta_\mathrm{osc}\simeq 0.0104$, and corresponding to $N_\mathrm{osc}-N_\mathrm{end}=-13.3$, to be compared with the exact time found numerically, $N_\mathrm{osc}-N_\mathrm{end}=-12.89$.

As noticed above, the form of $\omega_\mathrm{h}^2(t)$ is complicated. However, depending on the regimes considered, some useful approximations can be found. For instance, in the regime dominated by the light field, the Hubble parameter simplifies and is given by Eq.~(\ref{eq:hubblelightphase}), $H^2=2m_\ell^2(N_\mathrm{end}-N)/3$. Moreover, given that $\dd N=H\dd t$, the relationship between the number of e-folds and the cosmic time can be expressed as [this equation was already derived, see Eq.~(\ref{eq:linktN}), and we reproduce it here for convenience]
\begin{align}
\label{eq:linktNapp}
    t-t_\mathrm{p}=-\frac{\sqrt{6}}{m_\ell}\left[\left(N_\mathrm{end}-N\right)^{1/2}-\left(N_\mathrm{end}-N_\mathrm{p}\right)^{1/2}\right].
\end{align}
As a consequence, inserting those equations in Eq.~(\ref{eq:defwh}), one finds that $\omega^2_\mathrm{h}(t)$ reduces to a quadratic polynomial in time, namely
\begin{align}
\label{eq:quadraticw2}
\omega^2_\mathrm{h}(t)\simeq \bar{c}
+\bar{b}\, m_\ell\left(t-t_\mathrm{p}\right)    
+\bar{a}\, m_\ell^2\left(t-t_\mathrm{p}\right)^2,
\end{align}
with
\begin{align}
    \bar{a}&=-\frac14 m_\ell^2 \,,
    \\
    \bar{b}&= \frac14 m_\ell
    \Gamma_\mathrm{hm} +  \sqrt{\frac32} m_\ell^2 \sqrt{N_\mathrm{end}-N_\mathrm{p}} \,, 
    \\
    \bar{c} &= m_\mathrm{h}^2- \frac12 m_\ell^2  -\frac{\Gamma^2_\mathrm{hm}}{16}
    -\sqrt{\frac38}\Gamma_{\mathrm{hm}} m_\ell \sqrt{N_\mathrm{end}-N_\mathrm{p}}-\frac32 m_\ell^2\left(N_\mathrm{end}-N_\mathrm{p}\right).
\end{align}
This approximation is represented and compared to the exact $\omega^2_\mathrm{h}(t)$ in Fig.~\ref{fig:omegas} (solid orange line).

\subsection{The WKB approximation for the heavy field}
\label{subapp:wkb}

\begin{figure}
\begin{center}
    \includegraphics[width=0.75\linewidth]{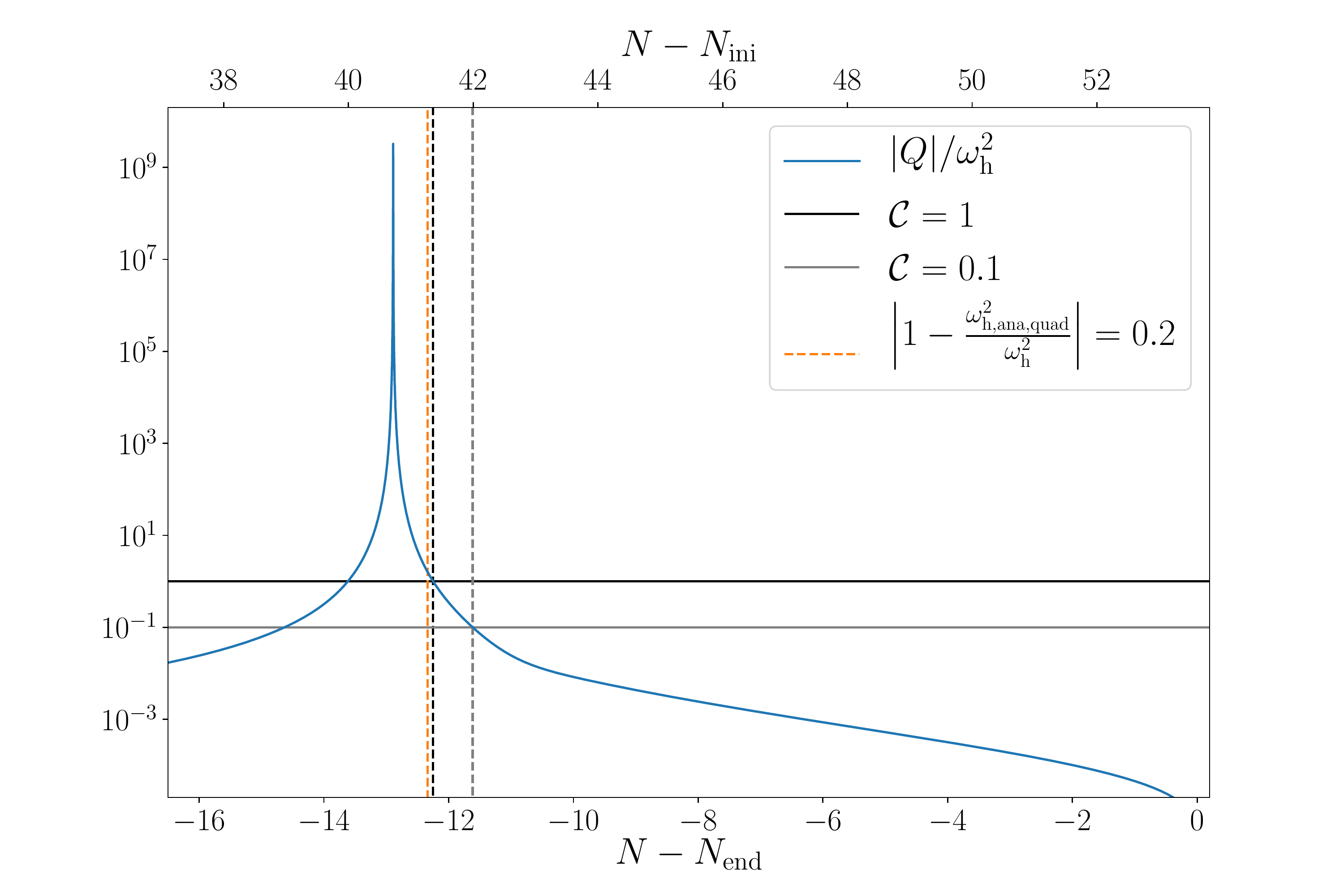}
    \caption{Time evolution of the quantity $Q/\omega^2_\mathrm{h}$ (blue line) with $Q$ defined in Eq.~\eqref{eq:defQwkb}, before, around and after the time ``$\mathrm{osc}$".
    The smallness of this quantity represents a criterion for the regime of validity of the WKB approximation, whose strictness is parameterized by the number $\mathcal{C}$ which is either $1$ (horizontal black line) or $0.1$ (horizontal gray line) in this work.
    The corresponding time from which the WKB approximation becomes valid are represented by vertical dashed lines (respectively black and gray ones).
    The time from which the quadratic approximation for the pulsation $\omega_\mathrm{h}^2$ becomes roughly valid is also represented by the orange vertical dashed lines.
    }
    \label{fig:Q over w2}
\end{center}
\end{figure}

As discussed earlier, Eq.~(\ref{eq:g(t)}) cannot be solved exactly due to the complex analytical time dependence of $\omega_\mathrm{h}^2(t)$. However, the WKB approximation can be used to find an approximate solution. Indeed, if one represents the WKB criterion, $Q/\omega^2_\mathrm{h}$, where 
\begin{align}
\label{eq:defQwkb}
    Q=\frac34 \frac{\dot \omega^2_\mathrm{h}}{\omega^2_\mathrm{h}}
    -\frac{\ddot{\omega_\mathrm{h}}}{2\omega_\mathrm{h}},
\end{align}
versus time, see Fig.~\ref{fig:Q over w2}, we notice that, quite quickly, on the left and right hand sides of the turning point (where, by definition, $Q$ blows up), this quantity tends to zero, signaling that WKB should be a very good approximation. Usually, the WKB approximation is used in the following way. It is first used on the left side of the turning point (region I) where the initial conditions are specified and, as consequence, where the solution is completely specified. The WKB approximation is also used on the right side (region III) where the solution, being a solution of a second order differential equation, depends on two arbitrary integration constants. These two constants can be fixed if the solution in region III is matched to the solution in region I. However, as noticed above, the solution in between (region II) is \textit{a priori} not known since the WKB approximation fails in this region. What is usually done to deal with this problem is to linearize the effective frequency in region II which allows us to find an approximation of the solution in this regime  which, then, permits to connect the solution in region I to the solution in region III. In this way, one can specified the two unknown integration constants in region III and have a complete knowledge of the solution.

Here, we closely follow this WKB procedure with, however, some differences in order to adapt our method to the problem at hand. In region I, the solution is known and is given by the slow-roll approximation. So we do not need to employ the WKB approximation in this regime (although it might be interesting to study how WKB is connected to slow-roll in this regime, see also Ref.~\cite{Martin:2002vn}). As a consequence, we will ``start the evolution" not deep in region I (on the left) but at the onset of region II. In fact, the time of transition, $t_\mathrm{trans}$ turns out to be located precisely at the beginning of this region and, therefore, we fix the ``initial conditions" (those relevant for our problem here; they have obviously nothing to do with the physical initial conditions, discussed before, and chosen at the onset of inflation) at this time. In addition, they are completely known from the previously studied slow-roll solution and read
\begin{equation}
\label{eq:iniwkb}
    g_\mathrm{h}(t_\mathrm{trans})= 1 \,, \quad \dot{g}_\mathrm{h}(t_\mathrm{trans})= - \frac{m_\mathrm{h}^2}{3  H(t_\mathrm{trans})} + \frac{3H(t_\mathrm{trans})}{2 } + \frac{ \Gamma_{\mathrm{hm}}}{4} \,,
\end{equation}
where we used the slow-roll expression of $\dot{\phi}_\mathrm{h}$. 
Numerically, with our fiducial parameters, given that $H_\mathrm{trans}=4.3 \times 10^{-5}$ (obtained analytically within the slow-roll approximation), we find that $\dot{g}_\mathrm{trans}=4.8 \times 10^{-5}$.
Moreover, one can compute the difference in cosmic time between the transition [$\theta_\mathrm{trans}=\mathrm{arctan}(1/R)$] and the onset of the oscillations [$\theta_\mathrm{osc}=\mathrm{arctan}(\sqrt{0.0104})$], with use of Eq.~\eqref{eq:thypergeo} which is indeed relevant around the transition time, and one finds $m_\ell(t_\mathrm{osc}-t_\mathrm{trans})=0.32$.

Let us now turn to the question of calculating $g_\mathrm{h}(t)$ in region II. As explained before, in this region, the WKB approximation breaks down and we need to use another technique which consists in linearizing the effective frequency around the turning point. Concretely, we write
\begin{align}
\label{eq:linearw2}
    \omega^2_\mathrm{h}=-\alpha (t-t_\mathrm{osc}) +\cdots
\end{align}
with $\alpha \equiv -\dd \omega ^2_\mathrm{h}(t=t_\mathrm{osc})/\dd t$. In our case, $\alpha $ is negative. The linear approximation~(\ref{eq:linearw2}) is represented in Fig.~\ref{fig:omegas}, see the solid green line. We have seen before that, in the vicinity of the turning point,  $\omega^2_\mathrm{h} \simeq m_\mathrm{h}^2 - 9H^2/4$, and this  implies that
$\alpha=-9/2\times (\epsilon_1 H^3)_\mathrm{osc}$. With the fiducial parameters, we find analytically $\alpha=-3.0 \times 10^{-14}$ in $\Mp^3$ units, where one has used the slow-roll formula for $\epsilon_1$, namely
\begin{equation}
\label{eq:eps1srappendix}
    \epsilon_1 = \frac{1}{2s} \frac{1+R^4 \mathrm{tan}^2\theta}{\cos^2\theta \left( 1+ R^2 \tan^2 \theta \right)^2} \,.
\end{equation}
Then, plugging the expression~(\ref{eq:linearw2}) into the equation~(\ref{eq:g(t)}), and introducing a new variable $y$ defined by the expression $y\equiv \alpha^{1/3}(t-t_\mathrm{osc})$, the equation for $g_\mathrm{h}(t)$ takes the form
\begin{align}
    \frac{\dd ^2g_\mathrm{h}}{\dd y^2}-yg_\mathrm{h}=0.
\end{align}
The solution to this equation is well-known and can be expressed in terms of the Airy functions of first and second kind, $\mathrm{Ai}$ and $\mathrm{Bi}$~\cite{Gradshteyn:1965aa,Abramovitz:1970aa}. As a consequence, one can write 
\begin{align}
\label{eq:airyII}
    g_\mathrm{h}(t)=B_1 \mathrm{Ai}(y)+B_2\mathrm{Bi}(y),
\end{align}
where $B_1$ and $B_2$ are two integration constants. These constants can be determined using the values of $g_\mathrm{h}$ and its derivative at the transition point, see Eqs.~(\ref{eq:iniwkb}). This leads to
\begin{align}
    B_1&=\pi\left[\frac{\dd}{\dd s}\mathrm{Bi}(y_\mathrm{trans})-\frac{1}{\alpha ^{1/3}}\dot{g}_\mathrm{trans}\mathrm{Bi}(y_\mathrm{trans})\right],
    \\
    B_2&=-\pi\left[\frac{\dd}{\dd s}\mathrm{Ai}(y_\mathrm{trans})-\frac{1}{\alpha ^{1/3}}\dot{g}_\mathrm{trans}\mathrm{Ai}(y_\mathrm{trans})\right].
\end{align}
Numerically, with our fiducial parameters, one finds, $B_1\simeq 1.98+3.90i$ and $B_2\simeq 3.75-2.26i$. 
An intermediate step has been to use $y_\mathrm{trans}=e^{i\pi/3}|\alpha|^{1/3}(t_\mathrm{trans}-t_\mathrm{osc})$, with the value of $(t_\mathrm{trans}-t_\mathrm{osc})$ that was found in the previous section.\footnote{These results and the following ones (until the numerical determination of the constants $C_\pm$) are derived with the choice of a complex cubic root for the negative number $\alpha=|\alpha| e^{i\pi}$ which is  $\alpha^{1/3}=e^{i \pi/3} |\alpha|^{1/3}$.
Note that in this case the time coordinate $y$ has a constant phase $\pi/3$, as indeed $y=e^{i \pi/3} |\alpha|^{1/3} (t-t_\mathrm{osc})$, a fact that is crucial in order to derive consistently the asymptotic limit shown in Eq.~\eqref{eq: airy asymptotic}.
This choice should not matter as long as it is consistent during the whole calculation, and indeed we have checked that using the different cubic root $\alpha^{1/3}=-|\alpha|^{1/3}$, leads to the same final results (but intermediate ones like the values for $y_\mathrm{trans}$ or $B_{1,2}$ are therefore obviously different).}
This completes our calculation of $g_\mathrm{h}$ (and, therefore, of $\phi_\mathrm{h}$) in region II which is now completely known.

The final step consists in writing the solution in region III. This region is the most interesting one for us since it corresponds to the phase dominated by the light field. Obtaining the solution for $g_\mathrm{h}(t)$ in this regime allows us to track the behavior of $\phi_\mathrm{h}(t)$ when the light field drives inflation and the heavy field oscillates and decays. As explained before, in this region, $Q/\omega_\mathrm{h}^2\ll 1$ and the WKB approximation is satisfied. As a consequence, the solution can be written as
\begin{align}
\label{eq:wkbregionIII}
    g_\mathrm{h}(t)=\frac{C_+}{\omega_\mathrm{h}^{1/2}}e^{i\int _{t_\mathrm{osc}}^t \omega_\mathrm{h} (\tau)\dd \tau}
+\frac{C_-}{\omega_\mathrm{h}^{1/2}}e^{-i\int _{t_\mathrm{osc}}^t 
\omega_\mathrm{h} (\tau)\dd \tau},
\end{align}
where $C_+$ and $C_-$ are two integration constants. This solution is valid only in region III, namely only for times $t$ where $Q/\omega_\mathrm{h}^2\ll 1$ (on the right hand side of the turning point). Notice, however, that the lower bounds in the integrals have been taken at the turning point. We first remark that changing the lower bound is in fact equivalent to redefining the constants $C_+$ and $C_-$. So choosing the lower bound to be the turning point is not in contradiction with the WKB approximation and just amounts to a particular (convenient) definition of $C_+$ and $C_-$. At this stage, our first goal is to determine these two quantities. In order to carry out this calculation, we must check that there is an overlap between region II and region III, that is to say a patching region where the linear approximation of the effective frequency is not too bad and where, at the same time, the WKB regime is satisfied. The linear approximation is acceptable if the second order term remains smaller than the first order one which means $t-t_\mathrm{osc}\lesssim 2(\dd \omega_\mathrm{h}^2/\dd t)/(\dd ^2 \omega^2_\mathrm{h}/\dd t^2)$ or 
\begin{align}
    m_\ell (t-t_\mathrm{osc})\lesssim \frac{3}{R(\epsilon_1-3\epsilon_2)\vert_\mathrm{osc}}.
\end{align}
On the other hand, if the linear approximation is accurate, the WKB approximation is satisfied provided $\vert \dot{\omega} _\mathrm{h}/\omega_\mathrm{h}^2\vert <1$, namely
\begin{align}
    m_\ell (t-t_\mathrm{osc})\gtrsim \left(\frac{m_\ell^3}{4\vert \alpha\vert}\right)^{1/3}.
\end{align}
This is possible if
\begin{align}
    \vert \alpha \vert> \frac{m_\mathrm{h}^3}{108}(\epsilon_1-3\epsilon_2)^3\vert_\mathrm{osc},
\end{align}
which is satisfied since, numerically, one finds $|\alpha|>2.5 \times 10^{-14}$ to be compared with the value found here, $|\alpha|=3.0 \times 10^{-14}$.
In this patching region, one can use at the same time the solution~(\ref{eq:airyII}) and the WKB solution~(\ref{eq:wkbregionIII}). Since both formula must lead to the same solution, this allows us to identify the coefficients $C_+$ and $C_-$ in terms of $B_1$ and $B_2$. Concretely, if one goes away from the turning point, Eq.~(\ref{eq:airyII}) leads to 
\begin{align}
\label{eq: airy asymptotic}
    g_\mathrm{h}(t)&\simeq 
\frac{1}{2\sqrt{\pi}}e^{-i\pi/12}
    \vert \alpha\vert^{-1/12}\left(t-t_\mathrm{osc}\right)^{-1/4}
    \biggl[(B_1+iB_2)\, e^{-\frac{2i}{3}\vert\alpha\vert^{1/2}(t-t_\mathrm{osc})^{3/2}}
    \nonumber \\ &
    +2B_2\, e^{\frac{2i}{3}\vert\alpha\vert^{1/2}(t-t_\mathrm{osc})^{3/2}}
\biggr],
     \end{align}
while the WKB solution~(\ref{eq:wkbregionIII}) in the linear part of the time-dependent frequency can be written as
\begin{equation}
    g_{\mathrm{h}}(t) \simeq |\alpha|^{-1/4} (t-t_\mathrm{osc})^{-1/4} \left[ C_+ e^{\frac{2i}{3}|\alpha|^{1/2}(t-t_\mathrm{osc})^{3/2}} +  e^{-\frac{2i}{3}|\alpha|^{1/2}(t-t_\mathrm{osc})^{3/2}} \right] \,.
\end{equation}
Requiring these two expressions to represent the same solution, this immediately leads to the relationship
\begin{align}
    C_+ &= \frac{B_2}{\sqrt{\pi}} |\alpha|^{1/6} e^{-i \pi/12}  \,, \quad 
    C_- = \frac{B_1+iB_2}{2\sqrt{\pi}} |\alpha|^{1/6} e^{- i\pi/12}  \,, 
\end{align}
Numerically, one finds $C_+\simeq 0.0096 - 0.01i$ and $C_-=C_+^*$, as required since $g_\mathrm{h}(t)$ is a real function. Therefore, we have reached our goal, namely expressing $g_\mathrm{h}(t)$ in the region where inflation is driven by the light field. The solution is \textit{a priori} entirely specified since the constants $C_+$ and $C_-$ are now related to what happens at the transition time. However, we must also calculate the WKB phase which requires the integration of $\omega_\mathrm{h}(t)$. It can be expressed as
\begin{align}
\label{eq:phasewkb}
    \int _{t_\mathrm{osc}}^t\omega(\tau)\dd \tau =\int _{t_\mathrm{osc}}^{t_{\mathrm{match}}}\omega(\tau)\dd \tau+
    \int _{t_\mathrm{match}}^t\omega(\tau)\dd \tau=\Phi 
    +\int _{t_\mathrm{match}}^t\omega(\tau)\dd \tau,
\end{align}
where we have introduced the time $t_\mathrm{match}$. This time is a time at which the quadratic approximation of $\omega_\mathrm{h}^2(t)$, see Eq.~(\ref{eq:quadraticw2}), becomes valid. In principle, the phase $\Phi $ can be calculated but, in our case, this would require the integration of the exact time-dependent frequency which is complicated given its form~(\ref{eq:defwh}) (and which would undermine the interest of having an analytical approximation). A rough approximation is to use the linear expression in the vicinity of the turning point which results in 
\begin{align}
\Phi \simeq  \frac{2}{3} |\alpha|^{1/2} 
(t_\mathrm{match}-t_\mathrm{osc})^{3/2}.  
\end{align}
However, this result does not give a very accurate estimation of the phase while the accuracy of the WKB approximation quite sensitively depends on $\Phi$. A more careful calculation, using $\omega_\mathrm{h}^2\simeq m_\mathrm{h}^2-9H^2/4$ gives, after the change of variable defined in Eq.~\eqref{eq: dt to dtheta} 
\begin{equation}
    \Phi=-\sqrt{6 C} \frac{R}{R^2-1} \int_{\theta_\mathrm{osc}}^{\theta_\mathrm{match}} 
\frac{\sqrt{1+R^2 \tan^2 \theta}}{\sin \theta \times \cos \theta}
(\tan \theta)^{\frac{1}{R^2-1}}  \left[1-\frac{3C}{2R^2} \frac{\left(\mathrm{tan} \theta \right)^\frac{2}{R^2-1}}{\mathrm{cos}^2\theta}\right]^{1/2} \dd \theta.
\end{equation}
Unfortunately, this integral cannot be computed analytically, but it is straightforward to compute it numerically. It gives  $\Phi=1.093$ for the fiducial parameters of the model, a value that should be compared with the numerical result $\Phi = 2.039$.  Admittedly, the result is still not very accurate and, therefore, the determination of $\Phi$ constitutes an additional source of error. Nevertheless, taking into account the phase determined according to the considerations presented above significantly improves the agreement 
between the exact and WKB solutions.

Let us now turn to the calculation of the second term in Eq.~(\ref{eq:phasewkb}). As already discussed above, this can be carried out with the help of the quadratic approximation for $\omega_\mathrm{h}^2(t)$, see Eq.~(\ref{eq:quadraticw2}). It is especially convenient to redefine the coefficients $\bar a$, $\bar b$ and $\bar c$ according to $a=-\bar{a}/m_\ell=1/4$, $b=\bar{b}/m_\ell^2$ and $c=\bar{c}/m_\ell^2$. In fact, since $c=R^2 -1/2 - b^2$, there remains only one free coefficient which can be chosen to be $b$, a positive definite quantity.
Then, defining 
\begin{align}
\label{eq:defu}
u= \frac{m_\ell(t-t_\mathrm{trans})-2b}{2 \sqrt{R^2-1/2}}\, ,
\end{align}
the frequency $\omega_\mathrm{h}^2(t)$ takes the form
\begin{align}
\label{eq:omegavsu}
    \omega_\mathrm{h}(t)=m_\ell\left(R^2-\frac12\right)^{1/2}\left(1-u^2\right)^{1/2}.
\end{align}
Notice that, in the regime we are interested in, $u^2$ is always smaller than one since $\omega _\mathrm{h}^2(t)$ is always real and positive. Then, it follows that
\begin{align}
\label{eq:intw}
    \int_{t_\mathrm{match}}^t \omega(\tau) \dd \tau  =  2 \left(R^2-\frac12\right) \int_{u_\mathrm{match}}^u \dd u \sqrt{1-u^2}=\left(R^2-\frac12\right)G(u),
\end{align}
with
\begin{align}
\label{eq:defbigG}
    G(u)=\left[u\sqrt{1-u^2} + \mathrm{arcsin}(u) \right]^u_{u_\mathrm{match}}.
\end{align}
This completes the calculation of $g_\mathrm{h}(t)$ since all the terms of Eq.~(\ref{eq:wkbregionIII}) are now explicitly known.\footnote{It is also interesting to notice that, in region III, if the quadratic approximation for the effective frequency is used, then an exact solution is available. Let us indeed define $z=2^{1/2}a^{1/4}[x-b/(2a)]$ and $A=-1/(2\sqrt{a})[b^2/(4a)+c]=1/2-R^2$. Concretely, $z$ can be expressed as 
\begin{align}
    z &= - \frac12\frac{\Gamma_\mathrm{hm}}{m_\ell}-\sqrt{6}(N_\mathrm{end}-N)^{1/2} .
\end{align}
Then, it follows that Eq.~(\ref{eq:g(t)}) can be written as
\begin{align}
    \frac{\dd ^2g}{\dd z^2}-\left(\frac{z^2}{4}+A\right)g=0,
\end{align}
which is the canonical form of the equation defining the Parabolic cylinder functions~\cite{Gradshteyn:1965aa,Abramovitz:1970aa}. As a consequence, the solution in region III takes the following form~\cite{Gradshteyn:1965aa,Abramovitz:1970aa}
\begin{align}
    g(z)=C_1 \mathrm{U}(A,z)+C_2 \mathrm{V}(A,z),
\end{align}
with
\begin{align}
    \mathrm{U}(A,z)&=\mathrm{D}_{-A-1/2}(z), 
    \\
    \mathrm{V}(A,z)&=\frac{1}{\pi}\Gamma\left(\frac12+A\right)
    \left[\mathrm{D}_{-A-1/2}(-z)+\sin (\pi A) \mathrm{D}_{-A-1/2}(z)\right]
\end{align}
where $C_1$ and $C_2$ are two integration constant and $\mathrm{D}_\nu(z)$ are parabolic functions. Notice that another canonical form which is also convenient is~\cite{Gradshteyn:1965aa,Abramovitz:1970aa}
\begin{align}
    \frac{\dd ^2g}{\dd z^2}+\left(\nu+\frac12 -\frac{z^2}{4}\right)g=0,
\end{align}
where $A=-\nu -1/2$ or $\nu=R^2-1$. In that case, the solution can be written as 
\begin{align}
    g(z)=C_1 \mathrm{D}_\nu(z)+C_2 \mathrm{D}_\nu(-z),
\end{align}
since $\mathrm{D}_\nu(\pm z)$ are linearly independent provided $\nu\neq 0, \pm 1, \cdots$ (we have used the same notations for the integration constants but obviously they differ from the ones already introduced before).}

\begin{figure}
    \centering
    \includegraphics[width=0.75\linewidth]{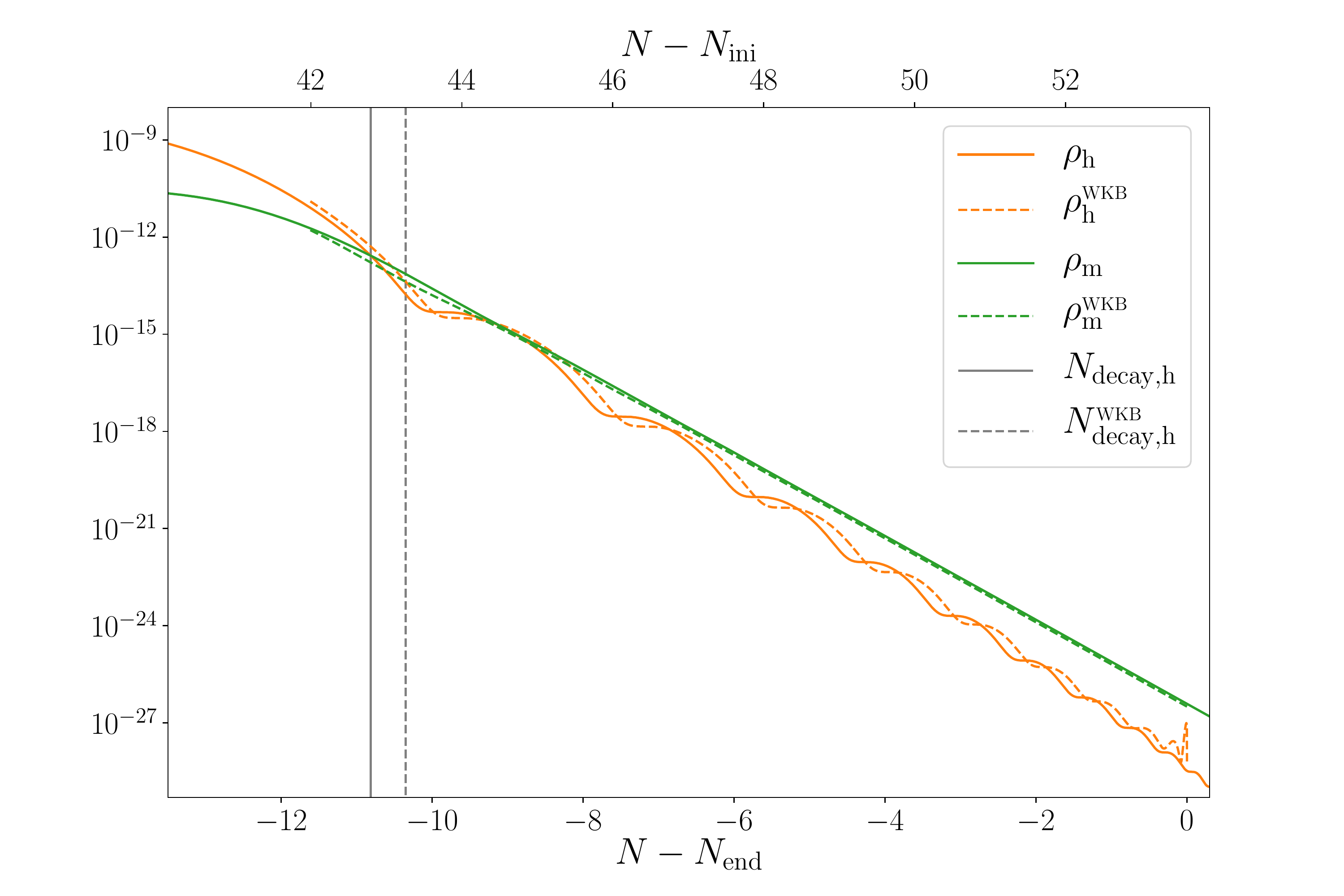}
    \caption{Comparison of $\rho_\mathrm{h}$ and $\rho_\mathrm{m}$ (in $\Mp^4$ units) during the oscillations and the decay of the heavy field, between their exact numerical results (respectively the orange and green solid lines), and their analytical approximations inferred from the WKB approach (respectively the orange and green dashed lines), see Eq.~\eqref{eq:rhohwkb} and Eq.~\eqref{eq:rhomwkb}.
    The time of decay of the heavy field, defined as the time when $\rho_\mathrm{h}=\rho_\mathrm{m}$ is shown with the vertical gray lines both from the numerical approach (solid line, $N_\mathrm{decay,h}=-10.81$) and the analytical one (dashed, $N_\mathrm{decay,h}^{_\mathrm{WKB}}=-10.35$).
    This approach clearly gives much better results than the one presented in body of the paper, compared to Fig.~\ref{fig:rhonaive} and the previous estimate $N_\mathrm{decay,h}^\mathrm{sd}=-8.39$ , and enables one to derive reliably and analytically the time of decay $N_\mathrm{decay,h}$.
    }
    \label{fig: rho decay WKB}
\end{figure}

The result of the WKB approximation is represented in Fig.~\ref{fig: rho decay WKB} where the evolution of the heavy field energy density
\begin{align}
\label{eq:rhohwkb}
    \rho_{\mathrm{h}}(t)=\frac12 \dot{\phi}_\mathrm{h}^2+\frac12 m_\mathrm{h}^2 
    \phi_{\mathrm{h}}^2
\end{align}
is represented and compared to the numerical solution. Evidently, the result is excellent and the WKB solution follows quite accurately the exact result.

\subsection{The evolution of the matter energy density}
\label{subapp:matter}

Having calculated the evolution of the heavy field, we now turn to the behavior of the matter energy density $\rho_\mathrm{m}$ whose equation of motion is Eq.~(\ref{eq:rhogammam}). As it was already noticed before, its solution can be expressed as
\begin{align}
\label{eq:solrhom2}
    a^3(t)\rho_\mathrm{m}(t)=a^3(t_\uini)
    \rho_\mathrm{m}(t_\uini)
    +\frac{\Gamma_{\mathrm{hm}}}{2}\int _{t_\uini}^t a^3(\tau)\dot{\phi}_\mathrm{h}^2(\tau)\dd \tau .
\end{align}
We see that this requires the calculation of the integral of the kinetic energy density of the field, ${\cal I}=\int _{t_\uini}^t a^3(\tau)\dot{\phi}_\mathrm{h}^2(\tau)\dd \tau $. In the slow-roll regime, $\dot \phi_\mathrm{h}^2(t)$ evolves slowly and could be put outside the integral, immediately leading to an explicit solution, see Sec.~\ref{subsubsec:mgammasr}. Here, however, it is no longer possible to use this trick and we must insert the WKB expression of $\phi_\mathrm{h}(t)$ in the above integral and calculate it explicitly. This leads to
\begin{align}
    {\cal I} =&
    a_\mathrm{p}^3\phi_\mathrm{p}^2\int_{t_\mathrm{match}}^t \mathrm{d}\tau e^{-\Gamma_{\mathrm{hm}}(\tau-t_\mathrm{p})/2}\biggl(\frac94 H^2g^2_\mathrm{h}+\frac34 H\Gamma_\mathrm{hm}g^2_\mathrm{h}-3H g_\mathrm{h}\dot{g}_\mathrm{h}
    -\frac{\Gamma_\mathrm{hm}}{2}g_\mathrm{h}\dot{g}_\mathrm{h}
    \nonumber \\ &
    +\frac{\Gamma_\mathrm{hm}^2}{16}g^2_\mathrm{h} +\dot{g}^2_\mathrm{h}\biggr).
\end{align}
In the present case, the initial time has been taken to be $t_\mathrm{match}$, the time from which the quadratic assumption becomes satisfied. Using the WKB solution~(\ref{eq:wkbregionIII}) for $g_\mathrm{h}(t)$, the above integral takes the form
\begin{align}
\label{eq:integralrhom}
{\cal I} &=\phi_\mathrm{p}a_\mathrm{p}^3\int_{t_\mathrm{match}}^t \mathrm{d}\tau e^{-\Gamma_{\mathrm{hm}}(\tau-t_\mathrm{p})/2}\Biggl[
C_+^2\left(\frac{{\cal C}}{\omega_\mathrm{h}}-i\right)^2
\omega_\mathrm{h} \, e^{2i\int_{t_\mathrm{osc}}^\tau \omega_\mathrm{h} \dd \tau}
\nonumber \\ &
+C_-^2\left(\frac{{\cal C}}{\omega_\mathrm{h}}+i\right)^2
\omega_\mathrm{h} \,e^{-2i\int_{t_\mathrm{osc}}^\tau \omega_\mathrm{h} \dd \tau}
+2C_+C_-\omega_\mathrm{h}\left(\frac{{\cal C}}{\omega_\mathrm{h}}-i\right)
\left(\frac{{\cal C}}{\omega_\mathrm{h}}+i\right)\Biggr],
\end{align}
with
\begin{align}
    \frac{{\cal C}}{\omega_\mathrm{h}}=\frac32 \frac{H}{\omega_\mathrm{h}}+\frac14 
    \frac{\Gamma_\mathrm{hm}}{\omega_\mathrm{h}}+\frac{\dot \omega_\mathrm{h}}{2\omega ^2_\mathrm{h}}\simeq \frac32 \frac{H}{\omega_\mathrm{h}}+\frac14 
    \frac{\Gamma_\mathrm{hm}}{\omega_\mathrm{h}}.
\end{align}
We see that we have three terms to calculate and we write them as ${\cal I}={\cal I}_1+{\cal I}_2+{\cal I}_3$. The next step consists in inserting the form of $\omega _\mathrm{h}(t)$ in the above expression. Since we are interested in the behavior of $\rho_\mathrm{m}$ in region III, one can use the quadratic approximation of $\omega^2_\mathrm{h}(t)$, see Eq.~(\ref{eq:omegavsu}) expressed in terms of the variable $u$, see Eq.~(\ref{eq:defu}). Then, the  first term reads
\begin{align}
\label{eq:calI1}
    {\cal I}_1 =a_\mathrm{p}^3\phi_\mathrm{p}
    C_+^2e^{i2\Phi}(2R^2-1)e^{-b\Gamma_\mathrm{hm}/m_\ell}
    \int _{u_\mathrm{match}}^u\dd u \, F(u) \, e^{i(2R^2-1)G(u)},
\end{align}
with
\begin{align}
    F(u)=e^{-\Gamma_\mathrm{hm}(R^2-1/2)^{1/2}u/m_\ell}(1-u^2)^{-1/2}
    \left[u-i\left(1-u^2\right)^{1/2}\right]^2,
\end{align}
and where the calculation of the WKB phase~(\ref{eq:intw}) has been used, see Eq.~(\ref{eq:defbigG}) for the definition of the function $G(u)$. The second term is then trivially found since this is just the complex conjugate of the first one, ${\cal I}_2={\cal I}_1^*$. Unfortunately, the integral~(\ref{eq:calI1}) cannot be performed analytically. However, we notice that the integrand has the form of a smooth function $F(u)$ multiplied by an \textit{a priori} rapid oscillatory phase (since $2R^2-1$ is \textit{a priori} quite a large number). As is well-known, several techniques (Filon method, Levin collocation method etc., see Ref.~\cite{olver2008numerical}) have been developed to deal with this type of integrals. The simplest one just consists in integrating by part leading to
\begin{align}
    \int _{u_\mathrm{match}}^u\dd u \, F(u) \, e^{i(2R^2-1)G(u)}&=
    \frac{1}{i(2R^2-1)}\biggl[\frac{F(u)}{G'(u)}e^{i(2R^2-1)G(u)}
    \nonumber \\ &
 -\frac{F(u_\mathrm{match})}{G'(u_\mathrm{match})}e^{i(2R^2-1)G(u_\mathrm{match})}\biggr]
   \nonumber \\ &
    -\frac{1}{i(2R^2-1)}\int _{u_\mathrm{match}}^u \dd u \frac{\dd}{\dd u}
    \left(\frac{F}{G'}\right)e^{i(2R^2-1)G(u)}.
    \end{align}
In terms of the (inverse of the) large parameter $2R^2-1$, the second term is of higher order and, therefore, can be neglected thus leading to an approximation of the integral. If one uses this approximation to estimate ${\cal I}_1$, this gives
\begin{align}
\label{eq:calI1approx}
    {\cal I}_1 &\simeq -a_\mathrm{p}^3\phi_\mathrm{p}
    iC_+^2e^{i2\Phi}e^{-b\Gamma_\mathrm{hm}/m_\ell}
    \left[\frac{F(u)}{G'(u)}e^{i(2R^2-1)G(u)}
 -\frac{F(u_\mathrm{match})}{G'(u_\mathrm{match})}e^{i(2R^2-1)G(u_\mathrm{match})}\right].
\end{align}
In principle, additional (successive) integrations by parts could lead to an even more accurate approximation but, here, we will restrict ourselves to first order.

Finally, the third term remains to be calculated. Straightforward manipulations show that it can be written as
 \begin{align}
\label{eq:calI3}
{\cal I}_3&=2a_\mathrm{p}^3\phi_\mathrm{p}C_+C_-
 (2R^2-1)e^{-b\Gamma_\mathrm{hm}/m_\ell}
    \int _{u_\mathrm{match}}^u\dd u\,  e^{-\aleph u}\, (1-u^2)^{-1/2}.
\end{align}
with
\begin{align}
    \aleph=\frac{\Gamma_\mathrm{hm}}{m_\ell}
    \left(R^2-\frac{1}{2}\right)^{1/2}.
\end{align}
The integral~(\ref{eq:calI3}) ${\cal J}=\int _{u_\mathrm{match}}^u\dd u\,  e^{-\aleph u}\, (1-u^2)^{-1/2}$ cannot be performed exactly. But it is however relatively easy to derive an accurate analytical approximation of ${\cal J}$. One can proceed as follows. Let us first introduce the new variable $\psi$ such that $u=\cos \psi$. Then the integral to be calculated reduces to 
\begin{align}
    {\cal J}
    =-\int ^{\psi}_{\psi_\mathrm{match}} e^{-\aleph\cos \Omega}\, \dd \Omega.
\end{align}
The integrand can easily be written as Fourier series, namely
\begin{align}
\label{eq:seriesexpcos}
    e^{-\aleph \cos \Omega}=I_0(-\aleph)+2\sum_{n=1}^{+\infty}I_n(-\aleph)
    \cos(n\Omega),
\end{align}
where $I_n$ is a modified Bessel function~\cite{Gradshteyn:1965aa,Abramovitz:1970aa}. Each term of the series~(\ref{eq:seriesexpcos}) can easily be integrated exactly and it follows that the integral ${\cal J}$ can be expressed as 
\begin{align}
    {\cal J}=\left[-I_0(-\aleph)\psi -2 \sum_{n=1}^{+\infty}\frac{1}{n}I_n(-\aleph)\sin(n\psi)\right]^\psi_{\psi_\mathrm{match}}
\end{align}
In practice, in order to reach a good accuracy, it is sufficient to keep only the first two or three terms in the above sum. We conclude that
\begin{align}
\label{eq:calI3final}
{\cal I}_3&=-2a_\mathrm{p}^3\phi_\mathrm{p}C_+C_-
 (2R^2-1)e^{-b\Gamma_\mathrm{hm}/m_\ell}
    \Biggl[I_0(-\aleph)\arccos u 
    \nonumber \\ &
    +2 \sum_{n=1}^{+\infty}\frac{1}{n}I_n(-\aleph)\sin(n\arccos u)\Biggr]^u_{u_\mathrm{match}}.
\end{align}

The final step consists in collecting the three terms calculated before in the final expression of the matter energy density. One obtains the rather lengthy, but explicit, following equation [using the fact that $G(u_\mathrm{match})=0$ by definition]
\begin{align}
\label{eq:rhomwkb}
   \rho_\mathrm{m}(t) &\simeq \rho_\mathrm{m}^\mathrm{match}\left[\frac{a_\mathrm{match}}{a(t)}\right]^3
    +\phi_\mathrm{p}\left[\frac{a_\mathrm{p}}{a(t)}\right]^3\frac{\Gamma_\mathrm{hm}}{2}e^{-b\Gamma_\mathrm{hm}/m_\ell}  
    \nonumber \\ & \times
    \Biggl(2 \mathrm{Re} \biggl\{ -iC_+^2e^{i2\Phi}\biggl[\frac{F(u)}{G'(u)}e^{i(2R^2-1)G(u)}
    -\frac{F(u_\mathrm{match})}{G'(u_\mathrm{match})}  \biggr] \biggr\}
    \nonumber \\ &
     -2\left\vert C_+\right \vert^2(2R^2-1)\Biggl[I_0(-\aleph)\arccos u +2 \sum_{n=1}^{+\infty}\frac{1}{n}I_n(-\aleph)\sin(n\arccos u)\Biggr]^u_{u_\mathrm{match}}\Biggr). 
\end{align}
This expression is represented in Fig.~\ref{fig: rho decay WKB} together with the exact (numerical) solution of $\rho_\mathrm{m}$. Obviously, it matches quite well the numerical result.

\subsection{Determination of the time of decay}
\label{subapp:decay}

Having determined the WKB form of $\rho_\mathrm{h}(t)$, see Eq.~(\ref{eq:rhohwkb}) and $\rho_\mathrm{m}(t)$, see Eq.~(\ref{eq:rhomwkb}), we can now calculate the time of decay. Recall that it is defined by the condition $\rho_\mathrm{h}(t_\mathrm{decay, h})=\rho_\mathrm{m}(t_\mathrm{decay, h})$. Given the complexity of Eqs.~(\ref{eq:rhohwkb}) and~(\ref{eq:rhomwkb}), it is clear that it is hopeless to solve $\rho_\mathrm{h}(t_\mathrm{decay, h})=\rho_\mathrm{m}(t_\mathrm{decay, h})$ analytically. Here, we solve this transcendental equation numerically and the solution is denoted $N_\mathrm{decay, h}^{_\mathrm{WKB}}$. For the fiducial parameters, one finds $N_\mathrm{decay,h}^{_\mathrm{WKB}}=-10.35$ to be compared with the exact value $N_\mathrm{decay,h}=-10.81$. The quantity $N_\mathrm{decay,h}^{_\mathrm{WKB}}$ is used in Table.~\ref{table:Smr} for the analytical estimation of the amplitude of non-adiabatic perturbations after double inflation.

\bibliographystyle{JHEP}
\bibliography{double-reh}

\providecommand{\href}[2]{#2}\begingroup\raggedright\begin{thebibliography}{10}

\bibitem{Starobinsky:1980te}
A.~A. Starobinsky, \emph{{A New Type of Isotropic Cosmological Models Without
  Singularity}},
  \href{https://doi.org/10.1016/0370-2693(80)90670-X}{\emph{Phys. Lett.}
  {\bfseries B91} (1980) 99}.

\bibitem{Guth:1980zm}
A.~H. Guth, \emph{{The Inflationary Universe: A Possible Solution to the
  Horizon and Flatness Problems}},
  \href{https://doi.org/10.1103/PhysRevD.23.347}{\emph{Phys.Rev.} {\bfseries
  D23} (1981) 347}.

\bibitem{Linde:1981mu}
A.~D. Linde, \emph{{A New Inflationary Universe Scenario: A Possible Solution
  of the Horizon, Flatness, Homogeneity, Isotropy and Primordial Monopole
  Problems}},
  \href{https://doi.org/10.1016/0370-2693(82)91219-9}{\emph{Phys.Lett.}
  {\bfseries B108} (1982) 389}.

\bibitem{Albrecht:1982wi}
A.~Albrecht and P.~J. Steinhardt, \emph{{Cosmology for Grand Unified Theories
  with Radiatively Induced Symmetry Breaking}},
  \href{https://doi.org/10.1103/PhysRevLett.48.1220}{\emph{Phys.Rev.Lett.}
  {\bfseries 48} (1982) 1220}.

\bibitem{Linde:1983gd}
A.~D. Linde, \emph{{Chaotic Inflation}},
  \href{https://doi.org/10.1016/0370-2693(83)90837-7}{\emph{Phys.Lett.}
  {\bfseries B129} (1983) 177}.

\bibitem{Starobinsky:1979ty}
A.~A. Starobinsky, \emph{{Spectrum of relict gravitational radiation and the
  early state of the universe}}, {\emph{JETP Lett.} {\bfseries 30} (1979) 682}.

\bibitem{Mukhanov:1981xt}
V.~F. Mukhanov and G.~Chibisov, \emph{{Quantum Fluctuation and Nonsingular
  Universe.}}, {\emph{JETP Lett.} {\bfseries 33} (1981) 532}.

\bibitem{Hawking:1982cz}
S.~Hawking, \emph{{The Development of Irregularities in a Single Bubble
  Inflationary Universe}},
  \href{https://doi.org/10.1016/0370-2693(82)90373-2}{\emph{Phys.Lett.}
  {\bfseries B115} (1982) 295}.

\bibitem{Guth:1982ec}
A.~H. Guth and S.~Pi, \emph{{Fluctuations in the New Inflationary Universe}},
  \href{https://doi.org/10.1103/PhysRevLett.49.1110}{\emph{Phys.Rev.Lett.}
  {\bfseries 49} (1982) 1110}.

\bibitem{Bardeen:1983qw}
J.~M. Bardeen, P.~J. Steinhardt and M.~S. Turner, \emph{{Spontaneous Creation
  of Almost Scale - Free Density Perturbations in an Inflationary Universe}},
  \href{https://doi.org/10.1103/PhysRevD.28.679}{\emph{Phys.Rev.} {\bfseries
  D28} (1983) 679}.

\bibitem{Martin:2015dha}
J.~Martin, \emph{{The Observational Status of Cosmic Inflation after Planck}},
  \href{https://doi.org/10.1007/978-3-319-44769-8_2}{\emph{Astrophys. Space
  Sci. Proc.} {\bfseries 45} (2016) 41}
  [\href{https://arxiv.org/abs/1502.05733}{{\ttfamily 1502.05733}}].

\bibitem{Martin:2013tda}
J.~Martin, C.~Ringeval and V.~Vennin, \emph{{Encyclop\ae{}dia Inflationaris}},
  \href{https://doi.org/10.1016/j.dark.2014.01.003}{\emph{Phys. Dark Univ.}
  {\bfseries 5-6} (2014) 75} [\href{https://arxiv.org/abs/1303.3787}{{\ttfamily
  1303.3787}}].

\bibitem{Martin:2013nzq}
J.~Martin, C.~Ringeval, R.~Trotta and V.~Vennin, \emph{{The Best Inflationary
  Models After Planck}},
  \href{https://doi.org/10.1088/1475-7516/2014/03/039}{\emph{JCAP} {\bfseries
  03} (2014) 039} [\href{https://arxiv.org/abs/1312.3529}{{\ttfamily
  1312.3529}}].

\bibitem{Chowdhury:2019otk}
D.~Chowdhury, J.~Martin, C.~Ringeval and V.~Vennin, \emph{{Assessing the
  scientific status of inflation after Planck}},
  \href{https://doi.org/10.1103/PhysRevD.100.083537}{\emph{Phys. Rev. D}
  {\bfseries 100} (2019) 083537}
  [\href{https://arxiv.org/abs/1902.03951}{{\ttfamily 1902.03951}}].

\bibitem{Polarski:1992dq}
D.~Polarski and A.~A. Starobinsky, \emph{{Spectra of perturbations produced by
  double inflation with an intermediate matter dominated stage}},
  \href{https://doi.org/10.1016/0550-3213(92)90062-G}{\emph{Nucl. Phys. B}
  {\bfseries 385} (1992) 623}.

\bibitem{Peter:1994dx}
P.~Peter, D.~Polarski and A.~A. Starobinsky, \emph{{Confrontation of double
  inflationary models with observations}},
  \href{https://doi.org/10.1103/PhysRevD.50.4827}{\emph{Phys. Rev. D}
  {\bfseries 50} (1994) 4827}
  [\href{https://arxiv.org/abs/astro-ph/9403037}{{\ttfamily
  astro-ph/9403037}}].

\bibitem{Langlois:1999dw}
D.~Langlois, \emph{{Correlated adiabatic and isocurvature perturbations from
  double inflation}},
  \href{https://doi.org/10.1103/PhysRevD.59.123512}{\emph{Phys. Rev. D}
  {\bfseries 59} (1999) 123512}
  [\href{https://arxiv.org/abs/astro-ph/9906080}{{\ttfamily
  astro-ph/9906080}}].

\bibitem{Gordon:2000hv}
C.~Gordon, D.~Wands, B.~A. Bassett and R.~Maartens, \emph{{Adiabatic and
  entropy perturbations from inflation}},
  \href{https://doi.org/10.1103/PhysRevD.63.023506}{\emph{Phys. Rev. D}
  {\bfseries 63} (2000) 023506}
  [\href{https://arxiv.org/abs/astro-ph/0009131}{{\ttfamily
  astro-ph/0009131}}].

\bibitem{GrootNibbelink:2001qt}
S.~Groot~Nibbelink and B.~J.~W. van Tent, \emph{{Scalar perturbations during
  multiple field slow-roll inflation}},
  \href{https://doi.org/10.1088/0264-9381/19/4/302}{\emph{Class. Quant. Grav.}
  {\bfseries 19} (2002) 613}
  [\href{https://arxiv.org/abs/hep-ph/0107272}{{\ttfamily hep-ph/0107272}}].

\bibitem{Tsujikawa:2002qx}
S.~Tsujikawa, D.~Parkinson and B.~A. Bassett, \emph{{Correlation - consistency
  cartography of the double inflation landscape}},
  \href{https://doi.org/10.1103/PhysRevD.67.083516}{\emph{Phys. Rev. D}
  {\bfseries 67} (2003) 083516}
  [\href{https://arxiv.org/abs/astro-ph/0210322}{{\ttfamily
  astro-ph/0210322}}].

\bibitem{Wands:2002bn}
D.~Wands, N.~Bartolo, S.~Matarrese and A.~Riotto, \emph{{An Observational test
  of two-field inflation}},
  \href{https://doi.org/10.1103/PhysRevD.66.043520}{\emph{Phys.Rev.} {\bfseries
  D66} (2002) 043520} [\href{https://arxiv.org/abs/astro-ph/0205253}{{\ttfamily
  astro-ph/0205253}}].

\bibitem{Wands:2007bd}
D.~Wands, \emph{{Multiple field inflation}},
  \href{https://doi.org/10.1007/978-3-540-74353-8_8}{\emph{Lect. Notes Phys.}
  {\bfseries 738} (2008) 275}
  [\href{https://arxiv.org/abs/astro-ph/0702187}{{\ttfamily
  astro-ph/0702187}}].

\bibitem{Choi:2008et}
K.-Y. Choi, J.-O. Gong and D.~Jeong, \emph{{Evolution of the curvature
  perturbation during and after multi-field inflation}},
  \href{https://doi.org/10.1088/1475-7516/2009/02/032}{\emph{JCAP} {\bfseries
  0902} (2009) 032} [\href{https://arxiv.org/abs/0810.2299}{{\ttfamily
  0810.2299}}].

\bibitem{Achucarro:2010jv}
A.~Achucarro, J.-O. Gong, S.~Hardeman, G.~A. Palma and S.~P. Patil, \emph{{Mass
  hierarchies and non-decoupling in multi-scalar field dynamics}},
  \href{https://doi.org/10.1103/PhysRevD.84.043502}{\emph{Phys. Rev. D}
  {\bfseries 84} (2011) 043502}
  [\href{https://arxiv.org/abs/1005.3848}{{\ttfamily 1005.3848}}].

\bibitem{Kaiser:2015usz}
D.~I. Kaiser, \emph{{Nonminimal Couplings in the Early Universe: Multifield
  Models of Inflation and the Latest Observations}},
  \href{https://doi.org/10.1007/978-3-319-31299-6_2}{\emph{Fundam. Theor.
  Phys.} {\bfseries 183} (2016) 41}
  [\href{https://arxiv.org/abs/1511.09148}{{\ttfamily 1511.09148}}].

\bibitem{Schimmrigk:2017jwa}
R.~Schimmrigk, \emph{{Multifield Reheating after Modular $j$-Inflation}},
  \href{https://doi.org/10.1016/j.physletb.2018.04.065}{\emph{Phys. Lett. B}
  {\bfseries 782} (2018) 193}
  [\href{https://arxiv.org/abs/1712.09961}{{\ttfamily 1712.09961}}].

\bibitem{Braglia:2020fms}
M.~Braglia, D.~K. Hazra, L.~Sriramkumar and F.~Finelli, \emph{{Generating
  primordial features at large scales in two field models of inflation}},
  \href{https://doi.org/10.1088/1475-7516/2020/08/025}{\emph{JCAP} {\bfseries
  08} (2020) 025} [\href{https://arxiv.org/abs/2004.00672}{{\ttfamily
  2004.00672}}].

\bibitem{Amendola:2016saw}
L.~Amendola et~al., \emph{{Cosmology and fundamental physics with the Euclid
  satellite}}, \href{https://doi.org/10.1007/s41114-017-0010-3}{\emph{Living
  Rev. Rel.} {\bfseries 21} (2018) 2}
  [\href{https://arxiv.org/abs/1606.00180}{{\ttfamily 1606.00180}}].

\bibitem{Dore:2014cca}
O.~Dor\'{e}, J.~Bock, P.~Capak, R.~de~Putter, T.~Eifler et~al., \emph{{SPHEREx:
  An All-Sky Spectral Survey}},
  \href{https://arxiv.org/abs/1412.4872}{{\ttfamily 1412.4872}}.

\bibitem{Matsumura:2013aja}
T.~Matsumura et~al., \emph{{Mission design of LiteBIRD}},
  \href{https://doi.org/10.1007/s10909-013-0996-1}{\emph{J. Low Temp. Phys.}
  {\bfseries 176} (2014) 733}
  [\href{https://arxiv.org/abs/1311.2847}{{\ttfamily 1311.2847}}].

\bibitem{Allen:2005ye}
L.~E. Allen, S.~Gupta and D.~Wands, \emph{{Non-gaussian perturbations from
  multi-field inflation}},
  \href{https://doi.org/10.1088/1475-7516/2006/01/006}{\emph{JCAP} {\bfseries
  01} (2006) 006} [\href{https://arxiv.org/abs/astro-ph/0509719}{{\ttfamily
  astro-ph/0509719}}].

\bibitem{Huston:2013kgl}
I.~Huston and A.~J. Christopherson, \emph{{Isocurvature Perturbations and
  Reheating in Multi-Field Inflation}},
  \href{https://arxiv.org/abs/1302.4298}{{\ttfamily 1302.4298}}.

\bibitem{Byrnes:2006fr}
C.~T. Byrnes and D.~Wands, \emph{{Curvature and isocurvature perturbations from
  two-field inflation in a slow-roll expansion}},
  \href{https://doi.org/10.1103/PhysRevD.74.043529}{\emph{Phys. Rev. D}
  {\bfseries 74} (2006) 043529}
  [\href{https://arxiv.org/abs/astro-ph/0605679}{{\ttfamily
  astro-ph/0605679}}].

\bibitem{Schutz:2013fua}
K.~Schutz, E.~I. Sfakianakis and D.~I. Kaiser, \emph{{Multifield Inflation
  after Planck: Isocurvature Modes from Nonminimal Couplings}},
  \href{https://doi.org/10.1103/PhysRevD.89.064044}{\emph{Phys. Rev. D}
  {\bfseries 89} (2014) 064044}
  [\href{https://arxiv.org/abs/1310.8285}{{\ttfamily 1310.8285}}].

\bibitem{Chen:2009we}
X.~Chen and Y.~Wang, \emph{{Large non-Gaussianities with Intermediate Shapes
  from Quasi-Single Field Inflation}},
  \href{https://doi.org/10.1103/PhysRevD.81.063511}{\emph{Phys. Rev. D}
  {\bfseries 81} (2010) 063511}
  [\href{https://arxiv.org/abs/0909.0496}{{\ttfamily 0909.0496}}].

\bibitem{Chen:2009zp}
X.~Chen and Y.~Wang, \emph{{Quasi-Single Field Inflation and
  Non-Gaussianities}},
  \href{https://doi.org/10.1088/1475-7516/2010/04/027}{\emph{JCAP} {\bfseries
  04} (2010) 027} [\href{https://arxiv.org/abs/0911.3380}{{\ttfamily
  0911.3380}}].

\bibitem{Baumann:2011nk}
D.~Baumann and D.~Green, \emph{{Signatures of Supersymmetry from the Early
  Universe}}, \href{https://doi.org/10.1103/PhysRevD.85.103520}{\emph{Phys.
  Rev. D} {\bfseries 85} (2012) 103520}
  [\href{https://arxiv.org/abs/1109.0292}{{\ttfamily 1109.0292}}].

\bibitem{Assassi:2012zq}
V.~Assassi, D.~Baumann and D.~Green, \emph{{On Soft Limits of Inflationary
  Correlation Functions}},
  \href{https://doi.org/10.1088/1475-7516/2012/11/047}{\emph{JCAP} {\bfseries
  11} (2012) 047} [\href{https://arxiv.org/abs/1204.4207}{{\ttfamily
  1204.4207}}].

\bibitem{Noumi:2012vr}
T.~Noumi, M.~Yamaguchi and D.~Yokoyama, \emph{{Effective field theory approach
  to quasi-single field inflation and effects of heavy fields}},
  \href{https://doi.org/10.1007/JHEP06(2013)051}{\emph{JHEP} {\bfseries 06}
  (2013) 051} [\href{https://arxiv.org/abs/1211.1624}{{\ttfamily 1211.1624}}].

\bibitem{Gong:2013sma}
J.-O. Gong, S.~Pi and M.~Sasaki, \emph{{Equilateral non-Gaussianity from heavy
  fields}}, \href{https://doi.org/10.1088/1475-7516/2013/11/043}{\emph{JCAP}
  {\bfseries 11} (2013) 043} [\href{https://arxiv.org/abs/1306.3691}{{\ttfamily
  1306.3691}}].

\bibitem{Arkani-Hamed:2015bza}
N.~Arkani-Hamed and J.~Maldacena, \emph{{Cosmological Collider Physics}},
  \href{https://arxiv.org/abs/1503.08043}{{\ttfamily 1503.08043}}.

\bibitem{Lee:2016vti}
H.~Lee, D.~Baumann and G.~L. Pimentel, \emph{{Non-Gaussianity as a Particle
  Detector}}, \href{https://doi.org/10.1007/JHEP12(2016)040}{\emph{JHEP}
  {\bfseries 12} (2016) 040}
  [\href{https://arxiv.org/abs/1607.03735}{{\ttfamily 1607.03735}}].

\bibitem{Garcia-Saenz:2018ifx}
S.~Garcia-Saenz, S.~Renaux-Petel and J.~Ronayne, \emph{{Primordial fluctuations
  and non-Gaussianities in sidetracked inflation}},
  \href{https://doi.org/10.1088/1475-7516/2018/07/057}{\emph{JCAP} {\bfseries
  1807} (2018) 057} [\href{https://arxiv.org/abs/1804.11279}{{\ttfamily
  1804.11279}}].

\bibitem{Garcia-Saenz:2018vqf}
S.~Garcia-Saenz and S.~Renaux-Petel, \emph{{Flattened non-Gaussianities from
  the effective field theory of inflation with imaginary speed of sound}},
  \href{https://doi.org/10.1088/1475-7516/2018/11/005}{\emph{JCAP} {\bfseries
  1811} (2018) 005} [\href{https://arxiv.org/abs/1805.12563}{{\ttfamily
  1805.12563}}].

\bibitem{Fumagalli:2019noh}
J.~Fumagalli, S.~Garcia-Saenz, L.~Pinol, S.~Renaux-Petel and J.~Ronayne,
  \emph{{Hyper non-Gaussianities in inflation with strongly non-geodesic
  motion}},  \href{https://arxiv.org/abs/1902.03221}{{\ttfamily 1902.03221}}.

\bibitem{Bjorkmo-Ferreira-Marsh_2019}
T.~Bjorkmo, R.~Z. Ferreira and M.~D. Marsh, \emph{Mild non-gaussianities under
  perturbative control from rapid-turn inflation models},
  \href{https://doi.org/10.1088/1475-7516/2019/12/036}{\emph{Journal of
  Cosmology and Astroparticle Physics} {\bfseries 2019} (2019) 036–036}.

\bibitem{Wang:2019gok}
D.-G. Wang, \emph{{On the inflationary massive field with a curved field
  manifold}}, \href{https://doi.org/10.1088/1475-7516/2020/01/046}{\emph{JCAP}
  {\bfseries 01} (2020) 046}
  [\href{https://arxiv.org/abs/1911.04459}{{\ttfamily 1911.04459}}].

\bibitem{Garcia-Saenz:2019njm}
S.~Garcia-Saenz, L.~Pinol and S.~Renaux-Petel, \emph{{Revisiting
  non-Gaussianity in multifield inflation with curved field space}},
  \href{https://arxiv.org/abs/1907.10403}{{\ttfamily 1907.10403}}.

\bibitem{Ferreira:2020qkf}
R.~Z. Ferreira, \emph{{Non-Gaussianities in models of inflation with large and
  negative entropic masses}},
  \href{https://doi.org/10.1088/1475-7516/2020/08/034}{\emph{JCAP} {\bfseries
  08} (2020) 034} [\href{https://arxiv.org/abs/2003.13410}{{\ttfamily
  2003.13410}}].

\bibitem{Pinol:2020kvw}
L.~Pinol, \emph{{Multifield inflation beyond $N_\mathrm{field}=2$:
  non-Gaussianities and single-field effective theory}},
  \href{https://arxiv.org/abs/2011.05930}{{\ttfamily 2011.05930}}.

\bibitem{Albrecht:1982mp}
A.~Albrecht, P.~J. Steinhardt, M.~S. Turner and F.~Wilczek, \emph{{Reheating an
  Inflationary Universe}},
  \href{https://doi.org/10.1103/PhysRevLett.48.1437}{\emph{Phys. Rev. Lett.}
  {\bfseries 48} (1982) 1437}.

\bibitem{Turner:1983he}
M.~S. Turner, \emph{{Coherent Scalar Field Oscillations in an Expanding
  Universe}}, \href{https://doi.org/10.1103/PhysRevD.28.1243}{\emph{Phys. Rev.
  D} {\bfseries 28} (1983) 1243}.

\bibitem{Shtanov:1994ce}
Y.~Shtanov, J.~H. Traschen and R.~H. Brandenberger, \emph{{Universe reheating
  after inflation}},
  \href{https://doi.org/10.1103/PhysRevD.51.5438}{\emph{Phys. Rev.} {\bfseries
  D51} (1995) 5438} [\href{https://arxiv.org/abs/hep-ph/9407247}{{\ttfamily
  hep-ph/9407247}}].

\bibitem{Bassett:2005xm}
B.~A. Bassett, S.~Tsujikawa and D.~Wands, \emph{{Inflation dynamics and
  reheating}}, \href{https://doi.org/10.1103/RevModPhys.78.537}{\emph{Rev. Mod.
  Phys.} {\bfseries 78} (2006) 537}
  [\href{https://arxiv.org/abs/astro-ph/0507632}{{\ttfamily
  astro-ph/0507632}}].

\bibitem{Allahverdi:2010xz}
R.~Allahverdi, R.~Brandenberger, F.-Y. Cyr-Racine and A.~Mazumdar,
  \emph{{Reheating in Inflationary Cosmology: Theory and Applications}},
  \href{https://doi.org/10.1146/annurev.nucl.012809.104511}{\emph{Ann. Rev.
  Nucl. Part. Sci.} {\bfseries 60} (2010) 27}
  [\href{https://arxiv.org/abs/1001.2600}{{\ttfamily 1001.2600}}].

\bibitem{Amin:2014eta}
M.~A. Amin, M.~P. Hertzberg, D.~I. Kaiser and J.~Karouby,
  \emph{{Nonperturbative Dynamics Of Reheating After Inflation: A Review}},
  \href{https://doi.org/10.1142/S0218271815300037}{\emph{Int. J. Mod. Phys.}
  {\bfseries D24} (2014) 1530003}
  [\href{https://arxiv.org/abs/1410.3808}{{\ttfamily 1410.3808}}].

\bibitem{Martin:2010kz}
J.~Martin and C.~Ringeval, \emph{{First CMB Constraints on the Inflationary
  Reheating Temperature}},
  \href{https://doi.org/10.1103/PhysRevD.82.023511}{\emph{Phys. Rev. D}
  {\bfseries 82} (2010) 023511}
  [\href{https://arxiv.org/abs/1004.5525}{{\ttfamily 1004.5525}}].

\bibitem{Martin:2014nya}
J.~Martin, C.~Ringeval and V.~Vennin, \emph{{Observing Inflationary
  Reheating}},
  \href{https://doi.org/10.1103/PhysRevLett.114.081303}{\emph{Phys. Rev. Lett.}
  {\bfseries 114} (2015) 081303}
  [\href{https://arxiv.org/abs/1410.7958}{{\ttfamily 1410.7958}}].

\bibitem{Martin:2016oyk}
J.~Martin, C.~Ringeval and V.~Vennin, \emph{{Information Gain on Reheating: the
  One Bit Milestone}},
  \href{https://doi.org/10.1103/PhysRevD.93.103532}{\emph{Phys. Rev. D}
  {\bfseries 93} (2016) 103532}
  [\href{https://arxiv.org/abs/1603.02606}{{\ttfamily 1603.02606}}].

\bibitem{Bassett:1998wg}
B.~A. Bassett, D.~I. Kaiser and R.~Maartens, \emph{{General relativistic
  preheating after inflation}},
  \href{https://doi.org/10.1016/S0370-2693(99)00478-5}{\emph{Phys. Lett. B}
  {\bfseries 455} (1999) 84}
  [\href{https://arxiv.org/abs/hep-ph/9808404}{{\ttfamily hep-ph/9808404}}].

\bibitem{Bassett:1999mt}
B.~A. Bassett, F.~Tamburini, D.~I. Kaiser and R.~Maartens, \emph{{Metric
  preheating and limitations of linearized gravity. 2.}},
  \href{https://doi.org/10.1016/S0550-3213(99)00495-2}{\emph{Nucl. Phys. B}
  {\bfseries 561} (1999) 188}
  [\href{https://arxiv.org/abs/hep-ph/9901319}{{\ttfamily hep-ph/9901319}}].

\bibitem{Bassett:1999ta}
B.~A. Bassett, C.~Gordon, R.~Maartens and D.~I. Kaiser, \emph{{Restoring the
  sting to metric preheating}},
  \href{https://doi.org/10.1103/PhysRevD.61.061302}{\emph{Phys. Rev. D}
  {\bfseries 61} (2000) 061302}
  [\href{https://arxiv.org/abs/hep-ph/9909482}{{\ttfamily hep-ph/9909482}}].

\bibitem{Tsujikawa:2000ab}
S.~Tsujikawa and B.~A. Bassett, \emph{{A New twist to preheating}},
  \href{https://doi.org/10.1103/PhysRevD.62.043510}{\emph{Phys. Rev. D}
  {\bfseries 62} (2000) 043510}
  [\href{https://arxiv.org/abs/hep-ph/0003068}{{\ttfamily hep-ph/0003068}}].

\bibitem{Tsujikawa:2002nf}
S.~Tsujikawa and B.~A. Bassett, \emph{{When can preheating affect the CMB?}},
  \href{https://doi.org/10.1016/S0370-2693(02)01813-0}{\emph{Phys. Lett. B}
  {\bfseries 536} (2002) 9}
  [\href{https://arxiv.org/abs/astro-ph/0204031}{{\ttfamily
  astro-ph/0204031}}].

\bibitem{Lyth:2003im}
D.~H. Lyth and D.~Wands, \emph{{Conserved cosmological perturbations}},
  \href{https://doi.org/10.1103/PhysRevD.68.103515}{\emph{Phys. Rev. D}
  {\bfseries 68} (2003) 103515}
  [\href{https://arxiv.org/abs/astro-ph/0306498}{{\ttfamily
  astro-ph/0306498}}].

\bibitem{Finelli:1998bu}
F.~Finelli and R.~H. Brandenberger, \emph{{Parametric amplification of
  gravitational fluctuations during reheating}},
  \href{https://doi.org/10.1103/PhysRevLett.82.1362}{\emph{Phys. Rev. Lett.}
  {\bfseries 82} (1999) 1362}
  [\href{https://arxiv.org/abs/hep-ph/9809490}{{\ttfamily hep-ph/9809490}}].

\bibitem{Jedamzik:2010dq}
K.~Jedamzik, M.~Lemoine and J.~Martin, \emph{{Collapse of Small-Scale Density
  Perturbations during Preheating in Single Field Inflation}},
  \href{https://doi.org/10.1088/1475-7516/2010/09/034}{\emph{JCAP} {\bfseries
  1009} (2010) 034} [\href{https://arxiv.org/abs/1002.3039}{{\ttfamily
  1002.3039}}].

\bibitem{Martin:2019nuw}
J.~Martin, T.~Papanikolaou and V.~Vennin, \emph{{Primordial black holes from
  the preheating instability}},
  \href{https://doi.org/10.1088/1475-7516/2020/01/024}{\emph{JCAP} {\bfseries
  2001} (2020) 024} [\href{https://arxiv.org/abs/1907.04236}{{\ttfamily
  1907.04236}}].

\bibitem{Martin:2020fgl}
J.~Martin, T.~Papanikolaou, L.~Pinol and V.~Vennin, \emph{{Metric preheating
  and radiative decay in single-field inflation}},
  \href{https://doi.org/10.1088/1475-7516/2020/05/003}{\emph{JCAP} {\bfseries
  05} (2020) 003} [\href{https://arxiv.org/abs/2002.01820}{{\ttfamily
  2002.01820}}].

\bibitem{Finelli:2000ya}
F.~Finelli and R.~H. Brandenberger, \emph{{Parametric amplification of metric
  fluctuations during reheating in two field models}},
  \href{https://doi.org/10.1103/PhysRevD.62.083502}{\emph{Phys. Rev. D}
  {\bfseries 62} (2000) 083502}
  [\href{https://arxiv.org/abs/hep-ph/0003172}{{\ttfamily hep-ph/0003172}}].

\bibitem{DeCross:2015uza}
M.~P. DeCross, D.~I. Kaiser, A.~Prabhu, C.~Prescod-Weinstein and E.~I.
  Sfakianakis, \emph{{Preheating after Multifield Inflation with Nonminimal
  Couplings, I: Covariant Formalism and Attractor Behavior}},
  \href{https://doi.org/10.1103/PhysRevD.97.023526}{\emph{Phys. Rev. D}
  {\bfseries 97} (2018) 023526}
  [\href{https://arxiv.org/abs/1510.08553}{{\ttfamily 1510.08553}}].

\bibitem{DeCross:2016fdz}
M.~P. DeCross, D.~I. Kaiser, A.~Prabhu, C.~Prescod-Weinstein and E.~I.
  Sfakianakis, \emph{{Preheating after multifield inflation with nonminimal
  couplings, II: Resonance Structure}},
  \href{https://doi.org/10.1103/PhysRevD.97.023527}{\emph{Phys. Rev. D}
  {\bfseries 97} (2018) 023527}
  [\href{https://arxiv.org/abs/1610.08868}{{\ttfamily 1610.08868}}].

\bibitem{DeCross:2016cbs}
M.~P. DeCross, D.~I. Kaiser, A.~Prabhu, C.~Prescod-Weinstein and E.~I.
  Sfakianakis, \emph{{Preheating after multifield inflation with nonminimal
  couplings, III: Dynamical spacetime results}},
  \href{https://doi.org/10.1103/PhysRevD.97.023528}{\emph{Phys. Rev. D}
  {\bfseries 97} (2018) 023528}
  [\href{https://arxiv.org/abs/1610.08916}{{\ttfamily 1610.08916}}].

\bibitem{Jiang:2018uce}
J.~Jiang, Q.~Liang, Y.-F. Cai, D.~A. Easson and Y.~Zhang, \emph{{Numerical
  study of inflationary preheating with arbitrary power-law potential and a
  realization of curvaton mechanism}},
  \href{https://doi.org/10.3847/1538-4357/ab189e}{\emph{Astrophys. J.}
  {\bfseries 876} (2019) 136}
  [\href{https://arxiv.org/abs/1812.08220}{{\ttfamily 1812.08220}}].

\bibitem{Nguyen:2019kbm}
R.~Nguyen, J.~van~de Vis, E.~I. Sfakianakis, J.~T. Giblin and D.~I. Kaiser,
  \emph{{Nonlinear Dynamics of Preheating after Multifield Inflation with
  Nonminimal Couplings}},
  \href{https://doi.org/10.1103/PhysRevLett.123.171301}{\emph{Phys. Rev. Lett.}
  {\bfseries 123} (2019) 171301}
  [\href{https://arxiv.org/abs/1905.12562}{{\ttfamily 1905.12562}}].

\bibitem{vandeVis:2020qcp}
J.~van~de Vis, R.~Nguyen, E.~I. Sfakianakis, J.~T. Giblin and D.~I. Kaiser,
  \emph{{Time scales for nonlinear processes in preheating after multifield
  inflation with nonminimal couplings}},
  \href{https://doi.org/10.1103/PhysRevD.102.043528}{\emph{Phys. Rev. D}
  {\bfseries 102} (2020) 043528}
  [\href{https://arxiv.org/abs/2005.00433}{{\ttfamily 2005.00433}}].

\bibitem{Malik:2002jb}
K.~A. Malik, D.~Wands and C.~Ungarelli, \emph{{Large scale curvature and
  entropy perturbations for multiple interacting fluids}},
  \href{https://doi.org/10.1103/PhysRevD.67.063516}{\emph{Phys. Rev. D}
  {\bfseries 67} (2003) 063516}
  [\href{https://arxiv.org/abs/astro-ph/0211602}{{\ttfamily
  astro-ph/0211602}}].

\bibitem{Malik:2004tf}
K.~A. Malik and D.~Wands, \emph{{Adiabatic and entropy perturbations with
  interacting fluids and fields}},
  \href{https://doi.org/10.1088/1475-7516/2005/02/007}{\emph{JCAP} {\bfseries
  0502} (2005) 007} [\href{https://arxiv.org/abs/astro-ph/0411703}{{\ttfamily
  astro-ph/0411703}}].

\bibitem{Berera:1995ie}
A.~Berera, \emph{{Warm inflation}},
  \href{https://doi.org/10.1103/PhysRevLett.75.3218}{\emph{Phys. Rev. Lett.}
  {\bfseries 75} (1995) 3218}
  [\href{https://arxiv.org/abs/astro-ph/9509049}{{\ttfamily
  astro-ph/9509049}}].

\bibitem{Yokoyama:1998ju}
J.~Yokoyama and A.~D. Linde, \emph{{Is warm inflation possible?}},
  \href{https://doi.org/10.1103/PhysRevD.60.083509}{\emph{Phys. Rev. D}
  {\bfseries 60} (1999) 083509}
  [\href{https://arxiv.org/abs/hep-ph/9809409}{{\ttfamily hep-ph/9809409}}].

\bibitem{Mukhanov:1990me}
V.~F. Mukhanov, H.~Feldman and R.~H. Brandenberger, \emph{{Theory of
  cosmological perturbations. Part 1. Classical perturbations. Part 2. Quantum
  theory of perturbations. Part 3. Extensions}},
  \href{https://doi.org/10.1016/0370-1573(92)90044-Z}{\emph{Phys. Rept.}
  {\bfseries 215} (1992) 203}.

\bibitem{Bardeen:1980kt}
J.~M. Bardeen, \emph{{Gauge Invariant Cosmological Perturbations}},
  \href{https://doi.org/10.1103/PhysRevD.22.1882}{\emph{Phys. Rev.} {\bfseries
  D22} (1980) 1882}.

\bibitem{Malik:2008im}
K.~A. Malik and D.~Wands, \emph{{Cosmological perturbations}},
  \href{https://doi.org/10.1016/j.physrep.2009.03.001}{\emph{Phys. Rept.}
  {\bfseries 475} (2009) 1} [\href{https://arxiv.org/abs/0809.4944}{{\ttfamily
  0809.4944}}].

\bibitem{Visinelli:2014qla}
L.~Visinelli, \emph{{Cosmological perturbations for an inflaton field coupled
  to radiation}},
  \href{https://doi.org/10.1088/1475-7516/2015/01/005}{\emph{JCAP} {\bfseries
  1501} (2015) 005} [\href{https://arxiv.org/abs/1410.1187}{{\ttfamily
  1410.1187}}].

\bibitem{Martin:2004um}
J.~Martin, \emph{{Inflationary cosmological perturbations of quantum-mechanical
  origin}}, \href{https://doi.org/10.1007/11377306_7}{\emph{Lect. Notes Phys.}
  {\bfseries 669} (2005) 199}
  [\href{https://arxiv.org/abs/hep-th/0406011}{{\ttfamily hep-th/0406011}}].

\bibitem{Martin:2007bw}
J.~Martin, \emph{{Inflationary perturbations: The Cosmological Schwinger
  effect}}, \href{https://doi.org/10.1007/978-3-540-74353-8_6}{\emph{Lect.
  Notes Phys.} {\bfseries 738} (2008) 193}
  [\href{https://arxiv.org/abs/0704.3540}{{\ttfamily 0704.3540}}].

\bibitem{Martin:2015qta}
J.~Martin and V.~Vennin, \emph{{Quantum Discord of Cosmic Inflation: Can we
  Show that CMB Anisotropies are of Quantum-Mechanical Origin?}},
  \href{https://doi.org/10.1103/PhysRevD.93.023505}{\emph{Phys. Rev. D}
  {\bfseries 93} (2016) 023505}
  [\href{https://arxiv.org/abs/1510.04038}{{\ttfamily 1510.04038}}].

\bibitem{Kodama:1985bj}
H.~Kodama and M.~Sasaki, \emph{{Cosmological Perturbation Theory}},
  \href{https://doi.org/10.1143/PTPS.78.1}{\emph{Prog. Theor. Phys. Suppl.}
  {\bfseries 78} (1984) 1}.

\bibitem{Weinberg:2008zzc}
S.~Weinberg, \emph{{Cosmology}}. Oxford Univ. Press, 2008.

\bibitem{Press1996}
W.~H. Press, S.~a. Teukolsky, W.~T. Vetterling and B.~P. Flannery,
  \emph{{Numerical Recipes in Fortran 77: the Art of Scientific Computing.
  Second Edition}}, vol.~1. 1996.

\bibitem{Leach:2002ar}
S.~M. Leach, A.~R. Liddle, J.~Martin and D.~J. Schwarz, \emph{{Cosmological
  parameter estimation and the inflationary cosmology}},
  \href{https://doi.org/10.1103/PhysRevD.66.023515}{\emph{Phys. Rev. D}
  {\bfseries 66} (2002) 023515}
  [\href{https://arxiv.org/abs/astro-ph/0202094}{{\ttfamily
  astro-ph/0202094}}].

\bibitem{Gradshteyn:1965aa}
I.~S. Gradshteyn and I.~M. Ryzhik, \emph{Table of Integrals, Series, and
  Products}. Academic Press, New York and London, 1965.

\bibitem{Abramovitz:1970aa}
M.~Abramowitz and I.~A. Stegun, \emph{Handbook of mathematical functions with
  formulas, graphs, and mathematical tables}. National Bureau of Standards,
  Washington, US, ninth~ed., 1970.

\bibitem{Gonzalez:2018jax}
P.~Gonz\'alez, G.~A. Palma and N.~Videla, \emph{{Covariant evolution of
  perturbations during reheating in two-field inflation}},
  \href{https://doi.org/10.1088/1475-7516/2018/12/001}{\emph{JCAP} {\bfseries
  12} (2018) 001} [\href{https://arxiv.org/abs/1805.10360}{{\ttfamily
  1805.10360}}].

\bibitem{Martin:2002vn}
J.~Martin and D.~J. Schwarz, \emph{{WKB approximation for inflationary
  cosmological perturbations}},
  \href{https://doi.org/10.1103/PhysRevD.67.083512}{\emph{Phys. Rev. D}
  {\bfseries 67} (2003) 083512}
  [\href{https://arxiv.org/abs/astro-ph/0210090}{{\ttfamily
  astro-ph/0210090}}].

\bibitem{olver2008numerical}
S.~Olver, \emph{Numerical approximation of highly oscillatory integrals}, Ph.D.
  thesis, University of Cambridge, 2008.

\end{thebibliography}\endgroup
%
\end{document}